\documentclass[School=Princeton]{Dissertate}
\pdfoutput=1

\usepackage{comment}
\usepackage{amssymb}
\usepackage{lipsum}
\usepackage{nomencl}
\usepackage{pstricks}
\usepackage{tablefootnote}
\usepackage{siunitx}
\usepackage{array, makecell}
\usepackage{multibib}
\makenomenclature

\usepackage{lipsum}
\usepackage{units}
\usepackage[utf8]{inputenc}
\usepackage{multirow}
\usepackage{dsfont}
\usepackage{setspace}

\begin{document}

\title{Modeling Multiphase Flow Through and Around Multiscale Deformable Porous Materials}
\author{Francisco J. Carrillo}
\advisor{Dr. Ian C. Bourg}

\degree{Doctor of Philosophy}
\field{Chemical and Biological Engineering}
\degreeyear{2021}
\degreemonth{September}
\department{Chemical and Biological Engineering}

\pdOneName{B.S.}
\pdOneSchool{University of Texas at Austin}
\pdOneYear{2016}

\pdTwoName{M.A.}
\pdTwoSchool{Princeton University}
\pdTwoYear{2018}
\maketitle
\copyrightpage

	\phantomsection
	\addcontentsline{toc}{chapter}{Abstract}
	\newpage
	\pagenumbering{roman}
	\setcounter{page}{3}
	\pagestyle{fancy}
	\renewcommand{\headrulewidth}{0.0pt}
	\vspace*{35pt}
	\begin{center}
	\scshape Abstract \\ \rm
	\end{center}

Detailed understanding of the coupling between fluid flow and solid deformation in porous media is crucial for the development biomedical devices and novel energy technologies relating to a wide range of geological and biological processes. Well established models based on poroelasticity theory exist for describing coupled fluid-solid mechanics. However, these models are not adapted to describe systems with multiple fluid phases or ``hybrid-scale" systems containing both solid-free regions and porous matrices. To address this problem, we present a novel computational fluid dynamics approach based on a unique set of volume-averaged partial differential equations that asymptotically approach the Navier-Stokes Volume-of-Fluid equations in solid-free-regions and Biot's Poroelasticity Theory in porous regions. Unlike existing multiscale multiphase solvers, it can match analytical predictions of capillary, relative permeability, and gravitational effects at both the pore and Darcy scales. Through careful consideration of interfacial dynamics and extensive benchmarking, we show that the resulting model accurately captures the strong two-way coupling that is often exhibited between multiple fluids and deformable porous media during processes such as swelling, compression, cracking, and fracturing. The versatility of the approach is illustrated through studies that 1) quantified the effects of microporosity on sedimentary rock permeability, 2) identified the governing non-dimensional parameters that predict capillary and viscous fracturing in porous media, 3) characterised the effects of cracking on hydraulic fracture formation, and 4) described wave absorption and propagation in poroelastic coastal barriers. The approach's open-source numerical implementation \href{https://github.com/Franjcf}{``hybridBiotInterFoam"}, effectively marks the extension of computational fluid dynamics simulation packages into deformable, multiphase, multiscale porous systems.

	\vspace*{\fill}
	\newpage \lhead{} \rhead{}
	\cfoot{\thepage}

\tableofcontents
\listoffigures

	\newpage \thispagestyle{fancy} \vspace*{\fill}
	\scshape \noindent 

To my beautiful family, who I love with all my heart.
	\vspace*{\fill} \newpage \rm

	\chapter*{Acknowledgments}
	\noindent

\newthought{I am deeply grateful to} God and the Blessed Mother for giving me the strength and faith to work with joy and purpose. To my wife, Marisabel, for following me to Princeton, challenging me to be a better person everyday, cheering me up when I'm down, and for loving me unconditionally. I want to thank my parents, Francia and Federico, for their continuous encouragement and undying support. Thank you to Danilo and Ileana, for welcoming me into their home and family for the last 2 years. 

I would also like to thank my advisor, Dr. Ian Bourg, for believing in me and giving me the opportunity to work, learn, and grow at this great university. I am deeply thankful to Dr. Cyprien Soulaine for his continued friendship and guidance, and to my thesis committee (Doctors Howard Stone, Sujit Datta, and Michael Celia) for their support and productive discussions. 

Lastly, I want to thank and acknowledge the U.S. Department of Energy, Office of Science, Office of Basic Energy Sciences (DOE-BES) EFRC program under Award DE-AC02-05CH11231 (through the Center for Nanoscale Controls on Geologic CO2), the National Science Foundation Division of Earth Sciences (NSF EAR) CAREER program under Award EAR-1752982, and the Mary and Randall Hack ‘69 Research Fund provided by the High Meadows Environmental Institute at Princeton University for making my research possible. 
	\vspace*{\fill} \newpage
	\setcounter{page}{1}
	\pagenumbering{arabic}

\doublespacing
\hypersetup{linkcolor=SchoolColor} 


\begin{savequote}[75mm]
Research is what I'm doing when I don't know what I'm doing.
\qauthor{Wernher von Braun}
\end{savequote}

\chapter{Introduction}
\label{Sect:Intro}

\newthought{Fluid flow in deformable porous media} is a ubiquitous phenomenon with important implications in many energy and environmental technologies including geologic ${\mathrm{CO}}_{2}$ sequestration, soil bioremediation, water treatment, enhanced biochemical production, nuclear waste disposal, and fuel cell design \citep{Bacher2019,Bock2010,Cunningham2003,Rass2018,Towner1987}. It also underlies iconic geophysical features at many scales, from coastal, riparian, and volcanic landforms to fractures in subsurface reservoirs, cracks in clay soils, and bubbles in soft sediments. 

Whereas single-phase flow in porous media is relatively well understood from atomistic to continuum scales, the dynamics of systems containing multiple phases remain challenging to describe at all scales \citep{Gray2015,Li2018}. Multiphase flow involves strong feedback between inertial, viscous, capillary, and interfacial forces \citep{Meakin2009,Datta2014}. This coupling is intrinsically multiscale, as inertial and viscous forces dominate in large pores or fractures while capillary forces and interfacial energetics dominate within smaller porous or microporous structures. The complex linkage between microscopic geometric heterogeneities and macroscopic processes makes it necessary to consider multiple scales across porous media in order to create truly predictive models, from the scale of microscopic interfaces ($\sim\mu$m), to pore networks and lab columns ($\sim$cm), all the way up to the field scale ($\sim$km).

\begin{figure}[htb]
\begin{center}
\includegraphics[width=0.9\textwidth]{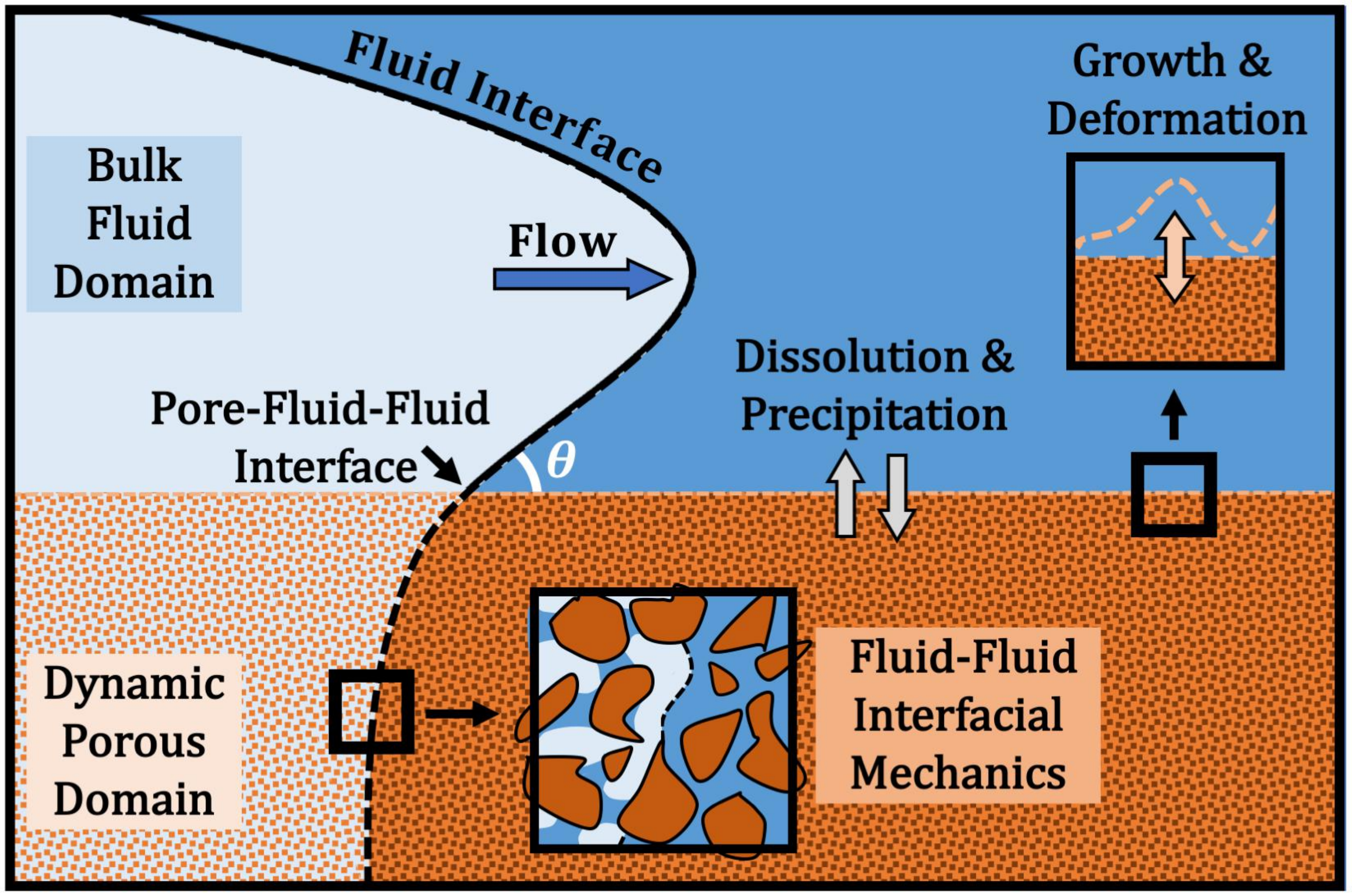}
\caption[Conceptual representation of a multiscale deformable porous medium and its related processes]{Conceptual representation of a multiscale deformable porous medium and its related processes. The porous domain is shown in the lower half (orange), the free-fluid domain is shown in the upper half (blue), two immiscible fluids (left and right) are shown in different shades of blue and are separated by an interface (black). $\theta$ is the contact angle.  \label{fig:MDBB}}
\end{center} 
\end{figure}

\begin{figure}[htb!]
\begin{center}
\includegraphics[width=0.97\textwidth]{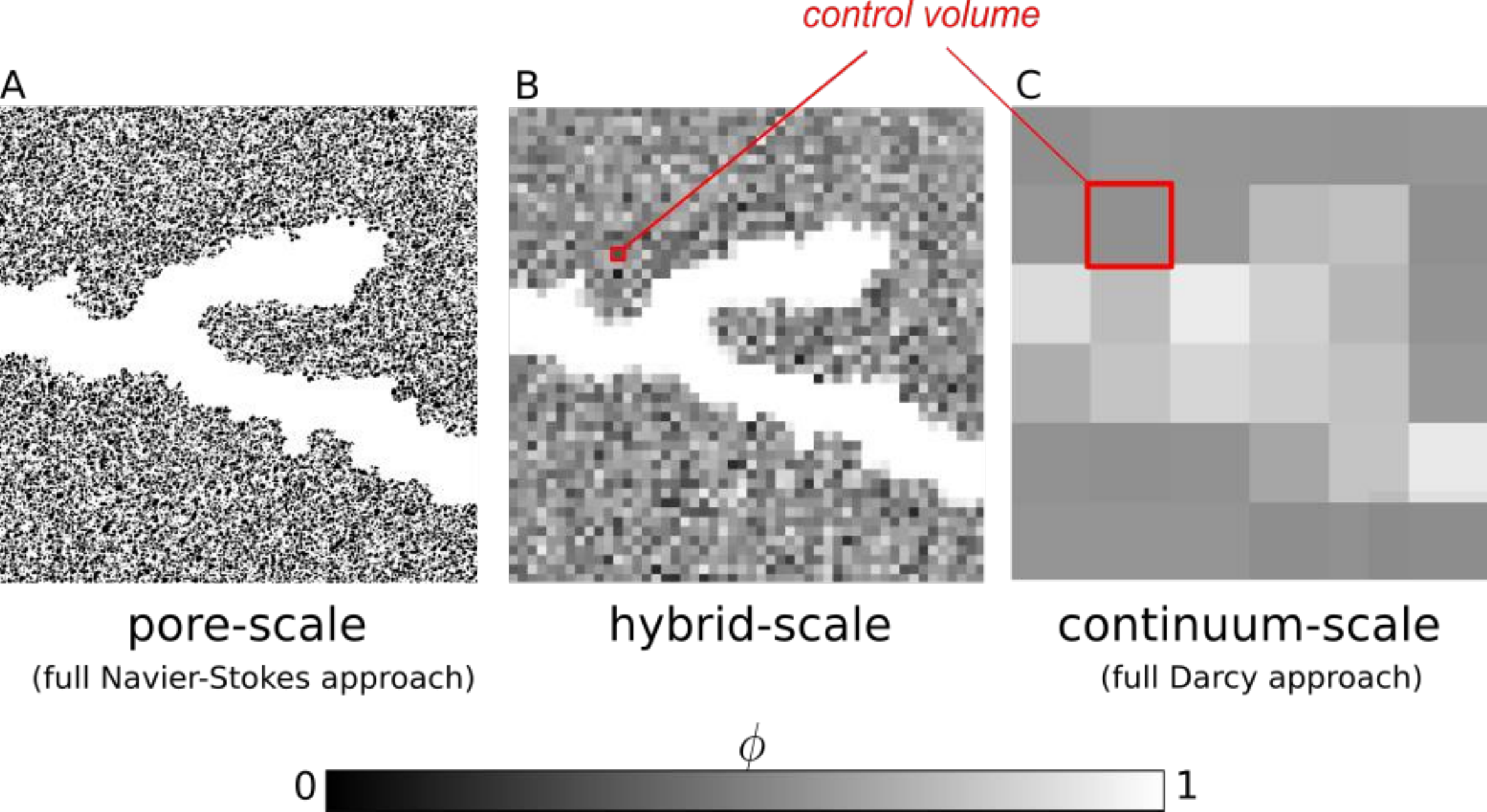}
\caption[Schematic representations of a porous medium with two characteristic pore sizes]{Schematic representations of a porous medium with two characteristic pore sizes depending on the scale of resolution: (a) full pore scale (Navier-Stokes), (b) intermediate or hybrid scale, and (c) full continuum scale (Darcy). Our objective is to derive a framework that can describe multiphase flow at all three scales described in the figure based on a single set of equations resolved throughout the entire system.   \label{fig:Figure1}}
\end{center} 
\end{figure}

An important and largely unresolved challenge in the areas outlined above is the difficulty of describing situations where multiple fluids interacts with a deformable porous material. For example, when modeling flow through biofilms or membranes it is imperative to understand how fluid flow behaves inside the microporous medium (in pores with length scales of $\sim10^{-6} \ \unit{m}$) while simultaneously understanding how the deformation of this medium affects the overall flow field (often controlled by much larger flow paths with length scales on the order of $\sim10^{-2} \ \unit{m}$) \citep{Bottero2010}. Similarly, the propagation of flow-driven fractures in porous materials and the propagation of waves in coastal barriers involve feedbacks between flow and mechanics in systems with characteristics pore widths that differ by three or more orders of magnitude. In this dissertation, we develop a framework capable of representing multiphase flow and solid mechanics in systems with two characteristic pore length scales, as required to simulate many of the aforementioned phenomena (see Figs. \ref{fig:MDBB} and \ref{fig:Figure1}). 

The starting point for our study is based on the present ample understanding of multiphase flow dynamics within and around \textit{static} porous materials, from viscous and capillary fingering \citep{Ferer2004,Lenormand1989,Lenormand1988} to temperature and surface tension driven flows \citep{Shih1996}, all the way to turbulent multiphase flows \citep{Colombo2015,Soulaine2014}. This knowledge, in conjunction with numerical techniques such as the Lattice Boltzmann Method, the Finite Volume Method, Homogenization Theory, and Averaging Theory, forms the basis of fast and accurate models that are routinely applied to help design and improve hydrocarbon production \citep{Burrus1991,Mehmani2019}, ${\mathrm{CO}}_{2}$ sequestration \citep{Hassan2012}, and even nuclear reactors \citep{Tentner2008}. However, the study of multiphase flow across different scales remains limited as shown by the absence of well-established approaches to describe how bubbles or waves propagate into an unsaturated porous medium or how a multiphase fluid mixture is pushed out of a porous medium into open space. 

A similar situation pertains with regard to computational models that couple fluid flow and solid mechanics. Theoretical and numerical approaches based on Biot's Theory of poroelasticity \citep{Biot1941}, Terzaghi's effective stress principle \citep{Terzaghi1943}, and Mixture Theory \citep{Siddique2017d} have been successful at modeling systems with flow in deformable porous media including arteries, biofilms, boreholes, hydrocarbon reservoirs, seismic systems, membranes, soils, swelling clays, and fractures \citep{Auton2017a,Barry1997,Jha2014,Lo2005,Lo2002,Macminn2016b,Mathias2017,Santillan2017}. However, as mentioned above, we still have very little understanding of how flow-induced deformation of these solid materials affects the macroscopic flow around them (and thus their boundary conditions) or how fluid-fluid interfaces behave when pushed against a soft porous medium and vice-versa. 

Three major approaches have been proposed to resolve the challenge posed by fluid flow in porous media containing both solid-free regions and microporous domains (hereafter referred to as multiscale systems). The most straightforward of these involves performing direct numerical simulations (DNS) throughout the entire multiscale domain, both within and outside the porous medium \citep{Breugem2004,Hahn2002,Krafczyk2015}. Although rigorous, this technique is impractical in situations with a large difference in length scales between the largest and smallest pores, where it requires exceedingly fine grids and tremendous computational resources. 

To save time and resources, other studies have relied on hybrid DNS-Darcy approaches, where fluid and solid mechanics within a porous medium are modeled as averaged quantities through Darcy's law, pore-network models, or Biot theory \citep{Weishaupt2019,Ehrhardt2010}. One such approach relies on the use of the Beavers-Joseph (BJ) boundary condition to couple fluid flow in solid-free domains (simulated using the Navier-Stokes equations) and in microporous domains (simulated using Darcy's law) for \textit{single} phase flow and \textit{static} porous media \citep{Beavers1967a,Fetzer2016}. Recent studies have extended this BJ approach to allow multiphase flow in the solid-free domain \citep{Baber2016} or to include the effects of poroelasticity within the porous medium \citep{Lacis2017,Zampogna2019}. However, to the best of our knowledge, no BJ based technique has yet been developed to couple solid mechanics with multiphase flow simultaneously within the solid-free and porous domains. 

The Darcy-Brinkman (DB) approach --also referred to as Darcy-Brinkman-Stokes (DBS) equations-- \citep{Brinkman1947} presents a well-known alternative to the BJ interface matching technique. These equations arise from volume averaging the Stokes (or Navier-Stokes) equations in a control volume that contains both fluids and solids \citep{Vafai1981,Hsu1990,Bousquet-Melou2002,Goyeau2003}. It consists in a Stokes-like momentum equation that is weighted by porosity and contains an additional drag force term that describes the mutual friction between the fluids and solids within said control volume. Unlike standard continuum scale equations for flow and transport in porous media such as Darcy's law, the DB equation remains valid in solid-free regions (see Figure~\ref{fig:Figure1}A) where the drag force term vanishes and the DB equation turns into the Stokes (or Navier-Stokes) equation. In porous regions (see Figure~\ref{fig:Figure1}C), in contrast, viscous dissipation effects are negligible compared with the drag force exerted onto the pore walls and the DB momentum equation tends asymptotically towards Darcy's law \citep{Tam1969,Whitaker1986,Auriault2009}. Therefore, the ``micro-continuum" DB equation has the ability to simultaneously solve flow problems through porous regions and solid-free regions \citep{Neale1974}, paving the path to hybrid scale modeling (see Figure~\ref{fig:Figure1}B). In the case of single phase flow, it is known to be analogous (in fact, formally equivalent) to the previously mentioned and well-established Beavers-Joseph boundary conditions \citep{Beavers1967,Neale1974}.

The resulting so-called ``micro-continuum'' approach has been extensively used to solve \textit{single} phase flow through \textit{static} multiscale porous media. A prime example is the modelling of fluid flow in three-dimensional images of rock samples that contain unresolved sub-voxel porosity \citep{Knackstedt_2006,Apourvari2014, Scheibe2015,Soulaine2016,Kang2019,Singh2019}. It also has been used to simulate dissolution wormholing during acid stimulation in cores by updating the weighting porosity field through geochemical reactions \citep{Liu1997,Golfier2002,Soulaine2016a,Tomin_2018}. Moreover, it has been shown that whenever low-porosity low-permeability porous regions are present, the velocity within these regions drops to near zero, such that the micro-continuum DB framework can be used as a penalized approach to map a solid phase onto a Cartesian grid with a no-slip boundary at the solid surface \citep{Angot1999,Khadra2000,Soulaine2016a}. Therefore, this approach can be used to move fluid-solid interfaces efficiently in a Cartesian grid without a re-meshing strategy. For example, \cite{Soulaine2017} used a micro-continuum framework to predict the dissolution kinetics of a calcite crystal and successfully benchmarked their model against state-of-the-art pore scale dissolution solvers with evolving fluid-solid interfaces \citep{Molins2019}.

In the present thesis, we propose the Multiphase Darcy-Brinkman-Biot (DBB) approach, a fully coupled multiscale model for two-phase flow in deformable porous media based on the micro-continuum approach, rooted in elementary physical principles and rigorously derived using the method of the volume averaging \citep{Whitaker1999}. We show that there exists a single set of partial differential equations that can be applied in pore, continuum, and hybrid scale representation of multiphase flow in porous media. Particular attention is paid to the derivation of gravity and capillary effects in the porous domain for both fluid and solid mechanics. The resulting two-phase micro-continuum framework is validated using an extensive series of test cases where reference solutions exist. We verify that the multiscale solver converges to the standard Darcy/Biot scale solutions (Buckley-Leverett, capillary-gravity equilibrium, drainage in a heterogeneous reservoir, Terzaghi consolidation tests, clay swelling experiments) when used at the continuum scale in porous media and to the two-phase Navier-Stokes solutions (droplet on a flat surface, capillary rise, drainage with film deposition, two-phase flow in a complex porous structure) when used at the pore scale. The fully implemented numerical model, along with the aforementioned verification and tutorial cases, is provided as an accompanying open-source solver: \href{https://github.com/Franjcf}{\textit{hybridBiotInterFoam}}.

This thesis is organized as follows. Chapter \ref{FullDerivation} introduces the concept of volume averaging and describes the derivation of the governing equations for coupled fluid and solid mechanics. Chapter \ref{numerical_impl} explains the numerical implementation and algorithm development for the coupled mass and momentum equations and introduces the resulting open-source solver. Then, Chapters \ref{chp:singlePhaseDBB} to \ref{chp:MDD_Applications} focus on verifying and showcasing the model's ability to capture coupled fluid-solid mechanics in multiscale porous media. This was done in an incremental manner, where Chapter \ref{chp:singlePhaseDBB} focuses on \textit{single} phase flow through \textit{deformable} porous media, Chapter \ref{chp:hybridPhase} on \textit{multiphase} flow through \textit{static} porous media, and Chapter \ref{chp:MDD_Applications} on \textit{multiphase} flow through \textit{deformable} porous media. After that, Chapter \ref{chapter:PRL} uses the developed model to identify the governing non-dimensional parameters that predict capillary and viscous fracturing in porous media. Lastly, Chapter \ref{chapter:clogging} presents a slight detour through the use of Machine Learning approaches to predict and simulate stochastic clogging process in heterogeneous porous media. Chapter \ref{conclusion} concludes with a summary of this thesis and a discussion on future work.

\begin{savequote}[75mm]
The difference between screwing around and science is writing it down.
\qauthor{Adam Savage}
\end{savequote}

\chapter{Derivation of the PDEs for Multiphase Flow in Deformable Porous Media}
\label{FullDerivation}

\begin{figure}[htb!] 
\begin{center}
\includegraphics[width=0.9\textwidth]{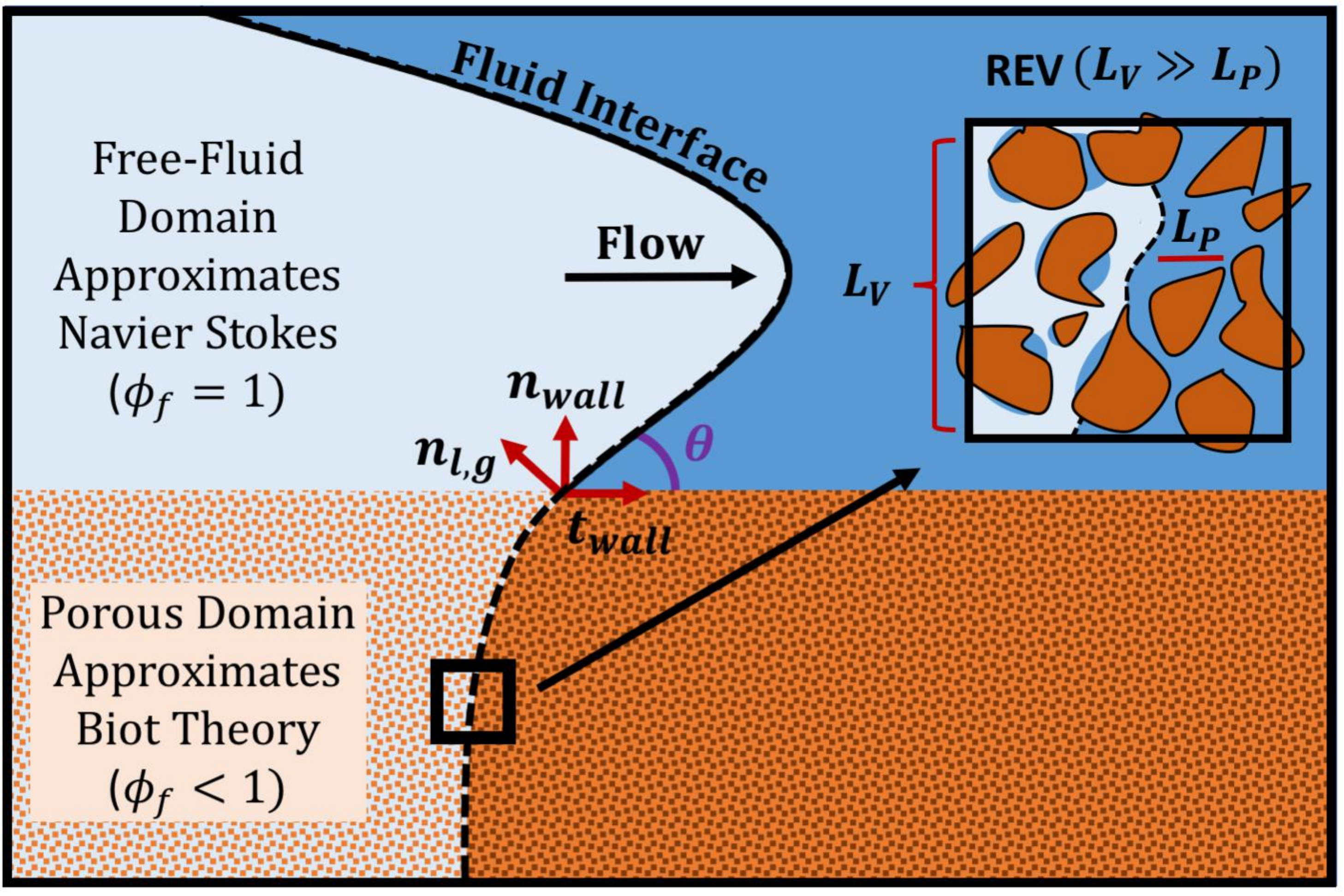}
\caption[Conceptual representation of the Multiphase DBB modelling framework]{\label{fig:conceptual_derivation} Conceptual representation of the Multiphase DBB modelling framework. The porous domain is shown in the lower half (orange), the free-fluid domain is shown in the upper half (blue), the two immiscible fluids (left and right) are shown in different shades of blue and are separated by an interface (black), $\theta$ is the contact angle, and ${\phi }_f$ is the porosity. REV is the ``Representative Elementary Volume'' over which all conservation equations are averaged. Note that the stated relation between the averaging volume’s length scale $L_V$ and the porous length scale $L_P$ is required for the creation of a REV.}
\end{center}
\end{figure}

\newthought{In this chapter,} we present the complete derivation of the micro-continuum equations for multiphase flow in static and deformable porous media. This Computational Fluid Dynamics (CFD) approach allows for simulation of two-phase flow in hybrid systems containing solid-free regions and porous matrices, as illustrated schematically in Fig. \ref{fig:conceptual_derivation}. Our approach consists of a unique set of five volume-averaged partial differential equations that asymptotically approach the Navier-Stokes Volume-of-Fluid equations in solid-free-regions and multiphase Biot Theory in porous regions. This set of equations consists of \hyperref[Sect:SolidMass]{1)} a solid mass conservation equation, \hyperref[Sect:FluidMass]{2)} a fluid mass conservation equation, \hyperref[Sect:saturationEqn]{3)} a fluid saturation conservation equation, \hyperref[Sect:FluidMomeqn]{4)} a fluid momentum conservation equation, and \hyperref[Sect:generalSolidEqn]{5)} a solid momentum conservation equation. This work is adapted from \cite{Carrillo2019}, \cite{Carrillo2020}, and \cite{Carrillo2020MDBB}.

\section{Volume Averaging}
\label{Sect:Vol_Avg}

We start by introducing the concept of volume averaging. This technique forms the basis of the micro-continuum equations, as it allows the classical mass and momentum conservation equations to account for the coexistence of solids ($s$), wetting fluids ($w$), and non-wetting fluids ($n$) within a given control volume. This method is well suited for use in conjunction with the Finite Volume Method \citep{Patankar1980}, as the latter method's numerical grid provides an intuitive and straightforward numerical interpretation of what we will define as the averaging volume ($V$).  We start by defining the volume averaging operator:

\begin{equation}{\overline{\beta}}_i=\frac{1}{V}\int_{V_i}{{\beta}_idV}\end{equation} 

\noindent where ${\beta}_i$ is a function defined in each phase's respective volume $V_i\left(i=w,n,s\right).$ We also define the phase averaging operator: 

\begin{equation}{\overline{\beta}}^i_i=\frac{1}{V_i}\int_{V_i}{{\beta}_idV}\end{equation} 

Each fluid's ($f$) phase averaged variables are thus intrinsically related by the porosity ${\phi}_f=V_f/V=(V_w+V_n)/V$ and saturation ${\alpha}_i=V_i/V_f$ fields (where $V_f \equiv V_w +  V_n$) such that ${\overline{\beta}}_i={\phi}_f{\alpha}_i{\overline{\beta}}^i_i\ \left(i=w,n\right)$. For phase-averaged solid variables, the equivalent relationship only involves the solid fraction ${\phi}_s$, such that:  ${\overline{\beta}}_s=\ {\phi}_s{\overline{\beta}}^s_s$. We note that ${\phi}_f+{\phi}_s=1$ and ${\alpha}_1+{\alpha}_2=1$; thus knowledge of one of the $\phi$ or $\alpha$ variables always leads to knowledge of the other. Volume averaging then allows for the definition of several regions within a hybrid-scale numerical simulation: 

\begin{equation}\label{phi_def}
\phi_f=
    \begin{cases}
        1, & \textnormal{in solid-free regions}\\
        \left]0;1\right[, &  \textnormal{in porous regions}
    \end{cases}
\end{equation} 

\begin{equation}\label{alpha_def}
\alpha_w=
    \begin{cases}
        0, & \textnormal{in regions saturated with non wetting fluid}\\
        \left]0;1\right[, &  \textnormal{in unsaturated regions} \\
        1, &  \textnormal{in regions saturated with wetting fluid} \\
    \end{cases}
\end{equation}

The application of an averaging transformation to mass and momentum conservation equations will result in variables and equations that are weighted differently in each of the regions identified in Eqns. \ref{phi_def}-\ref{alpha_def}. However, the averaging of differential equations is not straightforward, which is why we introduce the following spatial averaging theorems for volumes containing three distinct phases \citep{Howes1985,Whitaker1999}:

\begin{equation}\label{surf_integral_1}\overline{\frac{\partial {\beta}_i}{\partial t}}=\ \frac{\partial \overline{{\beta}_i}}{\partial t}-\frac{1}{V}\int_{A_{i,j}}{{\beta}_i{\boldsymbol{v}}_{i,j} \cdot {\boldsymbol{n}}_{i,j}dA}-\frac{1}{V}\int_{A_{i,k}}{{\beta}_i{\boldsymbol{v}}_{i,k} \cdot {\boldsymbol{n}}_{i,k}dA}\end{equation} 
\begin{equation}\label{eq:vol_avg_scalar}\overline{\nabla {\beta}_i}=\nabla\overline{{\beta}_i}+\frac{1}{V}\int_{A_{i,j}}{{\beta}_i{\boldsymbol{n}}_{i,j}dA}+\frac{1}{V}\int_{A_{i,k}}{{\beta}_i{\boldsymbol{n}}_{i,k}dA}\ \ \ \end{equation} 
\begin{equation}\label{surf_integral_3}\overline{\nabla \cdot {\boldsymbol{\beta}}_i}=\nabla\overline{\boldsymbol{\beta}_i}+\frac{1}{V}\int_{A_{i,j}}{{\boldsymbol{\beta}}_i\cdot {\boldsymbol{n}}_{i,j}dA}+\frac{1}{V}\int_{A_{i,k}}{{\boldsymbol{\beta}}_i{ \cdot \boldsymbol{n}}_{i,k}dA}\ \ \ \end{equation} 

\noindent where $A_{i,j}$ represents the interfacial area between phase $i$ and $j$, $\boldsymbol{n}_{i,j}$ is a vector normal to the interface and oriented toward phase $j$, and ${\boldsymbol{v}}_{i,j}$ is the velocity of the interface. The notation for symbols with subscript pair $i,k$ is equivalent, and the symbols $i,j,k$ represent any combination of the solid, wetting, and non-wetting phases. The surface integrals in Eqns. \ref{surf_integral_1} - \ref{surf_integral_3} are crucial components of the following derivations as they convert the interfacial conditions at the fluid-fluid and fluid-solid interfaces into body forces within the averaged partial differential equations. 

The following properties will also become useful, which follow directly from the basic averaging theorem in a system with a single fluid and solid phase \citep{Whitaker1986}.
\begin{equation}\ \frac{1}{V}\int_{A_{i,s}}{{\boldsymbol{n}}_{f,s}dA}\ =\ -\nabla{\phi}_f\ \end{equation} 

However, if the integral is over a fluid-solid surface in a multiphase system, the previous equation needs to be modified to account for the fact that fluid phase $i$ only partially covers the full solid surface,

\begin{equation}\ \frac{1}{V}\int_{A_{i,s}}{{\boldsymbol{n}}_{i,s}dA}\ =\ -{\alpha}_{i}\nabla{\phi}_f\ \end{equation} 

The original property can then be easily recovered by integrating over the solid surface spanned by both fluid phases:

\begin{equation}\label{Eq:porosity_grad}\frac{1}{V}\int_{A_{i,s}}{{\boldsymbol{n}}_{i,s}dA}+\ \frac{1}{V}\int_{A_{j,s}}{{\boldsymbol{n}}_{j,s}dA}\ =\ -{\alpha}_{i}\nabla{\phi}_f-{\alpha}_{j}\nabla{\phi}_f=\ -\nabla{\phi}_f\ \end{equation} 

\begin{figure}[htb!]
\begin{center}
\includegraphics[width=1.0\textwidth]{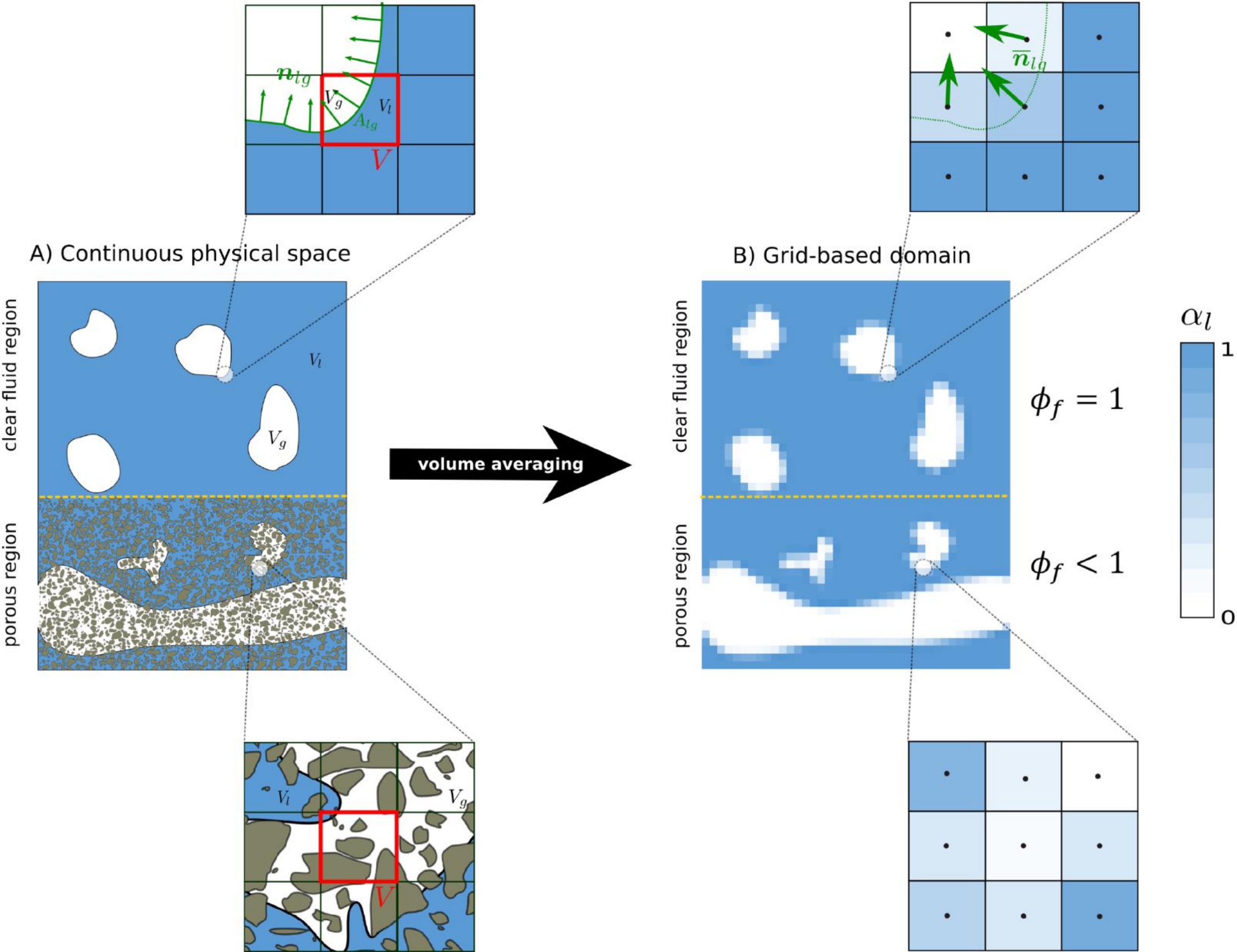}
\caption[Distribution of the fluid phases in a continuous physical domain and a discrete Eulerian grid]{Distribution of the fluid phases in (a) the continuous physical domain, (b) a discrete Eulerian grid. In this figure, the subscripts $g$ and $l$ represent gaseous and liquid phases, respectively. \label{fig:FigureVOF}}
\end{center} 
\end{figure}

\section{Solid Mass Conservation Equation} \label{Sect:SolidMass}

We start the derivation of our solid mass conservation equation with an elementary mass conservation equation:

\begin{equation}
    \frac{\partial{\rho}_s}{\partial t}+\nabla\cdot\left({\rho}_s{\boldsymbol{U}}_s\right)=0
\end{equation} 

\noindent where ${\rho}_s$ and $\boldsymbol{U}_s$ are the solid's density and velocity, respectively. We then apply the averaging operator within a volume containing the solid ($s$) and two immiscible fluid phases ($w,n$) such that:
\begin{multline}
\overline{\frac{\partial {\rho}_s}{\partial t}}+\overline{\nabla\cdot\left({\rho}_s{\boldsymbol{U}}_s\right)} = \\
\frac{\partial \overline{{\rho}_s}}{\partial t}-\frac{1}{V}\int_{A_{s,w}}{{\rho}_s{\boldsymbol{U}}_s \cdot {\boldsymbol{n}}_{s,w}dA}-\frac{1}{V}\int_{A_{s,n}}{{\rho}_s{\boldsymbol{U}}_s \cdot {\boldsymbol{n}}_{s,n}dA} \\
+\nabla\cdot\left(\overline{{\rho}_s{\boldsymbol{U}}_s}\right)+\frac{1}{V}\int_{A_{s,w}}{{\rho}_s{\boldsymbol{U}}_s\cdot {\boldsymbol{n}}_{s,w}dA} +\frac{1}{V}\int_{A_{s,n}}{{\rho}_s{\boldsymbol{U}}_s\cdot {\boldsymbol{n}}_{s,n}dA}
\end{multline}

It then becomes clear that the integral operators that arise from the transient term cancel with the ones that arise from the convection term such that:

\begin{equation}\frac{\partial \overline{{\rho}_s}}{\partial t}+\nabla\cdot\left(\overline{{\rho}_s{\boldsymbol{U}}_s}\right)=0\end{equation} 

We can now decompose the equation in terms of the intrinsic phase averages,

\begin{equation}\frac{\partial {\phi}_s\overline{{\rho}^s_s}}{\partial t}+\nabla\cdot\left({\phi}_s{\overline{\rho}}^s_s\ \overline{{\boldsymbol{U}}^s_s}\right)=0\end{equation} 

Assuming the solid density is constant and given that intrinsic phase averages of constants are equal to themselves, we can state the following:

\begin{equation}\frac{\partial {\phi}_s}{\partial t}+\nabla\cdot\left({\phi}_s\overline{{\boldsymbol{U}}^s_s}\right)=0\end{equation} 

This finalizes the derivation of the solid mass conservation equation.

\section{Fluid Mass Conservation Equation} \label{Sect:FluidMass}

Just as before, we start with a mass conservation equation for a given fluid $"i"$ $\left(i=w,n\right)$,

\begin{equation}\frac{\partial {\rho}_i}{\partial t}+\nabla\cdot\left({\rho}_i{\boldsymbol{U}}_i\right)=0\end{equation} 

\noindent where ${\rho}_i$ and $\boldsymbol{U}_i$ are the fluid's density and velocity, respectively. We then follow the same procedure outlined in the Section \ref{Sect:SolidMass} (also shown in Appendix \ref{App:Skipped_Fluid_Continuity}) in order to obtain the following equation: 

\begin{equation}\label{eq:fluid_mass}\frac{\partial \overline{{\rho}_i}}{\partial t}+\nabla\cdot\left(\overline{{\rho}_i{\boldsymbol{U}}_i}\right)=0\end{equation} 

Just as before, we want to decompose the averaged terms into their intrinsically averaged components. This time, however, we do so by noting that
${\overline{\beta}}_i={\phi}_f{\alpha}_i{\overline{\beta}}^i_i$. 

\begin{equation}\frac{\partial {\phi}_f{\alpha}_i\overline{{\rho}^i_i}}{\partial t}+\nabla\cdot\left({\phi}_f{\alpha}_i\overline{{\rho}^i_i}\overline{{\boldsymbol{U}}^i_i}\right)=0\end{equation} 

Assuming the fluid density is constant and given that intrinsic phase averages of constants are equal to themselves, we can show that:

\begin{equation}\label{Eq:sat_conservation}\frac{\partial {\phi}_f{\alpha}_i}{\partial t}+\nabla\cdot\left({\phi}_f{\alpha}_i\overline{{\boldsymbol{U}}^i_i}\right)=0\end{equation} 

This step completes the derivation of the mass conservation equation for a given fluid ``$i$''. We now sum together the mass conservation equations for both wetting ($w)$ and non-wetting ($n)$ fluids, such that:

\begin{equation}\frac{\partial {\phi}_f{\alpha}_w}{\partial t}+\frac{\partial {\phi}_f{\alpha}_n}{\partial t}+\nabla\cdot\left({\phi}_f{\alpha}_w\overline{{\boldsymbol{U}}^w_w}\right)+\nabla\cdot\left({\phi}_f{\alpha}_n\overline{{\boldsymbol{U}}^n_n}\right)=\end{equation} 

\begin{equation}\frac{\partial {\phi}_f({\alpha}_w+{\alpha}_n)}{\partial t}+\nabla\cdot\left({\phi}_f{\alpha}_w\overline{{\boldsymbol{U}}^w_w}+{\phi}_f{\alpha}_n\overline{{\boldsymbol{U}}^n_n}\right)=0\end{equation} 

The result is the \textit{single-field} mass conservation equation in terms of the porosity and the single-field fluid velocity ${\boldsymbol{U}}_f={\phi}_f\left[{\alpha}_w{\overline{\boldsymbol{U}}}^w_w\ {+\ \alpha}_n{\overline{\boldsymbol{U}}}^n_n\right]$,

\begin{equation}\frac{\partial {\phi}_f}{\partial t}+\nabla\cdot{\boldsymbol{U}}_f=0\end{equation} 

\noindent where the phrase ``single-field" refers to averaged variables that depend on the properties of both fluids. This concludes the derivation of the volume averaged fluid mass conservation equation. 

\section{Saturation Convection Equation}\label{Sect:saturationEqn}

We start this derivation with the volume-averaged mass conservation equation for the wetting fluid phase (Equation \ref{Eq:sat_conservation} in Section \ref{Sect:FluidMass}):

\begin{equation}\frac{\partial {\phi}_f{\alpha}_w}{\partial t}+\nabla\cdot\left({\phi}_f{\alpha}_w\overline{{\boldsymbol{U}}^w_w}\right)=0\end{equation} 

This equation can be used to advect the saturation function ${\alpha}_w.$ However, for practical reasons that will be made clear later, we wish to express this as a function of the single-field velocity ${\boldsymbol{U}}_f$ (rather than $\overline{{\boldsymbol{U}}^w_w}$). We start by multiplying by 1 and by noting that ${1 = \alpha}_w+{\alpha}_n$,

\begin{equation}\frac{\partial {\phi}_f{\alpha}_w}{\partial t}+\nabla\cdot\left({\phi}_f{\alpha}_w({\alpha}_w+{\alpha}_n)\overline{{\boldsymbol{U}}^w_w}\right)=0\end{equation} 
\begin{equation}\frac{\partial {\phi}_f{\alpha}_w}{\partial t}+\nabla\cdot\left({\phi}_f{\alpha}_w{\alpha}_w\overline{{\boldsymbol{U}}^w_w}\right)+\nabla\cdot\left({\phi}_f{\alpha}_w{\alpha}_n\overline{{\boldsymbol{U}}^w_w}\right)=0\end{equation} 

We then add ${\phi}_f{\alpha}_w{\alpha}_n\overline{{\boldsymbol{U}}^n_n}$ to one side and subtract it from the other (essentially adding zero to the equation),

\begin{equation}\frac{\partial {\phi}_f{\alpha}_w}{\partial t}+\nabla\cdot\left({\alpha}_w\left({\phi}_f{\alpha}_w\overline{{\boldsymbol{U}}^w_w}{+\phi}_f{\alpha}_n\overline{{\boldsymbol{U}}^n_n}\right)\right)+\nabla\cdot\left({\phi}_f{\alpha}_w{\alpha}_n\left(\overline{{\boldsymbol{U}}^w_w}-\overline{{\boldsymbol{U}}^n_n}\right)\right)=0\end{equation} 
\begin{equation}\frac{\partial {\phi}_f{\alpha}_w}{\partial t}+\nabla\cdot\left({\alpha}_w{\boldsymbol{U}}_f\right)+\nabla\cdot\left({\phi}_f{\alpha}_w{\alpha}_n\left(\overline{{\boldsymbol{U}}^w_w}-\overline{{\boldsymbol{U}}^n_n}\right)\right)=0\end{equation} 

Finally, we define the relative velocity ${\boldsymbol{U}}_r= \overline{{\boldsymbol{U}}^w_w}-\overline{{\boldsymbol{U}}^n_n}$ in order to obtain the following saturation equation:

\begin{equation}\frac{\partial {\phi}_f{\alpha}_w}{\partial t}+\nabla\cdot\left({\alpha}_w{\boldsymbol{U}}_f\right)+\nabla\cdot\left({\phi}_f{\alpha}_w{\alpha}_n{\boldsymbol{U}}_r\right)=0\end{equation} 

The definition of the relative velocity is domain dependent (it is defined differently in porous and solid-free regions, \citep{Carrillo2020}). Its full analytical derivation and description is provided in Section \ref{matchingtoDarcy}. 

\section{Fluid Momentum Conservation Equation} \label{Sect:FluidMomeqn}

We start with the classical form of the momentum conservation equation for a single-phase fluid $i$,

\begin{equation}\frac{\partial {\rho}_i{\boldsymbol{U}}_i}{\partial t}+\nabla\cdot\left({\rho}_i{\boldsymbol{U}}_i{\boldsymbol{U}}_i\right)=-\nabla p_i+{\rho}_i\boldsymbol{g}+\nabla\cdot{\boldsymbol{S}}_i\end{equation} 

\noindent where ${\rho}_i$ is the density, ${\boldsymbol{U}}_i$ is the velocity, $p_i$ is the fluid pressure, $\boldsymbol{g}$ is gravity, and ${\boldsymbol{S}}_i=\ {\mu}_i\left(\nabla{\boldsymbol{U}}_i+{\left(\nabla{\boldsymbol{U}}_i\right)}^T\right)$ is the viscous stress tensor for a Newtonian fluid. For simplicity, we neglect the inertial and convective terms for the time being:

\begin{equation}0=-\overline{\nabla p_i}+\overline{{\rho}_i\boldsymbol{g}}+\overline{\nabla\cdot{\boldsymbol{S}}_i}\end{equation} 

Applying the averaging theorems in a volume containing an additional immiscible fluid ($j$) and a solid ($s$), we obtain:

\begin{multline}
0=-\nabla{\overline{p}}_i-\frac{1}{V}\int_{A_{i,j}}{p_i{\boldsymbol{n}}_{i,j}dA}-\frac{1}{V}\int_{A_{i,s}}{p_i{\boldsymbol{n}}_{i,k}dA}+\ \overline{{\rho}_i\boldsymbol{g}}+\nabla\cdot\overline{{\boldsymbol{S}}_i}+ \\ \frac{1}{V}\int_{A_{i,j}}{{\boldsymbol{S}}_i\cdot {\boldsymbol{n}}_{i,j}dA}+\frac{1}{V}\int_{A_{i,s}}{{\boldsymbol{S}}_i{ \cdot \boldsymbol{n}}_{i,s}dA}
\end{multline} 

For improved readability, we now group the terms integrating over the fluid-fluid interface into a single term ${\boldsymbol{D}}_{i,j}=\ \frac{1}{V}\int_{A_{i,j}}{{\boldsymbol{n}}_{i,j}\cdot (-\boldsymbol{I}p_i+{\boldsymbol{S}}_i)dA}$. Furthermore, we now convert the volume averages into their intrinsic phase averages such that:

\begin{equation}0=-\nabla{{({\phi}_f\alpha}_i p }^i_i)+{\phi}_f{\alpha}_i{\rho}_i\boldsymbol{g}+\nabla\cdot\overline{{\boldsymbol{S}}_i}-\frac{1}{V}\int_{A_{i,s}}{p_i{\boldsymbol{n}}_{i,k}dA}+\frac{1}{V}\int_{A_{i,s}}{{\boldsymbol{S}}_i{ \cdot \boldsymbol{n}}_{i,s}dA}+{\boldsymbol{D}}_{i,j}\end{equation} 

Before we continue, we note that we can separate the pressure into its intrinsic phase average and deviation terms \citep{Whitaker2013}, such that $p_i=\ {\overline{p}}^i_i+{\tilde{p}}_i$, 

\begin{multline}0=-\nabla{{({\phi}_f\alpha}_i\overline{p}}^i_i)+{\phi}_f{\alpha}_i{\rho}_i\boldsymbol{g}+\nabla\cdot\overline{{\boldsymbol{S}}_i} -\frac{1}{V}\int_{A_{i,s}}{{\hat{p}}^i_i{\boldsymbol{n}}_{i,k}dA}\ \\ -\frac{1}{V}\int_{A_{i,s}}{{\hat{p}}_i{\boldsymbol{n}}_{i,k}dA}\ +\frac{1}{V}\int_{A_{i,s}}{{\boldsymbol{S}}_i{ \cdot \boldsymbol{n}}_{i,s}dA}+{\boldsymbol{D}}_{i,j}\end{multline} 

Assuming that the intrinsic phase averages are constant along the integration surface and by expanding the terms inside the gradient operators we get the following,

\begin{multline}0=-{\phi}_f\nabla{{(\alpha}_i\overline{p}}^i_i)-{{\alpha}_i\overline{p}}^i_i\nabla{\phi}_f+{\phi}_f{\alpha}_i{\rho}_i\boldsymbol{g}+\nabla\cdot\overline{{\boldsymbol{S}}_i}- \\ {\overline{p}}^i_i\frac{1}{V}\int_{A_{i,s}}{{\boldsymbol{n}}_{i,k}dA}+ 
\frac{1}{V}\int_{A_{i,s}}{{\boldsymbol{n}}_{i,j}\cdot \left(-\boldsymbol{I}{\tilde{p}}_i+{\boldsymbol{S}}_i\right)dA}+{\boldsymbol{D}}_{i,j}\end{multline} 

We now apply the averaging theorem that relates the surface integral of the unit normal to the porosity gradient (Equation \ref{Eq:porosity_grad}),

\begin{multline}
0=-{\phi}_f\nabla{{(\alpha}_i\overline{p}}^i_i)-{{\alpha}_i\overline{p}}^i_i\nabla{\phi}_f+{\phi}_f{\alpha}_i{\rho}_i\boldsymbol{g}+\nabla\cdot\overline{{\boldsymbol{S}}_i} \\
+{{\alpha}_i\overline{p}}^i_i\nabla{\phi}_f\ +\frac{1}{V}\int_{A_{i,s}}{{\boldsymbol{n}}_{i,j}\cdot \left(-\boldsymbol{I}{\tilde{p}}_i+{\boldsymbol{S}}_i\right)dA}+{\boldsymbol{D}}_{i,j}
\end{multline} 

We then cancel out like-terms and group the remaining integrals into a single term that represents the total momentum exchange across the solid-fluid interface ${\boldsymbol{D}}_{i,s}=\frac{1}{V}\int_{A_{i,s}}{\boldsymbol{n}}_{i,j}\cdot$ \\ $\left(-\boldsymbol{I}{\tilde{p}}_i+{\boldsymbol{S}}_i\right)dA$. The result is the averaged fluid momentum equation over a volume containing another immiscible fluid and a solid,

\begin{equation}\label{Eq:stokes_singlephase}0=-{\phi}_f\nabla{{(\alpha}_i\overline{p}}^i_i)+{\phi}_f{\alpha}_i{\rho}_i\boldsymbol{g}+\nabla\cdot\overline{{\boldsymbol{S}}_i}\ +{\boldsymbol{D}}_{i,s}+{\boldsymbol{D}}_{i,j}\end{equation} 

Adding the momentum equations for a wetting ($w)$ and a non-wetting ($n$) fluid we obtain the single-field averaged momentum conservation equation in terms of ${\boldsymbol{U}}_f$ and the single-field pressure $p = \alpha_w \overline{p}^w_w +\alpha_n \overline{p}^n_n $.

\begin{equation}\label{Eq_vol_avg_stokes}0=-{\phi}_f\nabla p+{\phi}_f{\rho}_f\boldsymbol{g}+\nabla\cdot\overline{\boldsymbol{S}}\ +{\boldsymbol{D}}_{w,s}+{\boldsymbol{D}}_{n,s}+{\boldsymbol{D}}_{w,n}+{\boldsymbol{D}}_{n,w}\end{equation} 

\noindent where $\overline{\boldsymbol{S}}={\mu}_f(\nabla{\boldsymbol{U}}_f+{(\nabla{\boldsymbol{U}}_f)}^T)$ is the averaged single-field viscous stress tensor (Appendix \ref{App:Viscous_Stress}), ${\mu}_f$ is the arithmetic average of each fluid's viscosity ${\mu}_f={\alpha}_w{\mu}_w+{\alpha}_n{\mu }_n$, and ${\rho}_f$ is the arithmetic average of each fluid's density ${\rho}_f={\alpha}_w{\rho}_w+{\alpha}_n{\rho}_n$. If necessary, we can now add the averaged single-field material derivative (Appendix \ref{App:material_derivative}) to obtain,

\begin{multline}\frac{\partial {\rho}_f{\boldsymbol{U}}_f}{\partial t}+\nabla\cdot\left(\frac{{\rho}_f}{{\phi}_f}{{\boldsymbol{U}}_f\boldsymbol{U}}_f\right)=-{\phi}_f\nabla p+{\phi}_f{\rho}_f\boldsymbol{g}+\nabla\cdot\overline{\boldsymbol{S}} + \\ {\boldsymbol{D}}_{w,s}+{\boldsymbol{D}}_{n,s}+{\boldsymbol{D}}_{w,n}+{\boldsymbol{D}}_{n,w}\end{multline} 

Finally, for this equation to be valid in both a porous medium and within a solid-free domain, we now need closure for the ${\boldsymbol{D}}_{i,j}$ interaction terms. This was done in \citet{Carrillo2020} and described in the following section. The results is shown here:
\begin{equation}{\boldsymbol{D}}_{w,s}+{\boldsymbol{D}}_{n,s}+{\boldsymbol{D}}_{w,n}+{\boldsymbol{D}}_{n,w}=-{\phi}_f\mu k^{-1}\left({\boldsymbol{U}}_f-{\overline{\boldsymbol{U}}}_s\right)+{\phi}_f{\boldsymbol{F}}_{c,1}+{\phi}_f{\boldsymbol{F}}_{c,2}\end{equation} 
where $\mu k^{-1}$ is the single-field mobility of the fluids (i.e. a representation for drag) and ${\boldsymbol{F}}_c$ represents the capillary forces arising from fluid-fluid interfacial dynamics. The analytical representations of these terms are derived in the next section (Section \ref{App:Multiphase_Terms}) and summarized in Section \ref{Sect: Parameter Closure}. The result is the complete single-filed fluid momentum equation for two-phase flow in deformable porous media (i.e. a re-formulated VOF approach for flow in dynamic porous media):

\begin{multline}\label{eq:complete_fluid}\frac{\partial {\rho}_f{\boldsymbol{U}}_f}{\partial t}+\nabla \cdot\left(\frac{{\rho}_f}{{\phi}_f}{\boldsymbol{U}}_{\boldsymbol{f}}{\boldsymbol{U}}_{\boldsymbol{f}}\right)=-{\phi}_f\nabla p+{\phi}_f{\rho}_f\boldsymbol{g}+\nabla\cdot\overline{\boldsymbol{S}}- \\ {\phi}_f\mu k^{-1}\left({\boldsymbol{U}}_f-{\overline{\boldsymbol{U}}}_s\right)+{\phi}_f{\boldsymbol{F}}_{c,1}+{\phi}_f{\boldsymbol{F}}_{c,2}\end{multline}

The equation presented above tend towards the standard VOF approach in solid-free regions (where the drag term becomes negligible) and towards the multiphase Darcy equations in porous regions. The latter can be explained by the fact that the viscous stress tensor $\nabla\cdot\overline{\boldsymbol{S}}$ becomes negligible under the scale-separation assumption, inertial terms become negligible under the assumption of low Reynold's number flow in the porous medium, and the $\boldsymbol{F}_c$ terms are set to fit standard multiphase Darcy's law \citep{Whitaker1986,Carrillo2020}. Therefore, Eqn. \ref{eq:complete_fluid} $\approx$

\begin{multline}
    \begin{cases}
        \frac{\partial {\rho }_f{\boldsymbol{U}}_f}{\partial t}+\nabla\cdot\left({\rho }_f{{\boldsymbol{U}}_f\boldsymbol{U}}_f\right)=-\nabla p+\nabla\cdot\overline{\boldsymbol{S}}+{\rho }_f\boldsymbol{g}+{\boldsymbol{F}}_{c,1} &  \textnormal{in solid-free regions} \\
        \left({\boldsymbol{U}}_f-{\overline{\boldsymbol{U}}}_s\right)=-\frac{k}{\mu }\left(\nabla p-{\rho }_f\boldsymbol{g}-{\boldsymbol{F}}_{c,1}-{\boldsymbol{F}}_{c,2}\right) & \textnormal{in porous regions}
    \end{cases}
\end{multline}


\section{Closure of the Fluids' Multiscale Interaction Terms} \label{App:Multiphase_Terms}

To form the multiscale momentum equation, we express the sum of the average shear stress at the fluid-solid and fluid-fluid interfaces as the sum of two independent terms, an Eulerian drag force $-{\phi}_f\mu k^{-1}\left({\boldsymbol{U}}_f-{\overline{\boldsymbol{U}}}_s\right)$ and a capillary force $\boldsymbol{F}_c$:

\begin{equation}{\boldsymbol{D}}_{w,s}+{\boldsymbol{D}}_{n,s}+{\boldsymbol{D}}_{w,n}+{\boldsymbol{D}}_{n,w}=-{\phi}_f\mu k^{-1}\left({\boldsymbol{U}}_f-{\overline{\boldsymbol{U}}}_s\right)+{\phi}_f{\boldsymbol{F}}_c\end{equation} 
these so called ``multi-scale parameters" reflect sub-grid scale variables and processes such as the location of the fluid-fluid interface and the hydrodynamic impact of the porous micro-structure.  They have different meanings and different formulations depending on whether their averaging volume contains solid material or not. Therefore we will present distinct derivations of these terms with the purpose of matching the solution of the two-phase Navier-Stokes equations in solid-free regions and of the two-phase Darcy equations in porous regions. 

\subsection{Matching the Volume-of-Fluid Model in Solid-Free Regions}

In CFD, the Volume of Fluid (VOF) method \citep{Hirt1981} is a standard approach to track the interface movement of two immiscible fluids in a fixed Eulerian grid.  In it, a phase indicator representing the volume of fluid in each grid block is used to track the distribution of the fluid phases in a computational domain as illustrated in the upper part of Figure \ref{fig:FigureVOF}B. This phase indicator has the same form as the saturation field $\alpha_w$ defined before, where $\alpha_w=1$ in cells saturated by the wetting phase and $\alpha_w=0$ in cells that contain the non-wetting phase only. Note that $0<\alpha_w<1$ in cells that contain the immiscible interface between both fluids. 

In standard VOF approaches, the computational cells do not contain solid ($\phi_f=1$). In these cases, the relative velocity $\boldsymbol{U}_r$ is used as a compression term to force the fluid-fluid interface to be as sharp as possible \citep{Rusche2002a}. This compression velocity acts in the direction normal to the interface. In the VOF framework, the normal to the fluid-fluid interface is computed using the gradient of the saturation. \citet{Rusche2002a} proposes a relative velocity oriented in the direction normal to interface with a value based on the maximum magnitude of $\boldsymbol{U}_f$:

\begin{equation}
   \boldsymbol{U}_r=C_\alpha \max \left(\left| \boldsymbol{U}_f \right|\right) \overline{\boldsymbol{n}_{w,n}} ,
\end{equation}
where $C_\alpha$ is a model parameter used to control the compression of the interface and $\overline{\boldsymbol{n}_{w,n}}$ is the mean normal vector. For low values of $C_\alpha$, the interface diffuses. For higher values, the interface is sharper, but excessive values are known to introduce parasitic velocities and lead to unphysical solutions. In practice, $C_{\alpha}$ is often chosen between 0 and 4. The mean normal vector $\boldsymbol{n}_{w,n}$ is computed by using the gradient of the phase indicator function $\alpha_w$. The relation between these two vectors can be obtained by applying Eq. \ref{eq:vol_avg_scalar} to the liquid phase indicator function $\mathds{1}_l$ (a function equal to 1 in $V_w$ and 0 elsewhere) in solid-free regions such that,
\begin{equation}\label{eq:lemmaBISA}
\nabla\alpha_w=-\frac{1}{V}\int_{A_{w,n}}\mathds{1}_l\boldsymbol{n}_{w,n}dA.
\end{equation}
Therefore, 

\begin{equation}\label{eq:meanNormalA}
\overline{\boldsymbol{n}_{w,n}}=-\frac{\nabla\alpha_w}{\left|\nabla\alpha_w\right|},
\end{equation}
is a unit vector defined at the cell centers that describes the mean normal to the fluid-fluid interface in a control volume.

Another consequence of the absence of solid in the VOF equations is that the forces describing the shear stresses of the fluids onto the solid surface are null, hence $\boldsymbol{D}_{w,s}=\boldsymbol{D}_{n,s}=0$. Therefore, the Darcy term in the momentum equation vanishes: 

\begin{equation}
{\phi}_f\mu k^{-1}\left({\boldsymbol{U}}_f-{\overline{\boldsymbol{U}}}_s\right) = 0.
\end{equation}

The integration of the shear boundary condition at the fluid-fluid interface, yields a relationship between the mutual shear between the two fluids and the surface integral of the surface tension effects: 
\begin{equation}
    \boldsymbol{D}_{w,n}+\boldsymbol{D}_{n,w} = \phi_f \boldsymbol{F}_c =\frac{1}{V}\int_{A_{lg}}\boldsymbol{n}_{lg}\cdot \gamma \kappa dA. 
    \label{eq:int_surf_force_A}
\end{equation}

This equation cannot be used directly, because the terms under the volume integral require knowledge of the location and curvature of the fluid-fluid interface within a grid block. This information is unknown in a grid-based formulation for which all the physical variables and forces are averaged on control volumes. In the VOF method, the curvature of the interface $\kappa$ is approximated by a mean interface curvature $\overline{\kappa}$. \citet{Brackbill1992} assumes that the mean curvature of the interface can be approximated by calculating the divergence of the mean normal vector,  $\overline{\kappa}=\nabla \cdot \overline{\boldsymbol{n}_{w,n}}$. Because $\overline{\kappa}$ and $\gamma$ are constant within a control volume, they can be extracted from the integral in Eq. \ref{eq:int_surf_force_A} to obtain (after applying Eq. \ref{eq:lemmaBISA}) the so-called Continuum Surface Force (CSF) formulation \citep{Brackbill1992}: 
\begin{equation}
    \boldsymbol{F}_c = \phi_f^{-1}\gamma \nabla \cdot\left(\frac{\nabla \alpha_w}{\left| \nabla\alpha_w \right|}\right)\nabla \alpha_w.
\end{equation}

\subsection{Matching the Standard Two-Phase Darcy Model in Porous Regions} \label{matchingtoDarcy}

In this subsection, we recall the formulation of the standard two-phase Darcy model that is classically used to describe two-phase flow in porous media at the continuum scale. The model can be derived by applying the volume averaging operators on a REV of the porous structure \citep{Whitaker1986a}. Unlike the present micro-continuum model, the two-phase Darcy model is a two-field model, meaning that instead of one velocity field describing the flow, there are two velocities ($\overline{\boldsymbol{U}_i}$ with  $i=w,n$) with separate pressure fields ($\overline{p_i}$ with  $i=w,n$). 

The incompressible, immiscible two-phase Darcy model in static porous media consists of a saturation equation for the wetting phase,
\begin{equation}
    \frac{\partial \phi \alpha_w}{\partial t}+\nabla \cdot \overline{\boldsymbol{U}_w} =0,
\end{equation}
a mass balance equation,
\begin{equation}
    \nabla \cdot \overline{\boldsymbol{U}_i} =0,
\end{equation}
and two momentum balance equations, one for each phase, 
\begin{align}
    \overline{\boldsymbol{U}_i} = \phi_f \alpha_i \overline{\boldsymbol{U}_i^i}  &=-\frac{k_0k_{r,i}}{\mu_i}\left( \nabla \overline{p_i}^i  - \rho_i \boldsymbol{g}\right),\quad i=w,n, \nonumber
\\
    &= -M_i\left( \nabla \overline{p_i}^i  - \rho_i \boldsymbol{g}\right),\quad i=w,n.
    \label{eq:Darcy-law_A}
\end{align}
These can also be written as,
\begin{equation}\label{Darcy-Mom}
    0 = - \nabla \overline{p_i}^i  + \rho_i \boldsymbol{g} - M_i^{-1}\overline{\boldsymbol{U}_i} ,\quad i=w,n,
\end{equation}
where $k_0$ is the absolute permeability of the porous structure, $k_{r,w}$ and $k_{r,n}$ are the relative permeabilities with respect to each fluid (classically represented here as functions of wetting fluid saturation; more complex formulations exist that account for viscous coupling between the two fluids or for the Klinkenberg effect in the non-wetting (often gaseous) phase \citep{Picchi2018}), and $M_i=\frac{k_0k_{r,i}}{\mu_i}$ are the fluid mobilities. Equations \ref{Darcy-Mom} arise from further simplification of the volume averaged Stokes equations, Eq. \ref{Eq:stokes_singlephase}, where the drag forces are combined and described as a single Darcy term. Moreover, \citet{Whitaker1986} showed that the viscous dissipative term, $\nabla \cdot \overline{\boldsymbol{S}}_i$ is negligible in comparison to the drag forces whenever the system's macroscopic length scale is significantly larger than the length scale of that averaging volume. This feature is a fundamental aspect of the multiscale framework described above because it means that even though the viscous dissipative term is retained in the single-field momentum equation, it naturally vanishes when the computational cells contain solid content. This allows the continuity of stresses between porous and solid-free domains \citep{Neale1974}.

Because it involves four equations and five unknown variables, the two-phase Darcy model is complemented by the definition of macroscopic capillary pressure $p_c$, which provides an additional relationship between the two averaged pressure fields:
\begin{equation}
    p_c\left(\alpha_w \right)=\left( \overline{p_g}^g - \overline{p_l}^l\right).
    \label{eq:capillary_pressure_A}
\end{equation}
This equation has been theoretically derived through homogenization techniques and is assumed to be independent of sub-volume fluid properties. \citep{Whitaker1986a,Torres1987}.

As the two-phase Darcy model explicitly represents the two phase-averaged velocities, it can be used to derive an expression for the relative velocity $\boldsymbol{U}_{r}$ in the porous region. Before going through the derivation, we note that the application of the gradient operator to the definition of the single-field pressure $p$, along with the definition of capillary pressure, Eq. \ref{eq:capillary_pressure_A}, results in:
\begin{align}
    \nabla  \overline{p_w}^w &= \nabla p - \nabla \left( \alpha_n p_c \right),
\nonumber
\\
    \nabla  \overline{p_n}^n &= \nabla p + \nabla \left( \alpha_w p_c \right).
    \label{eq:pressure_gradient_2}
\end{align}

We can then obtain an analytical form for $\boldsymbol{U}_{r}$ by using the multi-phase Darcy equations presented above (Eqn. \ref{eq:Darcy-law_A}):

\begin{align}
\boldsymbol{U}_{r} &=\left(\overline{\boldsymbol{U}_w^w}-\overline{\boldsymbol{U}_n^n}\right),
\nonumber
\\
&=-\frac{ M_w}{\phi_f \alpha_w}  \left( \nabla\overline{p_w^w} - \rho_w  \boldsymbol{g}\right)+\frac{M_n}{\phi \alpha_n} \left(\nabla \overline{p_n^n} - \rho_n  \boldsymbol{g}\right),
\nonumber
\\
&=\phi^{-1}_f \left[-\frac{ M_w}{\alpha_w} \nabla\overline{p_w^w}+ \frac{M_n}{ \alpha_n}\nabla\overline{p_n^n} +  \left(  \rho_w \frac{ M_w}{ \alpha_w}- \rho_n \frac{M_n}{ \alpha_n} \right)\boldsymbol{g}\right],
\nonumber
\\
&= \phi^{-1}_f \left[\begin{array}{c}
     -\left(\frac{ M_w}{\alpha_w}- \frac{M_n}{\alpha_n}\right)\nabla p + \left(  \rho_w \frac{ M_w}{\alpha_w}- \rho_n \frac{M_n}{\alpha_n} \right)\boldsymbol{g} + \\ \frac{ M_w}{\alpha_w}\nabla\left(\alpha_n p_c \right) +\frac{M_n}{\alpha_n}\nabla\left(\alpha_w p_c \right)\end{array}\right],
\nonumber
\\
&= {\phi_f}^{-1}\left[ \begin{array}{c}
-\left(M_w{\alpha }^{-1}_w-M_n{\alpha }^{-1}_n\right)\nabla p + \left({\rho }_wM_w{\alpha }^{-1}_w-{\rho }_wM_n{\alpha }^{-1}_n\right)\boldsymbol{g}+ \\ 
\left(M_w{\alpha }_n{\alpha }^{-1}_w +M_n{\alpha }_w{\alpha }^{-1}_n\right)\nabla p_c- \left(M_w{\alpha }^{-1}_w- M_n{\alpha }^{-1}_n\right)p_{c}\nabla{\alpha }_w \end{array}\right].
\label{eq:vr_porous_2}
\end{align}

A two-phase Darcy model for the single-field velocity $\boldsymbol{U}_f$ is then formed to derive the continuum scale formulation of the drag force $\mu k^{-1}\boldsymbol{U}_f$ and capillary force $\boldsymbol{F}_c$. This is achieved by summing both phase velocities, Eq. \ref{eq:Darcy-law_A}, and using the pressure gradient relationship, Eq. \ref{eq:pressure_gradient_2}. We obtain: 
\begin{align}
    \boldsymbol{U}_f &= \overline{\boldsymbol{U}_w}  + \overline{\boldsymbol{U}_n}, 
    \nonumber
    \\
    &=-M_n\nabla \overline{p_n^n} - M_w\nabla \overline{p_w^w}  +\left(   \rho_n  M_n+ \rho_w   M_w\right) \boldsymbol{g},
    \\
    \nonumber
    &= -\left(M_n+ M_w\right)\nabla  p  +\left(   \rho_n  M_n+ \rho_w   M_w\right) \boldsymbol{g} + \left[ M_w \nabla \left( \alpha_n p_c \right) -  M_n\nabla \left( \alpha_w p_c \right)\right],
\end{align}
The previous equation can be recast into:
\begin{equation}
    0 = -\nabla  p  +\rho_{M} \boldsymbol{g} - M^{-1}\boldsymbol{U}_f+M^{-1}\left[ M_w \nabla \left( \alpha_n p_c \right) -  M_n\nabla \left( \alpha_w p_c \right)\right],
    \label{eq:single-field-darcy-momentum_2}
\end{equation}
where $M= M_w+M_n$ is the total mobility and $\rho_{M}=\left( \rho_w  M_w +  \rho_n M_n\right)/\left( M_w + M_n\right)$ is a mobility-weighted average fluid density. This single-field two-phase Darcy equation matches the two-phase micro-continuum momentum equation, Eq. \ref{Eq_vol_avg_stokes}, if this equation's drag coefficient and the capillary force equal
\begin{equation}
    \mu k^{-1} = M^{-1}=k_0^{-1}\left(\frac{\mu_w}{k_{r,w}} +\frac{\mu_n}{k_{r,n}}\right)^{-1},
    \label{eq:single-field-kr_2}
\end{equation}
and
\begin{align}
    \boldsymbol{F}_c & = M^{-1}\left[ M_w \nabla \left( \alpha_n p_c \right) -  M_n\nabla \left( \alpha_w p_c \right)\right], \nonumber\\
    & = \left[M^{-1}\left(  M_w \alpha_n - M_n \alpha_w \right)\left(\frac{\partial p_c}{\partial \alpha_w}\right) - p_c \right]\nabla \alpha_w,
    \label{eq:Fc-1_2}
\end{align}
respectively. Here, the single-field relative permeability, Eq. \ref{eq:single-field-kr_2}, is a harmonic average of the two-phase mobilities, in agreement with the proposal of \citet{Wang1993} and \citet{Soulaine2019}. 

Finally, we note that in Eq. \ref{eq:single-field-darcy-momentum_2}, the single-field fluid density $\rho_{M}$ in the buoyant term is a weighted average based on the fluid mobilities, or more exactly, the fractional flow functions, $M_iM^{-1}$. This is a classic concept in multiphase flow in porous media. As shown in \citet{Carrillo2020} a strictly equivalent solution can be derived where $\rho_{M}$ is replaced by $\rho_f$ in Eq. \ref{eq:single-field-darcy-momentum_2} and the capillary force expression is replaced by:
\begin{equation}
    \boldsymbol{F}_c = \boldsymbol{F}_{c,1} + \boldsymbol{F}_{c,2} = M^{-1}\left(  M_w \alpha_n - M_n \alpha_w \right) [( \rho_w  -  \rho_n )\boldsymbol{g} + \nabla p_c ] - p_c \nabla \alpha_w
\end{equation}
this is the expression shown in Sections \ref{Sect:FluidMomeqn} and \ref{Sect: Parameter Closure}.

\section{Momentum Conservation Equation for a Linear Elastic Solid} \label{Sect:elastic}

We start with the differential form of the force balance for a solid body in terms of the displacement vector ${\boldsymbol{d}}_s$. Fundamentally, it consists of an elastic stress tensor ($\boldsymbol{\sigma})$ balanced by a Terzaghi effective stress tensor ($\boldsymbol{\tau})$ \citep{Carrillo2019,Jasak2000}.

\begin{equation}\frac{{\partial }^2\left({\rho}_s{\boldsymbol{d}}_s\right)}{\partial t^2}-\nabla \cdot \boldsymbol{\sigma}=\nabla \cdot \boldsymbol{\tau}+{\rho}_s\boldsymbol{g}\end{equation} 

\noindent where $\boldsymbol{\tau}$ is a function of the confining pressure $(P_{conf})$, the fluid pressure $(p)$, and the swelling pressure $(p_{swell})$ such that: $\boldsymbol{\tau}=\ {\boldsymbol{P}}_{conf}-\boldsymbol{I}p-\boldsymbol{I}p_{swell}$. Applying the volume averaging operators for a solid ($s$) in contact with a wetting ($w$) and non-wetting fluid ($n$) we obtain:

\begin{equation}\overline{\frac{{\partial }^2\left({\rho}_s{\boldsymbol{d}}_s\right)}{\partial t^2}}-\overline{\nabla \cdot \boldsymbol{\sigma}}=\overline{\nabla \cdot \boldsymbol{\tau}}\ +\overline{{\rho}_s\boldsymbol{g}}\end{equation}

Expanding the first term from the left we get:
\begin{equation}
\begin{split}
=\frac{{\partial }^2\left(\overline{{\rho}_s{\boldsymbol{d}}_s}\right)}{\partial t^2}-\frac{\partial }{\partial t}\frac{1}{V}\int_{A_{s,w}}{{{\rho}_s{\boldsymbol{d}}_s\boldsymbol{U}}_s \cdot {\boldsymbol{n}}_{s,w}dA}-\frac{1}{V}\int_{A_{s,w}}{\frac{\partial \left({\rho}_s{\boldsymbol{d}}_s\right)}{\partial t}{\boldsymbol{U}}_s \cdot {\boldsymbol{n}}_{s,w}dA} \\
-\frac{\partial }{\partial t}\frac{1}{V}\int_{A_{s,n}}{{{\rho}_s{\boldsymbol{d}}_s\boldsymbol{U}}_s \cdot {\boldsymbol{n}}_{s,n}dA}-\frac{1}{V}\int_{A_{s,n}}{\frac{\partial \left({\rho}_s{\boldsymbol{d}}_s\right)}{\partial t}{\boldsymbol{U}}_s \cdot {\boldsymbol{n}}_{s,n}dA}
\end{split}
\end{equation} 

Given that linear elasticity theory only deals with infinitesimal deformations, we can safely assume that the velocity of the solid-fluid interface $({\boldsymbol{U}}_s)$ is very close to zero. This approximation yields:

\begin{equation}\overline{\frac{{\partial }^2\left({\rho}_s{\boldsymbol{d}}_s\right)}{\partial t^2}}=\frac{{\partial }^2\left(\overline{{\rho}_s{\boldsymbol{d}}_s}\right)}{\partial t^2}=\frac{{\partial }^2\left({\phi}_s{\rho}_s\overline{{\boldsymbol{d}}^s_s}\right)}{\partial t^2}\end{equation} 

We now continue by expanding the stress terms,
\begin{equation}
\begin{split}
\frac{{\partial }^2\left({\phi}_s{\rho}_s\overline{{\boldsymbol{d}}^s_s}\right)}{\partial t^2}-\nabla\cdot \overline{\boldsymbol{\sigma}}-\frac{1}{V}\int_{A_{s,n}}{\boldsymbol{\sigma}{ \cdot \boldsymbol{n}}_{s,w}dA}-\frac{1}{V}\int_{A_{s,n}}{\boldsymbol{\sigma}{ \cdot \boldsymbol{n}}_{s,n}dA} \\
-\nabla\cdot \overline{\boldsymbol{\tau}}-\frac{1}{V}\int_{A_{s,w}}{\boldsymbol{\tau}{ \cdot \boldsymbol{n}}_{s,w}dA}-\frac{1}{V}\int_{A_{s,n}}{\boldsymbol{\tau}{ \cdot \boldsymbol{n}}_{s,n}dA}=0
\end{split}
\end{equation} 

Just as before, we can write the Terzaghi stress tensor (i.e. the fluid pressure) as the sum of its intrinsic phase average and its sub-volume deviations. Furthermore, we also expand the averaged Terzaghi term into its phase averaged form,

\begin{equation}
\begin{split}
\frac{{\partial }^2\left({\phi}_s{\rho}_s\overline{{\boldsymbol{d}}^s_s}\right)}{\partial t^2}-\nabla\cdot \overline{\boldsymbol{\sigma}}-\frac{1}{V}\int_{A_{s,w}}{{\boldsymbol{\sigma} \cdot \boldsymbol{n}}_{s,i}dA}-\frac{1}{V}\int_{A_{s,n}}{\boldsymbol{\sigma}{ \cdot \boldsymbol{n}}_{s,n}dA}-\nabla\cdot {(\phi}_s{\overline{\boldsymbol{\tau}}}^s) \\
-\frac{1}{V}\int_{A_{s,w}}{\left(\overline{{\boldsymbol{\tau}}^{\boldsymbol{s}}}+\widetilde{\boldsymbol{\tau}}\right){ \cdot \boldsymbol{n}}_{s,w}dA}-\frac{1}{V}\int_{A_{s,n}}{(\overline{{\boldsymbol{\tau}}^{\boldsymbol{s}}}+\widetilde{\boldsymbol{\tau}}){ \cdot \boldsymbol{n}}_{s,n}dA}=0
\end{split}
\end{equation} 

Assuming that the intrinsic phase averages of the stress tensors are constant along the integration surfaces and using the geometric relation between said surfaces and the porosity gradient (Equation \ref{Eq:porosity_grad}), we obtain:

\begin{multline}\frac{{\partial }^2\left({\phi}_s{\rho}_s\overline{{\boldsymbol{d}}^s_s}\right)}{\partial t^2}-\nabla\cdot \overline{\boldsymbol{\sigma}}-\nabla \cdot {(\phi}_s\overline{{\boldsymbol{\tau}}^{\boldsymbol{s}}})+ {\overline{{\boldsymbol{\tau}}^{\boldsymbol{s}}}\cdot\nabla\phi}_s - \\ \frac{1}{V}\int_{A_{s,w}}{{\left(\boldsymbol{\sigma}+\widetilde{\boldsymbol{\tau}}\right) \cdot \boldsymbol{n}}_{s,w}dA}- \frac{1}{V}\int_{A_{s,n}}{(\boldsymbol{\sigma}+\widetilde{\boldsymbol{\tau}}){ \cdot \boldsymbol{n}}_{s,n}dA}=0\end{multline} 

This expression can be simplified into the following by the use of the chain rule,

\begin{multline}\frac{{\partial }^2\left({\phi}_s{\rho}_s\overline{{\boldsymbol{d}}^s_s}\right)}{\partial t^2}-\nabla \cdot \overline{\boldsymbol{\sigma}}-{\phi}_s\nabla \cdot \overline{{\boldsymbol{\tau}}^{\boldsymbol{s}}}- \\ \frac{1}{V}\int_{A_{s,w}}{{\left(\boldsymbol{\sigma}+\widetilde{\boldsymbol{\tau}}\right) \cdot \boldsymbol{n}}_{s,w}dA}-\frac{1}{V}\int_{A_{s,n}}{(\boldsymbol{\sigma}+\widetilde{\boldsymbol{\tau}}){ \cdot \boldsymbol{n}}_{s,n}dA}=0\end{multline} 

Finally we represent the integral terms as two separate interaction terms defined as ${\boldsymbol{B}}_{s,i}=\frac{1}{V}\int_{A_{s,i}}{{\left(\boldsymbol{\sigma}+\widetilde{\boldsymbol{\tau}}\right) \cdot \boldsymbol{n}}_{s,i}dA}$:

\begin{equation}\frac{{\partial }^2\left({\phi}_s{\rho}_s\overline{{\boldsymbol{d}}^s_s}\right)}{\partial t^2}-\nabla \cdot \overline{\boldsymbol{\sigma}}={\phi}_s\nabla \cdot \overline{{\boldsymbol{\tau}}^{\boldsymbol{s}}}+{{\phi}_s\rho}_s\boldsymbol{g}+{\boldsymbol{B}}_{s,w}+{\boldsymbol{B}}_{s,n}\label{elasticEqn}\end{equation} 

In the case of a linear elastic solid the stress tensor is a function of the two lame coefficients $({\mu}_s,\ {\lambda}_s)$ and takes the following form: 

\begin{equation}\overline{\boldsymbol{\sigma}}={\phi}_s\mu_s\nabla \overline{{\boldsymbol{d}}^s_s}+{{\phi}_s{\mu}_s\left(\nabla \overline{{\boldsymbol{d}}^s_s}\right)}^T+{\phi}_s\lambda_str(\nabla \overline{{\boldsymbol{d}}^s_s})\boldsymbol{I}\end{equation} 

\section{Momentum Conservation Equation for a Viscoplastic Solid} \label{sect:viscoplastic_eqn}

The elementary momentum conservation equation for a plastic solid can be written as:
\begin{equation}\frac{\partial {\rho}_s{\boldsymbol{U}}_s}{\partial t}+\nabla\cdot\left({\rho}_s{\boldsymbol{U}}_s{\boldsymbol{U}}_s\right)-\nabla\cdot\boldsymbol{\sigma}=\nabla\cdot\boldsymbol{\tau}+{\rho}_s\boldsymbol{g}\end{equation}

\noindent where $\boldsymbol{\sigma}$ represents the plastic viscous stress tensor and $\boldsymbol{\tau}$ is the Terzaghi effective stress tensor. Note how in this case, the solid is essentially a viscous fluid, albeit a non-Newtonian one. Fortunately, the averaging procedure for each term within this equation has already been shown in Sections \ref{Sect:FluidMomeqn} and \ref{Sect:elastic}. Putting it all together, the averaged momentum equation for a plastic solid is as follows:

\begin{equation}\frac{\partial {{\phi}_s\rho}_s{\overline{\boldsymbol{U}}}^s_s}{\partial t}+\nabla\cdot\left({\phi}_s{\rho}_s{\overline{\boldsymbol{U}}}^s_s{\overline{\boldsymbol{U}}}^s_s\right)-\nabla\cdot\overline{\boldsymbol{\sigma}}={\phi}_s\nabla\cdot{\overline{\boldsymbol{\tau}}}^s+{{\phi}_s\rho}_s\boldsymbol{g}+{\boldsymbol{B}}_{s,w}+{\boldsymbol{B}}_{s,n}\label{plasticEq} \end{equation} 

For a viscoplastic solid the stress tensor is a function of the effective plastic viscosity  $({\mu}^{eff}_s)$ and takes the following form:

\begin{equation}\overline{\boldsymbol{\sigma}}={\phi}_s\ {\mu}^{eff}_s\left(\ \nabla {\overline{\boldsymbol{U}}}^s_s+{\ \left(\nabla {\overline{\boldsymbol{U}}}^s_s\right)}^T-\frac{2}{3}\nabla \cdot \left({\overline{\boldsymbol{U}}}^s_s\boldsymbol{I}\right)\right)\end {equation} 

\section{General Solid Momentum Conservation Equation} \label{Sect:generalSolidEqn}

If we assume low solid velocities ($\mathrm{Re}_s\lesssim 1)$ we can reasonably neglect the inertial terms and time derivatives in both the elastic and plastic solid momentum equations (Eqns. \ref{plasticEq} and \ref{elasticEqn} in order to arrive to a general equation for solid mechanics:

\begin{equation}\label{eq:avg_solid_mech}-\nabla\cdot\overline{\boldsymbol{\sigma}}={\phi}_s\nabla\cdot{\overline{\boldsymbol{\tau}}}^s+{{\phi}_s\rho}_s\boldsymbol{g}+{\boldsymbol{B}}_{s,w}+{\boldsymbol{B}}_{s,n}\end{equation} 

\noindent where ${\overline{\boldsymbol{\sigma}}}^s_s$\textbf{ }is either the solid's elastic or viscoplastic stress tensor. The only thing left to do is to close the ${\boldsymbol{B}}_{s,i}$ interaction terms. This was done in \citet{Carrillo2020} and shown in the next section (Section \ref{App:Biot_Terms}), where: 

\begin{equation}{\boldsymbol{B}}_{s,w}+{\boldsymbol{B}}_{s,n} = {\phi}_f\mu k^{-1}\left({\boldsymbol{U}}_f-{\overline{\boldsymbol{U}}}_s\right)-{\phi}_f{\boldsymbol{F}}_{c,1}+{\phi}_s{\boldsymbol{F}}_{c,2}\end{equation} 

\noindent here, $\mu k^{-1}$ is the single-field mobility of the fluids (i.e. a representation for drag) and ${\boldsymbol{F}}_{c,i}$ represents the capillary forces arising from fluid-fluid interfacial dynamics within the porous medium. Putting everything together we now obtain the averaged momentum equation for a solid containing two immiscible phases:
\begin{equation}\label{complete_solid}-\nabla\cdot\overline{\boldsymbol{\sigma}}={\phi}_s\nabla\cdot{\overline{\boldsymbol{\tau}}}^s+{{\phi}_s\rho}_s\boldsymbol{g}+{\phi}_f\mu k^{-1}\left({\boldsymbol{U}}_f-{\overline{\boldsymbol{U}}}_s\right)-{\phi}_f{\boldsymbol{F}}_{c,1}+{\phi}_s{\boldsymbol{F}}_{c,2}\end{equation}

\section{Closure of the Solid's Multiscale Interaction Terms} \label{App:Biot_Terms}

Just as we did for the fluid equations in Section \ref{App:Multiphase_Terms}, we will assume that the sum of the averaged stresses at the solid-fluid interface can be expressed as the sum of two independent terms: a drag force that captures shear-induced momentum exchange $\left({\boldsymbol{B}}_{drag}\right)$ and a capillary force originating from capillary pressure jumps across the integrated solid surfaces within the porous media $\left({\boldsymbol{B}}_{cap}\right)$. 

\begin{equation}
\begin{aligned}
{\boldsymbol{B}}_{drag}+{\boldsymbol{B}}_{cap}={\boldsymbol{B}}_{s,w}+{\boldsymbol{B}}_{s,n}
\end{aligned}
\end{equation}

We now seek closure of these two coupling terms. By conservation of momentum, we know that any drag-induced momentum lost by the fluid must be gained by the solid. Thus, by Eqn. \ref{eq:complete_fluid},

\begin{equation}
\begin{aligned}
{\boldsymbol{B}}_{drag}={\phi }_f\mu k^{-1}\left({\boldsymbol{U}}_f-{\overline{\boldsymbol{U}}}_s\right)
\end{aligned}
\end{equation}

Closure of the capillarity-induced interaction term ${\boldsymbol{B}}_{cap}$ is obtained by adding the solid and fluid momentum equations (Eqns. \ref{eq:complete_fluid} and \ref{eq:avg_solid_mech}) within the porous medium at low Reynold numbers and low permeability, which yields:
\begin{equation*}0=-{\phi}_f\nabla p+{\phi}_f{\rho}_f\boldsymbol{g}-{\phi}_f\mu k^{-1}\left({\boldsymbol{U}}_f-{\overline{\boldsymbol{U}}}_s\right)+{\phi}_f{\boldsymbol{F}}_{c,1}+{\phi}_f{\boldsymbol{F}}_{c,2}\ \end{equation*} 
\begin{equation*}+\end{equation*} 
\begin{equation*}-\nabla\cdot\overline{\boldsymbol{\sigma}}={\phi}_s\nabla\cdot{\overline{\boldsymbol{\tau}}}^s+{{\phi}_s\rho}_s\boldsymbol{g}+{\phi}_f\mu k^{-1}\left({\boldsymbol{U}}_f-{\overline{\boldsymbol{U}}}_s\right)+{\boldsymbol{B}}_{cap}\end{equation*}
\begin{equation*}=\end{equation*} 
\begin{equation}\label{added}
-\nabla\cdot\overline{\boldsymbol{\sigma }}={\phi }_s\nabla\cdot\overline{{\boldsymbol{\tau }}^{\boldsymbol{s}}}-{\phi }_f\nabla p+\left({{\phi }_s\rho }_s+{\phi }_f{\rho }_f\right)\boldsymbol{g}+{\phi }_f{\boldsymbol{F}}_{c,1}+{\phi }_f{\boldsymbol{F}}_{c,2}+{\boldsymbol{B}}_{cap}\boldsymbol{\ }
\end{equation}

In multiphase porous systems with incompressible grains and no swelling pressure (i.e. $\nabla\cdot\overline{{\boldsymbol{\tau }}^{\boldsymbol{s}}}=-\nabla p)$, Biot Theory states that $\nabla\cdot\overline{\boldsymbol{\sigma }}=\nabla p-{\rho }^*\boldsymbol{g}+p_c\nabla{\alpha }_w$, where ${\rho }^*=({{\phi }_s\rho }_s+{\phi}_f{\rho}_f)$ \citep{Jha2014,Kim2013}. This expression is satisfied by the previous equation in the absence of capillary forces \citep{Carrillo2019}. In multiphase systems, however, it imposes the following equality,

\begin{equation}
\begin{aligned}
{\boldsymbol{B}}_{cap}=-({\phi }_f{\boldsymbol{F}}_{c,1}+{\phi }_f{\boldsymbol{F}}_{c,2}+p_c\nabla{\alpha }_w) 
\end{aligned}
\end{equation}

Given that ${\boldsymbol{F}}_{c,1} = M^{-1}\left(M_w{\alpha}_n-M_n{\alpha}_w\right)\left(\nabla p_c+\left({\rho}_w-{\rho}_n\right)\boldsymbol{g}\right)$ and that ${\boldsymbol{F}}_{c,2} = -p_c\nabla{\alpha }_w$ (see Eqns. \ref{Fc1_def} and \ref{Fc2_def}), the previous equation can be rearranged to obtain:

\begin{equation}\label{Eq:fc_closure}
\begin{aligned}
{\boldsymbol{B}}_{cap}=-{\phi }_f{\boldsymbol{F}}_{c,1}+{\phi }_s{\boldsymbol{F}}_{c,2}\  
\end{aligned}
\end{equation}

Equation \ref{Eq:fc_closure} gives closure to the last coupling parameter and marks the end of this derivation. The result is the solid momentum conservation equation shown in Sections \ref{Sect:generalSolidEqn} and \ref{Sect: Parameter Closure}.

\section{Interfacial Conditions between Solid-Free Regions and Porous Regions} \label{sec:interfacial_condition}

One of the most important features allowed by the equations presented above is the existence of an interface between solid-free and microporous domains. Although the creation of a rigorous un-averaged description of this interface is still an open question, we approximate a solution to it by guaranteeing its necessary components within our fluid and solid averaged equations.  

An accurate description of fluid behavior at the interface requires three components: 1) mass conservation across the interface, 2) continuity of stresses across the interface, and 3) an interfacial wettability condition. Components 1 and 2 are intrinsically fulfilled by our solver due to its single-field formulation for velocity and pressure within the fluid conservation equations (Eqns. \ref{Eq:finalFluidMass} and \ref{Eq:Final_fluid}). As shown in \citet{Neale1974} and \citet{Carrillo2019} these two components are necessary and sufficient to model single-phase flow within a multiscale system. Furthermore, these conditions have also been used for closure when modelling multiphase flow in moving porous media \citep{Lacis2017,Zampogna2019,Carrillo2020}. The required wettability condition at the porous interface (Component 3) is included in our model through the implementation of a penalized contact angle condition (Eqn. \ref{nw_def}) following the steps outlined in \citet{Horgue2014} and \citet{Carrillo2020}.

The complementary solid conditions at the porous interface are very similar: 1) solid mass conservation across the interface, 2) continuity of fluid-induced stresses across the interface, and 3) a discontinuity of solid stresses at the interface. Just as before, the first two conditions are intrinsically fulfilled through the use of a single set of mass and momentum conservation equations across both domains and have also been used as closure conditions in previous studies \citep{Lacis2017,Zampogna2019}. The third condition is enforced by the use of volume-averaged solid rheology models that tend towards infinitely deformable materials in solid-free regions, as shown in \citet{Carrillo2019}. When volume-averaged, the behavior of the solid's stress tensor is domain dependent (i.e. solid fraction dependent). Thus, in solid regions, the elasticity and viscosity of the porous medium is determined by standard averaged rheological properties (the elastic and viscoplastic moduli). Contrastingly, in solid-free regions, the solid fraction tends to zero and, as such, said properties do as well. The result is a stress-free ``ghost" solid that does not apply resistance to the porous region, creating the required stress discontinuity at the porous interface. 

Although necessary, these conditions represent but an approximation to the complete description of fluid and solid mechanics at the porous interface. However, to the best of our knowledge, there does not exist an alternative set of boundary conditions that can or have been used to model multiphase flow in multiscale porous media. 

\section{Advection-Diffusion Equation for Tracer Particles}

The derivation of the following advection-diffusion equation for tracer particles flowing within a single fluid follows the same averaging procedure as all the previous mass conservation equations. 

\begin{equation} \label{eq:11} 
\frac{\partial \left({\phi }_f{\overline{C}}_{i,x}^i\right)}{\partial t}+\nabla \left({\overline{C}}_{i,x}^i{\boldsymbol{U}}_{f}\right)-\nabla\cdot \left({\phi }_{f\ }D_{\mathrm{eff,}x}\nabla{\overline{C}}_{i,x}^i\right)=0 
\end{equation} 

\noindent where ${\overline{C}}_{i,x}^i$ and $D_{\mathrm{eff,}x}$ are the molar concentration and effective diffusion coefficient of a given species \textit{x}, respectively. A reaction term can readily be added to Eqn. \ref{eq:11} to describe reactive solutes \citep{Soulaine2017}. The expansion of Eqn. \ref{eq:11} into multiphase flow is beyond the scope of our investigation, but can be found in \citet{Maes2020}.

\section{Summary of Equations and Multiscale Parameters} \label{Sect: Parameter Closure}

The final set of conservation equations used in this framework now follows. The combination of these solid and fluid conservation equations leads to a model that tends towards multiphase Navier-Stokes in solid-free regions and towards Biot Theory in porous regions (see Appendix \ref{App:Biot_Theory}).

\begin{equation}\label{Eq:finalFluidMass}\frac{\partial {\phi}_f}{\partial t}+\nabla \cdot{\boldsymbol{U}}_f=0\end{equation} 

\begin{equation}\label{Eq:finalSaturation}\frac{\partial {\phi}_f{\alpha}_w}{\partial t}+\nabla \cdot\left({\alpha}_w{\boldsymbol{U}}_f\right)+\nabla \cdot\left({\phi}_f{\alpha}_w{\alpha}_n{\boldsymbol{U}}_r\right)=0\end{equation} 

\begin{multline}
\label{Eq:Final_fluid}\frac{\partial {\rho}_f{\boldsymbol{U}}_f}{\partial t}+\nabla \cdot\left(\frac{{\rho}_f}{{\phi}_f}{\boldsymbol{U}}_f{\boldsymbol{U}}_f\right)=-{\phi}_f\nabla p+{\phi}_f{\rho}_f\boldsymbol{g}+\nabla\cdot\overline{\boldsymbol{S}}- \\ {\phi}_f\mu k^{-1}\left({\boldsymbol{U}}_f-{\overline{\boldsymbol{U}}}_s\right)+{\phi}_f{\boldsymbol{F}}_{c,1}+{\phi}_f{\boldsymbol{F}}_{c,2}
\end{multline} 

\begin{equation}\label{Eq:finalSolidMass}\frac{\partial {\phi}_s}{\partial t}+\nabla \cdot\left({\phi}_s{\overline{\boldsymbol{U}}}^s_s\right)=0\end{equation} 

\begin{equation}\label{Eq:Final_Solid}-\nabla\cdot\overline{\boldsymbol{\sigma}}={\phi}_s\nabla \cdot \overline{{\boldsymbol{\tau}}^{\boldsymbol{s}}}+{{\phi}_s\rho}_s\boldsymbol{g}+{\phi}_f\mu k^{-1}\left({\boldsymbol{U}}_f-{\overline{\boldsymbol{U}}}_s\right)-{\phi}_f{\boldsymbol{F}}_{c,1}+{\phi}_s{\boldsymbol{F}}_{c,2}\end{equation} 

The closed-form expressions of the multi-scale parameters $\mu k^{-1} \mathrm{, \ } \boldsymbol{U}_r \mathrm{\ ,and}\ {\boldsymbol{F}}_{c,i}$, which are defined differently in each region now follow. A full derivation and discussion of these parameters can be found in Sections \ref{App:Multiphase_Terms} and \ref{App:Biot_Terms} \citep{Carrillo2020}. 

\begin{equation}\label{uk_def}
\mu k^{-1}=
    \begin{cases}
        0 & \textnormal{in solid-free regions}\\
        k^{-1}_0{\left(\frac{k_{r,w}}{{\mu }_w}+\frac{k_{r,n}}{{\mu }_n}\right)}^{-1} &  \textnormal{in porous regions}
    \end{cases}
\end{equation}

\begin{equation}\label{Fc1_def}
{\boldsymbol{F}}_{c,1}=
    \begin{cases}
        -\frac{\gamma}{{\phi}_f}\nabla\cdot \left({\boldsymbol{n}}_{w,n}\right)\nabla{\alpha}_{w} & \textnormal{in solid-free regions}\\
        M^{-1}\left(M_w{\alpha}_n-M_n{\alpha}_w\right)\left(\nabla p_c+\left({\rho}_w-{\rho}_n\right)\boldsymbol{g}\right) &  \textnormal{in porous regions}
    \end{cases}
\end{equation}

\begin{equation}\label{Fc2_def}
{\boldsymbol{F}}_{c,2}=
    \begin{cases}
        0 & \textnormal{in solid-free regions}\\
        -p_c\nabla{\alpha }_w &  \textnormal{in porous regions}
    \end{cases}
\end{equation}

\begin{equation}\label{nw_def}
{\boldsymbol{n}}_{w,n}=
    \begin{cases}
        \frac{\nabla{\alpha}_{w}}{\left|\nabla{\alpha }_{w}\right|} & \textnormal{in solid-free regions}\\
        {cos \left(\theta \right)\ }{\boldsymbol{n}}_{wall}+{sin \left(\theta \right)\ }{\boldsymbol{t}}_{wall} &  \textnormal{at the interface between solid-free porous regions}
    \end{cases}
\end{equation}

\begin{equation}\label{Ur_def}
{\boldsymbol{U}}_{r}=
    \begin{cases}
        C_{\alpha \ }max\left(\left|{\boldsymbol{U}}_f\right|\right)\frac{\nabla{\alpha }_w}{\left|\nabla{\alpha }_w\right|} & \textnormal{in solid-free regions}\\
        {\phi }^{-1}\left[ \begin{array}{c}
-\left(M_w{\alpha }^{-1}_w-M_n{\alpha }^{-1}_n\right)\nabla p + \\ \left({\rho }_wM_w{\alpha }^{-1}_w-{\rho }_wM_n{\alpha }^{-1}_n\right)\boldsymbol{g}+ \\ 
\left(M_w{\alpha }_n{\alpha }^{-1}_w +M_n{\alpha }_w{\alpha }^{-1}_n\right)\nabla p_c- \\ \left(M_w{\alpha }^{-1}_w- M_n{\alpha }^{-1}_n\right)p_{c}\nabla{\alpha }_w \end{array}\right] &  \textnormal{in porous regions}
\end{cases}
\end{equation}

\noindent where $\gamma $ is the surface tension, $C_{\alpha}$ is an interface compression parameter (traditionally set to values between 1-4 in the Volume-of-Fluid method), $k_0$ is the absolute permeability, $k_{r,i}$ are the fluids' relative permeabilities, $M_i=k_{i,r}/{\mu}_i\ $are the fluids' mobilities, $M=M_w+M_n$ is the single-field mobility, and $p_c$ is the average capillary pressure between the two fluids in a given control volume. The phase-specific parameters can be readily calculated from closed-form relative permeability and capillary pressure models such as the Brooks-Corey \citep{RHBrooks1964} and Van Genutchen \citep{VanGenutchen1980} models. Lastly, $\theta $ is the imposed contact angle at the porous wall, and $\boldsymbol{n}_{wall}$  and $\boldsymbol{t}_{wall}$ are the normal and tangential directions relative to said wall, respectively. For the reader's convenience, a full implementation of this model, complete with rheological, relative permeability, and capillary pressure models, is included within the \href{https://github.com/Franjcf/hybridPorousInterFoam}{\textit{hybridPorousInterFoam}} (for static porous media) and \href{https://github.com/Franjcf/hybridBiotInterFoam}{\textit{hybridBiotInterFoam}} (for deformable porous media) open-source solvers. 

\section{Conclusion}

A complete set of partial differential equations required to model multiphase flow in multiscale deformable porous media is presented above. In the following chapters, we focus on the numerical implementation of these equations and their application to a wide range of different natural and engineered systems. As such, for the rest of this dissertation, whenever any partial differential equation is mentioned, we will be referring to the ones described in this chapter. The main model limitations stemming from the assumptions presented in this chapter are summarized and discussed in Chapter \ref{conclusion}.

\nomenclature{$\rho_i$}{Density of phase $i$}%
\nomenclature{$\rho_f$}{Single-field fluid density}%
\nomenclature{$\rho^*$}{Biot density}%
\nomenclature{$\overline{\boldsymbol{U}_i}$}{Superficial velocity of phase $i$ in the grid-based domain}%
\nomenclature{$\overline{\boldsymbol{U}_i^i}$}{Phase-averaged velocity of phase $i$ in the grid-based domain}%
\nomenclature{$\boldsymbol{U}_f$}{Single-field fluid velocity in the grid-based domain}%
\nomenclature{$\boldsymbol{U}_r$}{Relative velocity in the grid-based domain}%
\nomenclature{$\boldsymbol{U}_s$}{Solid velocity in the grid-based domain}%
\nomenclature{$\boldsymbol{v}_{i,j}$}{Velocity of the i-j interface in the continuous physical space}%
\nomenclature{$V$}{Volume of the averaging-volume}%
\nomenclature{$V_i$}{Volume of phase $i$ in the averaging-volume}%
\nomenclature{$A_{i,j}$}{Interfacial area between phase $i$ and $j$}%
\nomenclature{$\boldsymbol{n}_{i,j}$}{Normal vector to the i-j interface in the continuous physical space}%
\nomenclature{$p$}{Single-field fluid pressure in the grid-based domain}%
\nomenclature{$p_c$}{Capillary pressure}%
\nomenclature{$\boldsymbol{S}$}{Single-field fluid viscous stress tensor in the grid-based domain}%
\nomenclature{$\boldsymbol{\sigma}$}{Elastic (or plastic) solid stress tensor in the grid-based domain}%
\nomenclature{$\boldsymbol{\tau}$}{Terzaghi stress tensor in the grid-based domain}%
\nomenclature{$\gamma$}{Interfacial tension}%
\nomenclature{$I$}{Identity matrix}%
\nomenclature{$\phi_f$}{Porosity field}%
\nomenclature{$\phi_s$}{Solid fraction field}%
\nomenclature{$\alpha_w$}{Saturation of the wetting phase}%
\nomenclature{$\alpha_n$}{Saturation of the non-wetting phase}%
\nomenclature{$\boldsymbol{D}_{i,k}$}{Drag force exerted by phase $k$ on phase $i$}%
\nomenclature{$\boldsymbol{B}_{i,k}$}{Drag force exerted by phase $k$ on phase $i$}%
\nomenclature{$\mu_i$}{Viscosity of phase $i$ }%
\nomenclature{$\mu_f$}{Single-field viscosity }%
\nomenclature{$k$}{Apparent permeability}%
\nomenclature{$\boldsymbol{g}$}{Gravity vector}%
\nomenclature{$\boldsymbol{F}_{c,i}$}{Surface tension force in the grid-based domain}%
\nomenclature{$C_{\alpha}$}{Parameter for the compression velocity model}%
\nomenclature{$k_0$}{Absolute permeability}%
\nomenclature{$k_{r,i}$}{Relative permeability with respect to phase $i$ }%
\nomenclature{$M_i$}{Mobility of phase $i$}%
\nomenclature{$M$}{Total mobility}%
\nomenclature{$\theta$}{Surface contact angle}%
\nomenclature{$\boldsymbol{n}_{wall}$}{Normal vector to the porous surface}%
\nomenclature{$\boldsymbol{t}_{wall}$}{Tangent vector to the porous surface}%
\nomenclature{$m$}{Mass of fluid per control volume}%
\nomenclature{$b$}{Biot coefficient}%
\nomenclature{$\epsilon$}{Volumetric strain}%
\nomenclature{$\mu_s$}{First Lame coefficient}%
\nomenclature{$\lambda_s$}{Second Lame coefficient}%
\nomenclature{$\boldsymbol{d}_s$}{Solid displacement}%
\nomenclature{$\mu_s^{eff}$}{Effective plastic viscocity}%

\printnomenclature 
\begin{savequote}[75mm]
Computers are useless. They can only give you answers.
\qauthor{Pablo Picasso}
\end{savequote}

\chapter{Numerical Implementation}\label{numerical_impl}

\newthought{Having rigorously derived} the governing equations for multiphase flow in deformable porous media in Chapter \ref{FullDerivation}, we now turn to the task of implementing them into a computational framework apt for performing numerical simulations. This chapter will provide the computational tools required to perform all the simulations shown in subsequent chapters. The work presented here is adapted from \cite{Carrillo2019}, \cite{Carrillo2020}, and \cite{Carrillo2020MDBB}.

\section{Numerical Platform}

The implementation of the multiphase DBB model was done in OpenFOAM{\circledR}, a free, open-source, parallelizable, and widely used computational fluid mechanics platform. This C++ code uses the Finite Volume Method to discretize and solve partial differential equations in complex 3-D structured and unstructured grids. Its object-oriented structure and multitude of supporting libraries allows the user to easily customize each simulation's setup with different numerical discretization schemes, time-stepping procedures, matrix-solution algorithms, and supporting physical models. The implementation described below stems directly from OpenFOAM's{\circledR} standard two-phase incompressible flow solver ``\textit{interFoam}''.

\section{Solution Algorithm}

The solution of the governing equations is done in a sequential manner, starting with the fluid mechanics equations and following with the solid mechanics equations for every time step. Of particular importance is the handling and modification of the velocity-pressure coupling required for modeling incompressible fluids in conjunction with a moving solid matrix. For this step, we based our solution algorithm on the Pressure Implicit Splitting-Operator (PISO) \citep{Issa1986}. First, we explicitly solve the fluid saturation equation (Eqn. \ref{Eq:finalSaturation}) for ${\alpha }^{t+1}_w$ through the Multidimensional Universal Limiter of Explicit Solution (MULES) algorithm \citep{Marquez2013}. This allows for stable numerical advection of the saturation field by the application of Flux Corrected Transport Theory \citep{Rudman1997}. Then, we update the boundary values of ${\boldsymbol{U}}_f$ and ${\boldsymbol{U}}_r$ in addition to the cell-centered values of the permeability $k^{t+1}$, density ${\rho }_f^{t+1}$, and viscosity ${\mu_f}^{t+1}$ based on the newly calculated saturation field ${\alpha }^{t+1}_w$. The capillary forces ${\boldsymbol{F}}^{t+1}_{c,i}$ are also updated accordingly. After that, a preliminary value of the fluid velocity ${\boldsymbol{U}}^*_f$ is calculated by implicitly solving the algebraically discretized form of the fluid momentum equation used in the Finite Volume Method.

\begin{equation}\label{1}
\begin{aligned}
a_p{\boldsymbol{U}}^*_f=\boldsymbol{H}\left({\boldsymbol{U}}^*_f\right)+{\rho }^{t+1}_f\boldsymbol{g}+{\boldsymbol{F}}^{t+1}_{c,i}-\nabla p^t
\end{aligned}
\end{equation}

\noindent where $\boldsymbol{H}\left({\boldsymbol{U}}^*_f\right)$ contains inertial, convective, viscous, and drag source terms originating from neighboring cells and $a_p$ represents these same terms but at the volume of interest. Note that the ${\boldsymbol{U}}^*_f$ field does not follow mass conservation. To account for this, we use the fluid continuity equation (Eqn. \ref{Eq:finalFluidMass}) in conjunction with the previous equation (Eqn. \ref{1}) to update the velocity field ${\boldsymbol{U}}^{**}_f$ and calculate a preliminary mass-conservative pressure field $p^*$. In other words, these fields must satisfy,

\begin{equation}\label{discretized}
\begin{aligned}
{\boldsymbol{U}}^{**}_f=\frac{1}{a_p}\left(\boldsymbol{H}\left({\boldsymbol{U}}^*_f\right)+{\rho }^{t+1}_f\boldsymbol{g}+{\boldsymbol{F}}^{t+1}_{c,i}-\nabla p^*\right) 
\end{aligned}
\end{equation}

\begin{equation}\label{divergence}
\begin{aligned}
\nabla\cdot {\boldsymbol{U}}^{**}_f=-\ \frac{\partial {\phi }_f}{\partial t} 
\end{aligned}
\end{equation}

These equations can be recast into a single coupled equation which is then used to implicitly solve for pressure. This step can be done through several generalized matrix solvers that are standard in OpenFOAM{\circledR}. 

\begin{equation}\label{coupled_eqn}
\begin{aligned}
\nabla\cdot \left(\frac{1}{a_p}\left(\boldsymbol{H}\left({\boldsymbol{U}}^*_f\right)+{\rho }^{t+1}_f\boldsymbol{g}+{\boldsymbol{F}}^{t+1}_{c,i}\right)\right)-\nabla\cdot \left(\frac{1}{a_p}\nabla p^*\right)=-\ \frac{\partial {\phi }_f}{\partial t}
\end{aligned}
\end{equation}

After solving for pressure $p^*$, velocity can be re-calculated from Equation \ref{discretized}. This semi-implicit pressure-velocity correction step is repeated until the desired convergence is reached. It has been shown that at least two pressure-velocity correction loops are required to ensure mass conservation \citep{Issa1986}. 

At this point ${\boldsymbol{U}}^{t+1}_f$ and $p^{t+1}$ are set and used as input values for updating the drag and pressure source terms present in the solid mechanics momentum equation (Eqn. \ref{Eq:Final_Solid}). Then, in the case of visco-poro-plasticity, said equation is discretized in a similar way as the fluid momentum equation (Eqn. \ref{Eq:Final_fluid}) and used to implicitly solve for ${\boldsymbol{U}}^{t+1}_s$. In the case of poroelasticity, the solid mechanics equation is solved through the algorithm presented in \citet{Jasak2000}. Here, the solid's elastic equation (Eqn. \ref{elasticEqn}) is discretized and segregated into implicit and explicit components, after which it is iteratively solved until convergence is reached. This segregated method not only guarantees fast convergence but also memory efficiency. Finally, the updated solid velocity is used to ``advect'' the solid fraction field ${\phi }_s$ by solving the mass conservation equation (Eqn. \ref{Eq:finalSolidMass}). At this point the algorithm advances in time according to the imposed Courant-Friedrichs-Lewy (CFL) number. Further discussion regarding the discretization techniques and matrix-solution procedures can be found in \citet{Carrillo2019}, \citet{Carrillo2020}, \citet{Carrillo2020MDBB}, \citet{Jasak1996}, and \citet{Jasak2000}.

\section{Open-Source Implementation}

The complete set of governing equations and solution algorithms, along with the necessary rheology, relative permeability, and capillary pressure models (Appendix \ref{multiphase_models} and \ref{rheology_mod}) were implemented into the solvers \href{https://github.com/Franjcf}{\textit {hybridPorousInterFoam}} (for static porous media) and \href{https://github.com/Franjcf}{\textit{ thybridBiotInterFoam}} (for deformable porous media). These solvers, along with representative tutorial cases, automated compilation and running procedures, and all the simulated cases presented in this dissertation were incorporated into open-source CFD packages of the same names. OpenFOAM{\circledR} and our code are free to use under the GNU general public license and can be found at \url{https://openfoam.org/} and \url{https://github.com/Franjcf} \citep{Carrillo2020code,hybridBiotInterFoam_Code}, respectively. 

\begin{savequote}[75mm]
In God we trust, all others bring data.
\qauthor{William Edwards Deming}
\end{savequote}

\chapter{Applications to Single-Phase Flow in Multiscale Deformable Porous Media}\label{chp:singlePhaseDBB}

\newthought{In this chapter} we apply the equations derived in Chapter \ref{FullDerivation} and implemented in Chapter \ref{numerical_impl} to model single phase flow through deformable porous media. Special emphasis is given to the simulation of of swelling clays and their influence on the permeability of sedimentary rocks. The work presented in this chapter represents the first step towards verifying and showcasing the Multiphase DBB model. This chapter is adapted from \citet{Carrillo2019}.

\section{Introduction}

How does the permeability of sedimentary media depend on porosity, mineralogy, fluid chemistry, and stress history? This question has been a recurrent theme in subsurface hydrology for over half a century \citep{Berg1970,Bourg2017a,Brace1980}. It impacts a range of endeavors that shape humanity's energy landscape including hydrocarbon migration and recovery \citep{Alvarado2010}, geothermal energy production \citep{Barbier1997}, geologic carbon sequestration \citep{Klaus2003}, and radioactive waste storage \citep{Sellin2013}. The last two technologies have the potential to contribute up to half of the mitigation effort required to stabilize global CO${}_{2}$ emissions \citep{Metz2005,Socolow2004} but require the ability to accurately predict the permeability evolution of ductile fine-grained sedimentary rocks (shale, mudstone) over millennial time-scales in the presence of geochemical and geomechanical disturbances \citep{Bourg2017a,Neuzil2013}. 

\begin{figure} 
\begin{center}
\includegraphics[width=0.95\textwidth]{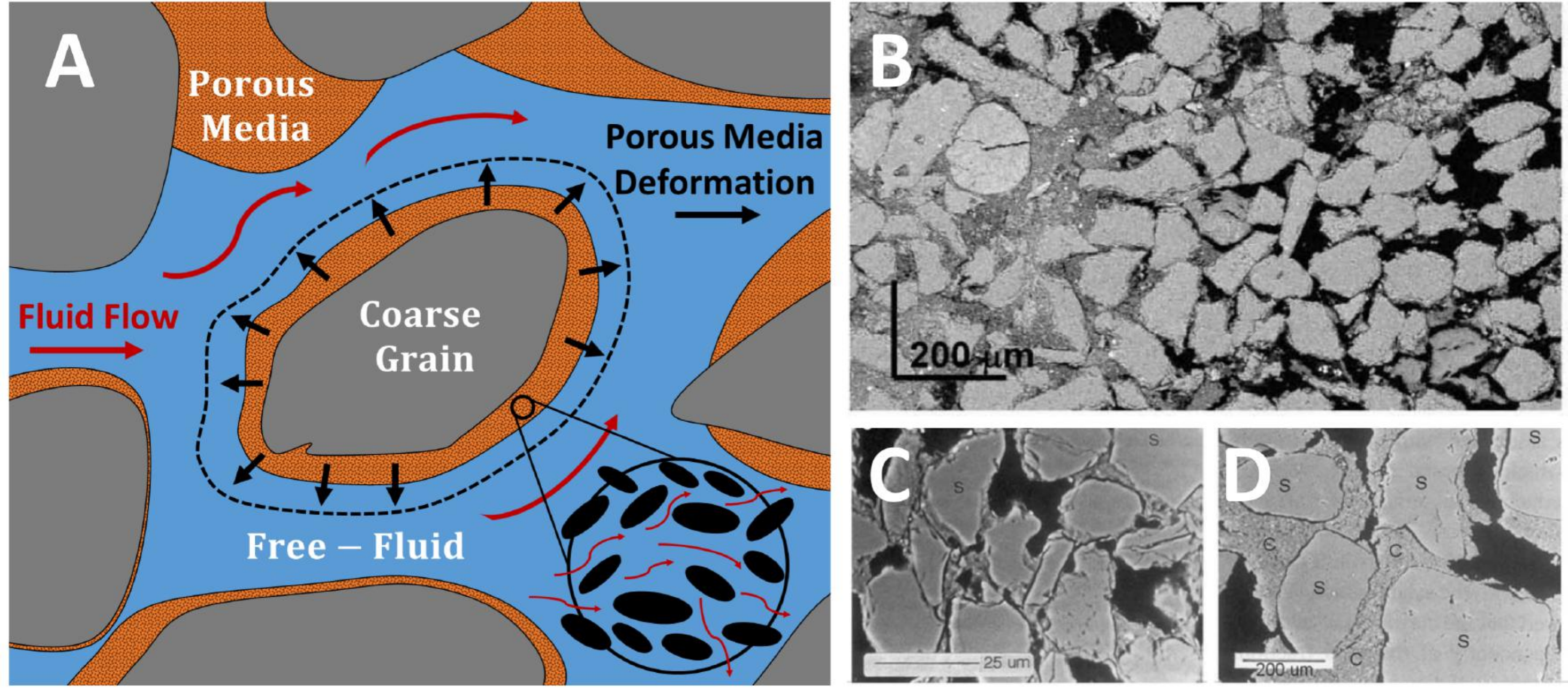}
\caption[Conceptual representation of the two key challenges posed by fluid flow in clay-rich sedimentary rocks]{(a) Conceptual representation of the two key challenges posed by fluid flow in clay-rich sedimentary rocks. The structure contains three regions: rigid coarse grains, macropores, and a deformable microporous clay matrix. Fluid flow (red arrows) occurs primarily in the bulk-fluid domain, but the boundaries of this domain are influenced by the deformation of the microporous clay domain (black arrows). Figures b-d are electron microscopy images of clayey media, specifically (b) a Canadian shale and (c,d) a mixture of clay and sand (denoted by the letters c and s, respectively) associated with reductive and expansive clay swelling states \citep{Fies1998,Peters2009}. Coarse grains, microporous clay, and macropores are shown in gray, orange, and blue in Fig. \ref{fig:1}a and in light gray, dark gray, and black in Figs. 1b-d. \label{fig:1}}
\end{center} 
\end{figure}

A major challenge associated with developing a predictive understanding of flow in fine-grained sedimentary media is that these structures have two characteristic length scales: a \textit{macroscale} defined by the assemblages of coarse grains of quartz, feldspar, or carbonate over distances of micrometers and a \textit{microscale} defined by the assemblages of clay minerals (primarily smectite and Illite) over distances of nanometers \citep{Bourg2017a}. Figure \ref{fig:1}a shows a conceptual model of the macroscale structure of sedimentary rocks as a mixture of rigid coarse grains, a deformable microporous clay matrix, and macropores \citep{Crawford2008a,Marion1992a,Revil1999b}. This model is consistent with electron microscopy observations \citep{Nadeev2013,Nole2016b,Tuller2003} as exemplified in Fig. \ref{fig:1}b-d \citep{Fies1998,Peters2009}. Experimental data indicate that permeability in fine-grained soils, sediments, and sedimentary rocks can be highly sensitive to the spatial distribution of the clay matrix \citep{Abichou2002,Nadeau1998}.

A second major challenge is that the microporous clay matrix is non-rigid: it can swell or shrink in response to changes in salinity and deform in response to fluid flow or external stresses in a manner that reflects the nanoscale colloidal interactions between negatively-charged clay particles \citep{Liu2013,Madsen1989,Suzuki2005,Teich-McGoldrick2015}. The resulting dynamics of the clay matrix give rise to significant couplings between the hydrologic, chemical, and mechanical (HCM) properties of clayey media \citep{Carey2014,Erol1977c,Murad1997a}. These couplings are particularly strong if the clay fraction contains significant amounts of smectite (i.e., swelling clay minerals) and if the pore water contains predominantly sodic salts. In these cases, swelling by more than 1400 \% and swelling pressures up to $\mathrm{\sim}$50 MPa have been reported \citep{Karnland2007,Norrish1954}. In short, fundamental predictions of the permeability of fine-grained sediments and sedimentary rocks require a model capable of describing coupled fluid flow in pores with very different sizes (intergranular macropores and clay micropores with pore widths on the order of micrometers and nanometers, respectively) while also predicting the deformations of the microporous clay matrix induced by flow, external stresses, and salinity changes \citep{Bourg2017a}.

Existing approaches to predicting the hydrology of fine-grained sedimentary media have focused on addressing either one of the two challenges outlined above, but not both simultaneously. Approaches focused on the existence of two length scales have generally used an ``ideal packing model'' representation of clay-rich sedimentary media as a microporous (clay) medium embedded within a network of coarse grains \citep{Crawford2008a,Marion1992a,Revil1999b}. On this model, a threshold naturally emerges at a clay mineral mass fraction of $\mathrm{\sim}$1/3 where the microporous clay matrix becomes the load-bearing phase, in agreement with experimental observations on the core-scale hydrologic and mechanical properties of sedimentary rocks \citep{Bourg2015,Crawford2008a}. However, existing models based on this framework invariably neglect the mechanics of the clay matrix by assuming either that the clay has a fixed porosity or that it uniformly occupies the space available within the network of coarse grains. One consequence of this approximation is that these models do not capture the influence of salinity on the permeability of clayey media \citep{Kwon2004,Quirk1986}.

Conversely, approaches focused on the HCM couplings such as Terzaghi's consolidation theory, Biot's theory of poroelasticity, and Mixture Theory (theories initially developed to describe clayey media and now widely applied to other deformable porous media including hydrogels and biological tissues) simplify the governing equations for fluid and solid dynamics into a single macroscopic equation by assuming that both phases are superimposed continua with negligible inertial forces \citep{Auton2017a,Barry1992,Jain2009,Santillan2018,Terzaghi1964a}. This approach results in an implicitly coupled momentum equation for deformable porous media that can be paired to different body forces or solid deformation constitutive relations to create system-specific models. A drawback of this approach in the context of sedimentary rocks, however, is that it does not reflect the existence of the two characteristic length scales illustrated in Fig. \ref{fig:1} or consequences such as the permeability threshold at a finite clay content noted above.

The advent of Digital Rock Physics (DRP) over the last decade---i.e., the combined use of X-ray computed tomography (XCT) and computational methods to generate numerical models of fluid flow in real rocks \citep{Adler1992,Fredrich1999,Mirabolghasemi2015,Raeini2017}---provides a potential route to simultaneously addressing both challenges outlined above. In the last few years, a handful of DRP studies have examined rocks with two characteristic length scales including fractured porous rocks, vuggy media, and clay-rich sedimentary rocks \citep{Keller2013,Li2016,Saif2017}. In these studies, the macroscale features (coarse grains, macropores, and microporous regions) are fully resolved, while the microscale features (i.e., particles and pores within the microporous regions) are below the resolution of the XCT measurements. Many of these studies used a computational framework based on a pore network model (PNM) with two characteristic length scales \citep{Mehmani2014,Mehmani2013,Prodanovic2015}. Alternatively, at least two used an approach based on computational fluid dynamics (CFD) where flow in the macropores and microporous regions were coupled through slip-flow boundary conditions and volume-averaging based on the Darcy-Brinkman formalism \citep{Bijeljic2018,Guo2018}. The resulting models demonstrate the possibility of characterizing porous media with two characteristic length scales as mixtures of three regions (as in Fig. \ref{fig:1}a) and in coupling fluid flow in the macropores and microporous regions \citep{Golfier2015,Soulaine2017}. However, these studies all assumed a static porous medium configuration, a significant limitation in the case of clay-rich media, as noted above.

\begin{figure} 
\begin{center}
\includegraphics[width=0.99\textwidth]{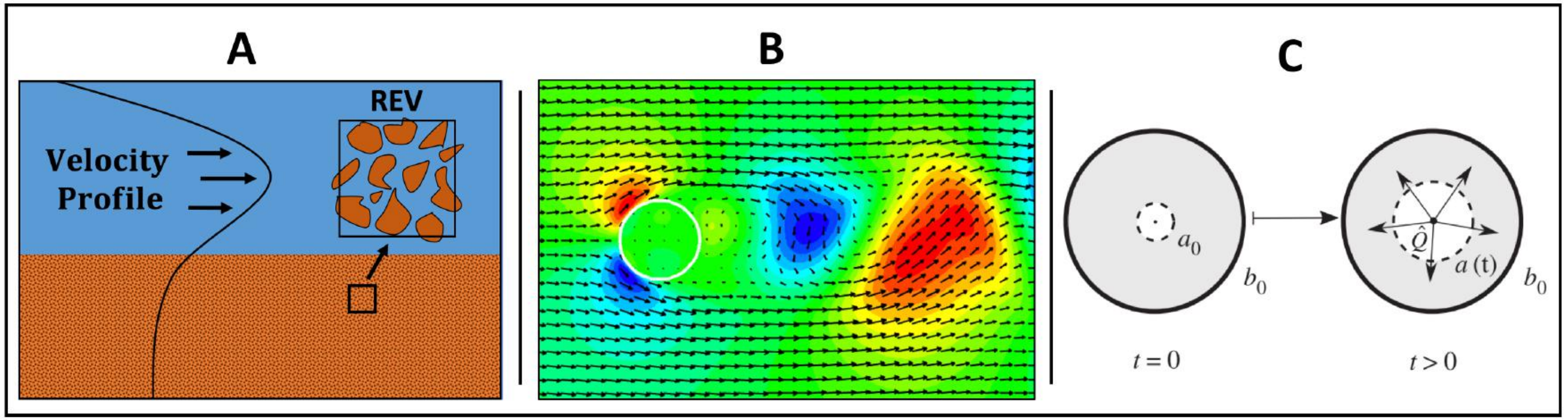}
\caption[Illustration of the three main theoretical frameworks combined in this study to model fluid flow in porous media with two characteristic length scales and a deformable microporous matrix]{Illustration of the three main theoretical frameworks combined in the present study to model fluid flow in porous media with two characteristic length scales and a deformable microporous matrix: (a) models based on the Darcy-Brinkman formulation that describe coupled fluid flow in macropores (clear region) and in a microporous medium (shaded region) \citep{Ochoa-Tapia1995a}; (b) boundary methods that simulate flow around moving impermeable solids \citep{Breugmem2013}; and (c) poromechanics model of a cylindrical conduit where the fluid and porous solid are in a single domain (grey) and forces are imposed through boundary conditions \citep{Auton2017a}. Here $Q$ represents the fluid flow rate, and $a(t)$ and $b_0$ represent the inner and outer radii of the cylinder, respectively. \label{fig:2}}
\end{center} 
\end{figure}

The main contribution of this chapter is the application of the system of differential equations derived in Chapter \ref{FullDerivation} to single phase flow in deformable porous media. The model is parameterized with a focus on viscoplastic clay-rich sedimentary media (Section \ref{Sect:Parametrization4}) and validated against experimental results on the properties of these media (Section \ref{Sect:Application5}). With proper parameterization, our framework should be applicable to other systems that involve coupled fluid flow in macropores and in a deformable microporous matrix such as soils \citep{Muradt1996,Vo2011}, hydrogels \citep{Datta2016}, biological tissues \citep{Dai2014}, and fractures \citep{Noiriel2007b}. To illustrate this versatility, we also formulate and apply the model for cases where the microporous matrix undergoes elastic rather than viscoplastic deformation (Section \ref{Sect:elastic6}). 

\section{Parametrization to Porous Media: Na-Smectite Clay}
\label{Sect:Parametrization4}

As noted in the introduction, a key challenge in subsurface hydrology is the strong influence of clay minerals on fluid flow in sedimentary rocks \citep{Bourg2017a}. This challenge is particularly profound in systems that contain smectite clay, a microporous material whose permeability, swelling pressure, and plastic rheology are highly sensitive to porosity and aqueous chemistry \citep{Aksu2015d,Mondol2008,Spearman2017}. Smectite is the predominant clay mineral in many unconsolidated sediments, in bentonite (a mixture of sand and clay used in geotechnical applications), and in soils formed in temperate weathering environments \citep{Abichou2002,Sposito2008a}. Smectite and Illite clay minerals constitute roughly half of the world's sedimentary rock mass, with Illite being generally more abundant on a mass basis but both clay minerals being roughly equally important on a surface area basis. The present section parameterizes the set of equations derived above for systems where the microporous matrix consists entirely of smectite by choosing the appropriate permeability, salt diffusion, swelling pressure, and rheology models. We focus on conditions where aqueous chemistry is dominated by sodic salts such as NaCl, where smectite swelling and rheology have been most extensively characterized. The result is a closed, fully coupled system of equations. 

\subsection{Parametrization of the Permeability Function} \label{permpara}

First, we define the permeability of the clay matrix ($k$), the key variable that determines the sub-REV-scale momentum interaction between water and clay. For simplicity, we use the well-known Kozeny-Carman relation: 

\begin{equation} \label{eq:20} 
k\left({\phi }_f\right)=\frac{1}{b\ t^2a^{2\ }_s}\ \frac{{\phi }^3_f}{{\left(1-{\phi }_f\right)}^2} 
\end{equation} 

Equation \ref{eq:20} describes the permeability of a bundle of capillary tubes of uniform radius within an impermeable solid, where $b$ is a pore shape factor, $t$ is the tortuosity of the pore network, and $a_s$ is the specific surface area expressed as the ratio of surface area to solid volume. The parameter values used in applying Eqn. \ref{eq:20} to Na-smectite and Illite are summarized in Table \ref{tbl:1}. The shape factor was set to \textit{b} = 3 $\mathrm{\pm}$ 1 based on empirical observations in a range of porous media \citep{Bourg2017a}. Tortuosity was set to  $t^2$ = 4.0 $\mathrm{\pm}$ 1.6 based on measurements of water tracer diffusion in compacted Na-smectite \citep{Bourg2006}. The specific surface area was set to 1.7 $\mathrm{\pm}$ 0.2 nm${}^{-1\ }$for smectite and 0.21 $\mathrm{\pm}$ 0.03 nm${}^{-1}$ for Illite based on experimental values reported in previous studies ($a_s$\textit{ }= 703 $\mathrm{\pm}$ 60 m${}^{2}$ g${}^{-1}$ for smectite and \textit{a}${}_{s}$ = 78 $\mathrm{\pm}$ 10 m${}^{2}$ g${}^{-1}$ for Illite, converted to units of area per volume using grain densities of 2400 $\mathrm{\pm}$ 140 kg m${}^{-3\ }$and 2700 $\mathrm{\pm}$ 70 kg m${}^{-3}$, respectively) \citep{Brooks1955,Diamond1956a,Mesri1971c,Orchiston1954,Quirk1955}. The values of specific surface area used here are based on methods that probe the entire water-accessible surface area of clay particles, including glycerol and ethylene glycol monoethyl ether (EGME) retention techniques but not standard N${}_{2}$ adsorption techniques \citep{Diamond1956a,Tournassat2015}.

A well-known limitation of Eqn. \ref{eq:20} is that it does not capture the influence of pore-size heterogeneity \citep{Bourg2017a,Dixon1999} and, therefore, does not accurately predict the macro-scale permeability of clay-rich soils or sedimentary rocks \citep{Mondol2008,Ren2016a}. The comparison with experimental data in Fig. \ref{fig:3}A, however, indicates that Eqn. \ref{eq:20} provides a reasonable permeability model for pure Na-smectite and Illite without the need for fitting parameters. 

Two important features of the permeability of pure clay are not described by Eqn. \ref{eq:20}. First, the permeability data in Fig. \ref{fig:3}A relate exclusively to sodium-exchanged clays. Other data indicate that smectite (and, to a smaller extent, Illite) has a higher permeability at the same porosity values if equilibrated with calcium or potassium electrolytes \citep{Mesri1971c}. This permeability difference likely reflects the less uniform pore-size distributions of K- and Ca-smectite (relative to Na-smectite) that results from the stronger charge-screening capacity of K and Ca ions at the clay surface. Second, clay-water mixtures exposed to one-dimensional consolidation are known to develop a textural anisotropy that makes their permeability tensor anisotropic \citep{Hicher2000}. Overall, Eqn. \ref{eq:20} is used here as a simple, yet reasonably accurate permeability model in the range of conditions of interest, i.e., fully-saturated Na-smectite or Illite in the absence of excessive anisotropy. Modifications to account for clay anisotropy, other clay counterions (such as calcium or potassium), multiphase flow effects (e.g., the Klinkenberg Effect), or to implement other permeability models \citep{Chapuis2003a,Samarasinghe1982} can be readily carried out within the present framework.

\subsection{Parametrization of the Swelling Pressure Term}

Second, we define the swelling pressure of the microporous clay matrix. For this, we use the well-known Derjaguin-Landau-Verwey-Overbeek (DLVO) theory of colloidal interactions. More precisely, we use a semi-empirical formulation proposed by \citet{Liu2013} for the disjoining pressure in a water film of thickness ``h'' between clay tactoids (i.e., stacks of clay particles separated by $\mathrm{\le}$ 3 water layers):

\begin{equation} \label{eq:21} 
\begin{split}
p_{swell}=2{\hat{C}}_{\mathrm{NaCl}} & R T \mathrm{cosh} \left(y_m\ -1\right) \\ &-\frac{A_h}{6\pi }\left(\frac{1}{h^3}-\frac{2}{{\left(h+D_p\right)}^3}+\frac{1}{{\left(h+2D_p\right)}^3}\right)+S_0{\mathrm{exp} \left(-\frac{h}{l}\right)\ }\ 
\end{split}
\end{equation}

In Eqn. \ref{eq:21}, the first term on the right side is Langmuir's model \citep{Kemper1972,Langmuir1938} of the osmotic swelling pressure caused by overlapping electrostatic double layers in a slit-shaped nanopore, where ${\hat{C}}_{\mathrm{NaCl}}$\textit{ }is the salt concentration in mol m${}^{-3}$, \textit{R} is the ideal gas constant (8.31446 J K${}^{\mathrm{-}1}$ mol${}^{\mathrm{-}1}$), \textit{T} is the temperature in Kelvin, and $y_m$ is the scaled electrostatic potential at the mid-plane of the nanopore, calculated here using the so-called ``compression approach'' for solving the Poisson-Boltzmann equation between two charged parallel plates in a symmetrical electrolyte solution \citep{Liu2008}. The salinity dependence of $y_m$ gives rise to the well-known salinity dependence of clay swelling. The second term is the contribution of London dispersion forces to the swelling pressure modeled through Hamaker's approach \citep{Hamaker1937}, where$\ D_p$ is the thickness of clay tactoids and $A_h$ is Hamaker's constant for clay tactoids separated by a water film. The third term is an empirical description of the short-range ``hydration repulsion'' between clay particles, where S${}_{0}$ and \textit{l }are empirical coefficients. Parameter values and the relation between ${\phi }_f$ and \textit{h} were taken from \citet{Liu2013} and are listed in Table \ref{tbl:1}. Equation \ref{eq:21}1 as parametrized by \citet{Liu2013} is used here as a convenient parametric fit to experimental data on smectite swelling as a function of ${\phi }_f$ and C.

We note that Eq. \ref{eq:21} makes significant simplifications and approximations. For example, the hydration repulsion term is purely phenomenological, while the optimal formulation and parameterization of the London dispersion term is still unsettled in the case of smectite clay \citep{Gilbert2015a,Tester2016}. In addition, the first two terms in Eqn. \ref{eq:21} are based on mean-field theories, i.e., they neglect short-range interactions between water, ions, and clay surfaces \citep{McBride1997,Missana2000}. In the case of Na-smectite, this last approximation is valid only at interparticle distances greater than 3 nm (i.e. ${\phi }_f>0.75$) \citep{Adair2001}. One consequence of this is that DLVO theory does not predict the existence of stable ``crystalline'' swelling states with interparticle distances $\mathrm{\le}$ 1 nm that predominate at high salinity, high compaction, or in aqueous chemistries dominated by divalent ions \citep{Bourg2017a,Pashley1984,shen2021}. Furthermore, this formulation assumes that clay swelling pressure is controlled by a single microstructural variable (interparticle distance \textit{h} or, equivalently, porosity ${\phi }_f$); more complex formulations have been proposed in the case of other microporous media with a deformable solid skeleton, such as activated carbon or zeolites \citep{Pijaudier-Cabot2011}. Finally, the numerical method used by \citet{Liu2013} to evaluate $y_m$ cannot be applied$\ $when C = 0; for simplicity we use C = 0.001 M as an approximation when calculating $p_{swell}$ in pure water (resulting in a $\mathrm{\sim}$0.05\% error in said value). 

\begin{figure} 
\begin{center}
\includegraphics[width=0.75\textwidth]{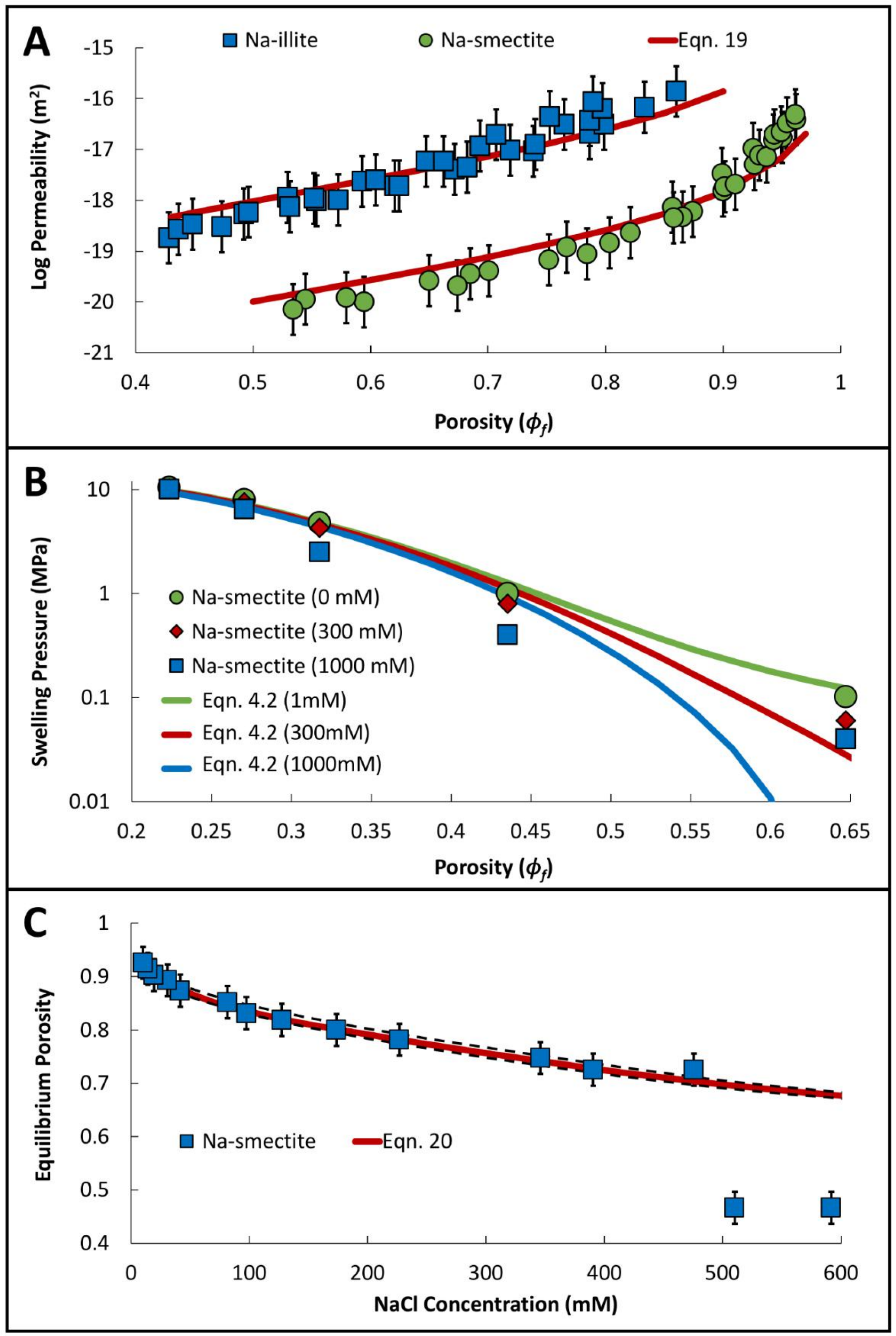}
\caption[Parameterization of clay models (Na-smectite)]{(a) Permeability of compacted Na-smectite and Illite as a function of porosity; (b) swelling pressure of compacted bentonite as a function of porosity at different salinities (error bars are smaller than the symbols); and (c) porosity of Na-smectite as a function of salinity at a small constant confining pressure. Symbols are experimental results \citep{Mesri1971c,Norrish1954,SKB2011}. Lines are model predictions obtained using Eqns. \ref{eq:20} and \ref{eq:21} with no fitting parameters in Fig. \ref{fig:3}A,B and with a single fitting parameter (the unspecified small confining pressure used in the experiments, which was set to 0.03 MPa; dashed lines show the sensitivity of the model predictions to $\pm$ 0.01 MPa differences in this value) in Fig. \ref{fig:3}C.}\label{fig:3}
\end{center} 
\end{figure}

To test the accuracy of Eqn. \ref{eq:21} as parameterized by \citet{Liu2013}, we used macroscopic data on the swelling pressure of confined water-saturated Na-smectite as a function of compaction and salinity (Fig. \ref{fig:3}B) as well as X-ray diffraction (XRD) measurements of \textit{h} vs. salinity (Fig. \ref{fig:3}C). Figure \ref{fig:3} shows that Eqn. \ref{eq:21} overestimates porosity at salinities $\ge$ 0.5 M NaCl, where crystalline swelling predominates, but accurately predicts experimental data at lower salinities.

\subsection{Parametrization of the Diffusion Coefficient of Dissolved Salts}

Third, we define the effective diffusion coefficient of dissolved salt (\textit{x}) in the microporous clay using a form of Archie's law \citep{Blum2007,Boving2001}:

\begin{equation} \label{eq:22} 
D_{\mathrm{e,}x}={\phi }^n_fD_x 
\end{equation} 

Although there does not exist a universal Archie's law exponent \textit{n} for all porous media, several studies have concluded that a value of 2.2 to 2.5 reasonably reproduces the diffusion of ions in pure compacted clay \citep{Revil2011,Shen2007,VanLoon2003,VanLoon2015}.

Equation \ref{eq:22} is known to oversimplify salt diffusion in clayey media, particularly by neglecting the salinity-dependence of $D_{\mathrm{e,}x}$ that arises from anion repulsion in the electrical double layer on charged clay surfaces \citep{Malusis2003,Sherwood2000a,VanLoon2007,Underwood2020}. A notable alternative to Eqn. \ref{eq:22} that accounts for this effect is Kemper's model \citep{Kemper1966,Kemper1966a}.

\begin{equation} \label{eq:23} 
D_{\mathrm{e,}x}=\frac{1-e^{-y_m}}{{\ t}^2}D_x 
\end{equation} 

\noindent where $t$ is tortuosity as in Eqn. \ref{eq:20}, \textit{y}${}_{m}$ is the same as in Eqn. \ref{eq:21}, and ``$1-e^{-y_m}$'' is the equilibrium ratio between anion concentration in the clay nanopores and in bulk liquid water. Equation \ref{eq:23} is less empirical than Eqn. \ref{eq:22} and more consistent with the permeability and swelling pressure models used above. However, its use would require a more complex treatment of the coupled advection and diffusion of ions in Eqn. \ref{eq:11}, where the impact of anion exclusion is neglected. We note that multiphase flow effects such as Knudsen diffusion are neglected here due to the fact that we focus exclusively in fully water-saturated media \citep{Malek2003}. Equation \ref{eq:22} is therefore used here for simplicity.

\subsection{Parametrization of Clay Rheology}

Lastly, we define the effective viscosity of the microporous clay in a manner that accounts for the impact of clay swelling and shearing on its plastic viscosity and critical shear stress \citep{Guven1993,Maciel2009a}. \citet{Spearman2017} developed a model based on floc fractal theory that predicts the rheological properties of a wide variety of clays when sheared at different solid-water ratios:

\begin{equation} \label{eq:24} 
{\mu }^{eff}_s={\left[{\mathrm{\tau }}^{\frac{1}{4-D}}_0+{\left(1+\frac{1}{\beta {\gamma }^m}{\left(\frac{{\tau }_0}{{\mu }_*}\right)}^m\right)}^{\frac{1}{4-D}}{\left({\mu }_*\gamma \right)}^{\frac{1}{4-D}}\right]}^{4-D}{\gamma }^{-1} 
\end{equation} 

\begin{equation} \label{eq:25} 
{\tau }_0={\tau }^*{\left[\frac{{\phi }_s/{\phi }^{max\ }_s}{(1-{\phi }_s/{\phi }^{max}_s)}\right]}^{4-D}\ \ \ \ \ \vdots \ \ \ \ \ \ {\mu }_*=\frac{{\mu }_f}{{\left(1-{\phi }_s/{\phi }^{max}_s\right)}^2}\ \  
\end{equation} 

In Eqns. \ref{eq:24}-\ref{eq:25}, ${\phi }^{max}_s$ is the maximum solid fraction of clay, ${\tau }^*$\textit{ }the critical shear stress at ${\phi }_s=0.5\ {\phi }^{max}_s$,  \textit{D} is the fractal dimension, $\beta $ is a structural parameter, \textit{m} is a structural break-up parameter, and $\gamma $ is the strain rate (calculated at each time step as the sum of symmetric components of $\nabla{\hat{U}}_s$). With proper tuning, this model was shown to successfully describe the rheologies of smectite, Illite, and kaolinite clays. Parameter values were taken from \citet{Spearman2017} based on fitting rheological data pertaining to a 90\% smectite-water mixture obtained by \citet{Coussot1993} and are listed in Table \ref{tbl:1}. 

We note that Eqns. \ref{eq:24} and \ref{eq:25} do not explicitly consider potential impacts of salinity or dynamic changes in solid fraction (as in the case of shrinking and swelling) on rheology. Instead, the Spearman model assumes that the clay is at an equilibrated solid fraction when calculating the effective viscosity. For simplicity, we assume that the resulting effective viscosity applies not only to shear, but also to expansion and contraction. 

The use of Eqns. \ref{eq:20}-\ref{eq:25} in the modeling framework derived in Chapter \ref{FullDerivation} provides closure for the coupled system of equations in the case of pure smectite in sodium electrolytes such as NaCl at salinities up to 0.5 M. As noted above, this closure relies on a number of assumptions and approximations. Notably, it neglects clay fabric anisotropy \citep{Hicher2000,Tessier1990} and clay dispersion into the bulk water phase. Furthermore, it relies on the assumption that DLVO theory, the Kozeny-Carman equation, Spearman's model, and the Fickian diffusion equation described above are accurate in conditions beyond their validation in Fig. \ref{fig:3} (for example, during dynamics shearing, shrinking, and swelling). The assumptions and approximations listed above are not intrinsic to the modeling framework derived in Chapter \ref{FullDerivation}. They can be readily addressed in future studies as additional information become available on the microscale permeability, swelling pressure, rheology, and dispersion of microporous clay.

\section{Model Validation and Application to Plastically Deformable Porous Media (Clay)}
\label{Sect:Application5}

\subsection{Model Verification by Comparison to the Fluid-Driven Deformation of a Clay Filter Cake}

A quantitative validation of our model was realized by testing its ability to predict the compaction of a clay plug caused by water flow through said plug. To do this, we compared our numerical results to predictions of an analytical model derived by \citet{Hewitt2016a} at equivalent experimental conditions. Said analytical model is based on Biot Theory and is able predict the 1-D solid deformation of the plug as a function its initial porosity, the deformation modulus of the solid, and the fluid pressure gradient across the plug. \citet{Hewitt2016a} validated their model against experimental results obtained with an elastically deformable microporous medium placed within a 12 cm by 25 cm container based on the assumptions that permeability follows the Kozeny-Carman equation, stresses follow the Terzaghi stress principle, and fluid flow (and thus the resulting deformation-inducing pressure gradient) is governed by Darcy's law. 

\begin{figure}[htb]
\begin{center}
\includegraphics[width=0.95\textwidth]{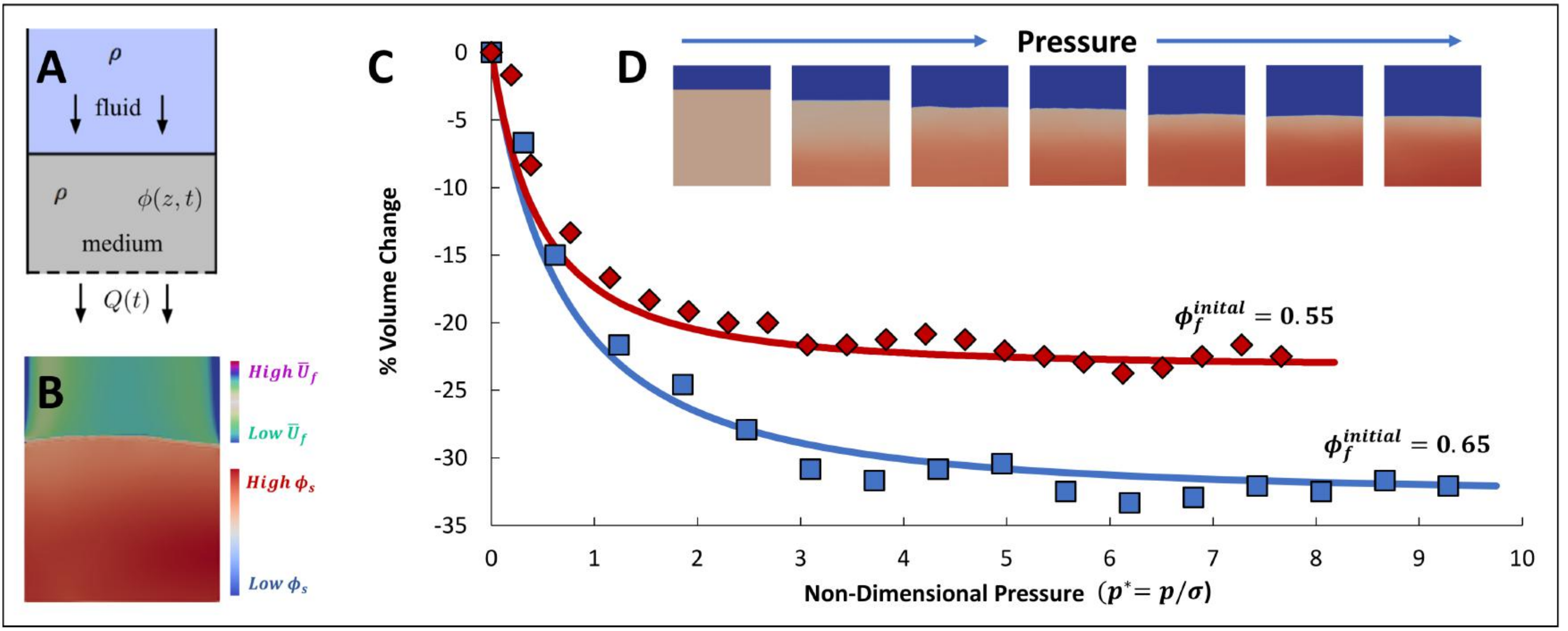}
\caption[Fluid-driven compaction of a clay filter cake]{Model prediction of the compaction associated with water flow through a clay plug as a function of pressure drop (from 0 to 200 kPa) across the clay plug: (a) schematic view of the simulated system; (b) equilibrated system as predicted by our simulation framework (colors show clay fraction ${\phi }_s$ in the microporous region and fluid velocity ${\overline{U}}_f$ in the free fluid region); (c) comparison between Hewitt's analytical model (solid lines) and the predictions of our numerical model (symbols) for initial clay porosities of 0.55 and 0.65; and (d) final configurations of the system as a function of increasing pressure drop. The simulations were carried out in a 1 by 300 grid (c) and a 240 by 300 grid (b, d). Results are reported as the non-dimensional volume change of the clay plug vs. non-dimensional pressure drop across the clay plug. Pressure in the simulations was non-dimentionalized using a deformation modulus calculated from the clay's critical stress at its initial porosity [\textit{p}* = \textit{p}/$\sigma$, where $\sigma ={\tau }^{initial}_{0\ }/({\phi }^{initial}_f/{\phi }^{final}_f-1)$]. For both simulations, mean average error (MAE) was about 0.04 times the final measured volume change. \label{fig:4}}
\end{center} 
\end{figure}

To replicate these conditions, our simulations where carried out on a 240 by 300 grid representing a 2-D container (12 cm by 15 cm) partially filled with non-swelling clay. We induced fluid flow through the clay by a constant pressure boundary condition at the top of the container, where fluid (and only fluid) was allowed to leave through the lower boundary. In Fig. \ref{fig:4}, we compare our steady-state, non-dimentionalized numerical model predictions for Na-smectite against \citet{Hewitt2016a} analytical model at two initial porosity values, showing good agreement between both models. Since \citet{Hewitt2016a} did not include a swelling pressure into their analytical framework, we set $p_{swell}=0$ in our numerical simulations to ensure proper comparison and verification of the flow-deformation mechanics. Please refer to Table \ref{tbl:1} for a detailed listing of the parameter used in this and all other subsequent simulations. 

The resulting steady-state deformation profiles are a result of the balance between the clay's structural forces (critical stress) and the forces imposed by the fluid (viscous drag): as compaction increases, the permeability of the porous medium decreases, which in turn increases the viscous stresses imposed on the medium. At the same time, given a constant pressure gradient, flow through the plug decreases as permeability decreases, which then reduces the magnitude of the viscous stresses. Steady-state is achieved once the viscous drag is balanced by the porous medium's structural forces, which explains why systems with a higher initial compaction deform less than their counterparts (Fig. \ref{fig:4}). The good agreement shown in Fig. \ref{fig:4}C is expected (both models rely on similar assumptions such as the validity of the Kozeny-Carman equation and Terzaghi and Biot's principles) but also provides some confidence in the ability of our model framework to represent the feedbacks between fluid flow and solid deformation in microporous media.   

\subsection{Model Verification by Comparison to Oedometric Clay Swelling Experiments}

\begin{figure}[htb!]
\begin{center}
\includegraphics[width=0.9\textwidth]{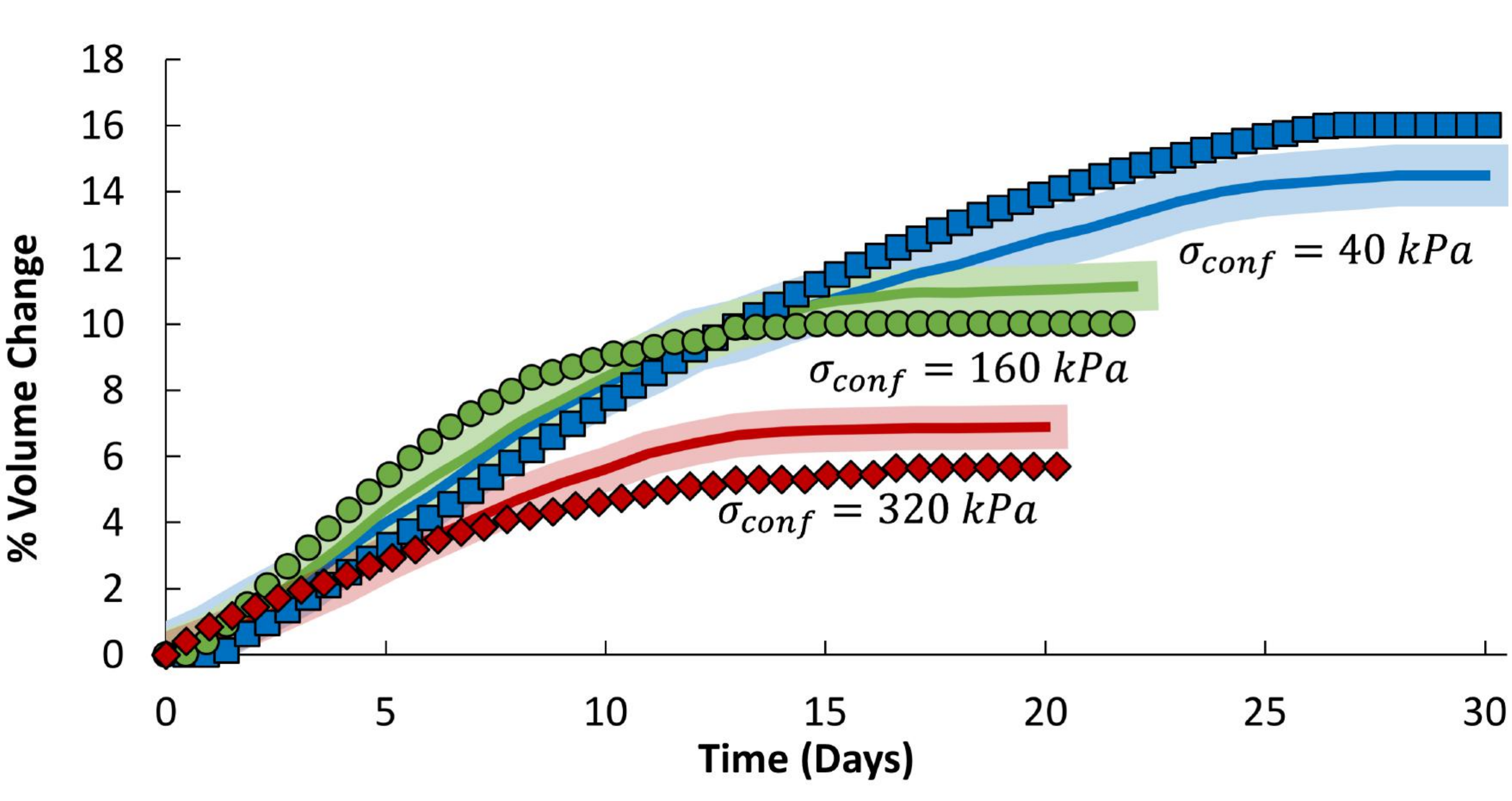}
\caption[Macroscopic swelling of NaCl-saturated smectite clay]{Macroscopic swelling of NaCl-saturated smectite clay (at confining pressures 40, 160, and 320 kPa) initially equilibrated with a 1M NaCl solution and then placed in contact with a reservoir of distilled water. Experimental results (solid lines) and approximated error (shaded regions) were obtained using data from \citet{DiMaio1996}. Simulation predictions (symbols) were obtained using two-dimensional 500 by 300 grids. For the three simulations, MAE was 0.06 to 0.09 times the measured final volume change. \label{fig:5}}
\end{center} 
\end{figure}

As a further quantitative validation of our model (and to verify the swelling pressure effects disregarded in the previous section) we tested model predictions against standard oedometric measurements of clay swelling driven by a salinity change. Specifically, we used measurements by \citet{DiMaio1996} of the volumetric swelling of water-saturated Na-smectite samples exposed to a salinity shock (from 1 to 0 M NaCl) at different confining pressures (Fig. \ref{fig:5}). The experimental conditions were straightforward: individual water-saturated clay samples were first compressed within a oedometer chamber to confining pressures of 40 kPa, 160 kPa, and 320 kPa. Subsequently, the samples were placed in contact with a 1 M NaCl solution through a porous boundary until equilibrated. Swelling was then induced by replacing the saltwater solution with distilled water while maintaining a constant confining pressure. 

The swelling portion of the experimentas was modeled by defining a 500 by 300 grid representing a 2-D container (5 cm by 3 cm) filled with 2 cm of 1M NaCl-equilibrated compacted smectite clay, where the clay was only allowed to swell in the positive y-direction as a result of salt diffusion out of the container at the lower boundary (where salt concentration was set to zero). All other boundary conditions on the container walls were set to replicate impermable surfaces with no-slip boundary conditions. For simplicity, the confining pressure was assumed uniform throughout the clay sample and was applied as a constant in the Terzaghi stress tensor (see Section \ref{Sect:elastic}).

As shown in Figure \ref{fig:5}, our model accurately captures the swelling kinetics of Na-smectite clay driven by osmotic uptake of liquid water. The agreement between our model and experimental data provides further confidence into the ability of our model framework to capture feedbacks between hydrology, mechanics, and salt transport in deformable microporous media. Overall, Figures \ref{fig:4} and \ref{fig:5} show that our model, with no parameter fitting, yields reasonably accurate predictions (with normalized MAEs below 0.09) of the overall extent of clay swelling or compaction and of the associated kinetics. Agreement is obtained regardless of whether the clay volume change is driven by fluid flow (Fig. \ref{fig:4}) or salinity changes (Fig. \ref{fig:5}). This good agreement suggests that our model captures key features of the coupled HCM behavior of Na-smectite despite the assumptions and simplifications noted above.

\subsection{Model Prediction of the Permeability of a Bead Pack Containing Na-Smectite as a Function of Salinity and Clay Content}

Having parametrized and tested our simulation framework, we used it to predict the permeability of an idealized model of fine-grained soils and sedimentary media: a bead pack containing coarse non-porous beads uniformly coated with Na-smectite clay. The idealized system simulated here has previously been used both as a conceptual model of the hydrology of sedimentary media and as an idealized experimental model of the properties of engineered clay barriers \citep{Abichou2002,Revil1999b,Tuller2003}. To the best of our knowledge, it has not been previously implemented in a numerical framework that accounts for the HCM couplings in microporous clay. As implemented here, this model system captures both key features of fine-grained soils and sedimentary media identified in the introduction: the co-existence of two characteristic length scales and the HCM coupled properties of the microporous clay. As noted above, Na-smectite is used here because reasonably accurate constitutive models exist to describe its swelling pressure, permeability, and rheology at the microscopic scale, even though the microporous regions in most soil and sedimentary media generally consist of more complex mixtures of Ca/Na-smectite, other clay minerals, and organic matter. 

Our simulations are based on a representative, yet simplified, rock geometry built upon the following assumptions: first, the medium's coarse-grained, load-bearing structure can be represented as a 2-D cross-section of a 3-D randomly distributed spherical packed bed with a porosity of 0.64 obtained from \citet{Finney1970}; second, the initial distribution of clay is modeled as a uniform film coating the coarse grains (a reasonable approximation given previous imaging studies) \citep{Abichou2002,Aksu2015d,Peters2009}, and third, a two-dimensional geometry is sufficient to capture major features of the simulated system (as in \citet{Quispe2005}). These are significant approximations; in particular, feedbacks between fluid flow and clay dynamics may create some degree of non-uniformity in the clay distribution within sedimentary rocks \citep{Song2015} and percolation thresholds associated with pore clogging may occur more readily in two-dimensional than in three-dimensional systems.

\begin{figure}[htb]
\begin{center}
\includegraphics[width=0.95\textwidth]{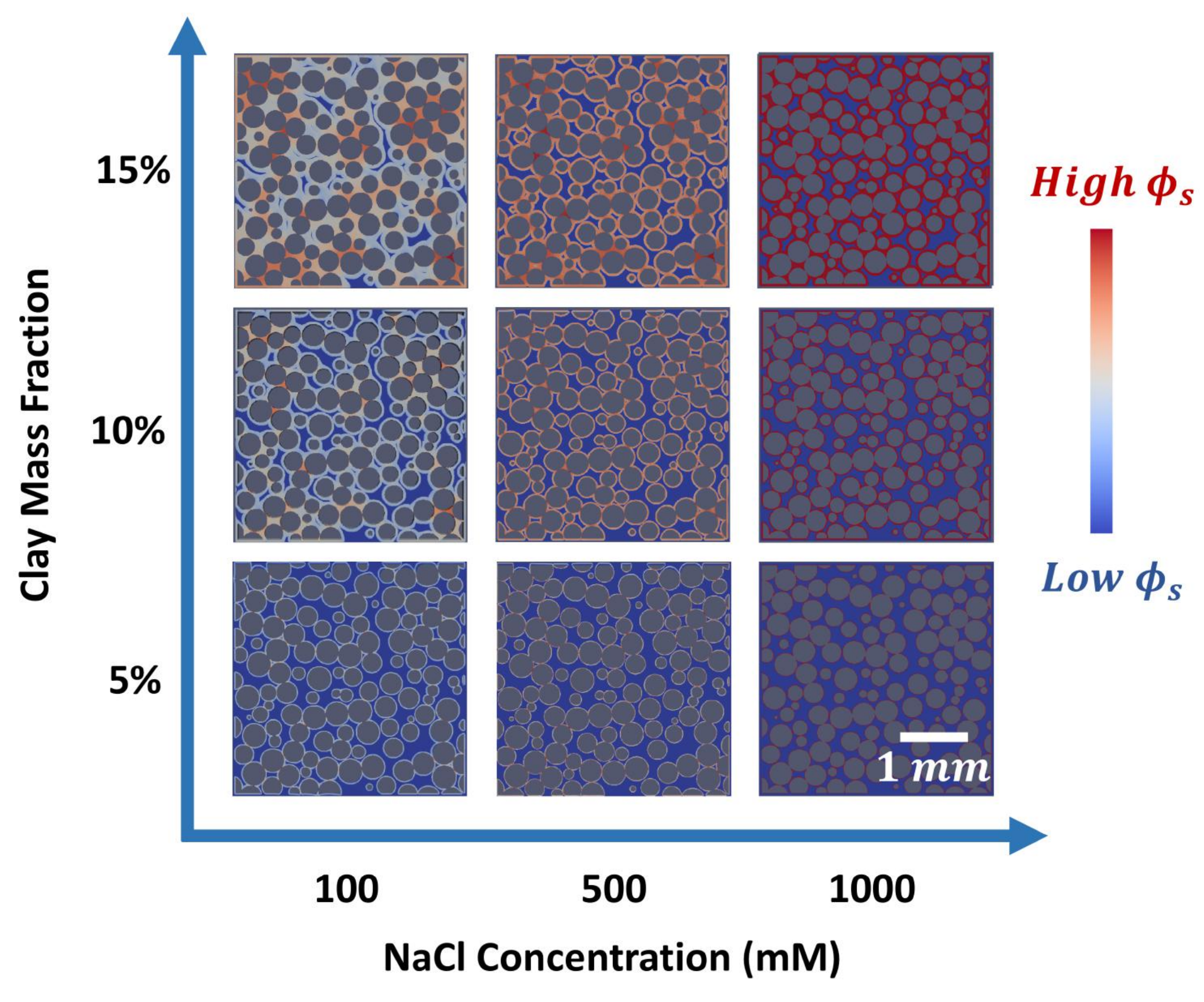}
\caption[Effect of clay mass fraction and NaCl concentration on the spatial distribution of smectite clay and available macropore space in a 2-D cross section of an ideal 3-D spherical packed bed]{Effect of clay mass fraction and NaCl concentration on the spatial distribution of smectite clay and available macropore space in a 2-D cross section of an ideal 3-D spherical packed bed \citep{Finney1970}. The simulated system is 3 by 3 mm. The grid resolution is 1020 by 1020 with a voxel size of 3 µm.\textbf{} \label{fig:6}}
\end{center} 
\end{figure}

\begin{figure}[htb]
\begin{center}
\includegraphics[width=0.8\textwidth]{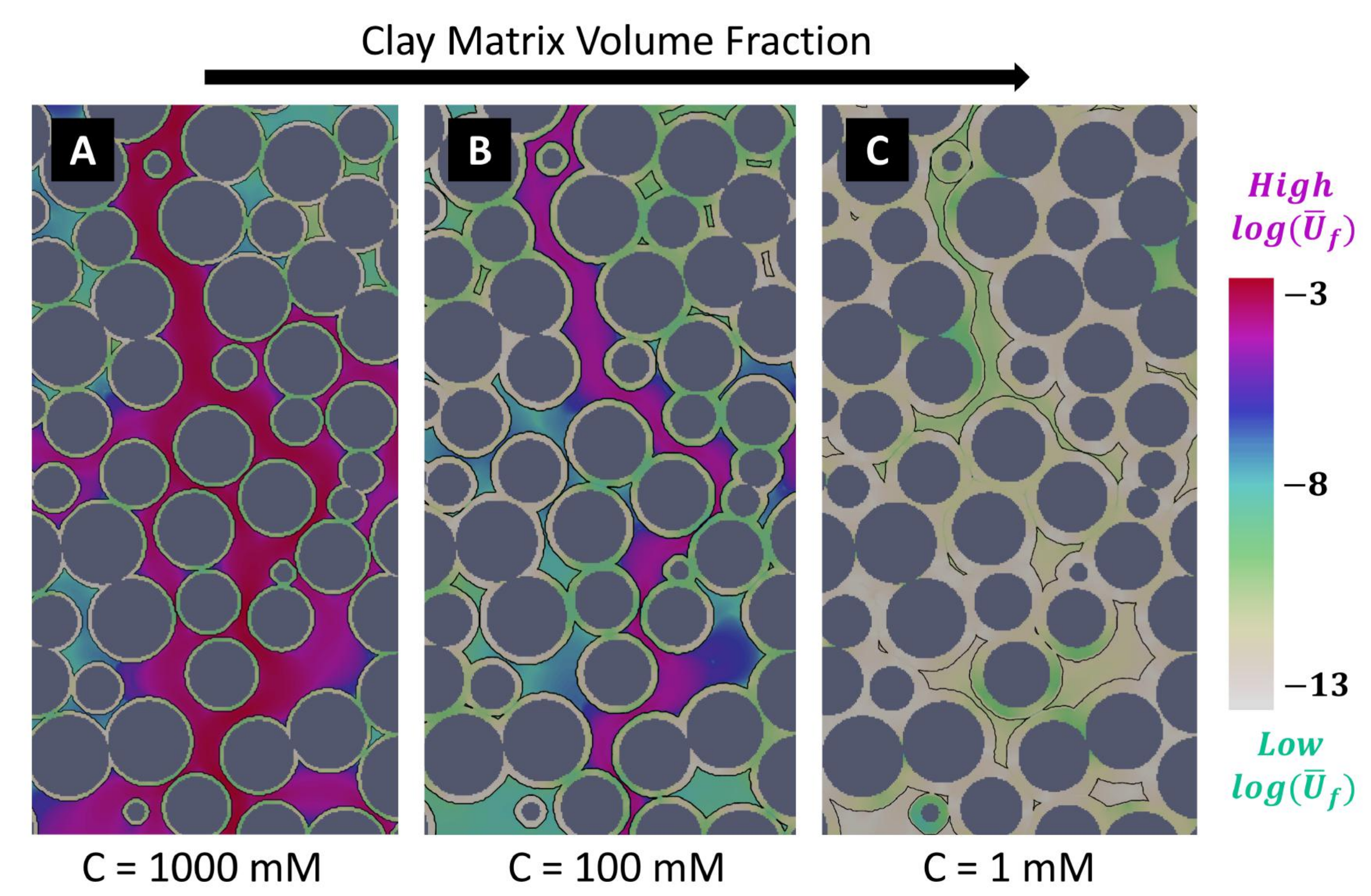}
\caption[Steady-state fluid velocity profiles with a 10 \% clay mass fraction at different salinity values]{Steady-state fluid velocity profiles with a 10 \% clay mass fraction at different salinity values. Thin black lines surrounding the solid grains represent the boundary between the clay matrix (within the lines) and the free fluid (outside the lines). At high salt concentrations (A), flow within the clay (greenish shades) does not control the overall permeability of the medium and can be considered negligible when compared to flow around the clay. Conversely, at lower salt concentrations and higher clay volumes (C), flow is controlled by the internal permeability of the clay matrix due to the absence of percolating flow paths through the macropores. The impact on the overall permeability of the system is shown in Fig. \ref{fig:8}.\label{fig:7}}
\end{center} 
\end{figure}

The system was simulated as a function of clay mass fraction (from 0 to 0.3) and salinity (from 1 to 1000 mM NaCl). We chose this parameter space because, as noted in the introduction, it represents the range over which micro and macro-pores coexist and over which the coarse grains are load bearing. Briefly, we initialized each set of simulations by populating the 3 by 3 mm$\mathrm{\ }$coarse grain structure with a uniform clay coating equilibrated at 1000 mM NaCl until we reached the desired clay mass fraction. Clay parametrization was consistent with the verification cases described in the previous sections. We then applied a constant hydrostatic pressure gradient in the y direction and a salinity step change across the sample and allowed the clay to swell until it reached its equilibrium volume fraction based on the new salinity value. A representative sample of the resulting steady-state configurations as a function of clay mass fraction and salinity is shown in Fig. \ref{fig:6}. Each sample was put through several swelling-shrinking cycles to ensure consistent results and to introduce heterogeneity approaching that present in natural systems. Although not evident in the equations, heterogeneous clay distributions can arise due to the combination of frictional forces imparted by the solid grains and the clay's non-Newtonian rheology. After we reached the steady state configuration, the permeability of each sample was evaluated by integration of the fluid velocities. Figures illustrating the predicted fluid velocity distribution in a region containing both macropores and microporous clay are shown in Fig. \ref{fig:7}.

\begin{figure}
\begin{center}
\includegraphics[width=0.8\textwidth]{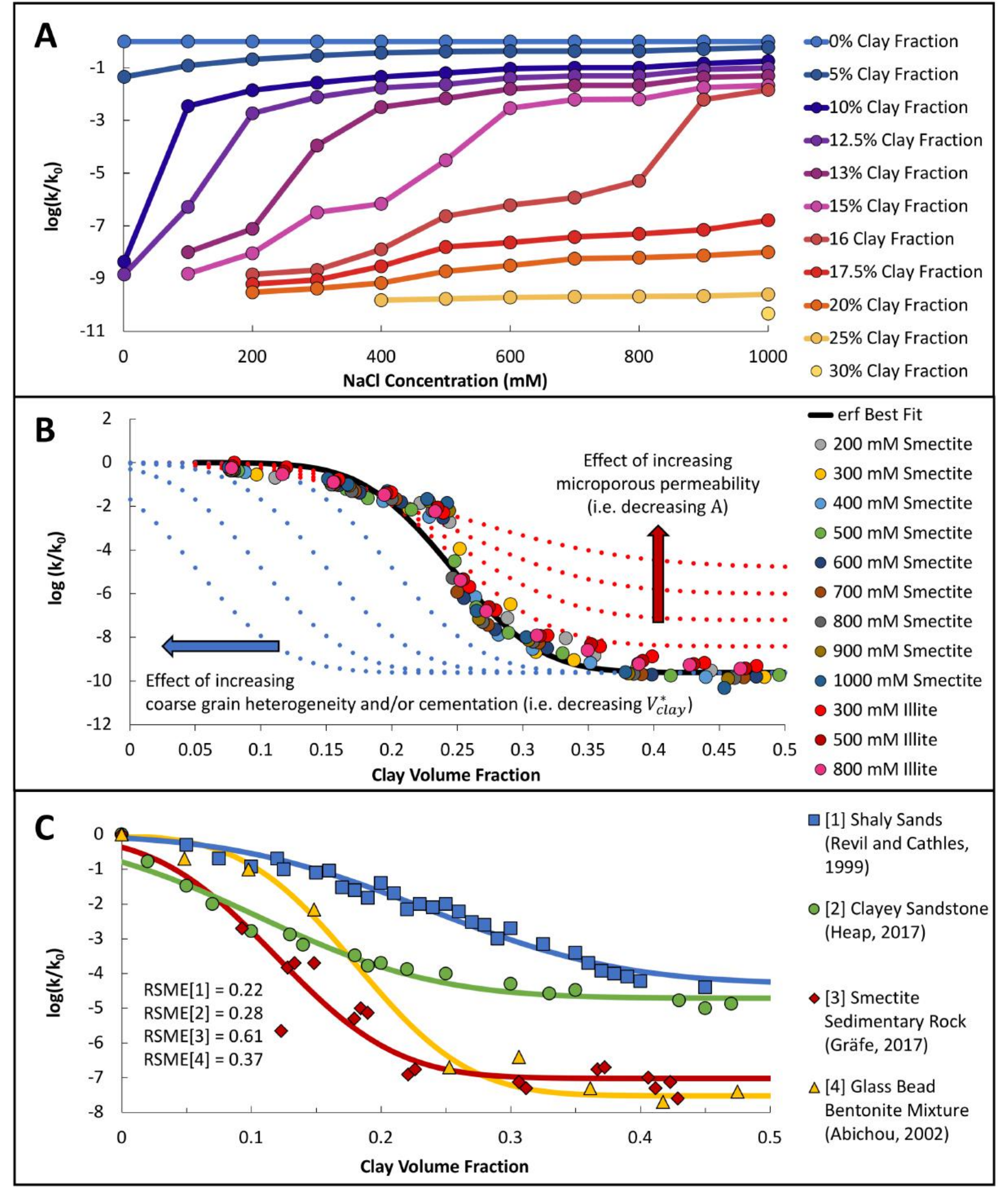}
\caption[Permeability of a simplified sedimentary rock model as a function NaCl concentration for a range of smectite clay mass fractions]{(a) Permeability of the simplified sedimentary rock model in Fig. \ref{fig:6} as a function NaCl concentration for a range of smectite clay mass fractions. The unprobed parameter space in the lower left corner represents conditions where clay more than fills the available space between the coarse grains (i.e., clay becomes load bearing). (b) Data from Fig. \ref{fig:8}A plotted as a function of clay matrix volume fraction. The permeability data collapse into a single curve that can be approximated with an error function (Eqn. \ref{eq:26} with A = 9.6, S = 1.38 $\times$ A, and $V^*_{clay}$ =0.25). Note that the figure also includes data obtained using Eqns. \ref{eq:20} and \ref{eq:21} parameterized for Illite instead of smectite. The smectite and Illite permeabilities coincide below the percolation threshold (where permeability is controlled by flow through macropores) and differ by up to two orders of magnitude above the threshold (where flow through the clay matrix predominates). (c) Comparison between Eqn. \ref{eq:26} (solid lines) and experimental datasets on the permeability of three different types of siliciclastic sedimentary rocks \citep{Grafe2017,Heap2017,Revil1999b} and a series of glass bead-smectite mixtures \citep{Abichou2002}. The curves' parameters are:$\ {\mathrm{A}}_{\mathrm{1}}\mathrm{=4.2\ \&\ }{\mathrm{V}}^{\mathrm{*}}_{\mathrm{1,\ clay}}\mathrm{=0.25,\ }{\mathrm{A}}_{\mathrm{2}}\mathrm{=4.8\ \&\ }{\mathrm{V}}^{\mathrm{*}}_{\mathrm{2,\ clay}}\mathrm{=0.10,\ }{\mathrm{A}}_{\mathrm{3}}\mathrm{=7\ \&\ }{\mathrm{V}}^{\mathrm{*}}_{\mathrm{3,\ clay}}\mathrm{=0.12,\ and\ }{\mathrm{A}}_{\mathrm{4}}=7.4\mathrm{\ \&\ }{\mathrm{V}}^{\mathrm{*}}_{\mathrm{4,clay}}\mathrm{=0.18}$ with S = 1.38 $\times$ A. \label{fig:8}}
\end{center} 
\end{figure}

Figure \ref{fig:8}A shows that permeability varies by ten orders of magnitude within the studied parameter space, with the largest permeability reduction (six orders of magnitude) coming from changing the salinity at clay mass fractions near 15\%. This large permeability range reflects the very low permeability of pure Na-smectite and the ability of the clay to block the main preferential fluid flow paths, effectively dividing the medium's macropores into isolated regions. Because of the clay's ability to swell, the location of the percolation threshold depends on both salinity and clay mass fraction.

Further analysis shows that the simulation predictions in Fig. \ref{fig:8}A collapse into a single relation between log(\textit{k/k${}_{0}$}) and the clay matrix volume fraction \textit{V}${}_{clay}$ (i.e., the ratio of volume occupied by the clay matrix, including its internal microporosity, to the total volume occupied by the clay and coarse grains). As shown in Fig. \ref{fig:8}B, the main preferential fluid flow path is consistently closed at \textit{V}${}_{clay}$ $\approx$ 25\% in our simulations, in reasonable agreement with previous predictions for spherical bead packs clogged either by clay or by cementation  \citep{MacArt2016,Tuller2006}. At clay volume fractions below this percolation threshold, clay swelling influences permeability by a relatively small (but still highly significant) two orders of magnitude. 

To compare the simulation predictions in Fig. \ref{fig:8}B with experimental data on the permeability of clayey media, we fitted the simulation predictions of log(\textit{k}/\textit{k}${}_{0}$) vs. \textit{V}${}_{clay}$ using an error function:

\begin{equation} \label{eq:26} 
{\mathrm{log} \left(\frac{k}{k_0}\right)\ }=\frac{A}{2}\left({\mathrm{erf} \left(\ S\ \left(V_{clay}-V^*_{clay}\right)\right)\ }-1\right) 
\end{equation} 

In Eqn. \ref{eq:26}, \textit{A} is the overall magnitude of the decrease in log \textit{k}, \textit{S} describes the sensitivity of permeability to $V_{clay}$ near the percolation threshold, and $V^*_{clay}$ describes the location of the threshold. Equation \ref{eq:26} provides a reasonable fit to the simulation predictions in Fig. \ref{fig:8}B with only three parameters (though it underestimates the sharpness of the threshold). To test the broader validity of Eqn. \ref{eq:26}, we identified four experimental datasets that reported the permeability of a series of porous media with similar mineralogy differing only in their clay content and that expressed clay content in the same manner as in Fig. \ref{fig:8}B \citep{Abichou2002,Grafe2017,Heap2017,Revil1999b}. As shown in Fig. \ref{fig:8}C, Eqn. \ref{eq:26} provides accurate descriptions of all four datasets. In the five parametric fits carried out with Eqn. \ref{eq:26} in Figs. \ref{fig:8}B,C, the values of \textit{S} and \textit{A} were consistently related (\textit{S}/\textit{A} = 1.38 $\pm$ 0.25). Fitted \textit{A} values ranged from 4.2 to 9.6, a range consistent with the reported permeability values of different clays, which span 4 to 5 orders of magnitude and increase in the order Na-smectite $\mathrm{<}$ Ca-smectite $\approx$ Illite $\mathrm{<}$ kaolinite \citep{Mesri1971c,Mondol2008}. Fitted values of $V^*_{clay}$ ranged from 0.10 to 0.25, in agreement with the expectation that the location of the percolation threshold depends on the porosity of the network of coarse grains \citep{Aksu2015d}, i.e., it should be lower for systems with greater cementation or with less uniform grain size distributions. The simulation predictions shown in Fig. \ref{fig:8}B, as expected, are at the upper end of the range of both \textit{A} values (because we used the lowest-permeability clay, Na-smectite) and $V^*_{clay}$ values (because we used a 2-D slice of a 3-D bead pack with a uniform bead size and no cementation).

The simulations described in this section are designed to approximate systems where the network of coarse grains is load-bearing and where clay influences permeability only through swelling and shrinking in response to salinity changes. Clay erosion, transport, and deposition are not included in our model, while viscous deformation of the clay by the flowing fluid is minimized by the use of a small fluid pressure gradient. Nevertheless, the good agreement between the results obtained with smectite and Illite and between the parametric fits to simulated and measured permeabilities suggest that the computational framework used here may be applicable to clayey media with a range of porosities, clay mineralogies, and configurations.

\section{Model Application to Fluid Injection Into a Poroelastic Medium}
\label{Sect:elastic6}

\begin{figure}[htb]
\begin{center}
\includegraphics[width=0.95\textwidth]{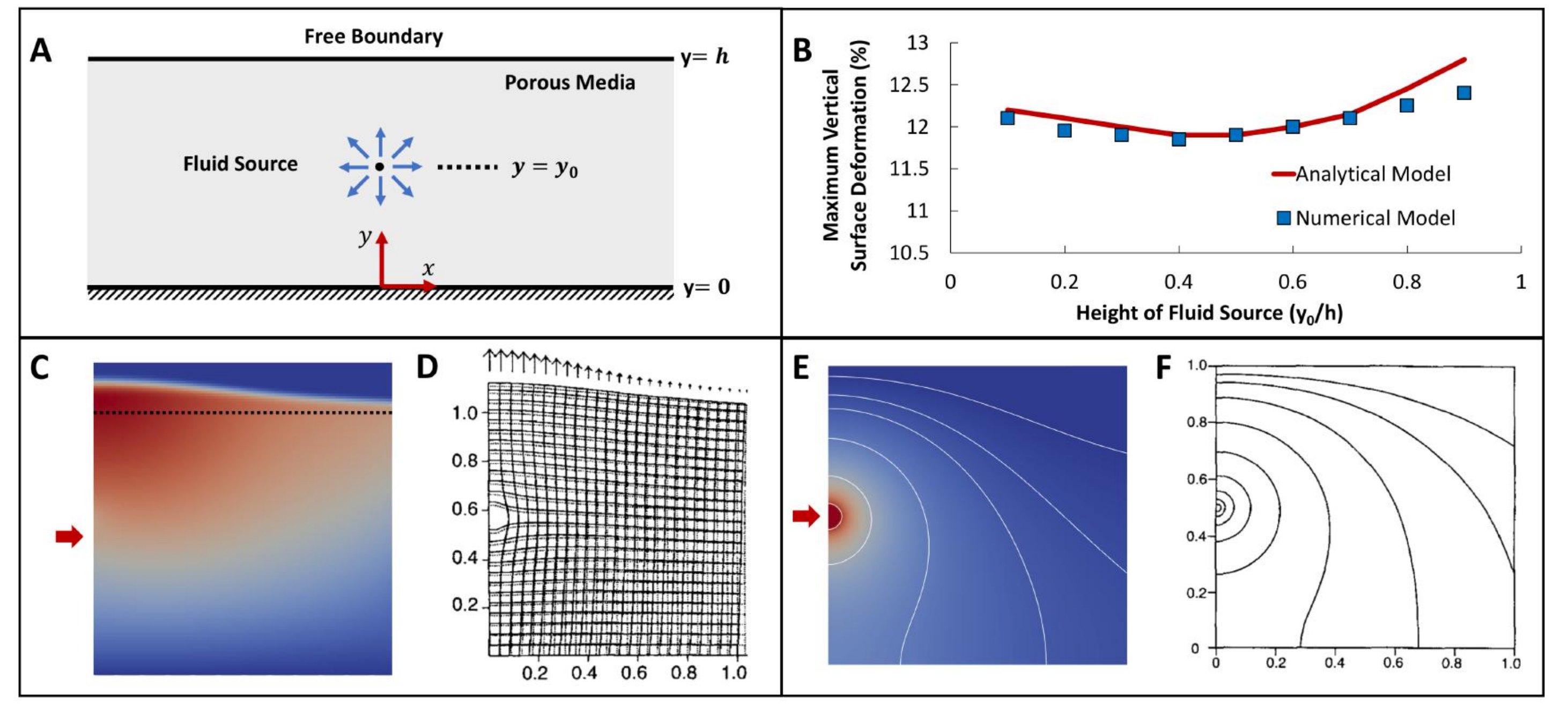}
\caption[Fluid-driven deformation of a poroelastic membrane]{(a) Representation of the modeled system. (b) Quantitative comparison of the maximum vertical deformation as a function of the point fluid source position ${(y}_0/h)$. Our model correctly predicts the vertical deformation of the material with a MAE equal to 0.01 times the overall deformation. One potential source of error is the fact that the fluid source in our numerical simulations was fixed in space and did not move upward with solid deformation as it does in the analytical solution. (c-f) Qualitative comparison of the deformation and pressure profiles obtained by our model on a 120 by 150 grid (c and e) and Barry's analytical solution (d and f), respectively. The red arrows in c and e represent the fluid source position in the numerical model. The parameters for these simulations can be found in Table \ref{tbl:1}. \label{fig:9}}
\end{center} 
\end{figure}

As a final illustration of the versatility of the modeling framework developed in Chapter \ref{FullDerivation}, we briefly apply and verify our model for a simple poroelastic system. More precisely, we simulate fluid injection into a symmetric poroelastic material, a model system used to mimic the properties of filters, biological tissues, membranes, and soils \citep{Auton2017a,Barry1995,Nagel2012}. \citet{Barry1997} used Biot theory and the assumption that material stress is directly proportional to the fluids' frictional forces (i.e. drag) to obtain an analytical solution for the pressure and deformation profiles brought about by a constant point fluid source placed at different depths in the system shown in Fig. \ref{fig:9}A. The system studied by Barry (and replicated in our simulations) is defined by a no-slip condition at the lower boundary (y = 0), a constant-pressure fluid point source placed at height $y_0$ on the axis of symmetry (x = 0), and zero-gradient boundary conditions at all other boundaries. Please refer to Table \ref{tbl:1} and \citet{Barry1997} for the complete parametrization of the simulation. 

The complete figure shows that our numerical framework replicates Barry's analytical solutions for the system's maximum deformation in the y-direction to a relatively high degree of accuracy for most values of $y_0$. As stated in the original paper, the minimum at $y_0/h\approx 0.4$ results from two phenomena: first, when the source is near $y/h=1\ $most of the flow leaves through the top boundary, increasing drag near this boundary and maximizing its deformation; second, the maximum vertical deformation is essentially the sum of all local deformations along the axis of symmetry ($x = 0$), meaning that at low values of $y_0$, fluid drag is able to act over more of the solid, thus increasing the overall accumulated vertical deformation. The sum of these two effects is minimized at intermediate values of $y_0,$ leading to the observed minimum. The tendency of our simulations to underestimate surface deformation at high values of $y_0$ likely reflects the fact that the location of the fluid source in our simulations is not affected by the solid's deformation, contrary to the analytical solution, a difference most significant at $y_0/h\approx 1$ because of the first phenomenon noted above.

One notable weakness of the simulations presented here is that the fluid-solid interface is not represented as a sharp step-function, but rather as a continuous interpolation of the stresses between both domains. This loss of sharpness is particularly noticeable as the magnitude of the deformations become large, as seen in Fig. \ref{fig:9}C, perhaps reflecting the breakdown of the validity of the elastic momentum equation in systems with large deformations. This loss of sharpness is much less apparent in the case of the swelling porous media modeled in the previous section, where the swelling pressure acts in favor of a sharp interface. Errors emanating from the lack of sharpness in elastic porous media could be addressed by the introduction of an artificial compression term (as used in VOF models) \citep{Rusche2002a}, higher resolution meshes, or the addition a suitable swelling pressure model. 

\begin{table}[htb]
\begin{center}
  \caption[Parameter values for experiments and simulations in Chapter 1]{Parameter values were obtained from \citet{Bourg2006}, \citet{Diamond1956a}, \citet{Liu2013}, \citet{Mesri1971c}, and \citet{Spearman2017} and are consistent across the simulations unless specified otherwise.}
  \label{tbl:1}
  \includegraphics[width=0.95\textwidth]{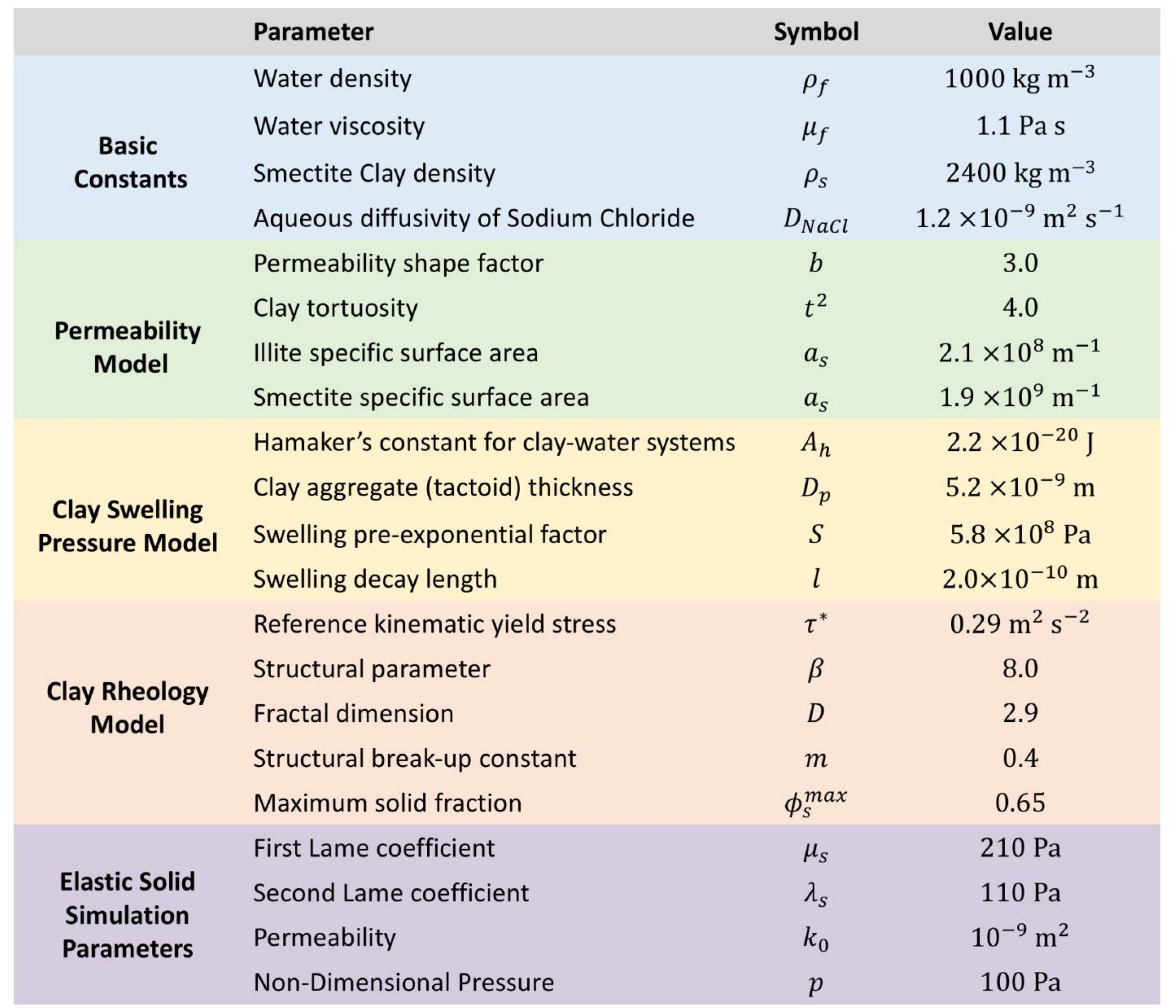}
\end{center}
\end{table}

\section{Conclusions}
\label{Sect:conclusion_DBB}

We have implemented, tested, and verified a multi-scale Hydro-Chemo-Mechanically coupled modeling framework for porous media containing both macropores and deformable microporous regions. We demonstrated the framework's accuracy and versatility by modeling HCM-coupled viscoplastic and poroelastic dynamics. 

The model was implemented and parametrized in particular to simulate the swelling behavior and fluid-induced deformation of water-saturated viscoplastic Na-smectite clay. Predicted couplings between hydrology, mechanics, and salinity were successfully validated against experimental data and numerical models by \citet{DiMaio1996} and \citet{Hewitt2016a} describing a broad range of simulated conditions. The model was then used to predict the permeability of a spherical bead pack as a function of clay content and salinity. For this simple model, a master parametric equation for permeability as a function of clay matrix volume fraction was extracted. This parametric equation was found to be consistent with experimental datasets on the permeability of smectite-coated glass bead packs and of different types of siliciclastic sedimentary rocks. Finally, the model's versatility was demonstrated by qualitatively predicting water-induced formation damage in a propped fracture in clayey rock and, also, by quantitatively predicting the pressure fields and deformation profiles resulting from fluid injection in an elastically-deformable axisymmetric porous medium, for which analytical solutions were developed by \citet{Barry1997}. 

Although the proposed framework has proven relatively accurate in the conditions examined, it comes with several limitations, all of which will be discussed in Chapter \ref{conclusion}. 
\begin{savequote}[75mm]
I have something good... but it's weird ...
\qauthor{Francisco J. Carrillo -- Cyprien Soulaine}
\end{savequote}

\chapter{Applications to Multiphase-Phase Flow in Static Multiscale Porous Media}\label{chp:hybridPhase}

\newthought{In this chapter} we focus exclusively on applying and verifying the multiphase aspect of our model. The objective is to show that the multiscale solver converges effectively towards its two asymptotic limits in static porous media, namely the two-phase Darcy model at the continuum scale and the Volume-Of-Fluid formulation at the pore scale. Additionally, we present two exemplary cases that show the potential and wide-range applicability of our model, where we simulate wave propagation in porous coastal barriers and drainage/imbibition in micro-fractures. The work shown in this chapter represents the second step toward verifying and showcasing the Multiphase DBB framework. This chapter is adapted from \citet{Carrillo2020}.

\section{Darcy Scale Validation \label{sec:Darcy_validation}}

The model’s ability to predict multiphase flow at the Darcy scale is validated against three well-known analytical and semi-analytical solutions. Together, these assessments test for the correct implementation of the relative permeability, gravity, and capillary terms derived in section 2.3. This validation follows the steps outlined in \cite{Horgue2015} for the development and validation of their own multiphase Darcy scale solver: \textit{impesFoam}.  A complete list of parameters used is provided in Tables \ref{table:F} and \ref{table:M}.

\begin{table}[htb!]

\begin{minipage}{.5\linewidth}
    \centering
    \setcellgapes{2pt}
    \makegapedcells
    \medskip
\begin{tabular}{||c  c||} 
\hline
 Property & Value \\ [0.5ex] 
 \hline\hline
 Water Density & \SI{1000}{kg.m^{-3}} \\
 \hline
 Water Viscosity &\SI{1e-3}{Pa.s}\\
 \hline
 Air Density & \SI{1}{kg.m^{-3}}    \\
 \hline
 Air Viscosity & \SI{1.76e-5}{Pa.s} \\ 
 \hline
 Oil Density & \SI{800}{kg.m^{-3}}  \\
 \hline
 Oil Viscosity & \SI{0.1}{Pa.s} \\
 \hline
  Gravity & \SI{9.81}{m.s^{-2}} \\
 \hline
\end{tabular}
 \caption[Table of Fluid Properties]{Table of Fluid Properties}
    \label{table:F}
    
\end{minipage}\hfill
\begin{minipage}{.5\linewidth}
 \centering 
 \setcellgapes{2pt}
 \makegapedcells
 \medskip
 \begin{tabular}{||c  c||} 
 \hline
 Model Parameter & Value \\ [0.5ex] 
 \hline\hline
 \(p_{c,0}\) & 100 Pa \\
 \hline
 m (Van Genuchten) & 0.5  \\ 
 \hline
 m (Brooks-Corey) & 3  \\ 
 \hline
 \(\beta \) (Brooks-Corey) & 0.5  \\ 
 \hline
\end{tabular}
 \caption[Table of Model Parameters]{Table of Model Parameters}
    \label{table:M}
\end{minipage}

\end{table}

\subsection{Buckley-Leverett}
\begin{figure}[htb]
\begin{centering}
\includegraphics[width=1\columnwidth]{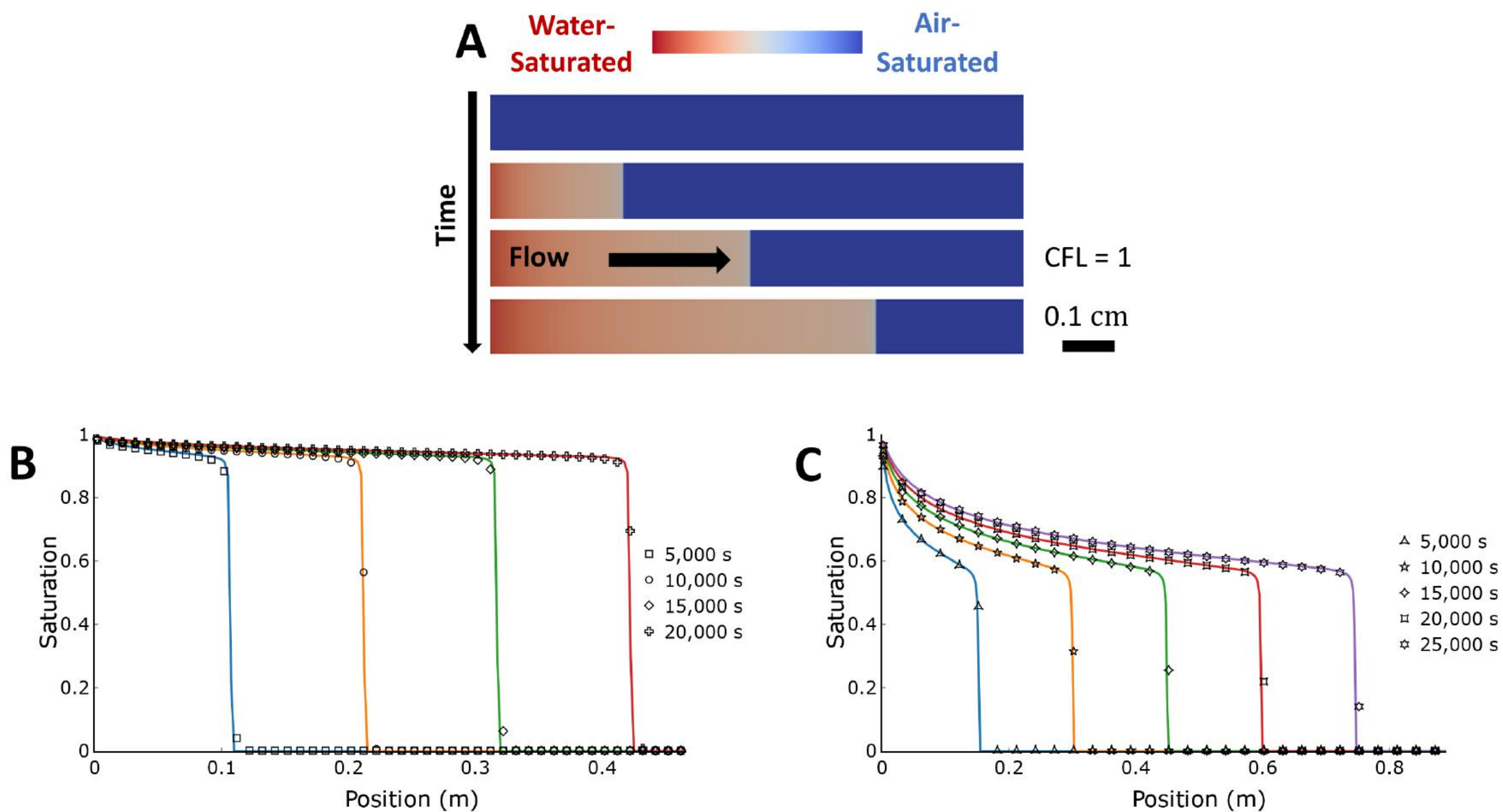}
\par\end{centering}
\caption[Comparison of predicted saturation profiles against Buckley-Leverett's solution]{Comparison of the time-dependent saturation profiles calculated from our numerical framework and Buckley-Leverett's semi-analytical solution for water injection into air-saturated (B) and oil-saturated (C) reservoirs. Figure A is a visual representation of the water saturation in the reservoir over time. Figures B and C show the semi-analytical (lines) and numerical (symbols) solutions of the system when using the Brooks-Corey and Van Genuchten relative permeability models, respectively.\label{fig:Buckley-Leverett}}
\end{figure}

We first consider the well-established Buckley-Leverett semi-analytical solution for two-phase flow in a horizontal one-dimensional system with no capillary effects (4 m long, 2000 cells, \( \phi_f = 0.5, k^{-1}_0=1\times10^{11}\) \si{m^{-2}}). In this case, water is injected into an air-saturated reservoir at a constant flow rate with the following boundary conditions: $\textbf{U}_{water}$ = \SI{1e-5}{\meter\per\second}, $\frac{\partial p_{inlet}}{\partial x} =0 \ \si{Pa.m^{-1}}$, and $ p_{outlet} = 0 \ \si{Pa}$. As water flows into the reservoir, it creates a saturation profile that is characterized by a water shock at its front, an effective shock velocity, and a saturation gradient behind said front. Figure \ref{fig:Buckley-Leverett} shows a good agreement between our numerical solutions and the semi-analytical solutions presented in \cite{Leverett1940} for all three features regardless of the chosen relative permeability model.

\subsection{Gravity dominated Buckley-Leverett }

\begin{figure}[htb]
\begin{centering}
\includegraphics[width=1\columnwidth]{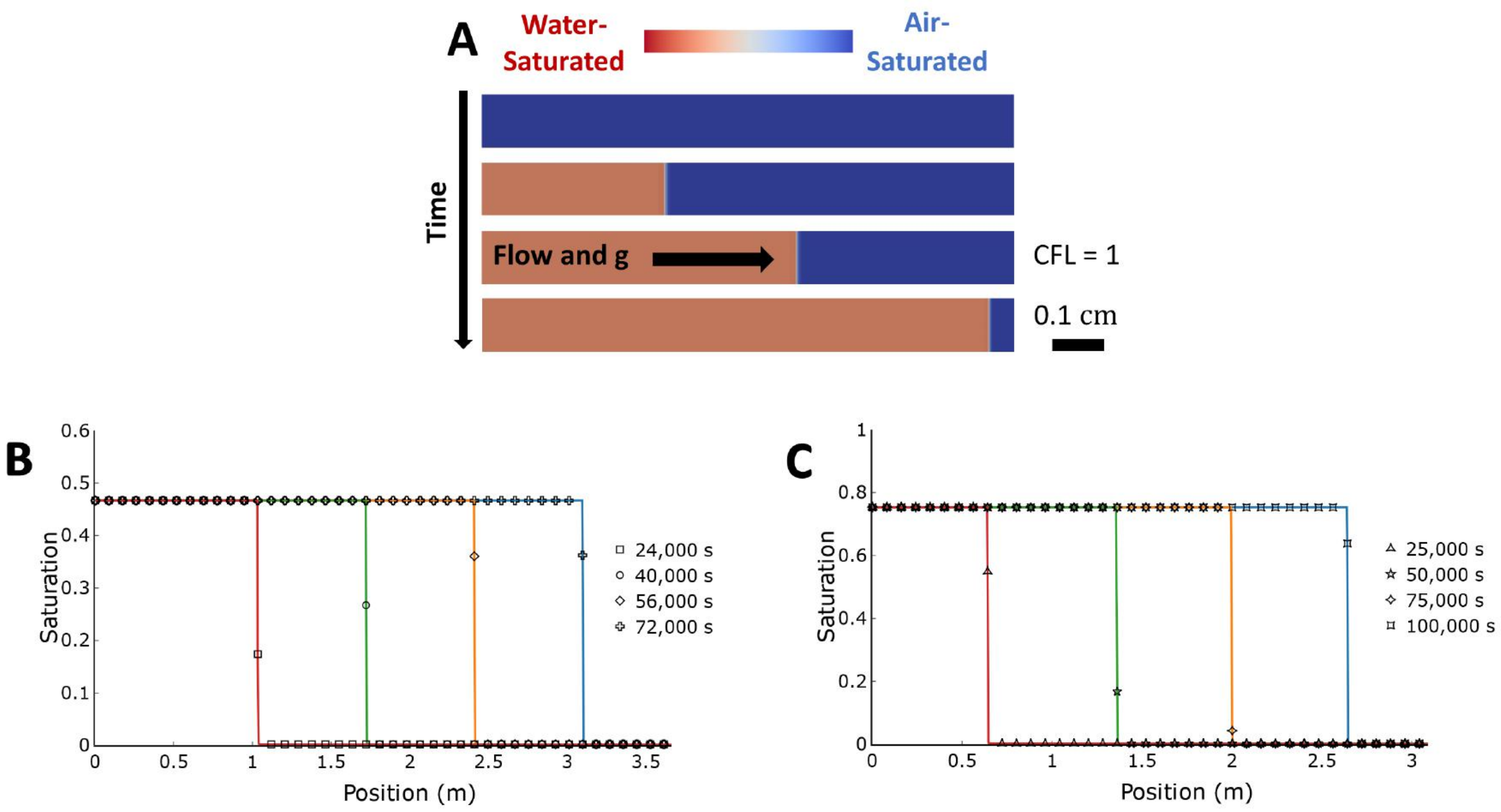}
\par\end{centering}
\caption[Comparison of predicted saturation profiles against Buckley-Leverett's solution with gravity]{Comparison of the time-dependent saturation profiles calculated from our numerical framework and the semi-analytical solution presented in section 4.1.2. Figure A is a visual representation of water saturation in the reservoir over time. Figures B and C show the semi-analytical (lines) numerical (symbols) solutions of the systems parameterized through the Brooks \& Corey and Van-Genuchten relative permeability models, respectively.\label{fig:Buckley-Leverett Gravity}}
\end{figure}

We then tested the exact same air-saturated system, but this time with the addition of gravity in the same direction of the water injection velocity (see Figure \ref{fig:Buckley-Leverett Gravity}). Under these conditions, gravity becomes the dominating driving force and the following equation can be used to calculate the water saturation at the front \citep{Horgue2015}:
\begin{equation}
 \overline{\boldsymbol{v}_l}^l - \frac{k_0 k_{r,l}(\alpha_l^{front})}{\mu_l}\rho_l \boldsymbol{g} = 0,
\label{eq:gravity-front}
\end{equation}
where the symbols are consistent with the ones presented in previous sections. Given the parameters presented in Tables \ref{table:F} and \ref{table:M}, Eq. \ref{eq:gravity-front} is solved iteratively to obtain \(\alpha_{l}^{front}  = 0.467  \)  and \(\alpha_{l}^{front}  = 0.753  \) when using the Brooks-Corey and Van Genuchten relative permeability $(k_{r,l})$ models, respectively (Appendix \ref{multiphase_models}). Figure \ref{fig:Buckley-Leverett Gravity} shows that our numerical solutions agree with the semi-analytical solutions.  

\subsection{Gravity-capillarity equilibrium}

\begin{figure}[htb!]
\begin{centering}
\includegraphics[width=0.9\columnwidth]{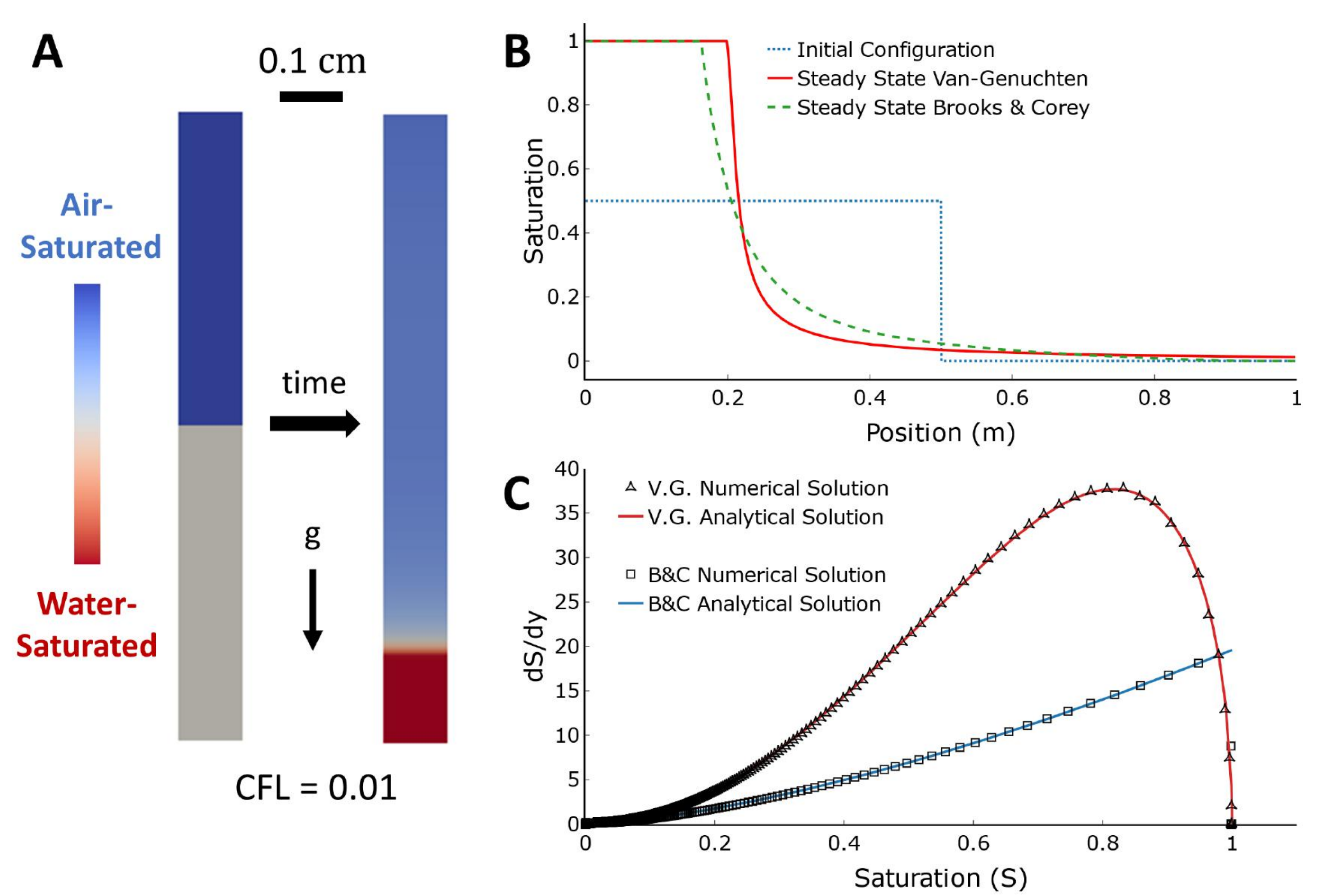}
\par\end{centering}
\caption[Gravity-capillary balance in an unsaturated column]{Comparison of the steady state water saturation profiles calculated from our numerical framework and the analytical solution shown in equation \ref{eq:gravity-capillarity2}.  Figure A is a visual representation of the initial and final water saturation profiles in the reservoir. Figures B and C show the steady state saturation profiles and the resulting equilibrium saturation gradients for both implemented capillary pressure models, respectively. \label{fig:Gravity-Capillarity Balance}}
\end{figure}

Lastly, we tested the validity of the capillary pressure term derived in Eqs. \ref{Ur_def},  \ref{Fc1_def}, and  \ref{Fc2_def} by solving for the steady state saturation profile of a one-dimensional porous column filled with water and air (1 m tall, 1500 cells, \( \phi_f = 0.5, \ k^{-1}_0=1\times10^{11}\) \si{m^{-2}}). Here, the initial water saturation of the column is set far from its thermodynamic equilibrium in a step-wise fashion: the lower half is partially saturated with water (\(\alpha_{water} = 0.5\)) while the upper half is initially dry as shown in Figure \ref{fig:Gravity-Capillarity Balance}A. To ensure proper equilibriation, both fluids are allowed to flow freely through the column's top boundary, but not through its lower one: $\frac{\partial \textbf{v}_{top}}{\partial y} =0 \ \si{m.s^{-1}.m^{-1}}, \frac{\partial p_{top}}{\partial y} = 0 \ \si{Pa.m^{-1}},
\ \textbf{U}_{bottom} = 0 \ \si{m.s^{-1}}, \ p_{bottom} = 0 \ \si{Pa}$. For this simplified case, the theoretical steady-state can be described as the balance between capillary and gravitational forces, where gravity pulls the heavier fluid (water) downwards while capillarity pulls it upwards. This behaviour can be described by the following equation \citep{Horgue2015}:
\begin{equation}
\frac{\partial p_c}{\partial y} = (\rho_g - \rho_l)g_y,
\label{eq:gravity-capillarity}
\end{equation}
which can be rearranged to yield: 
\begin{equation}
\frac{\partial \alpha_l}{\partial y} = \frac{(\rho_g - \rho_l)g_y}{\frac{\partial p_c}{\partial \alpha_l}}.
\label{eq:gravity-capillarity2}
\end{equation}
This last expression allows for the explicit calculation of the equilibrium water saturation gradient by using the closed-form Brooks-Corey or Van Genuchten capillary pressure models to obtain \(\frac{\partial p_c}{\partial \alpha_l}\) (Appendix \ref{multiphase_models}). Figure \ref{fig:Gravity-Capillarity Balance} shows that our numerical model accurately replicates the results obtained from Eq. \ref{eq:gravity-capillarity2} regardless of the chosen capillary pressure model.

\subsection{Darcy scale application: Oil drainage in a heterogeneous reservoir}

\begin{figure}
\begin{centering}
\includegraphics[width=0.6\columnwidth]{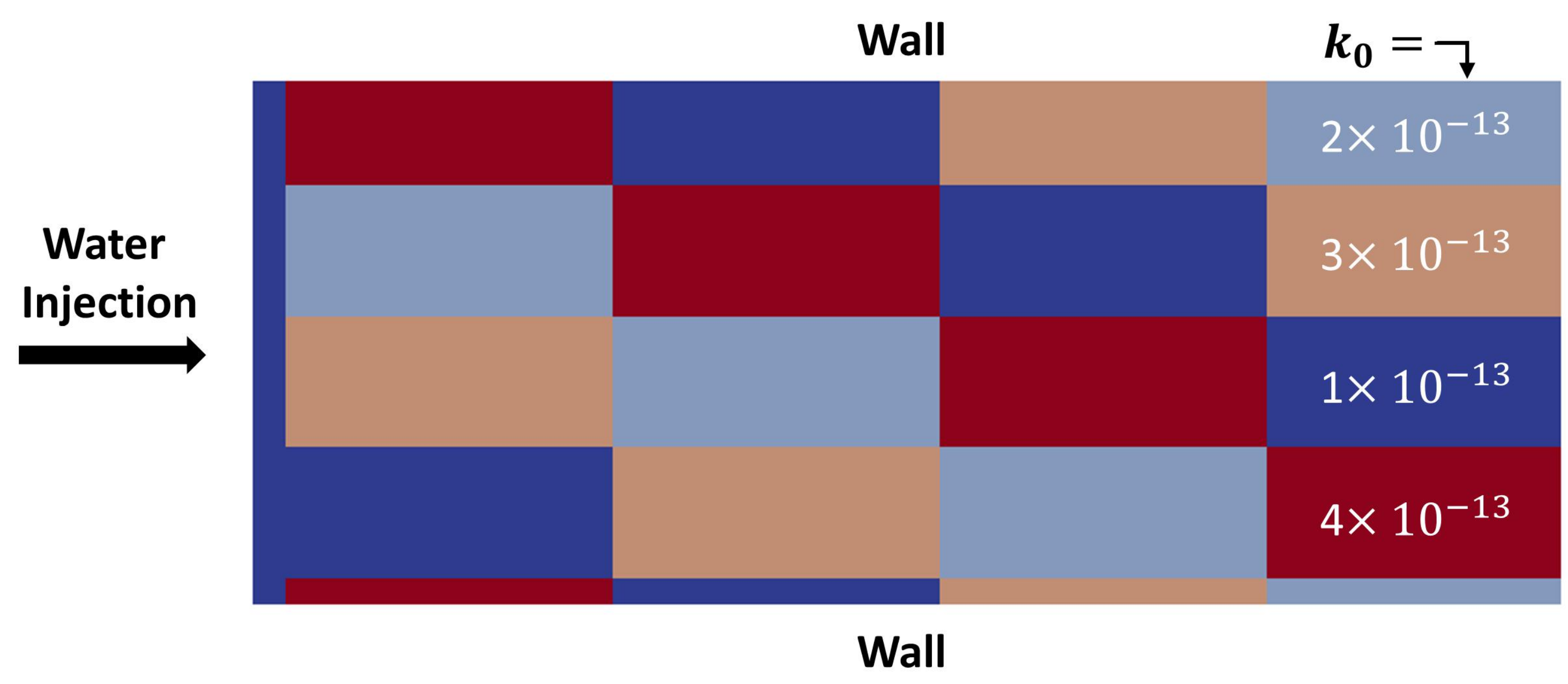}
\par\end{centering}
\caption[Simulation setup for oil drainage within a heterogeneous reservoir.]{Simulation setup for oil drainage within a heterogeneous reservoir. The different colored blocks represent the spatially variable permeability field. \label{fig:permeabilityField}}
\end{figure}

As an illustration of the applicability of our model to more complex systems at the Darcy scale, we simulate water injection into a heterogeneous oil-saturated porous medium (1 by 0.4 m, 2000 by 800 grid, water injection velocity = \SI{1e-4}{\meter \per \second}, \ $p_{outlet} = 0 \ \si{Pa}$). Oil drainage is commonly used in the energy sector, particularly as a form of enhanced oil recovery \citep{Alvarado2010}. Although analytical solutions such as the ones presented above are useful approximations for simple systems, they become greatly inaccurate when modeling complex multi-dimensional systems with spatially heterogeneous permeability. To illustrate this effect, we initialize our reservoir's permeability field as a grid of 0.25 by 0.1 m blocks with \(k_0\) values ranging from \SI{1e-13} to \SI{4e-13}{\meter^{2}} (see Figure \ref{fig:permeabilityField}). The relative permeabilities within the reservoir are modeled through the Van Genuchten model with negligible capillary effects  (Table \ref{table:M}). We note that this case was originally presented in \cite{Horgue2015} for the development of \textit{impesFoam}, a solver that uses the Implicit Pressure Explicit Saturation (IMPES) algorithm to solve the two-phase Darcy model, making it a convenient benchmark for comparison with \textit{hybridPorousInterFoam}. 

\begin{figure}[htb!]
\begin{centering}
\includegraphics[width=1\columnwidth]{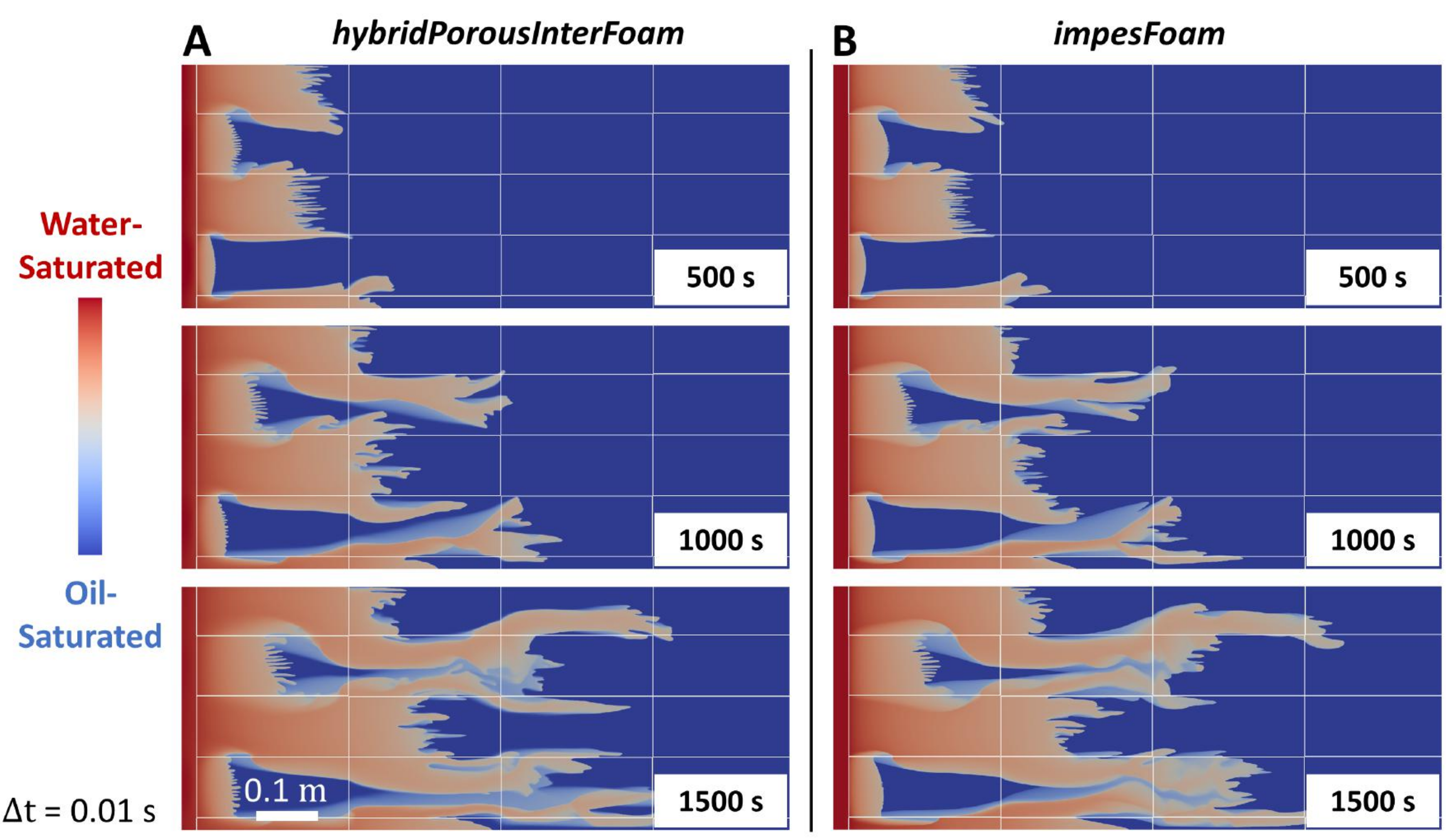}
\par\end{centering}
\caption[Oil drainage in a heterogeneous porous medium solved at the continuum scale]{Oil drainage in a heterogeneous porous medium solved at the continuum scale using \textit{hybridPorousInterFoam} or \textit{impesFoam}. The white rectangular grid represent the blocks with \(k_0\) values ranging from \SI{1e-13} to \SI{4e-13} {m^{2}} as shown in the previous figure.}   \label{fig:viscousFingering}
\end{figure}

Under the aforementioned parametric conditions and with equivalent numerical setups (i.e. same grid, time-stepping, and solver tolerances), Figure \ref{fig:viscousFingering} shows that the simulations performed with \textit{hybridPorousInterFoam} and \textit{impesFoam} develop very similar, yet not perfectly equivalent, saturation profiles. Of particular interest is the development of fingering instabilities that form due to the viscosity difference between the two fluids \citep{Saffman1958,Chen1985}. These instabilities are know to greatly reduce the efficiency of enhanced oil recovery, as they essentially trap residual oil behind the main water saturation front (Figure \ref{fig:viscousFingering}). Previous numerical studies have shown that the evolution of viscous fingering is highly dependent on the model's hyper-parameters, grid refinement, and/or solution algorithms \citep{Ferrari2013,Riaz2006,Horgue2015,Chen1998,Holzbecher2009}. This characteristic explains why \textit{hybridPorousInterFoam} and \textit{impesFoam} develop slightly different viscous fingering instabilities despite having virtually perfect agreement with the previously-presented analytical solutions: the two solvers rely on entirely distinct sets of governing equations, boundary conditions, discretization schemes, and pressure-solving algorithms (PISO vs IMPES). Nevertheless, this example application shows that our solver can readily simulate complex porous systems that have traditionally been modeled using conventional single-scale Darcy solvers.  

\section{Pore Scale Validation \label{sec:pore-scale-verification}}
Having validated all aspects of the model within the porous domain, we now test our model’s ability to recover known multiphase Navier-Stokes solutions within a non-porous domain. This validation follows the steps used in previous validations of multi-phase CFD solvers by \cite{Horgue2014}, \cite{Xu2017}, and \cite{Maes2020} and involves testing the implementation of the imposed contact angle boundary condition against several well-known numerical and analytical cases. Some of the simulation results obtained with our multi-scale solver are compared with simulations performed using \emph{interFoam}, the algebraic VOF solver of OpenFOAM\textregistered. In the following simulations, we implement a static contact angle as an approximate description of multiphase behaviour at solid interfaces, while noting the existence of more sophisticated formulations including dynamic contact angles with viscous bending or surface roughness \citep{Wenzel1936,Cassie1944,Voinov1976,Cox1986,Whyman2008,Meakin2009}.

\subsection{Contact angle on a flat plate}\label{flat_plate}

We first test the implementation of the penalized contact angle within \textit{hybridPorousInterFoam} by initializing several “square” water droplets on a 2-D flat porous plate with negligible permeability (6 by 2.4 mm, 480 by 192 cells, \(k_0^{-1} =\) \SI{1e20}{\meter^{-2}}) and allowing them to reach equilibrium for different prescribed contact angles (\(\theta_{water}\) = \ang{60},\ \ang{90}, \ \ang{150}). These tests are compared against equivalent droplets initialized on conventional non-porous boundaries and solved through \textit{interFoam}. Figure \ref{fig:penalizedContactAngle}A shows excellent agreement between the numerical simulations and the target equilibrium contact angle  \(\theta_{water}\). The lack of a perfectly sharp interface (an intrinsic feature of the VOF method) makes it difficult to accurately measure the contact angle at the solid interface. However, we can confidently state that all our results are within $\ang{5} $ of the target equilibrium contact angle. These tests are virtually identical to the ones shown in \cite{Horgue2014} and are consistent with their results. 

\subsection{Capillary rise}\label{cap_rise}

\begin{figure}[htb!]
\begin{centering}
\includegraphics[width=1\columnwidth]{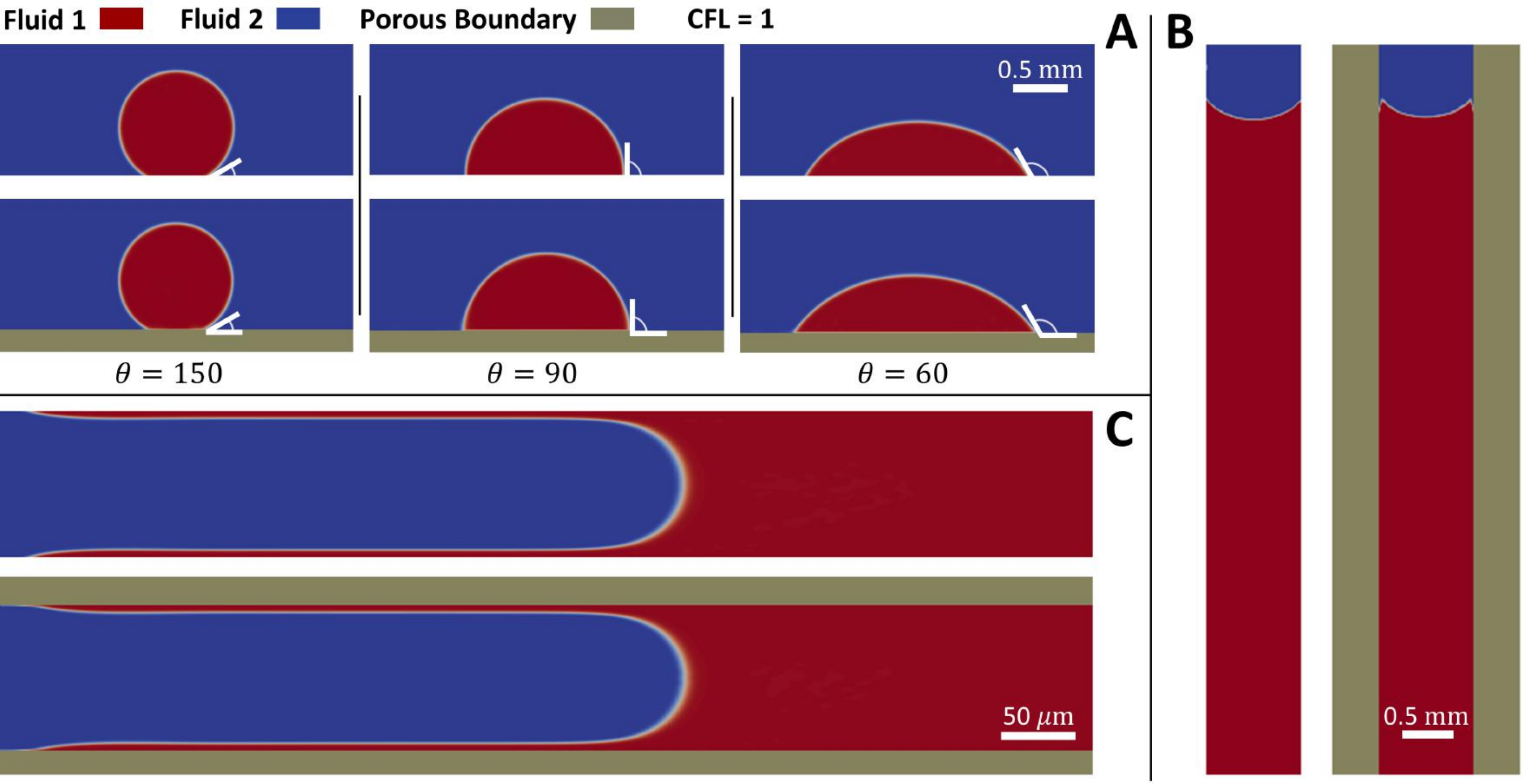}
\par\end{centering}
\caption[Compilation of all test cases performed for the verification of the solver within the Navier-Stokes domain]{Compilation of all test cases performed for the verification of the solver within the Navier-Stokes domain. Parts A, B, and C refer to the experiments described in Sections \ref{flat_plate}, \ref{cap_rise}, and \ref{Tay_film}, respectively. When present, the shaded walls show the porous boundaries used in \textit{hybridPorousInterFoam}, as opposed to the standard boundary (no-slip boundary condition at an impermeable wall) using in \textit{interFoam}. For reference and easy comparison, the white lines in Part A show the input equilibrium contact angle.  \label{fig:penalizedContactAngle}}
\end{figure}

As a second classic test for the correct implementation of multiphase flow at the pore-scale, we model water capillary rise in an air-filled tube (1 by 20 mm, 40 by 400 cells, \(\theta_{water}\) = \ang{45}) and measure the steady-state position of the water meniscus. To ensure a proper numerical setup, the tube's lower boundary is modeled as an infinite water reservoir and its upper boundary as open to the atmosphere. To prevent initialization bias, the meniscus is initialized  about 2 mm lower than the theoretical equilibrium height of 10 mm, which is given by the following equation \citep{Jup1719}:

\begin{equation}
h_{eq.} = \frac{\sigma cos(\theta)}{R \rho_l g_y },
\label{eq:capillaryHeight}
\end{equation}
where $R$ is the tube's radius. We then numerically simulate the system with \textit{hybridPorousInterFoam} and \textit{interFoam}, using impermeable porous boundaries with the former (\(k_0^{-1} =\) \SI{1e20}{\meter^{-2}}) and conventional sharp boundaries with the latter. Figure \ref{fig:penalizedContactAngle}B shows the steady state configurations of both cases, which have a meniscus height of 8.8 mm. According to Eq. \ref{eq:capillaryHeight}, this height is equivalent to an imposed contact angle of \ang{52}, a small yet significant difference to the imposed contact angle of \ang{45}. We are not the first to note that \textit{interFoam} (the standard pore scale multiphase flow solver in OpenFOAM\textregistered{}) presents minor inaccuracies in its ability to impose a prescribed contact angle \citep{Horgue2014, Grunding2019}. The comparisons presented here show that our solver's accuracy in this regard is similar to that of \textit{interFoam}. 

\subsection{Taylor film}\label{Tay_film}

We now model the drainage of ethanol (\(\mu_{eth.}\) =\SI{1.2e-3}{Pa.s}, \(\rho_{eth.}\) = \SI{789}{kg.m^{-3}}) by air in a 2-D micro-channel (800 by 100 $\mu$m, 280 by 116 cells, \(\theta_{eth.}\) = \ang{20}, injection velocity $U$ = 0.4 m/s, $p_{outlet}=0 \ \si{Pa}$). Under these circumstances, a  film forms at the channel's boundaries as a result of competing viscous and capillary forces at the solid interface (see Figure \ref{fig:penalizedContactAngle}C). The height of this film is given by the following analytical solution, which we use as a benchmark to verify our numerical simulations \citep{Aussillous2000},
\begin{equation}
\frac{h_{film}}{R} = \frac{1.34 Ca^{2/3}}{1+ 3.35 Ca ^{2/3}},
\label{eq:FilmHeight}
\end{equation}
where $Ca$ is the capillary number defined as $Ca=\frac{\mu_{eth.} U}{\sigma}$. We can solve Eq. \ref{eq:FilmHeight} with the given simulation parameters to obtain a film thickness of 4.35 $\mu$m. Simulations of this system performed using \textit{hybridPorousInterFoam} with impermeable porous boundaries ($k_0^{-1} =$ \SI{1e20}{\meter^{-2}}) and \textit{interFoam} with conventional boundaries yield a value of 4.50 $\mu$m, representing a relative error of about 3\% or 0.15 $\mu$m. These tests and their results are consistent with numerical simulations reported by \cite{Graveleau2017} and \cite{Maes2020} using \textit{interFoam}. 

\subsection{Pore scale application: Oil drainage in a complex pore network}

\begin{figure}[htb]
\begin{centering}
\includegraphics[width=1\columnwidth]{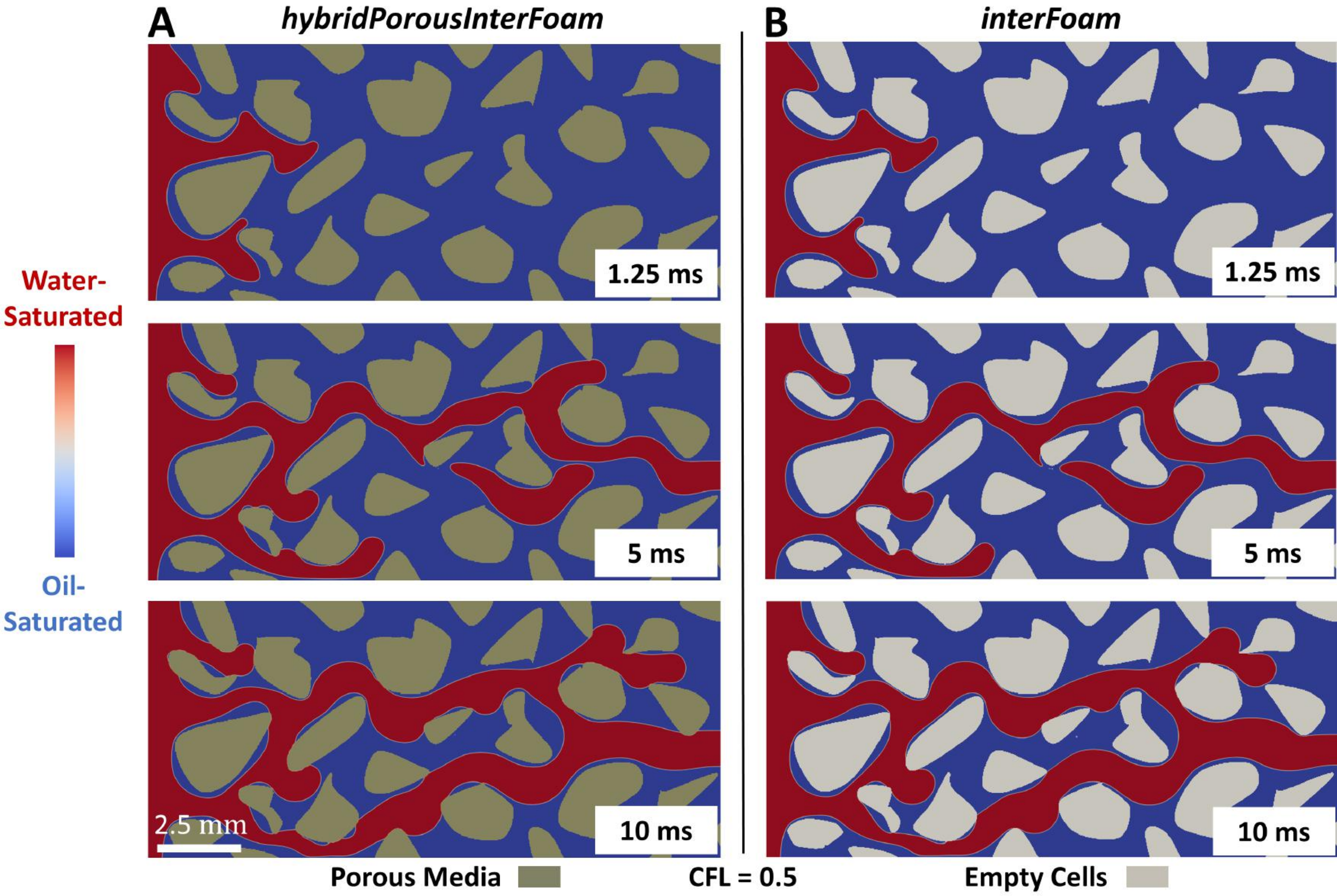}
\par\end{centering}
\caption[Oil drainage in a complex porous medium solved at the pore scale]{Oil drainage in a complex porous medium solved at the pore scale using \textit{hybridPorousInterFoam} and \textit{interFoam}. The shaded sections represent solid grains (modeled using $\phi_f$ = 0.001 and $k_0^{-1} =$ \SI{1e20}{\meter^{-2}} in \textit{hybridPorousInterFoam}) and the blue and red colors represent oil and water, respectively.\label{fig:drainageImbibition}}
\end{figure}

As we did at the end of the Darcy scale verification section, we now illustrate our model's applicability to more complex systems by presenting a simulation of oil drainage, this time at the pore scale. The relevance of the simulated system follows from our previous illustrative problem, as this is simply its un-averaged equivalent at a smaller scale. The complexity of the simulated system (1.7 by 0.76 mm, 1700 by 760 cells, water injection velocity = 0.1 m/s, \(\theta_{oil}\) = \ang{45}, $p_{outlet} = 0 \ \si{Pa}$) stems from the initialization of a heterogeneous porosity field as a representation of a cross-section of an oil-wet rock. Here, the porosity is set to one in the fluid-occupied space and close to zero in the rock-occupied space (See Fig. \ref{fig:drainageImbibition}A). This allows for the solid grains to act as virtually impermeable surfaces ($k_0^{-1} =$ \SI{1e20}{\meter^{-2}}) with wettability boundary condition \citep{Horgue2014}. To verify the accuracy of our solver, we solved an equivalent system with \textit{interFoam} by removing the rock-occupied cells from the mesh and imposing the same contact angle at these new boundaries through conventional methods. 

Figure \ref{fig:drainageImbibition} shows that the results of the two simulations are practically identical, down to the creation of the same preferential fluid paths and the same droplet snap-off at 5 ms. Nevertheless, there are minor differences in the results, where some interfaces are displaced at slightly different rates than their counterparts (see the upper right corner at 10 ms). We attribute these slight differences to the differing implementations of the contact angle at the solid boundaries. We invite interested readers to find this case in the accompanying toolbox and to refer to the extensive literature on this topic for further discussion on numerical and experimental studies of drainage and imbibition \citep{Lenormand1988,Ferrari2013,Datta2014,Roman2016,Zacharoudiou2018, Liu2019,Singh2019} .

\section{Hybrid Scale Applications \label{sec:applications}}

The complete body of work presented in the previous two sections verifies the capability of our model to perform simulations of multiphase flow in complex porous media at the pore and continuum scales.  We now show how \textit{hybridPorousInterFoam} makes the simulation of hybrid scale multiphase systems a fairly straightforward endeavor (a task that is challenging to perform with conventional methods). The main challenge when modeling these systems can be summarized by the following question: How can we rigorously model the porous interface between coupled Navier-Stokes and Darcy scale domains? Although this is still an open question, we attempt to approximate an answer by guaranteeing three of its necessary components in the present micro-continuum framework: first, mass conservation across the interface; second, continuity of stresses across the interface and; third, a wettability formulation at the interface. The first two components are intrinsic features of the solver which have been proven necessary and sufficient to accurately model single phase flow in hybrid scale simulations \citep{Neale1974} and have been used as closure conditions in previous multiphase models \citep{Lacis2017,Zampogna2019}. The latter, as explained in the pore scale validation section, is roughly approximated through a constant contact angle boundary condition. We recognize that these components represent an approximation to the complete description of the boundary. Nevertheless, to the best of our knowledge, there does not exist a better way to model this interfacial behavior, a testament to the novelty and potential of the proposed modeling framework. 

The following illustrative cases are used to show our model's capability to simulate multiphase systems at the hybrid scale. They are also included as tutorial cases in the accompanying toolbox. 

\subsection{Wave propagation in a coastal barrier}

Coastal barriers are used throughout the world to prevent flooding, regulate water levels, and protect against inclement weather \citep{Morton2002}. Accurate simulation of water interaction with these barriers is challenging as it requires predicting the behavior of open water at large scales (Navier-Stokes) while also resolving small-scale multiphase effects within the barrier itself (Darcy flow). 

\begin{figure}[htb!]
\begin{centering}
\includegraphics[width=1\columnwidth]{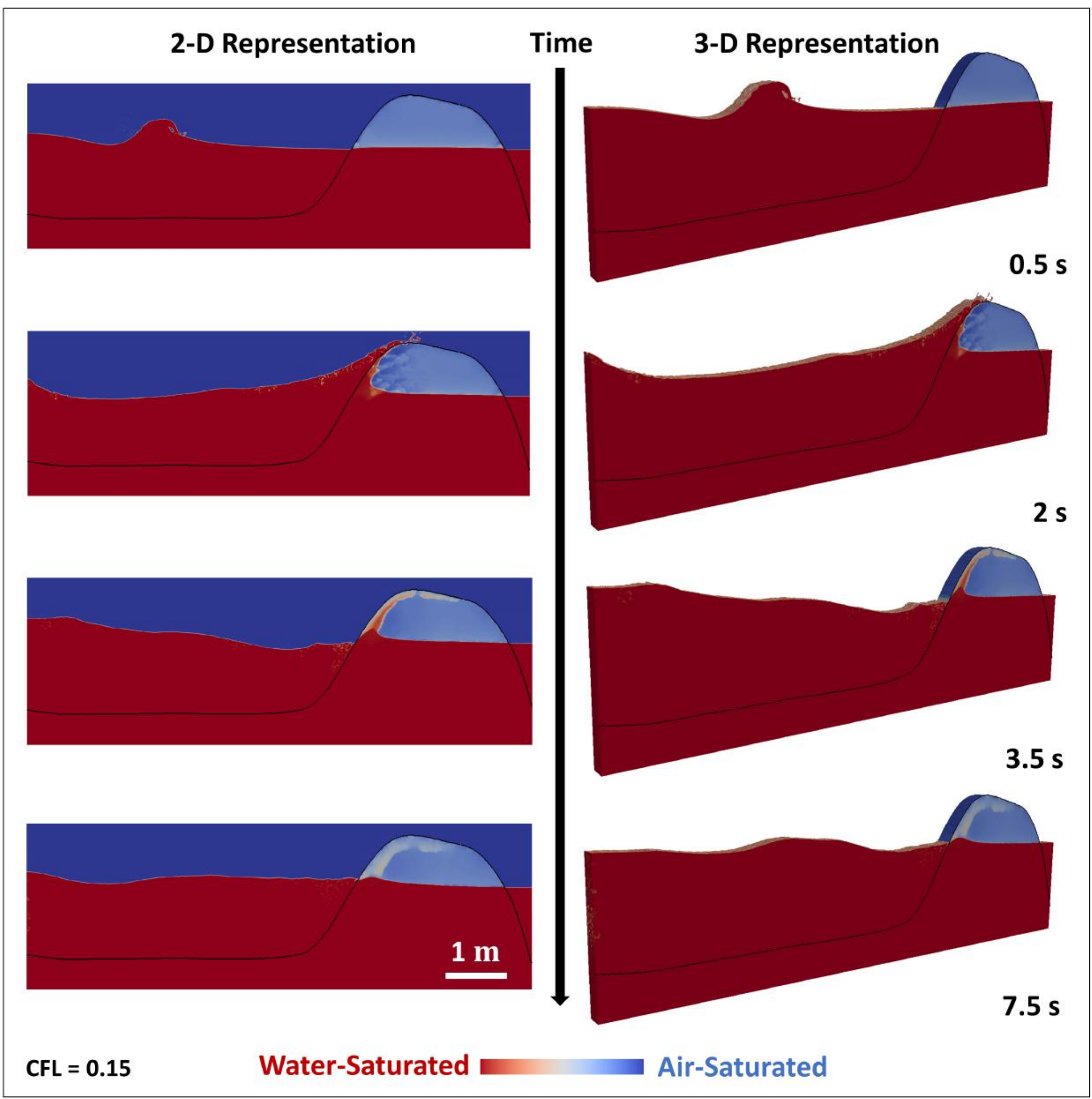}
\par\end{centering}
\caption[Wave dissipation in a porous coastal barrier]{Coastal barrier simulation at different times. The thin black line represents the boundary of the porous solid: $\phi_f$  and $k_0^{-1}$ are set to 0.5 and \SI{5e7}{\meter^{-2}} below said line and to 1 and 0 above it, respectively. The 2-D representation shows a plane that cuts through the middle of the 3-D simulation. \label{fig:coastalBarrier}}
\end{figure}

We created a three-dimensional coastal barrier (8.3 by 2.7 m by 0.25 m, 1660 by 540 by 50 cells) by initializing a heterogeneous porosity field in the shape of a barrier ($k_0^{-1} =$ \SI{5e7}{\meter^{-2}}, $\phi_{f,barrier}$ = 0.5) and by setting the water level such that it partially covers the barrier (see Figure \ref{fig:coastalBarrier}). In this particular case, we chose not to impose a contact angle at the barrier-water interface as its effects would be minimal when compared to macroscopic gravitational effects (Bond Number $= \frac{\Delta \rho (\textnormal{Length Scale})^2 g_y}{\sigma} >> 1$). To ensure proper initialization, we allowed the water saturation profile on the above-water section of coastal barrier to reach its capillary-induced steady state (similarly to the capillary rise simulation presented in Section \ref{cap_rise}). This process was modeled using the Van Genuchten relative permeability and capillary pressure models (m = 0.8, $p_{c,0}$ = 1000 Pa). After equilibration, we started the simulation by initializing a wave as a rectangular water extrusion above the water surface. To ensure proper wave propagation behavior, we tuned the simulation's numerical parameters (discretization schemes, linear solvers, time stepping strategy) according to the guidelines established in \cite{Larsen2019}.

The results from this simulation show that we can simultaneously model coupled wave and Darcy dynamics in three dimensions. The snapshots shown in Figure \ref{fig:coastalBarrier} illustrate how water saturation within the porous domain is controlled by the crashing of waves, gravity, and capillary effects. The associated wave absorption and dissipation cycle brought about by the porous structure is repeated every few seconds with lowering intensity until the initial configuration is eventually recovered. To the best of our knowledge, Figure \ref{fig:coastalBarrier} shows the first existing numerical simulation coupling multiphase flow with real capillary effects at two different scales without the use of different meshes, solvers, or complex interfacial conditions. Other models such as \textit{olaFlow} have been developed to simulate similar wave dynamics with coastal barriers \citep{Higuera2013}. Many of these models rely on the assumption that the pores within the coastal barrier are large ($>$10 cm), meaning that they can reasonably ignore capillary effects within the porous medium. Contrastingly, our model makes no such assumption, meaning it can also be used to model coastal barriers with arbitrarily small pores (such as in sand or gravel structures) and also should be applicable to other types of groundwater-surface water interaction \citep{Maxwell2014}.

\subsection{Drainage and imbibition in a fractured microporous matrix}

A second conceptually similar, yet completely different hybrid scale application of \textit{hybridPorousInterFoam} involves the injection of fluids into fractured porous materials. Accurately capturing the fluid behavior in these systems is especially challenging due to the fact that it requires accounting for multiphase effects simultaneously within the fracture (Navier-Stokes), in the surrounding microporous matrix (Darcy), and at the porous boundary (the contact angle implementation).

Here, we model drainage and imbibition in a water-wet fracture system, where we inject air into a 90\% water-saturated microfracture in the former and we inject water into a 90 \% air-saturated microfracture in the latter (1.2 by 0.5 mm, 1200 by 500 cells, \(\theta_{water}\) = \ang{45}, fluid injection velocity = 0.1 m s$^{-1}$, $p_{outlet} = 0 \ \si{Pa}$). The relative permeabilities and capillary pressures in the heterogeneously-initiated porous domain (\(\phi_f = 0.5,\)  $k_0^{-1} =$ \SI{4e12}{\meter^{-2}}) are modeled through the Brooks-Corey model with m = 3, $p_{c,0} = 100 \ \si{Pa}$, and $\beta=0.5$. 

\begin{figure}[htb]
\begin{centering}
\includegraphics[width=0.85\columnwidth]{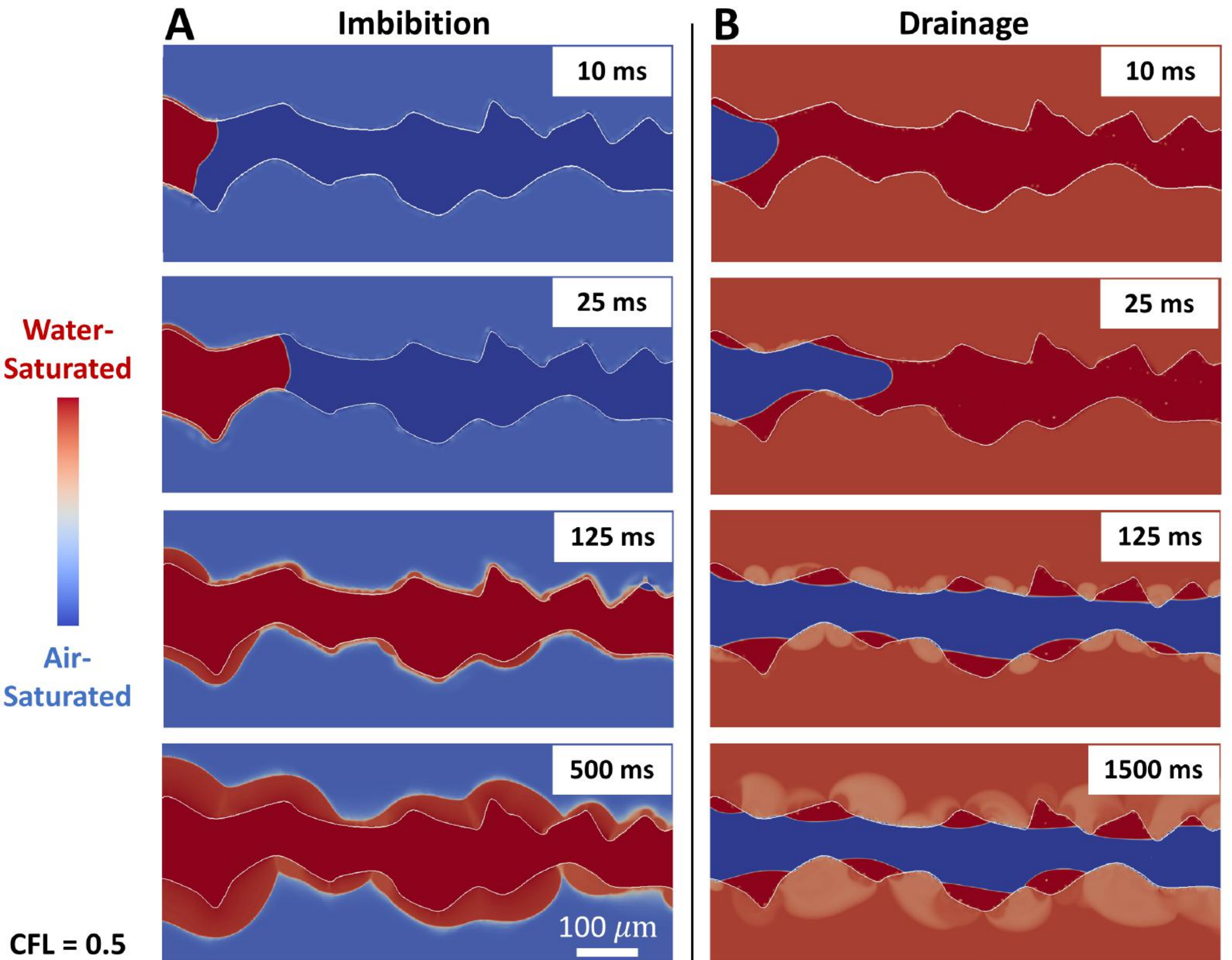}
\par\end{centering}
\caption[Drainage and imbibition in a microporous fracture]{Drainage and imbibition in a microporous fracture. Shades of blue and red represent of the degree of air and water saturation, respectively. The thin white line shows the fracture outline (i.e. the fluid-solid interface), which separates the open fracture (\(\phi_f =1, k_0^{-1} =0 \)) from the porous fracture walls (\(\phi_f =0.5, \) $k_0^{-1} =$ \SI{4e12}{\meter^{-2}}) located above and below it.   \label{fig:fracture}}
\end{figure}

Figure \ref{fig:fracture} presents the results of these simulations and illustrates how strongly multi-scale wettability effects can influence simulations results. In both cases, the injected fluid is able to invade the microporous matrix, but the mechanism through which it does is completely different. In the case of water injection (imbibition), the wetting contact angle boundary condition encourages complete water saturation of the whole fracture such that air is completely displaced by time = 125 ms. Furthermore, throughout the whole process, the microporous capillary pressure acts as an additional driving force for water invasion into the surrounding microporous matrix, leading to the almost complete saturation of the whole system by time = 500 ms. This process has some strong analogies to the flow of water in into hygroscopic materials\citep{Zhou2019}. 

The drainage case is slightly less intuitive, yet conceptually more interesting. Here, the contact angle and microporous capillary pressures act against the invasion of air into the fracture and into the surrounding porous material, respectively. The result is that the air cannot effectively displace water from the fracture, leading to the trapping of water droplets in fracture ridges. Initially, these droplets act as barriers that prevent air entry into the porous matrix (see time = 125 ms). However, as the flow-induced pressure gradient pushes air into the porous matrix, the water saturation in the pores surrounding the droplets decreases. The system then responds by increasing the capillary pressure at the porous interface, which eventually leads the water droplets to imbibe into the matrix. Lastly, we highlight the clear time scale separation between the imbibition and drainage cases, as the invading interface progresses about three times more slowly within the microporous matrix in the latter case.  

Several similar dual porosity models have been proposed to model the types of effects illustrated in Figure \ref{fig:fracture}, but never in this way or to this degree of detail \citep{Douglas1991,DiDonato2003}. Many of these models rely on a description of fractures as single-dimensional features with high porosity and permeability values within a pure Darcy scale simulation \citep{Nandlal2019,Yan2016}. Although very useful, many of these simulations ignore the geometric capillary effects and non-linear couplings presented above. Our approach can therefore be seen as the missing link between pore-scale modeling and Discrete Fracture Networks \citep{Karimi-Fard2016} and as a useful tool for refining the transfer functions used in these large scale models.

\section{Conclusions}

We have successfully tested and verified a multiscale model for two-phase flow in porous media. This modeling framework and its open-source implementation \textit{hybridPorousInterFoam} can be used to simultaneously model multiphase flow at two different length scales: a Darcy Scale where sub-voxel fluid-fluid interactions within a porous medium are modeled through relative permeability and capillary pressure constitutive models and a pore scale (or Navier-Stokes scale) where the solid material is non-porous and fluid-fluid interactions are modeled through a continuum representation of the Young-Laplace equation. Furthermore, our model is able to do this through the use of a single momentum conservation equation without the need to define different meshes, separate solvers/domains, or complex interfacial conditions. The proposed framework is an accurate and straightforward way to introduce the physics of two-phase flow in porous media in CFD softwares.

Through this study, we showed that our model can successfully simulate multiphase Darcy and Navier-Stokes flow to the same standard (and with the same assumptions and limitations) as conventional single-scale solvers \textit{impesFoam} and \textit{interFoam}. The coupling between the two scales at porous interfaces is handled by ensuring mass conservation and continuity of stresses at said boundary, as well as by implementing a constant contact angle wettability condition. We then leveraged all these features to show that our model can be used to model hybrid scale systems such as wave interaction with a porous coastal barrier and drainage and imbibition in a fractured porous matrix. 

Although the proposed formulation represents a significant advance in the simulation of multiscale multiphase systems, we note that further study is required in particular to properly and rigorously model the multi-scale porous interface. The implemented interface, as it stands, has been shown to accurately predict single phase flow into porous media \citep{Neale1974}, impose static contact angles over porous boundaries \citep{Horgue2014}, and approximate multiphase flow in porous media \citep{Lacis2017}. However, its accuracy when modeling multiphase flow at rough porous interfaces is still an open question, as there does not currently exist a rigorous formulation to model such behaviour. The derivation, implementation, and verification of such a boundary condition and the inclusion of erosion, chemical reactions \citep{Soulaine2017,Soulaine2018}, and solid mechanics \citep{Carrillo2019} into this framework will be the focus of subsequent chapters and investigations.  
\begin{savequote}[75mm]
All models are wrong, but some are useful \qauthor{George E. P. Box}
\end{savequote}

\chapter{Application to Multiphase Flow in Multiscale Deformable Porous Media}\label{chp:MDD_Applications}

\newthought{Most of the underlying components} of the Multiphase DBB approach described in Chapter \ref{FullDerivation} have been already tested and verified. Chapter \ref{chp:singlePhaseDBB} validated the momentum exchange terms as an effective coupling mechanism between a \textit{single} fluid phase and a \textit{deformable} plastic or elastic porous medium. The effects of confining and swelling pressures on porous media were also examined in said study. Then, Chapter \ref{chp:hybridPhase} extensively validated the extension of the Darcy-Brinkman equation into multiphase flow within and around \textit{static} porous media by comparison with reference test cases in a wide range of flow, permeability, capillarity, and wettability conditions. Therefore, the only thing left to validate is the ability of the Multiphase DBB model to accurately predict the behavior of multiscale systems that exhibit coupling effects between multiple fluids and a deformable porous matrix. This chapter focuses on addressing this last point by using our model to replicate known solutions to coupled problems, with a particular emphasis on fracture mechanics. Additionally, it presents two exemplary cases that show the potential and wide-range applicability of our coupled model, where we simulate wave propagation in poroelastic coastal barriers and surface cracking/subsidence as a result of subsurface hydraulic fracturing. The work presented here is adapted from \citet{Carrillo2020MDBB} and represents the third and last step in the journey of verifying and showcasing the Multiphase DBB model. 

\section{Model Validation}\label{Model_Val}

We begin with two validation cases relating to multiphase poroelasticity and the coupling between solid deformation and fluid pressure. Then, we proceed with two poroplastic cases that validate this framework for multiscale plastic systems. Finally, we conclude with two additional cases that verify the implementation of the capillary force interaction terms. These cases can be found in the accompanying CFD simulation package. 

\subsection{Terzaghi Consolidation Problem} \label{Terzaghi_Val}

\begin{figure}[htb!]
\begin{center}
\includegraphics[width=0.97\textwidth]{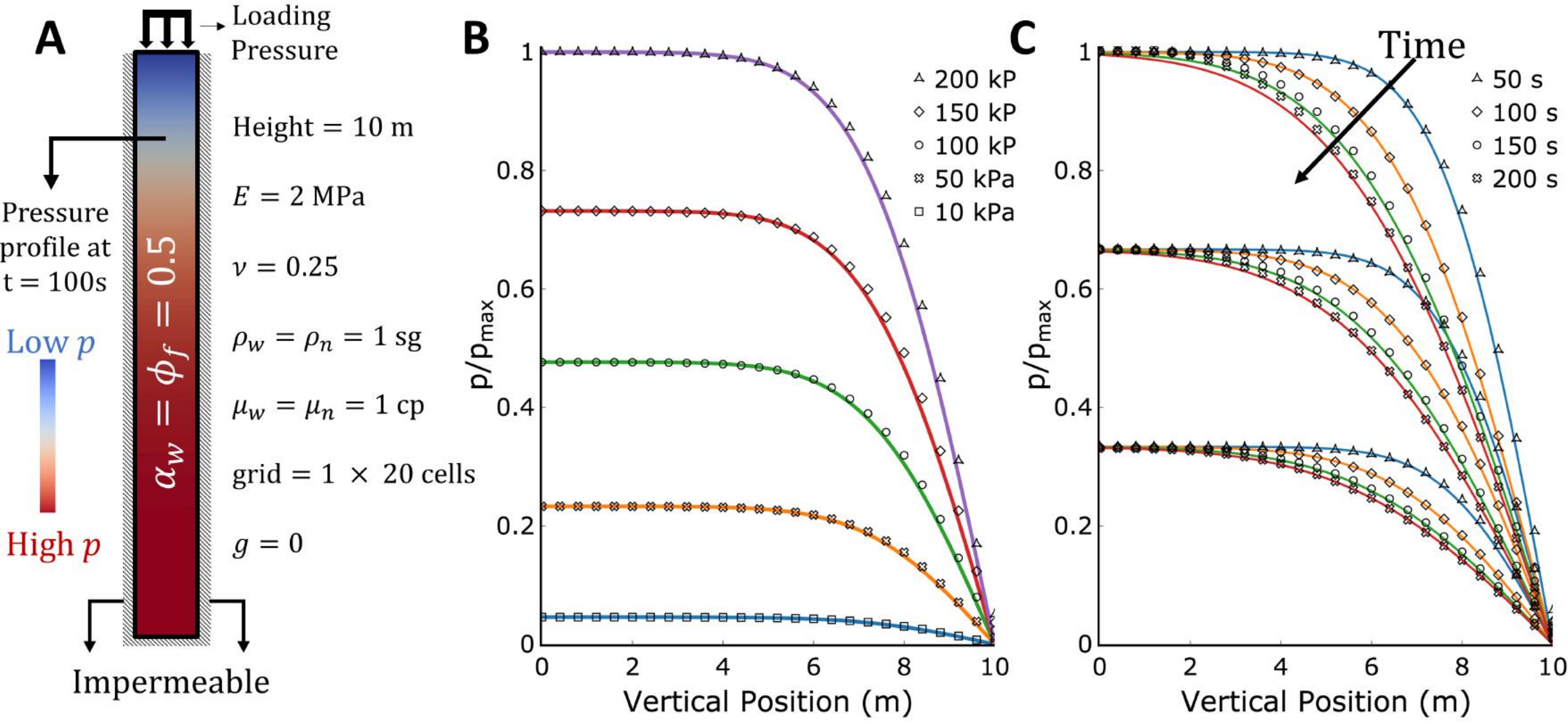}
\caption[One-dimensional Terzaghi consolidation problem]{One-dimensional Terzaghi consolidation problem. (A) Simulation setup. (B) Analytical (solid lines) and numerical (symbols) pressure profiles at $t = 100 \ \unit{s}$ for different loading pressure values. (C) Time-dependent pressure profiles for different column porosity values (From top to bottom: ${\phi}_s=0.75,\ 0.5,\ 0.25$).  \label{fig:Terzaghi}}
\end{center} 
\end{figure}

The Terzaghi uniaxial compaction test has been extensively used as a benchmark for the validation of numerical codes relating to poroelasticity \citep{Terzaghi1996}. Its main utility is to test the accuracy of the solid-fluid couplings that relate fluid pressure to solid deformation and \textit{vice versa}. The problem consists of a constrained saturated elastic porous medium that is abruptly compressed from its upper boundary by a constant uniaxial load (Figure \ref{fig:Terzaghi}). This creates a sudden increase in pore pressure, which is then dissipated by flow through the upper boundary (all other boundaries have impermeable boundary conditions). In the case of a one-dimensional porous medium, the resulting temporal and spatial evolution in fluid pressure can be described by the following simplified analytical solution \citep{Verruijt2013}.

\begin{equation}\label{Ver_Analytical}
\begin{aligned}
\frac{p}{p_{max}}={\mathrm{erf} \left(\frac{h-z}{2\sqrt{c_vt}}\right)\ }\ \ \ \mathrm{for\ }\ \ \frac{c_vt}{h^2}\ll 1\ 
\end{aligned}
\end{equation}

\noindent where ${c}_{v}=({k}_0E\ (\nu-1))/(\eta (2{\nu}^{2}+\nu-1))$ is the consolidation coefficient, ${k}_0$ is permeability, $E$ is Young's modulus, $\nu$ is Poisson's ratio,  $\eta$ is the fluid's unit weight, $h$ is the column height, and $z$ is the vertical coordinate. Our equivalent numerical setup is shown in Figure \ref{fig:Terzaghi}. The values of the relevant parameters in our simulations are $h=10 \ \unit{m}$, ${k}_0=5\times{10}^{-11} \ \unit{m^2}$, $E=2 \ \unit{MPa}$, and $\nu=0.25$. To show the accuracy of our model across different conditions, the loading pressure was varied from $10$ to $200 \ \unit{kPa}$ (Figure \ref{fig:Terzaghi}B) and the porosity from $0.25$ to 0.75 (Figure \ref{fig:Terzaghi}C). Lastly, the column was partially saturated $({\alpha}_{w}=0.5)$ with fluids with equal densities $({\rho}_{f}=1000 \ \unit{kg/m^3})$, viscosities $(\mu_f=1 \ \unit{cp})$, and negligible capillary effects. This last point allowed for testing the validity of the fluid-solid couplings irrespective of the simulated phases without violating any of the assumptions present in the analytical solution. Our numerical results show excellent agreement with Equation \ref{Ver_Analytical} for all tested conditions. 

\subsection{Pressure Oscillation in Poroelastic Core} \label{occilation}

\begin{figure}[htb!]
\begin{center}
\includegraphics[width=0.97\textwidth]{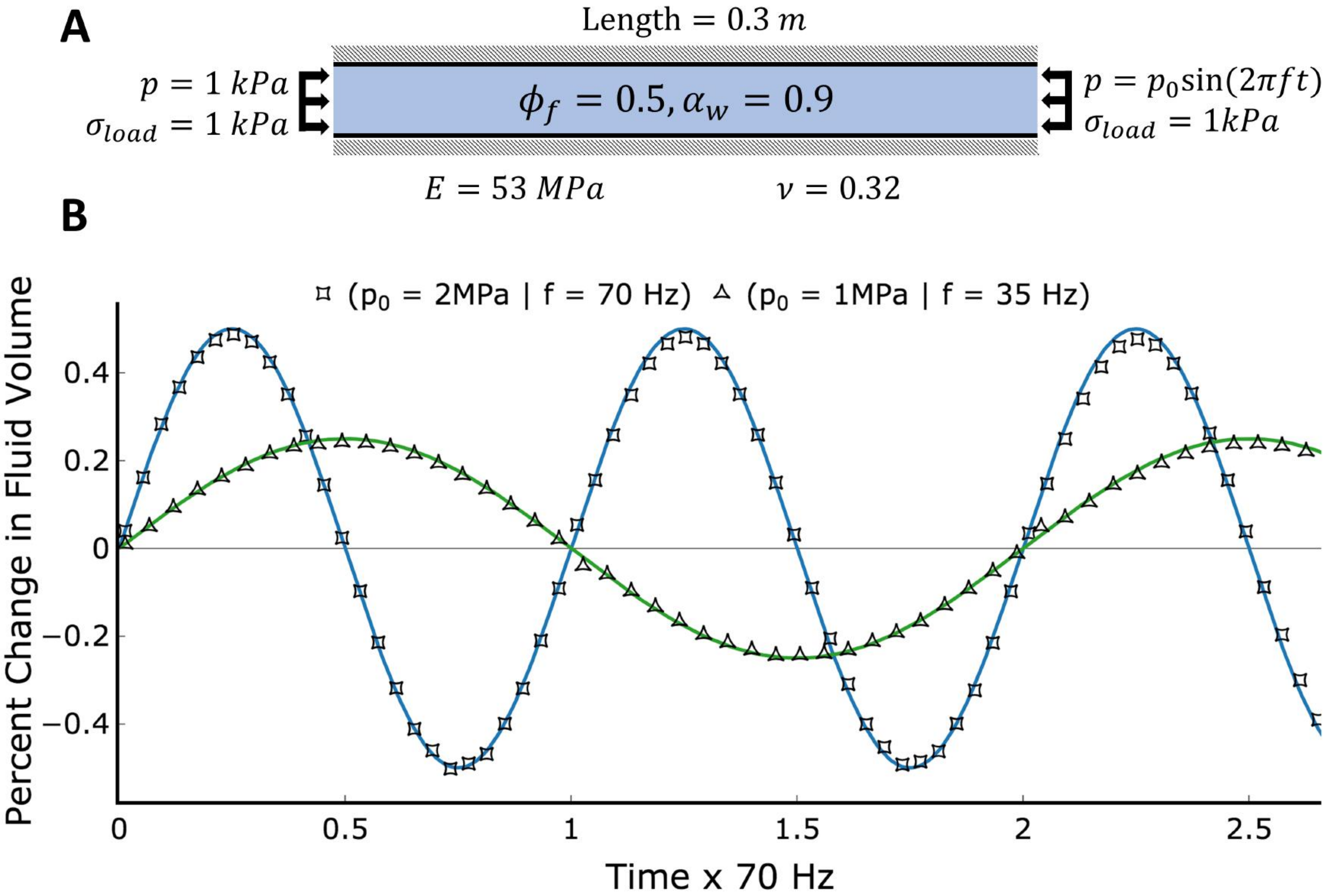}
\caption[Pressure oscillation in poroelastic core]{Change in fluid content of an oscillating poroelastic core. (A) Simulation setup. (B) Semi-analytical (solid lines) and numerical solutions (symbols) for the percent change in the core's fluid volume as a function of time. \label{fig:sposito}}
\end{center} 
\end{figure}

This verification quantifies the effects of the seismic stimulation of a poroelastic core saturated with water and trichloroethene (TCE). Our simulations follow the experimental and numerical set up described in \citet{Lo2012}, where a horizontal one-dimensional sand core ($0.3 \ \unit{m}$ long, $30\times 1$ grid cells, ${\phi}_f=0.5$, ${\alpha}_{w}=0.9$, ${k}_0=1.1\times{10}^{-11} \ \unit{m^2}$) is subjected to constant uniaxial compression and oscillatory pore pressure variations imposed by time-dependent boundary conditions (Figure \ref{fig:sposito}). In this case, flow is allowed through both boundaries, which results in a system that continuously undergoes a relaxation-compression cycle. The ensuing cyclical change in the core's fluid content as a function of time can be described by a semi-analytical solution first derived in \citet{Lo2012} and reproduced in Appendix \ref{analytical_occilation}.

For our matching simulations, the porous structure's Young's modulus was set to $E=53 \ \unit{MPa}$ and its Poisson ratio to $\nu=0.32$. Here, water density was ${\rho}_{w}=1000\ \unit{kg/m^3}$, water viscosity was $1 \ \unit{cp}$, TCE density was ${\rho}_{TCE}=1480\ \unit{kg/m^3}$, and TCE viscosity was ${\mu}_{TCE}=0.57 \ \unit{cp}$.  Furthermore, the pressure at the left boundary was held at $p=1 \ \unit{kPa}$ while the pressure at the right boundary was set by $p={p}_0{sin \left(2\pi ft\right)}$, with ${p}_0=1$ - $2 \ \unit{MPa}$ and $f=35$ - $70 \ \unit{Hz}$. Lastly, the core was uniaxially compressed through a constant stress of $1 \ \unit{kPa}$ applied at both boundaries. A comparison between our numerical solutions and Lo's semi-analytical solution is presented in Figure \ref{fig:sposito}, yielding excellent agreement for all tested cases.  

We note that the Multiphase DBB formulation should be able to describe `Slow" Biot pressure waves caused by the relative motion of the solid and fluid phases which occurs at much higher frequencies than the ones simulated here (i.e. 10 MHz). However, capturing these effects and modelling ``Fast/Compressional" pressure waves would require the implementation of a pressure-velocity coupling algorithm that allows for compressible flow \citep{Lo2012}. Such an endeavour is outside the scope of this dissertation. 

\subsection{Capillary Pressure Effects in a Poroelastic Column} \label{elastic_pc_Val}

\begin{figure}
\begin{center}
\includegraphics[width=0.97\textwidth]{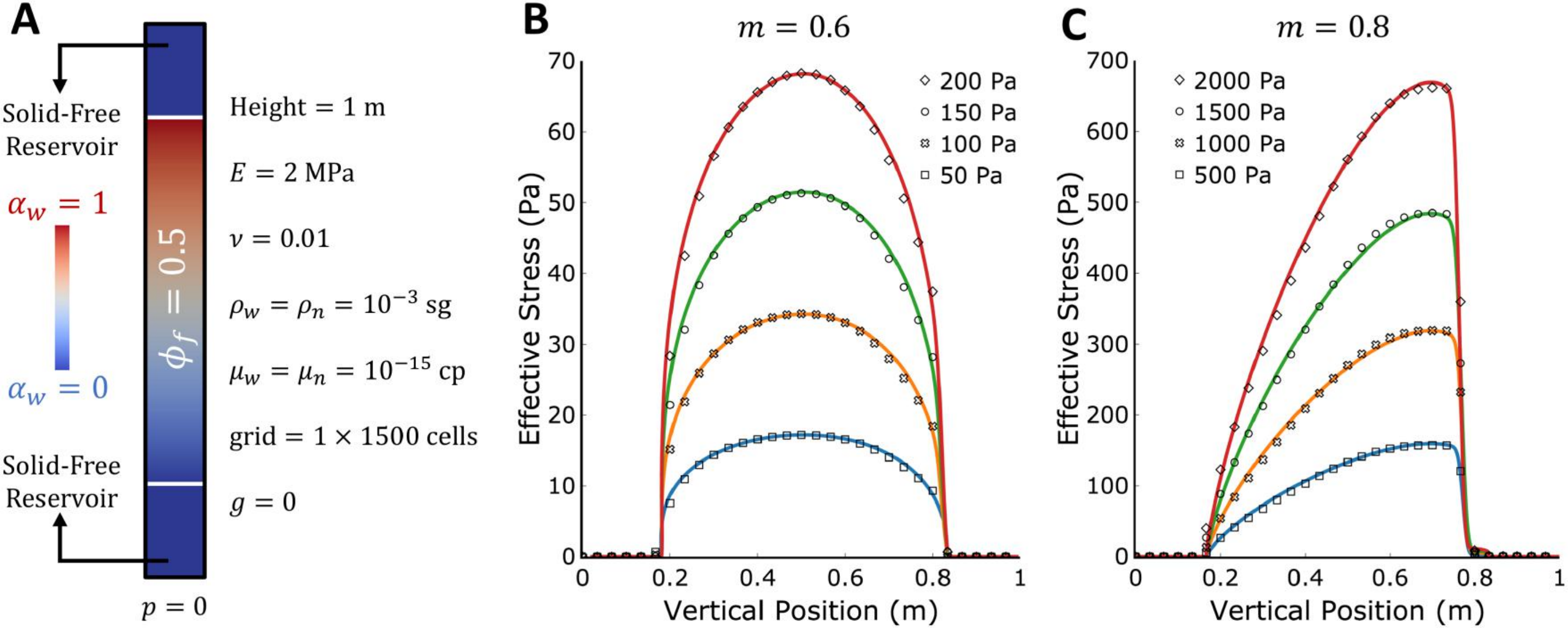}
\caption[Capillary effects in a poroelastic column]{Capillary effects in a poroelastic column. (A) Simulation setup. (B $\&$ C) Analytical (solid lines) and numerical (symbols) effective stress profiles for different capillary pressure values ($p_{c,0} = 50$ to $2000$ Pa) and Van Genuchten coefficients ($m = $ $0.6$ and $0.8$). \label{fig:elastic_pc}}
\end{center} 
\end{figure}

Having verified the two-way coupling between solid deformation and fluid pressure, we now verify the implementation of the capillary pressure terms within the solid mechanics equation. To do so, we simulate a poroelastic column ($1$ m tall, $1500$ cells, $\phi_f = 0.5$) bounded by two non-wetting fluid reservoirs at its upper and lower boundaries. The column is initialized with a linear saturation profile spanning from $\alpha_w = 0$ to $1$ (see Fig. \ref{fig:elastic_pc}). Fluid saturation is kept fixed by not solving Equation \ref{Eq:finalSaturation}, and the mobilities of both fluids are set to very high values ($M_i=1 \times 10 ^{10} \ \unit{m^3 /kg.s}$) to minimize drag-related effects. Under these conditions, the solid's effective stress is exclusively controlled by capillary effects and is described by the following analytical solution:

\begin{equation}
    \mathrm{Effective \ Stress} = \phi_s \times \alpha_w \times p_c
\end{equation}

We used the Van Genutchen capillary pressure model with $m = 0.6$ or $0.8$ and $p_{c,0} = 50$ to $2000$ Pa to calculate the solutions to said problem. The resulting agreement between the numerical and analytical solutions, shown in Fig. \ref{fig:elastic_pc}, confirms the accuracy of the fluid-solid capillary pressure coupling implemented in our model. Furthermore, the transitional behaviour of the effective stress at the macroscopic solid-fluid interface confirms the applicability of the interfacial condition described in Section \ref{sec:interfacial_condition}: as expected, solid stresses are dictated by standard elasticity theory in the porous region and become negligible in solid-free regions. 

Given that the fluid-solid couplings in a poroelastic solid are now verified, we proceed to verify said terms for poroplastic materials. 

\subsection{Fluid Invasion and Fracturing in a Hele-Shaw Cell }\label{HeleShaw_Val}

\begin{figure}[htb!]
\begin{center}
\includegraphics[width=0.8\textwidth]{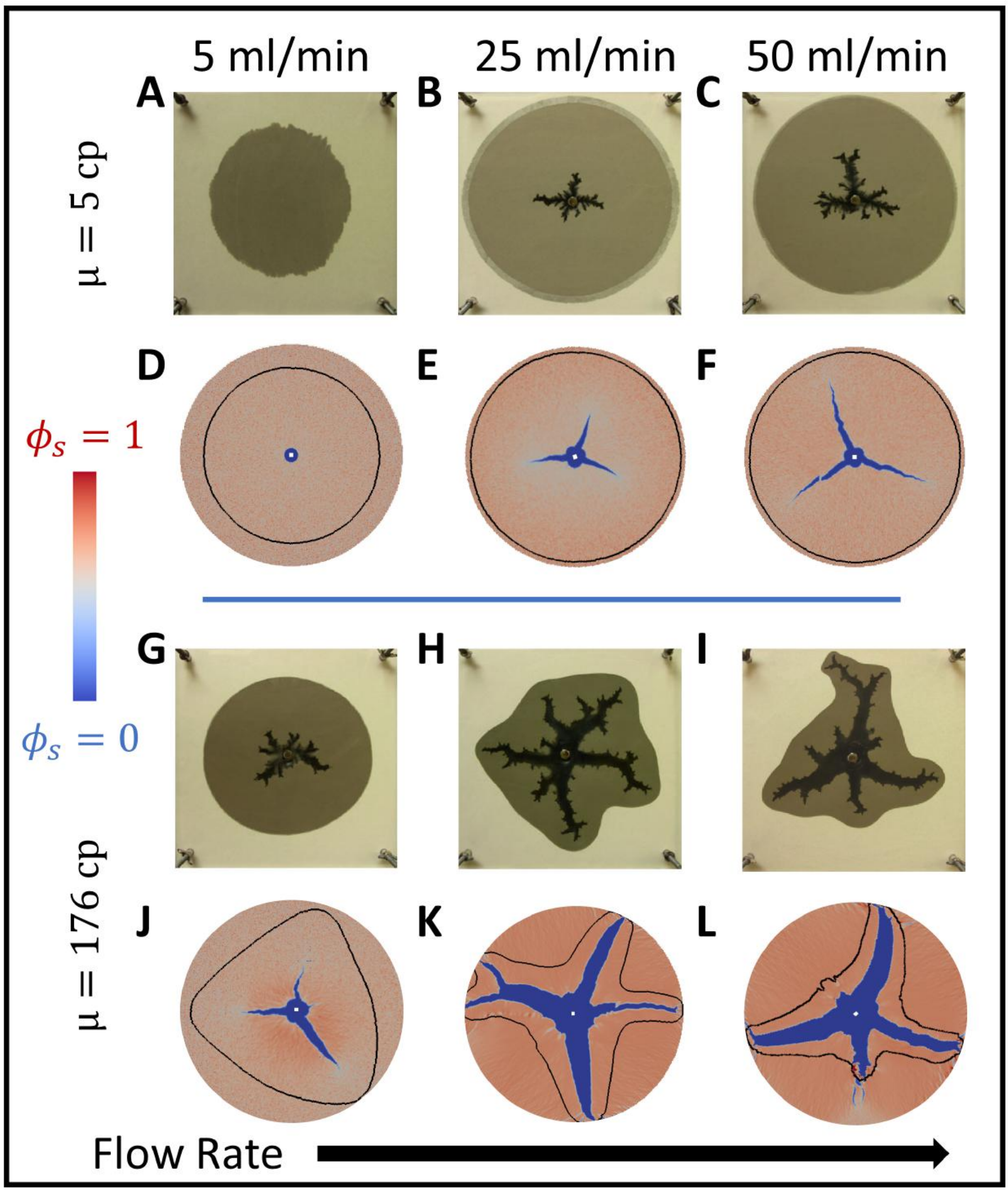}
\caption[Experimental and simulated fracturing in a Hele-Shaw cell]{Comparison of experimental (A, B, C, G, H, I) and simulated (D, E, F, J, K, L) fracturing in a Hele-Shaw cell. The color bar represents the solid fraction within the simulations (where red implies a pure solid and blue pure fluids) and the black lines represent the advancing glycerin saturation front. The experiments shown here are part of the results presented in \citet{Huang2012a}. \label{fig:hele-shaw}}
\end{center} 
\end{figure}

The third verification case (and the first poroplastic case) consists in the qualitative replication of a set of fracturing experiments that examined the injection of aqueous glycerin into dry sand within a 30 by 30 by 2.5 cm Hele-Shaw cell \citep{Huang2012a,Huang2012}. These experiments are inherently multiscale, in that the characteristic length scale of fractures in this system ($\sim \unit{cm}$) is orders of magnitude larger than that of pores within the microporous matrix ($\sim 100 \  \mathrm{\mu m} )$. They are also multiphysics, as they clearly exemplify the drag-controlled transition from Darcy flow within the porous medium to Stokes flow in the open fractures and the coupling between the hydrodynamics of fluid flow and the mechanics of fracture propagation (Fig. \ref{fig:hele-shaw}). 

The experimental setup involved the injection of aqueous glycerin at various flow rates $q$ between $5$ and $50$ ml/min while also varying the fluid's viscosity ${\mu}_{gly}$ between $5$ and $176 \ \unit{cp}$ for different experiments. Our numerical simulations were parameterized using measured values of the glycerin-air surface tension ($\gamma=0.063  \ \unit{kg/s^2}$), the density of pure glycerin (${\rho}_{gly}=1250 \ \unit{kg/m^3}$), the density of air (${\rho}_{air}=1 \ \unit{kg/m^3}$), the viscosity of air (${\mu}_{air}=0.017 \ \unit{cp}$), and the average radius and density of sand grains ($100\ \mathrm{\mu m}$ and $2650 \ \unit{kg/m^3}$, respectively). To mimic the sand's experimental configuration and permeability, the simulated solid fraction field was set to a random initial normal distribution such that ${\phi}_s=0.64\ \pm 0.05$ and the permeability was modelled as a function of the solid fraction through the Kozeny-Carman relation with ${k}_0=6.7 \times {10}^{-12} \ \unit{m^2}$. Relative permeabilities were calculated through the Van Genutchen model with the Van Genuchten coefficient $m$ set to $0.99$ (see Appendix \ref{multiphase_models}), while capillary pressures were deemed negligible (as $2\gamma{r}^{-1}\ll \mu{k}^{-1}{U}_{f}L)$. Finally, the porous medium was modeled as a continuous Hershel-Bulkley-Quemada plastic (Appendix \ref{rheology_mod}) with kinematic yield stress ${\tau}_0=16.02 \ \unit{m^2/s^2}$ \citep{Quemada1977}. Plasticity was used as the preferred mode of solid rheology due to its ability to account for the compressive and irreversible effects caused by fracturing within these experiments \citep{Abou-Sayed2004,VanDam2002}.

Numerically speaking, the simulations were carried out in a 30 by 30 cm 2-D grid (500 by 500 cells) with constant velocity and zero-gradient pressure boundary conditions at the inlet, zero-gradient velocity and zero pressure boundary conditions at the boundary walls, and a solid velocity tangential slip condition at all boundaries (i.e. the solid cannot flow across the boundaries, but the fluids can). Lastly, to enable a closer comparison between our 2D simulation and the 3D experiment we added an additional drag term to the fluid momentum equation equal to $12\mu {a}^{-2}{\boldsymbol{U}}_{f}$, which accounts for viscous dissipation through friction with the walls in a Hele-Shaw cell with aperture $a$ \citep{Ferrari2015}.

As shown in Figure \ref{fig:hele-shaw}, a dramatic transition in the mode of fluid invasion is observed with increasing fluid injection velocity and viscosity. At low flow rates and low viscosity $\left(q=5 \ \unit{ml/min},\ \mu=5 \ \unit{cp}\right)$, there is no discernible solid deformation and the main mode of fluid flow is through uniform invasion of the porous medium (Figure \ref{fig:hele-shaw}A). At intermediate flow rates and low viscosity ($q=25 \ \unit{ml/min}$ to $30 \ \unit{ml/min},\  \mu=5 \ \unit{cp}$), we still observe a uniform invasion front, but small fractures begin to appear (Figure \ref{fig:hele-shaw}B, C). At high viscosity $\left(\mu=176 \ \unit{cp}\right)$, we see clear fracturing patterns preceded by a non-uniform fluid invasion front (Figure \ref{fig:hele-shaw}H, I).

Figure \ref{fig:hele-shaw} shows that our simulation predictions are qualitatively consistent with the experiments presented in \citet{Huang2012a} with regard to both the stability of the capillary displacement front and the observed fracturing transition behavior. As suggested above, accurate prediction of this transition requires not only proper handling of fluid-fluid interactions (surface tension and relative permeability effects), but also accurate descriptions of their relationship with solid mechanics (drag) and the proper implementation of a solid rheological model that can replicate irreversible and unstable fracturing processes. We note that in our simulations, fracture initialization and propagation are predicted based on continuum-scale equations for the rheology and mechanics of the bulk microporous solid, with no specific treatment of grain-scale mechanics. Grid-level instabilities are brought about by the normally distributed porosity and permeability fields, as shown in Appendix \ref{fracturing_instabilities}. The microstructural differences between the experiments and our simulations (most clear in Figure \ref{fig:hele-shaw}C, F, and H, K) likely arise at least in part from the fact that the solid is modelled as a continuum rather than a granular material. 

This section demonstrates that the multiphase DBB model can be used to replicate and predict the main mode of fluid flow and solid deformation within fracturing systems. A comprehensive study of the controlling parameters for multiphase fracturing in the presence of both viscous and capillary stresses will be the focus of the next chapter (Chapter \ref{chapter:PRL}). 

\subsection{Modeling Fracturing Wellbore Pressure}\label{frac_validation}

\begin{figure}[htb!]
\begin{center}
\includegraphics[width=0.8\textwidth]{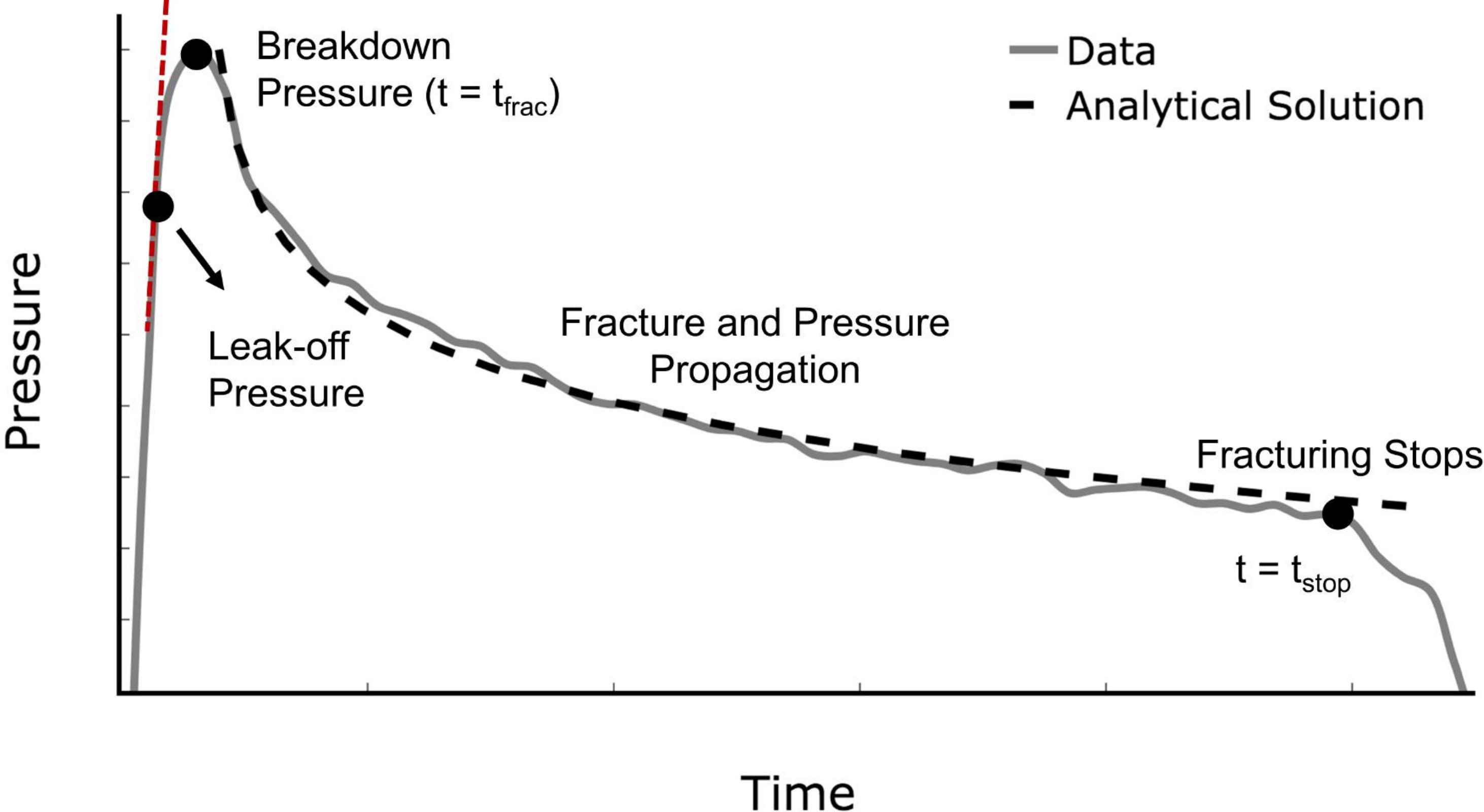}
\caption[Conceptual representation of wellbore pressure evolution during fluid-induced fracturing of low permeability rocks]{Conceptual representation of wellbore pressure evolution during fluid-induced fracturing of low permeability rocks. In this section, we are interested in modeling the behavior between $t_{frac}$ and $t_{stop}$. \label{fig:concept}}
\end{center} 
\end{figure}

Having shown that our model can qualitatively predict fracturing behavior, we now aim to determine whether it can do so in a quantitative matter. As depicted in Figure \ref{fig:concept}, fluid-induced fracturing of low-permeability rocks proceeds through the following well-established series of stages: First, fluid pressure increases linearly as fracturing fluid is injected into the wellbore. Second, as wellbore pressure increases and approaches the leak-off pressure, a small amount of pressure is propagated by fluid leakage into the rock. Third, fluid pressure continues to increase until it reaches the breakdown pressure, at which point it is high enough to fracture the rock. Fourth, a fracture is initiated and propagates; the wellbore pressure slowly decreases. Fifth, injection stops, fracture propagation stops, and wellbore pressure rapidly dissipates \citep{Abass1996,Abou-Sayed2004,Huang2012a,Papanastasiou2000,Santillan2017}.

In this section we aim to numerically replicate the time-dependent fracturing wellbore pressure during fracture propagation (i.e., the fourth stage outlined above) as described by an analytical solution presented in \citet{Barros-Galvis201}.

\begin{equation}\label{frac_Analytical}
p_{well}= p_0 - \frac{\mu q}{4\pi k_0 h} \left[ln\left(\frac{t k_0 \tau_0}{\phi_f \mu r_{well}^2}\right) + 0.81 \right]
\end{equation}

\noindent where $t$ is the time elapsed since fracture initialization, $q$ is the fluid injection rate, $p_{well}$ is the wellbore pressure, $p_0$ is the minimum pressure required for starting a fracture (a function of the solid's yield stress $\tau_0$), $h$ is the formation thickness, and $r_{well}$ is the wellbore radius. The remaining variables follow the same definitions described earlier. 

The general numerical setup is almost identical to the one presented in the previous section. The key difference is that we now inject aqueous glycerin into a strongly-non wetting (and thus almost impermeable) porous material. This is done to ensure an accurate replication of the analytical solution and its related assumptions, where fracturing is the main mode of fluid flow and there is virtually no fluid invasion into the porous matrix. The exact simulation parameters are $q=46$ to $110 \ \unit{ml/min}$, ${\tau}_0=0.2$ or $2 \ \unit{m^2/s^2}$, ${k}_0=6.7\times {10}^{-11}$ or $6.7\times{10}^{-12} \ \unit{m^2}$,  ${\mu}_{gly}=5 \ \unit{cp}$, and $m=0.05$. Note that low values of $m$ indicate that the porous formation is strongly non-wetting to the injected fluid. All other parameters are as in the previous section. 

\begin{figure}[htb!]
\begin{center}
\includegraphics[width=0.97\textwidth]{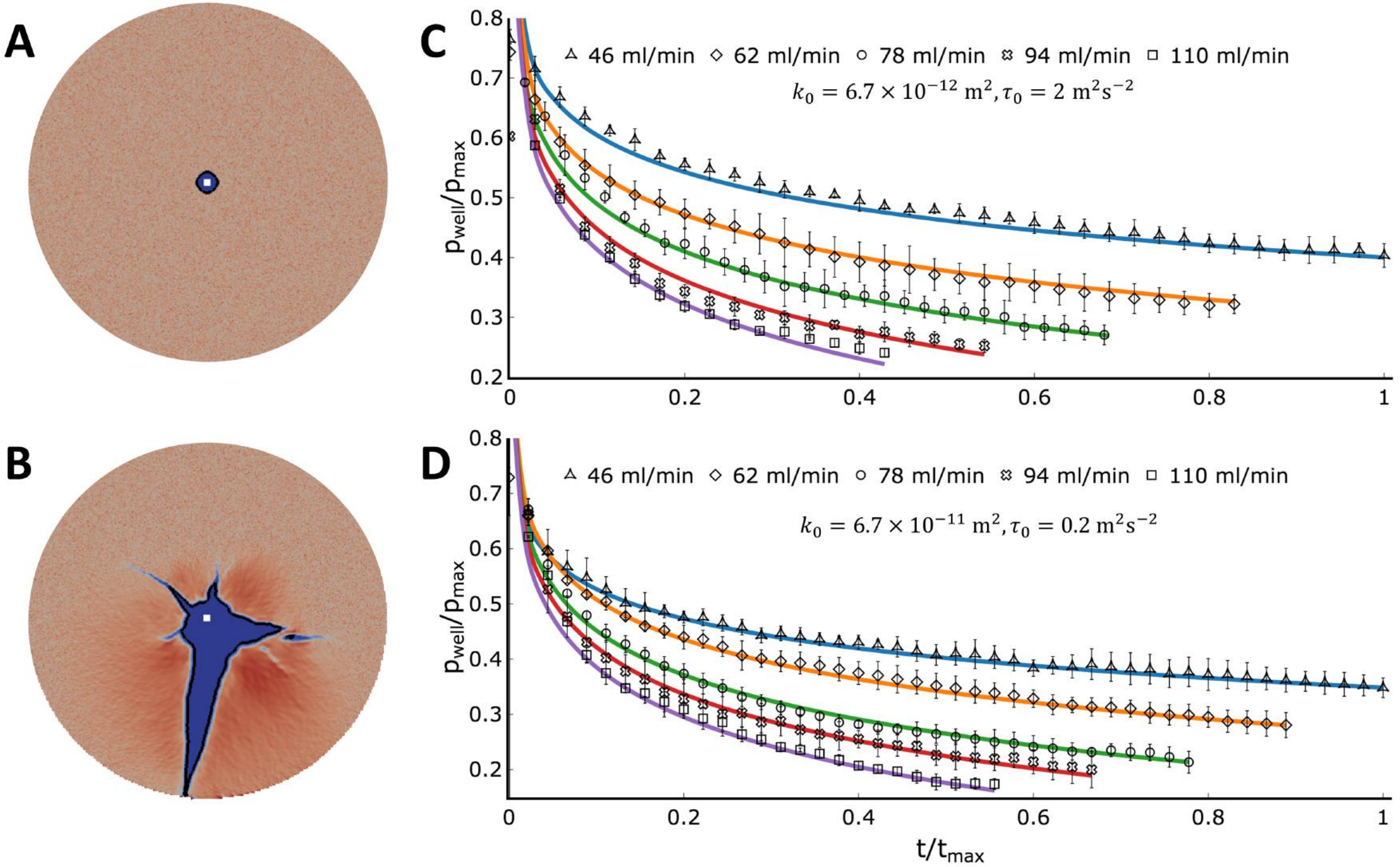}
\caption[Wellbore pressure as a function of injection rate and time]{Wellbore pressure as a function of injection rate and time. (A) The initial simulation setup showing the initial wellbore radius ${r}_{well}=1.3 \ \unit{cm}$, as well as the normally distributed solid fraction field. (B) The fractured system, where the thin black line represents the position of the advancing glycerin saturation front. C and D show the wellbore pressure as a function of time for different flow rates and different combinations of solid yield stress and permeability. Solid curves represent analytical solutions, while symbols represent simulation predictions. The color scheme in A and B is the same as in Figure \ref{fig:hele-shaw}, and ${p}_{max}$ is the maximum analytically-predicted pressure in each simulation. \label{fig:vis_frac}}
\end{center} 
\end{figure}

Lastly, as hinted at before, a notable characteristic of our model is that different normally-distributed solid fraction field initializations give different fracturing results (Appendix \ref{fracturing_instabilities}). For this reason, we performed four simulations for each parameter set. In Figure \ref{fig:vis_frac}, we present the average predicted wellbore pressure evolution with errors bar representing the 95\% confidence interval.

Figure \ref{fig:vis_frac} shows that our model can accurately and reliably predict the pressure and deformation behavior of a variety of fracturing systems, as all curves exhibit excellent agreement with their respective analytical solution. Note that the length of each curve relates inversely to the injection speed. This is because fractures at higher injection rates consistently reach the system's boundary faster than their counterparts, at which point there is a sharp decrease in pressure and the analytical solution no longer applies. Therefore, each curve's cutoff point represents the time at which the fracture effectively becomes an open channel between the wellbore and the outer boundary, normalized to the average value of that time for the slowest-moving fracture (i.e. $t = t_{max}$).

The successful replication of the analytical pressure profiles in this section verifies the model components pertaining to the pressure-velocity-deformation coupling and the two-way momentum transfer between the fluid and solid phases (drag). Therefore, the only model component left to verify is the implementation of the capillary force terms during fracturing of a plastic solid. 

\subsection{Capillary Effects on Fracturing Wellbore Pressure} \label{capillary_experiments}

Our fifth verification systematically varies the capillary entry pressure within non-wetting fracturing systems to quantify its effects on wellbore pressure. For this, we consider two different complementary cases: one where capillary forces are comparable to their viscous counterparts, and another where they are significantly larger than them. All parameters are the same as in the previous experiments (Section \ref{frac_validation}) unless otherwise specified. 

\begin{figure}[htb!]
\begin{center}
\includegraphics[width=0.97\textwidth]{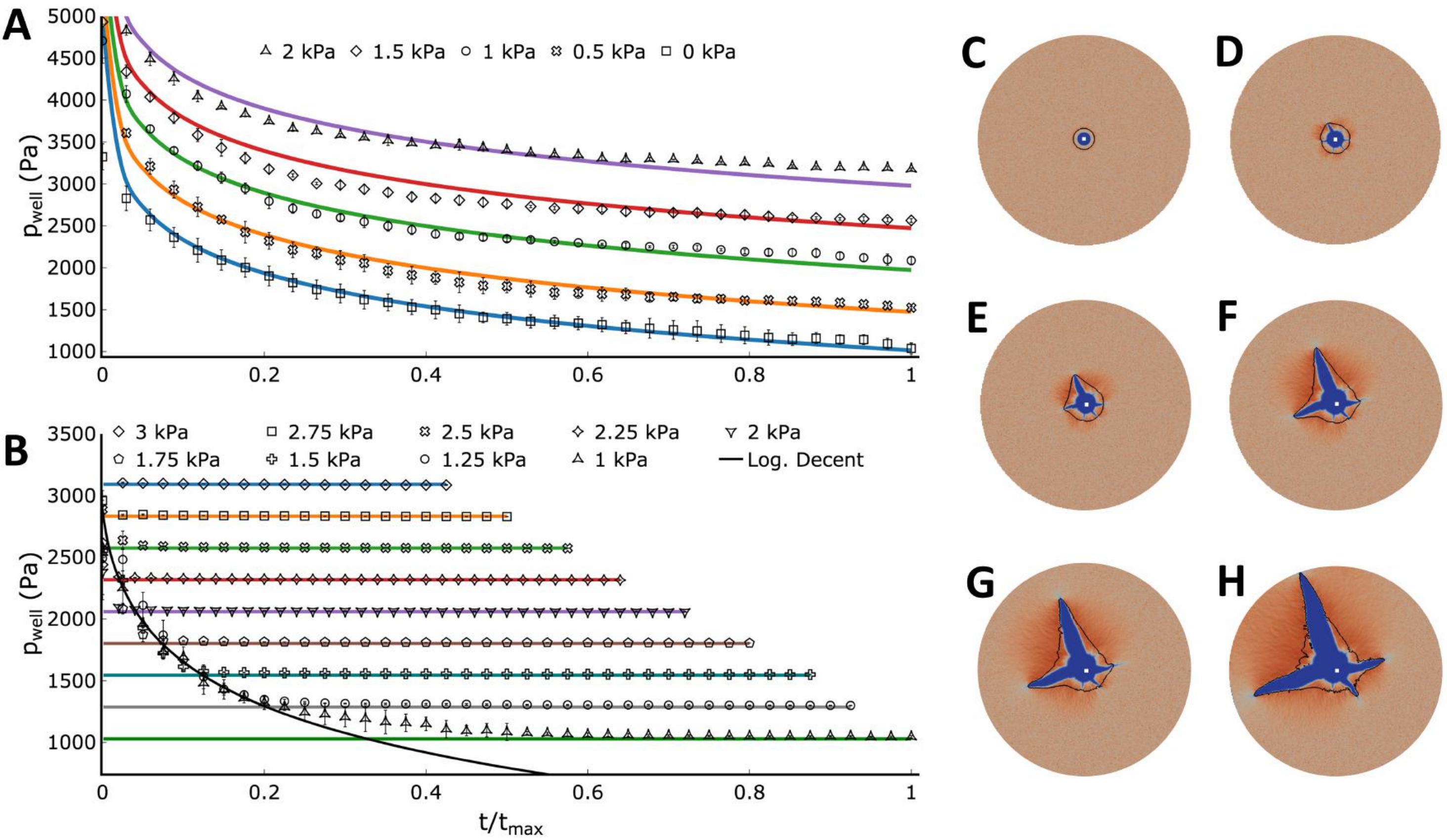}
\caption[Effect of capillary entry pressure on fracturing wellbore pressure]{Effect of capillary entry pressure on fracturing wellbore pressure. 
 (A-B) Wellbore pressure as a function of time and entry pressure for low and high permeability systems, respectively. In B, curves at increasingly high pressures were cut off for illustrative purposes and the solid line represents a fitted reference logarithmic pressure descent curve. (C-H) Time evolution of fractured system with a 1 kPa capillary entry pressure and high permeability. (C) Initial fluid invasion ($t/t_{max}<0$): at early times the wellbore pressure rises rapidly and becomes larger than the entry capillary pressure. The fluid invades the porous formation symmetrically. (D) Fracture initiation ($t/t_{max}=0$): The wellbore pressure continues to rise until it is larger than the breakdown pressure, at which point small fractures start to form. Fluid invasion continues. (E-F) Fracture propagation ($t/t_{max}>0 \mid$ $p_{well} > p_{c,0}$): the wellbore pressure drops as fractures propagate. Fluid invasion continues asymmetrically around said fractures. (G) Fluid invasion stops ($t/t_{max}>0 \mid$ $p_{well} \sim p_{c,0}$): As the wellbore pressure keeps dropping, the entry capillary pressure condition at the porous interface ensures that that wellbore pressure never goes below $p_{c,0}$, at which point fluid invasion stops. (H) Fracture reaches the simulation boundary ($t/t_{max}=1$).  The color convention in Figures C-H is the same as in Figure \ref{fig:hele-shaw}.
\label{fig:pc_frac}}
\end{center} 
\end{figure}

The first set of experiments expands the previous section's analysis into strongly non-wetting systems with the addition of a constant capillary pressure jump at the fracture interface imposed by a flat capillary pressure curve ($p_c=p_{c,0}=0$ to $2$ kPa, $ \tau_0=2 \ \unit{m^2/s^2}, \ {k}_0=6.7\times{10}^{-12} \ \unit{m^2},\ m=0.05$, and $q=78  \ \unit{ml/min}$ ). In this case, all the assumptions present in the fracturing analytical solution (Eqn. \ref{frac_Analytical}) are satisfied. However, said solution still does not account for capillarity. For constant flow in non-wetting systems, the addition of a constant capillary entry pressure jump at the fluid-solid interface would increase the calculated propagation pressure in Eqn. \ref{frac_Analytical} by said value such that ${p}^{new}_{well}={p}_{well}+{p}_{c}$. This effect is exemplified in Figure \ref{fig:pc_frac}A, where we present the updated analytical results in conjunction with our equivalent numerical results, demonstrating excellent agreement between them. Note that the predicted linear relationship between wellbore pressure and capillary entry pressure is not explicitly imposed in the numerical model. On the contrary, it arises naturally from the balance of viscous, capillary, and structural forces in Eqns. \ref{Eq:Final_fluid}-\ref{Eq:Final_Solid}.

The second set of experiments modifies the previous experiments by making the porous medium significantly more permeable, while still maintaining a constant capillary pressure jump at the fracture interface ($p_c=p_{c,0}=1$ to $3$ kPa, $ {\tau}_0=0.2 \ \unit{m^2/s^2},\ k_0=6.7\times{10}^{-11} \ \unit{m^2},\ m=0.99$, and $q=78  \ \unit{ml/min}$). This results in a set of cases where the wellbore pressure is increasingly controlled by the capillary pressure drop rather than by the viscous pressure drop across the fracture and porous formation.

Figure \ref{fig:pc_frac} demonstrates precisely this effect. Our simulations show that the wellbore pressure always decays towards the capillary entry pressure once viscous effects are dissipated by fracture growth, i.e., we observe a transition between viscous- and capillary-dominated regimes. At low values of ${p}_{c,0}$ $\left(<2500\ \unit{Pa}\right)$ the entry pressure is not high enough to prevent fluid flow into the surrounding porous matrix during fracturing (Figure \ref{fig:pc_frac}B-H). The resulting pressure drop cannot be modeled by the previously presented analytical solution (as it violates the no leak-off assumption), but still follows a logarithm-type curve that is characteristic of flow in fracturing systems. With increasing fracture propagation, the viscous pressure drop decreases until the wellbore pressure equals the entry pressure, which is, by definition, the minimum pressure drop required for fluid flow in highly permeable non-wetting systems. Finally, we note that in cases where capillary entry pressure is high relative to the pressure required to fracture the solid (i.e., at ${p}_{c,0}>2.250\ \unit{Pa}$ in the conditions simulated in Fig. \ref{fig:pc_frac}b), fracturing begins before the wellbore pressure can exceed ${p}_{c,0}$. This prevents essentially all flow into the porous formation, and the wellbore pressure is immediately stabilized at $\sim{p}_{c,0}$. For all cases, fractures continue to propagate until they reach the system boundary, at which point the pressure drops rapidly as noted in Section \ref{frac_validation}.

In this section we reduced the inherent complexity of the model's capillary force terms $F_{c,i}$ (Eqns. \ref{Fc1_def}-\ref{Fc2_def}) into a simple set of intuitive verifications. The quantitative agreement between these two analytical cases and their corresponding numerical simulations validate the implementation of the impact of capillary pressure effects on the mechanics of a ductile porous solid within our model. 

\section{Illustrative Applications}\label{illustrative_app}

Having verified and tested the model, we now proceed with two illustrations that demonstrate how \textit{hybridBiotInterFoam} enables the simulation of relatively complex coupled multiphase multiscale systems. The following cases serve as illustrative examples of our model's features and capabilities as well as tutorial cases within the accompanying toolbox.

\subsection{Elastic Failure in Coastal Barriers}

Coastal barriers are ubiquitous features in coastal infrastructure development. When designed appropriately, these structures can be very effective in regulating water levels and protecting against inclement weather \citep{Morton2002}. However, accurate prediction of the coupled fluid-solid mechanics of these structures (which can lead to barrier failure) is inherently challenging as it requires modeling large-scale features (waves) while also considering small-scale viscous and capillary interactions within the barrier. 

\begin{figure}[htb!]
\begin{center}
\includegraphics[width=0.97\textwidth]{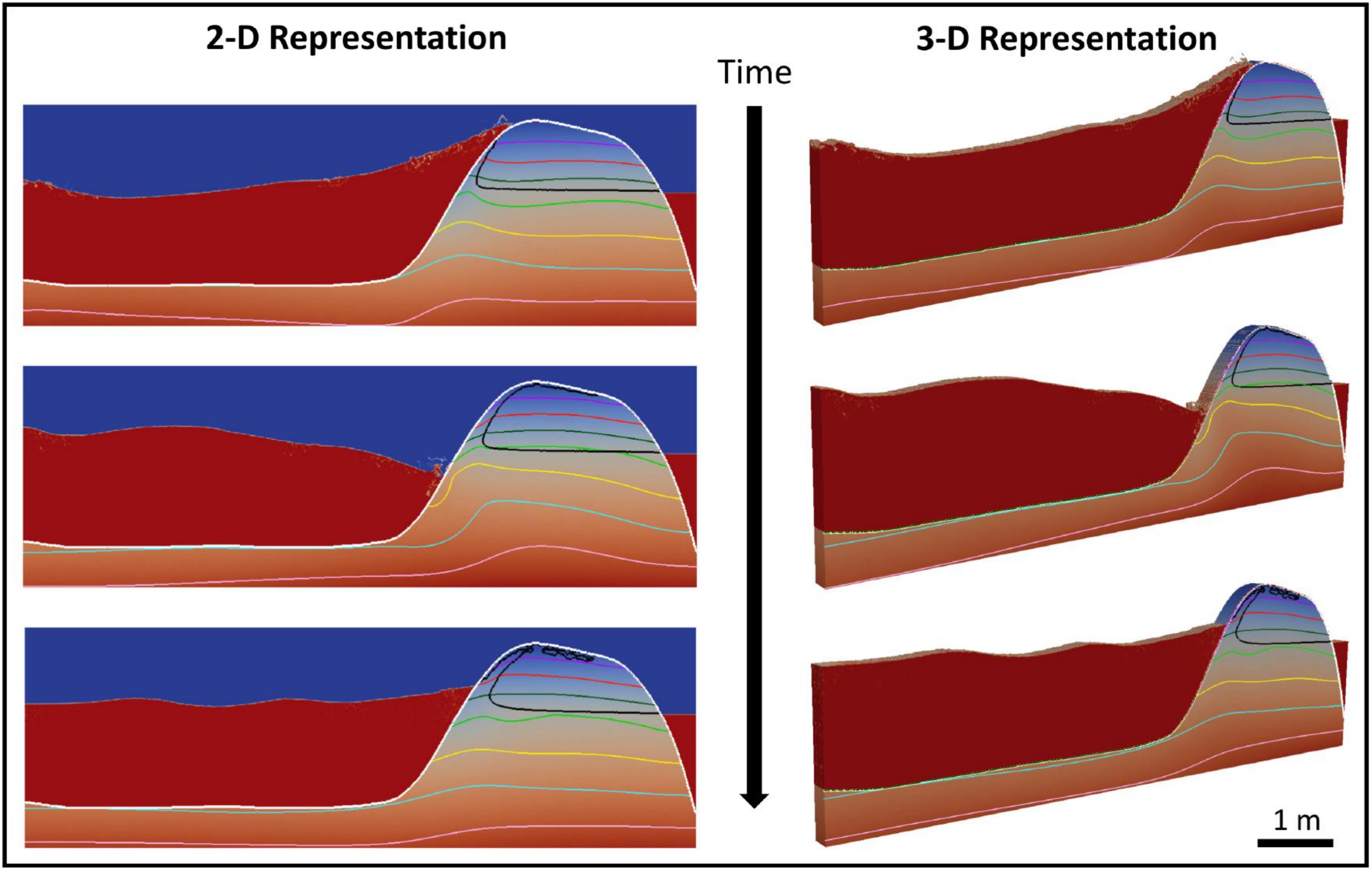}
\caption[Wave dissipation in a poroelastic coastal barrier]{Waves crashing against a poroelastic coastal barrier. Here, the thin black line represents the water-air interface $\left({\alpha }_w=0.5\right)$ and red-blue colors outside the coastal barrier represent water and air, respectively. Colored contours within the barrier are the calculated Von Mises stresses and are shown in 5 kPa increments in the general downwards direction. Note that the largest stresses are seen during the initial wave crash and increase towards the base of the barrier due to gravitational effects. \label{fig:barrier}}
\end{center} 
\end{figure}

The following case represents the continuation of the three-dimensional coastal barrier illustration presented in Chapter \ref{chp:hybridPhase} with the addition of linear-elastic poromechanics. As such, the simulation was created by initializing a heterogeneous porosity field (with $k_0=2\times {10}^{-8} \ \unit{m^2}$ and ${\phi }_f=0.5$) in the shape of a barrier within a 8.3 by 2.7 by 0.25 m rectangular grid (1600 by 540 by 50 cells). The relevant solid mechanics parameters were $E=5 \ \unit{MPa}$, $\nu =0.45$, and ${\rho }_s=2350 \ \unit{kg/m^3}$. Relative permeabilities and capillary pressures were evaluated through the Van Genuchten model with $m=0.8$ and $p_{c,0}=1 \ \unit{kPa}$. Before the start of the simulation, the water level was set to partially cover the barrier and then allowed to equilibrate. A single wave was then initialized at $t = 0$. This results in a simulation that exhibits a clear wave absorption cycle that gradually dissipates in time, as seen in Figure \ref{fig:barrier}. Detailed discussion on the fluid mechanics of this problem can be found in Chapter \ref{chp:hybridPhase} and \citet{Carrillo2020}.

Here, however, we are interested in evaluating the barrier's propensity to failure. We do this by applying the Von Mises yield criterion, which is commonly used to predict material failure in elastic systems. It states that if the second invariant of the solid's deviatoric stress (the Von Mises stress) is greater than a critical value (the yield strength) the material will begin to deform non-elastically \citep{VonMises1913}. Although we do not specify said critical value within our simulations, we can map the time-evolution of Von Misses stresses within the coastal barrier as a result of a wave absorption cycle (Figure \ref{fig:barrier}). Our results illustrate the potential utility of our simulation framework in predicting the location and time-of-formation of stress induced defects within coastal barrier as a function of wave characteristics, permeability, and barrier geometry.

\subsection{Flow-Induced Surface Deformation}

Surface deformation due to subsurface fluid flow is a common geological phenomenon occurring in strongly coupled systems and has clear implications in studies related to induced seismicity \citep{Shapiro2009}, ${\mathrm{CO}}_{2}$ injection in the subsurface \citep{Morris2011}, land subsidence \citep{Booker1986}, and the formation of dykes and volcanoes \citep{Abdelmalak2012,Mathieu2008}. In order to properly model these systems, it is necessary to be able to capture the time-evolution of surface uplift, cracks, and hydraulic fractures, as well as the effects that these features have on the overall flow field. Here, we use the terms hydraulic fracture vs. crack to refer to solid failure at vs. away from the injected fluid, respectively.

\begin{figure}[htb!]
\begin{center}
\includegraphics[width=0.97\textwidth]{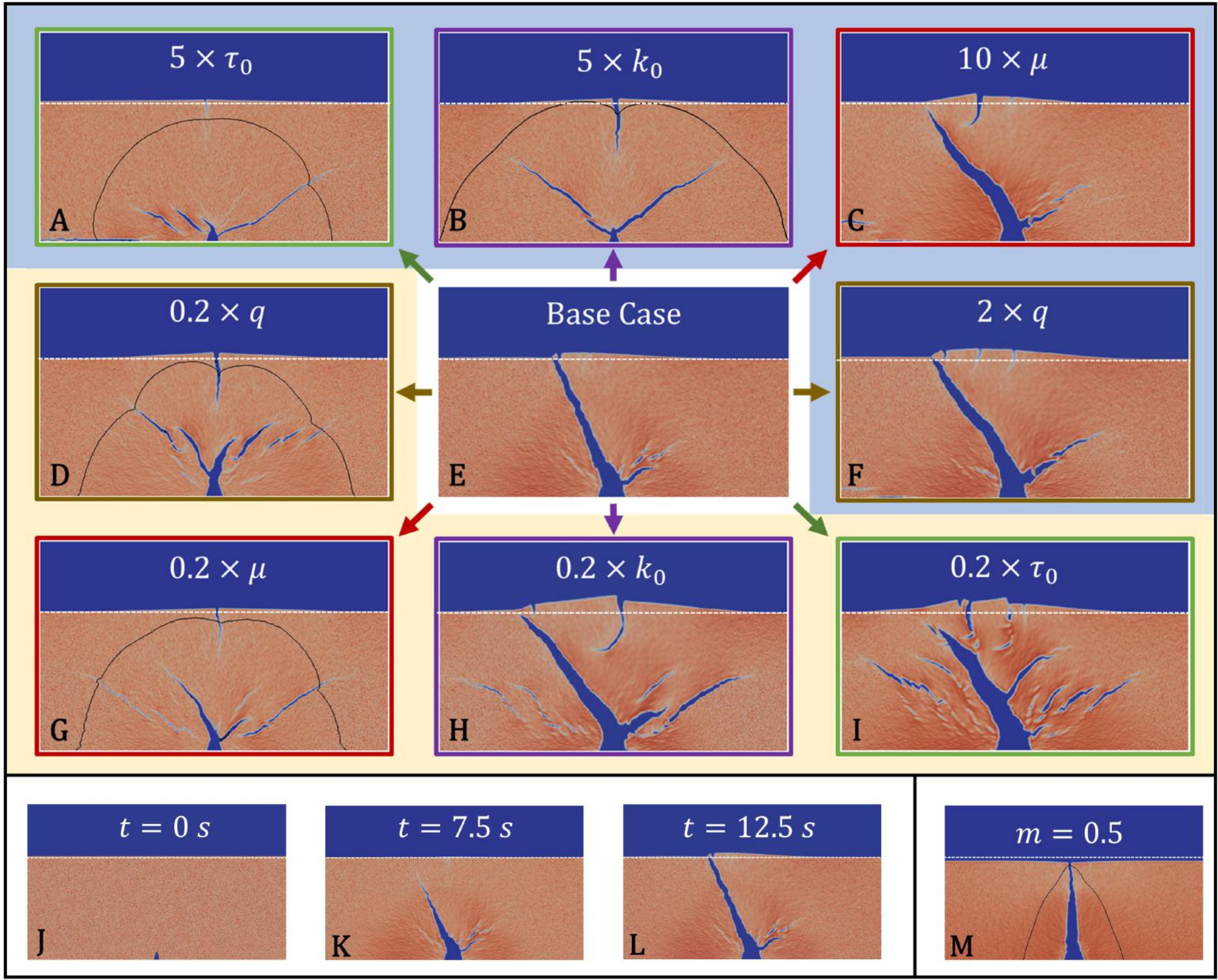}
\caption[Study of the effects of hydraulic fracturing and cracking on surface deformation]{Study of the effects of hydraulic fracturing and cracking on surface deformation. (A-I) Representative cases showing the effects of changing permeability $k_0$ (purple), solid yield stress $\tau_0$ (green), injection rate $q$ (brown), and injected fluid viscosity $\mu$ (red) on surface deformation. The blue and yellow subsections contain the results of increasing or decreasing the controlling parameters, respectively. (J-L) Time evolution of the fracturing base case. (M) Surface subsidence example. The difference between the base case (E) and all other simulations is shown in each case's legend. Dotted white lines represent the surface height of the initial solid fraction configuration. Note that the color scheme in all simulations is the same as in Figure \ref{fig:hele-shaw}. \label{fig:surface}}
\end{center} 
\end{figure}

This illustrative case was inspired by the experiments reported by \citet{Abdelmalak2012}, where the authors injected a highly viscous fluid into a dry silica powder in a Hele-Shaw cell in order to study the impact of hydraulic fractures on surface deformation, e.g., during the creation of volcanic structures. The system also bears some analogy to situations involving the injection of fluids into subsurface reservoirs, e.g., during geologic CO$_2$ sequestration \citep{Rutqvist2012}. The base case of our simulations consists of an impermeable rectangular container (50 by 30 cm, 500 by 300 cells) that is open to the atmosphere, is partially filled with a dry porous medium (${\phi}_s=0.6\ \pm 0.05$,\ ${\rho}_s=2650 \ \unit{kg/m^3},\ k_0=5 \times 10^{-11} \ \unit{m^2})$, and has an injection well at its lower boundary that injects water at $q=6.5 \ \unit{ml/s}$ (Fig. \ref{fig:surface}). To account for irreversible solid deformation, the porous medium is modeled as a plastic with a kinematic yield stress $\tau_0=0.22 \ \unit{m^2/s^2}.$ The solid is represented as impermeable to the invading fluid through the use of the Van Genuchten model with $m=0.05$ and $p_c=0$. Then, using this base case as a standard, we individually varied each of the main parameters ($q, \ k_0,\ \tau_0, \ m,\ {\phi}_s,\ {\mu }_{water})$ over several simulations in order to model the resulting solid deformation processes: fracturing, cracking, surface uplift, and subsidence (Figure \ref{fig:surface}).

The resulting cases demonstrate that cracking (solid failure away from the injected fluid) is strictly dependent on the number and orientation of existing hydraulic fractures, as it only occurs when there is more than one fracture branching off from the main injection point (Figure \ref{fig:surface}B, C, D, H, and I). This is likely because in cases presenting a single vertical fracture solid displacement is almost exclusively perpendicular to the fracturing direction, leading to virtually no surface deformation or cracking (Figure \ref{fig:surface}A, E and M). Contrastingly, the creation of inclined fractures exerts vertical forces on the solid, resulting in surface uplift and crack formation. The above diagram strongly suggests that deformation is controlled by the balance between viscous and structural forces: larger fractures occur within softer solids with higher momentum transfer, and smaller fractures occur in tougher solids with lower momentum transfer. As stated above, a comprehensive examination of the parameters that control solid fracturing will be the focus of Chapter \ref{chapter:PRL}.

In addition to the surface uplift presented above, subsurface subsidence is observed in the simulated system in conditions where the porous solid is rendered permeable to the invading fluid (i.e., $m \gg 0.05$). This phenomenon is not primarily controlled by momentum transfer, but rather by a gravitational effect whereby the displacement of air by water within the porous medium around the advancing hydraulic fracture renders the solid-fluid mixture heavier. Once it is heavy enough to overcome the plastic yield stress, the solid sinks and compresses around the fluid source (Figure \ref{fig:surface}M). 

\section{Conclusions}\label{conclusions}

Throughout this chapter and its predecessors, we have shown that our modeling framework is flexible and readily applicable to a large variety systems: from single-phase flow in static porous media, to elastic systems under compression, to viscosity- or capillarity-dominated fracturing systems, all the way up to multiscale wave propagation in poroelastic coastal barriers.  We invite the interested reader to tune, adapt, and expand the present illustrative simulations, which are included in the accompanying CFD toolbox. 

Lastly, we would like to note that this framework cannot be applied to every multiscale system, at it is based on strong assumptions of lengths-scale separation, sub-volume homogeneity, Eulerian descriptions, and relatively small elastic deformations. These assumptions come with their own limitations. A thorough discussion on the model's constraints and future work can be found in Chapter \ref{conclusion}.

\begin{savequote}[75mm]
The most exciting phrase to hear in science, the one that heralds the most discoveries, is not ‘Eureka!’ but ‘That’s funny…’\qauthor{Isaac Asimov}
\end{savequote}

\chapter{Capillary and Viscous Fracturing During Drainage in Porous Media} \label{chapter:PRL}

\newthought{In this chapter,} we use the Multiphase DBB framework to study the transition from uniform fluid invasion to capillary and viscous fracturing during drainage in porous media. To do so we examine multiphase flow in deformable porous media in a broad range of flow, wettability, and solid rheology conditions. We then demonstrate the existence of three distinct fracturing regimes controlled by two non-dimensional numbers that quantify the balance of viscous, capillary, and structural forces in a porous medium. We then use these parameters to establish a first-of-its-kind phase diagram for material failure caused by multiphase flow in poroplastic media. Lastly, we examine the effects of compaction on said dimensional numbers and the system's propensity to fracture. This chapter is adapted from \citet{Carrillo2021}.

\section{Introduction}

Multiphase flow in deformable porous media is a ubiquitous phenomenon in natural and engineered systems that underlies key processes in water and energy resource engineering and materials science, including membrane filtration, soil wetting/drying, unconventional hydrocarbon recovery, and geologic carbon sequestration \citep{Bacher2019,Rass2018,Towner1987}. A key obstacle to more accurate representations of this phenomenon is our limited understanding of the transition from fluid invasion to flow-induced fracturing, i.e., material failure caused by multiphase flow. In large part, this limitation is caused by a lack of computational approaches capable of representing multiphase flow in fractured deformable porous media. 

Previous work on \textit{multiphase} flow within \textit{static} porous media is extensive and includes detailed examinations of the influence of wettability, viscosity, and flow rate on flow in unsaturated porous media at multiple scales. In particular, existing studies have demonstrated how capillary forces give rise to differences between drainage and imbibition \citep{Lenormand1986}; how the ratio of fluid viscosities controls the stability of the invading fluid front \citep{JorgenMalby1985,Saffman1958,Stokes1976}; and how the magnitude of the capillary number delineates distinct flow regimes \citep{Ferer2004,Yortsos1997,Datta2013,Lu2020}. Each of the aforementioned controls is highly dependent on the system of interest. This complicates efforts to develop general relative permeability and capillary pressure models that apply to most systems of interest \citep{Picchi2018,Picchi2019,RHBrooks1964,VanGenutchen1980}.

Flow of a \textit{single fluid phase} through \textit{deformable} porous media also has been studied in depth. Numerical modeling studies are largely based on the work of Biot and Terzaghi \citep{Biot1941,Terzaghi1943} and have been used to reproduce the behavior of arteries, boreholes, swelling clays, and gels \citep{Auton2017a,Bertrand2016b,Carrillo2019,MacMinn2015a}. In the last decade, fundamental studies have generated detailed information on the dynamics that arise from fluid-solid couplings beyond the ideal poroelastic regime, including fracturing, granular fingering, and frictional fingering \citep{Campbell2017,Sandnes2011a,Zhang2013}. In particular, these studies have shown that the main parameters controlling the deformation of a porous solid by single phase flow are the material softness and the magnitude of the fluid-solid momentum transfer \citep{Sandnes2011a}.

The study of \textit{multiphase} flow in a \textit{deformable} porous medium is inherently more complex than the problems outlined above, as it requires simultaneous consideration of capillarity, wetting dynamics, fluid rheology, and solid deformation. Deformation modes associated with material failure (i.e., multiphase fracturing) are particularly challenging as they require simultaneous representation of multiphase flow in fractures and in the surrounding porous matrix. The existing detailed examinations of this phenomenon have focused exclusively on granular systems. Notably, Holtzman \& Juanes \citep{Holtzman2010a,Holtzman2012a} used experiments and discrete element models to demonstrate that the transitions between capillary fingering,
viscous fingering, and fracturing during multiphase flow in granular media reflect two non-dimensional numbers: a fracturing number (ratio of fluid
driving force to solid cohesive force) and a modified capillary number (the ratio between viscous and capillary pressure drops). Other discrete element approaches have shown that fracturing is highly dependent on the invading fluid's capillary entry pressure
\citep{Jain2009,Meng2020}. However, it is not clear how these conclusions translate to continuous non-granular systems.

To the best of our knowledge, no
experimental or numerical investigation has simultaneously explored the effects of flow rate, wettability, and
deformability during multiphase flow in deformable porous media at the \textit{continuum scale} and identified the controlling
parameters that relate \textit{all three} properties within a single phase diagram. Here, we use simulations carried out with our new Multiphase Darcy-Brinkman-Biot (DBB) framework \citep{Carrillo2020MDBB} to fill this knowledge gap and identify non-dimensional parameters that govern viscously-stable fluid drainage and fracturing in deformable porous media. We also find that the fracturing dynamics predicted by our continuum-scale framework is consistent with those observed or predicted for granular systems. In other words, in systems with a large length scale separation between pores and fractures, volume-averaged properties are sufficient to capture the onset and propagation of fractures at the continuum scale.

\section{Numerical Simulations}

\subsection{Crossover from Imbibition to Fracturing in a Hele-Shaw Cell } \label{subsect:Huang}

\begin{figure}[htb!]
\begin{center}
\includegraphics[width=0.8\textwidth]{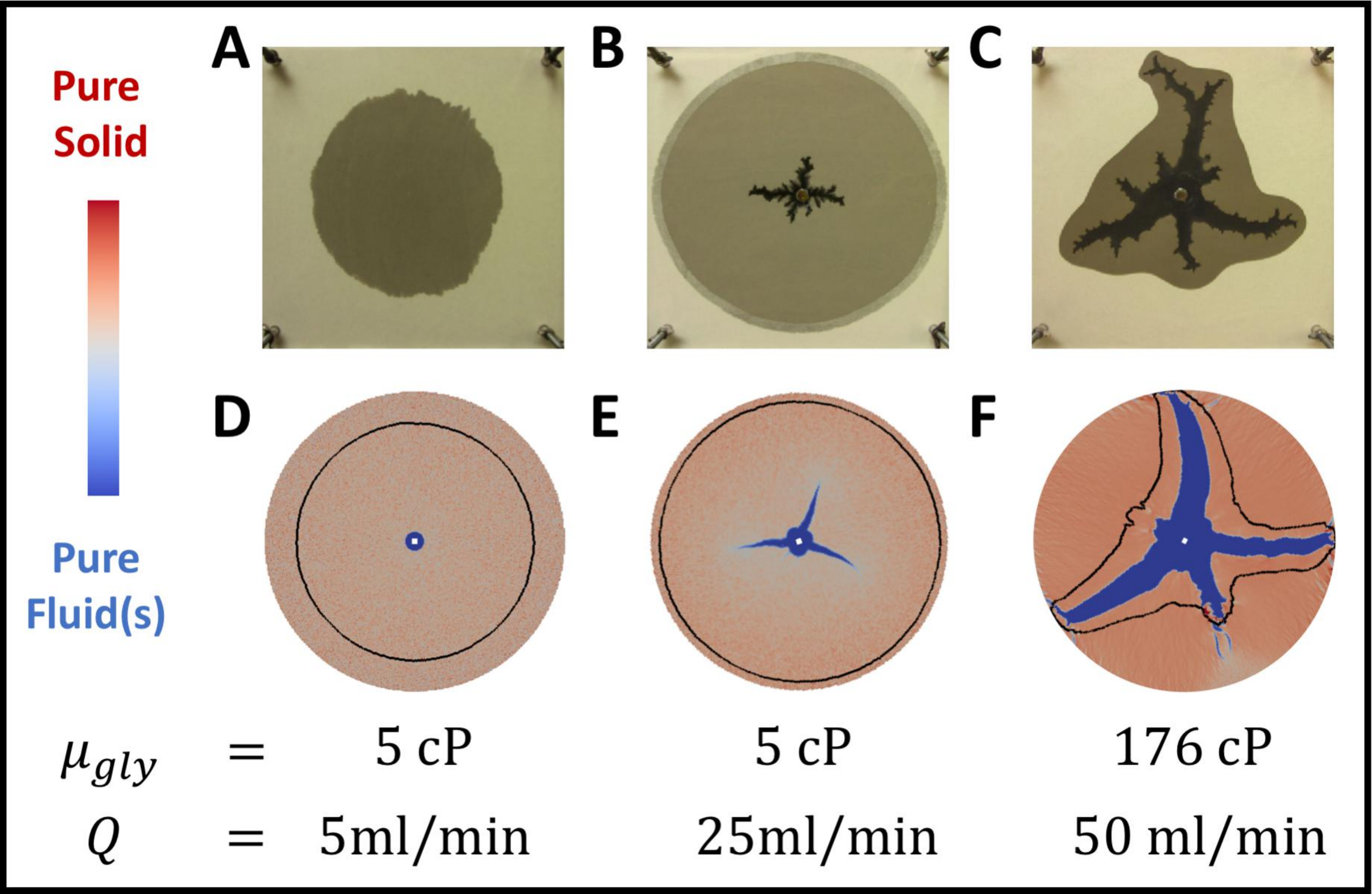}
\caption[Continuous transition from fluid imbibition to fracturing in a Hele-Shaw cell] {\label{fig:Huang_Frac} Continuous transition from fluid imbibition to fracturing in a Hele-Shaw cell. Experimental images (A, B, C) were taken from \citet{Huang2012a} and numerically replicated using equivalent conditions (D, E, F). Black lines represent the advancing saturation front. Additional cases can be found in Chapter \ref{chp:MDD_Applications}.}
\end{center}
\end{figure}

In addition to the derivation and extensive quantitative validation of the modeling framework, Chapter \ref{chp:MDD_Applications} included a qualitative validation of the ability of the Multiphase DBB model to predict the transition from invasion to fracturing during multiphase flow. Briefly, this validation replicated experiments by \citet{Huang2012a} involving the injection of aqueous glycerin into dry sand at incremental flow rates within a 30 by 30 by 2.5 cm Hele-Shaw cell. As shown in Fig. \ref{fig:Huang_Frac}, these experiments are inherently multiphysics as fluid flow is governed by Stokes flow in the fracture (aperture $\sim$cm) and by multiphase Biot Theory in the porous sand (pore width $\sim 100\mu$m).

As discussed in Chapter \ref{chp:MDD_Applications}, the similarities between our model and the experimental results are evident: as the viscous forces imposed on the solid increase, so does the system’s propensity to exhibit fracturing as the primary flow mechanism (as opposed to imbibition). Minor microstructural differences between our simulations and the
experiments reflect the manner in which the implemented continuum-scale rheology model approximates the solid's granular nature. It is clear, however, that both systems are controlled by the balance between viscous forces and solid rheology at the scale of interest \citep{Carrillo2020MDBB}. As such, these experiments present an ideal starting point for our investigation.

\subsection{Creation of Fracturing Phase Diagrams}

\begin{figure}
\begin{center}
\includegraphics[width=1\textwidth]{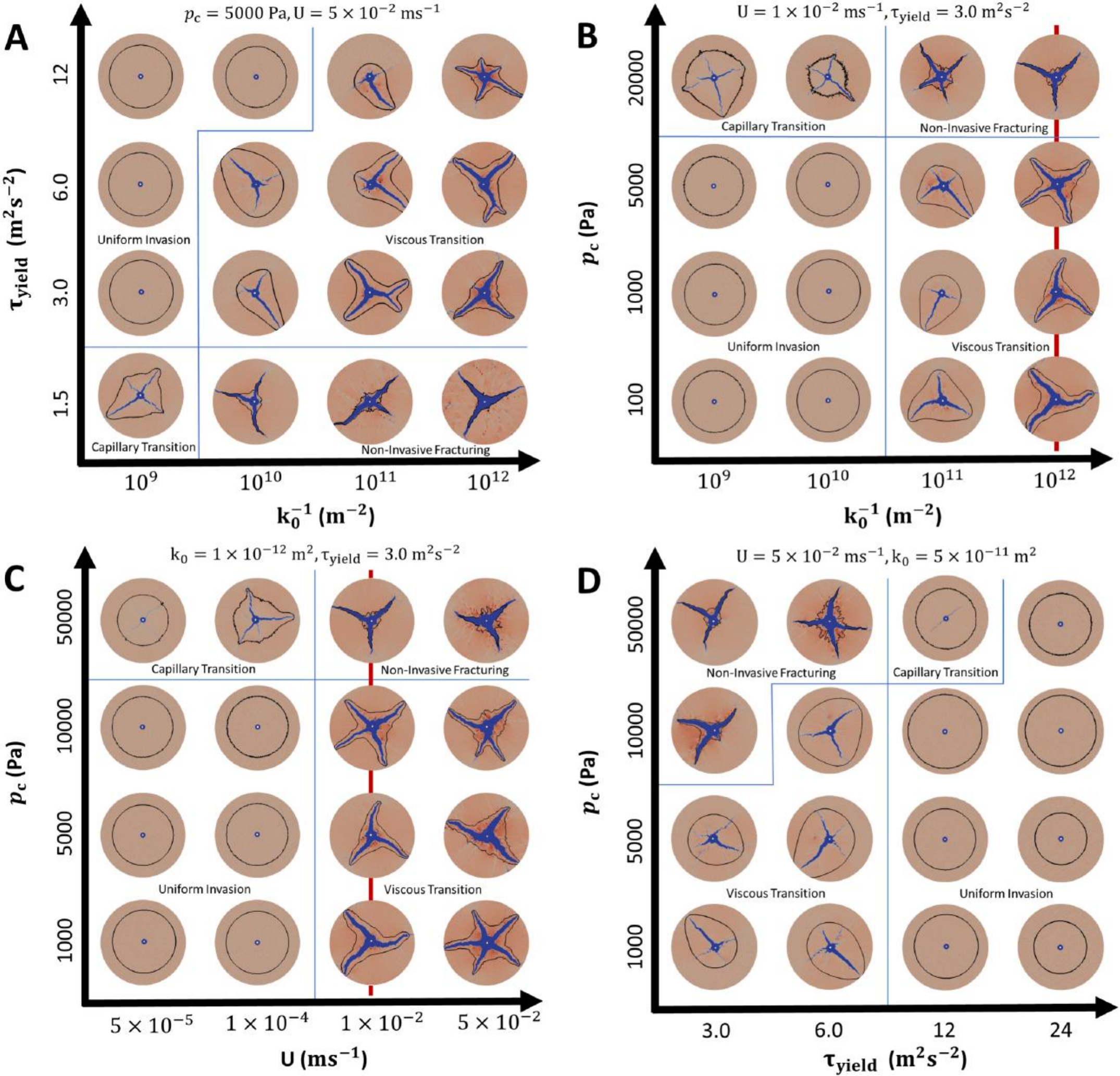}
\caption[Phase diagrams describing the impacts of permeability, plastic yield stress, fluid injection rate, and capillary entry pressure on the transition from fluid drainage to fracturing]{\label{fig:phase_diagrams}Phase diagrams describing the effects of varying permeability, plastic yield stress, fluid injection rate, and capillary entry pressure on the transition from fluid drainage to fracturing. All cases are at ${\phi }_s=0.60\pm 0.05$ and ${\mu }_{n}=5\ \mathrm{cP}$. The remaining parameters are case-specific and can be found in each figure's upper legend. The areas separated by thin blue lines highlight and label the four deformation regimes described in Section \ref{sect:fracture_mechanisms}. The vertical red lines represent where these diagrams intersect in 3-dimensional space. The color scheme is the same as in Fig. \ref{fig:Huang_Frac}.}

\end{center}
\end{figure}

Here, we use the same simulation methodology developed in \citep{Carrillo2020MDBB} and illustrated in Figure \ref{fig:Huang_Frac} to identify the general non-dimensional parameters that control the observed 
transitional behavior between invasion and fracturing in a plastic porous medium. To do so, we systematically vary
the solid's porosity ($\phi_f$ = 0.4 to 0.8), density-normalized plastic yield stress ($\tau_{yield}$ = $1.5$ to ${24\ \mathrm{m^2/s^2}}$), capillary entry pressure ($p_{c,0}$ = $100$ to $50,000\ \mathrm{Pa}$), and permeability ($k$ = $1\times10^{-13}$ and $5\times10^{-9} \
\mathrm{m^2}$) as well as the invading fluid's viscosity ($\mu_n$ = $0.5$ to $50\ \mathrm{cP}$) and injection rate ($\boldsymbol{U}_f$ = $1\times10^{-4}$ to $8\times10^{-2} \
\mathrm{m/s}$). As in our previous work, the solid's porosity was initialized as a normally-distributed field, the deformable solid was modeled as a Hershel-Bulkley-Quemada plastic \citep{Spearman2017,Quemada1977}, the porosity-dependence of permeability was modeled through the Kozeny-Carman relation, and relative permeabilities where calculated through the van-Genuchten model \citep{VanGenutchen1980}. Further details regarding the base numerical implementation of this model can be found in \citet{Carrillo2020MDBB} and the accompanying code \citep{hybridBiotInterFoam_Code}. The only major differences relative to our previous simulations are that we now include capillary effects and
represent viscously-stable drainage as opposed to imbibition (i.e., the injected glycerin is now non-wetting to the porous medium). A representative sample of the more than 400 resulting simulations is presented in the phase
diagrams shown in Fig. \ref{fig:phase_diagrams}.

Overall, the results make intuitive sense. Figure
\ref{fig:phase_diagrams}A shows that, \textit{ceteris-paribus}, less
permeable solids are more prone to fracturing. This is due
to the fact that, given a constant flow rate, lower permeability solids experience greater drag forces. Our results also show that solids
with lower plastic yield stresses fracture more readily, as their solid structure cannot
withstand the effects of relatively large viscous or capillary forces. The y-axis behavior of
Fig. \ref{fig:phase_diagrams}B further shows that systems with higher entry pressures are more likely to
fracture, i.e., the capillary stresses are more likely to overwhelm the solid's yield stress, in agreement with grain scale simulations \citep{Jain2009}. Finally, Fig.
\ref{fig:phase_diagrams}B also shows that higher injection rates lead to more fracturing, as these
increase viscous drag on the solid structure.   

\section{Characterization of Fracturing Mechanisms} \label{sect:fracture_mechanisms}

The deformation regimes observed in the previous experiments can be
delineated by defining two simple non-dimensional parameters that quantify the balance
between viscous pressure drop, solid softness, and capillary entry pressure.

\begin{equation}\label{eq:N_vis_radial}
    N_{vF}=\frac{\Delta p}{{\tau }_{yield}{\rho }_s}=\frac{{\mu }U{r}_{in}}{k{\tau }_{yield}{\rho }_s}{\mathrm{ln} \left(\frac{r_{out}}{r_{in}}\right)}\ \ \ \  
\end{equation}

\begin{equation} \label{Eq:N_cap} 
N_{cF}=\frac{p_{c,0}}{{\tau }_{yield}{\rho }_s}=\frac{2\gamma }{r_{pore}{\tau }_{yield}{\rho }_s} 
\end{equation} 

\begin{figure}[t!]
\begin{center}
\includegraphics[width=0.8\textwidth]{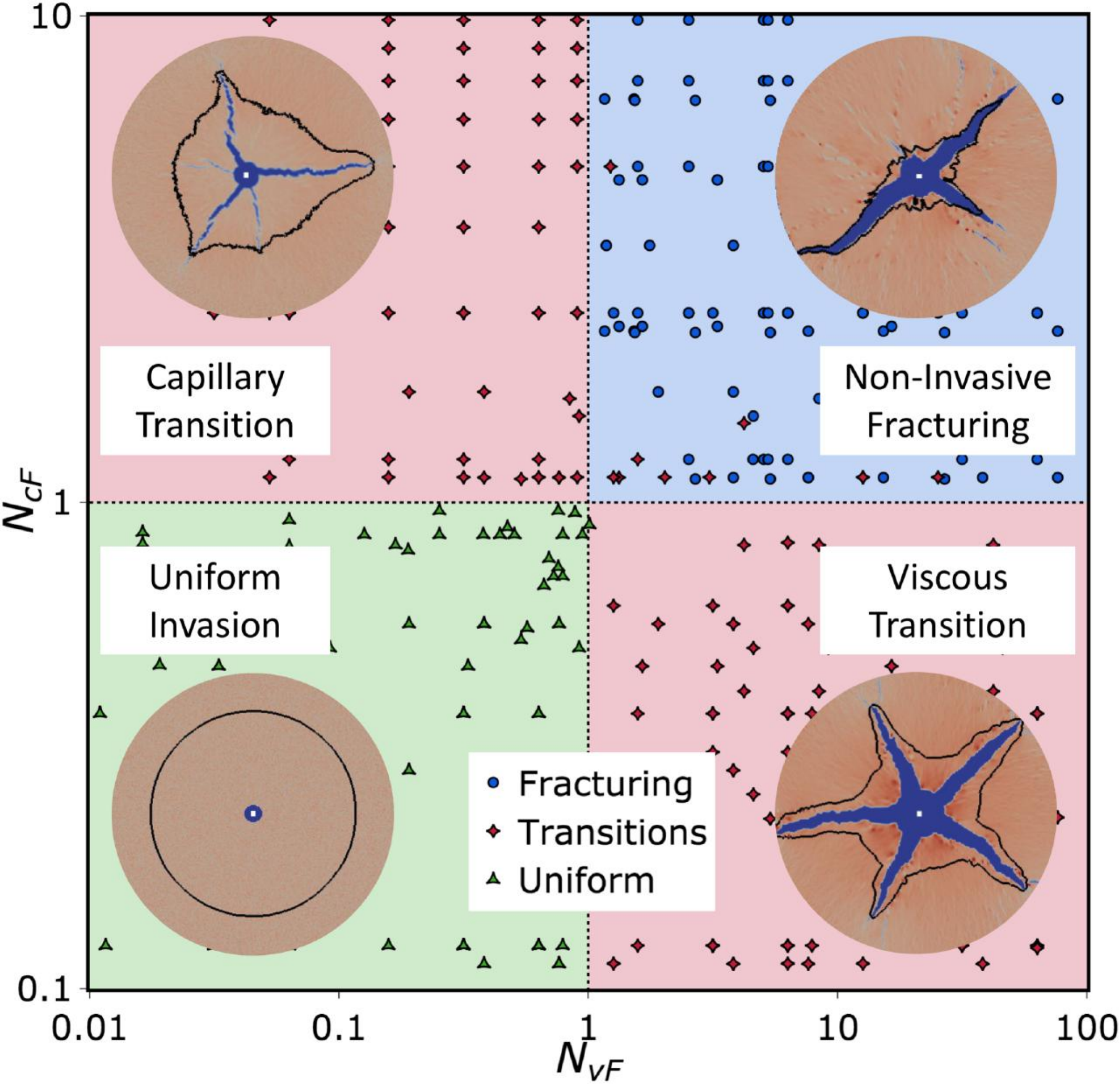}
\caption[Fluid invasion and fracturing in plastic porous media as a function of the viscous and capillary fracturing numbers]{\label{fig:Final_phase} Fluid invasion and fracturing in plastic porous media as a function of the viscous fracturing number $N_{vF}$ and the capillary fracturing number $N_{cF}$. Green triangles denote uniform invasion, red diamonds denote the transitional fracturing regimes, and blue circles denote non-invasive fracturing. The four image insets are representative samples of each fracturing regime.}
\end{center}
\end{figure}

Here, the viscous fracturing number ($N_{vF}$) represents the ratio between the viscous pressure drop and the solid's structural forces. It embodies the question: Does fluid flow generate sufficient friction to induce fracturing? As shown in Fig. \ref{fig:Final_phase}, the answer is no if $N_{vF}<1$ and yes if $N_{vF}>1.$ This number is the continuum scale analog to the fracturing number presented by \citet{Holtzman2012a} for granular solids. It also explains the experimental finding by \citet{Zhou2010} that fracture initiation is only a function of the resulting fluid pressure drop, irrespective of the injection rate or fluid viscosity used to create it. Furthermore, it illustrates why increasing the injection rate and decreasing the permeability have similar effects in Fig. \ref{fig:phase_diagrams}.

Complementarily, the capillary fracturing number ($N_{cF}$) represents the ratio between the capillary entry pressure and the solid's structural forces; it embodies the question: Does multiphase flow generate sufficient capillary stresses to fracture the solid? Figure \ref{fig:Final_phase} shows that when $N_{cF}<1$ drainage is the preferential flow mechanism and when $N_{cF}>1$ fracturing becomes the dominant phenomenon.

This analysis yields the rudimentary conclusion that fracturing should occur if either of the fracturing numbers is greater than unity, as confirmed by our simulations. However, our simulations further demonstrate the existence of three distinct fracturing regimes (Figs. \ref{fig:phase_diagrams}-\ref{fig:Final_phase}). The first regime, referred here as \textit{non-invasive fracturing} ($N_{vF}>1$ and $N_{cF}>1$) is characterized by fracturing of the porous solid with minimal fluid invasion, where fractures precede any invasion front. In the second regime, referred to here as the \textit{viscous fracturing transition} ($N_{vF}>1$ and $N_{cF}<1$), only the viscous stresses are sufficiently large to fracture the solid. This leads to the formation of relatively wide fractures enveloped and preceded by a non-uniform invasion front. Finally, in the third regime, referred to here as the \textit{capillary fracturing transition} ($N_{vF}<1$ and $N_{cF}>1$), only the capillary stresses are sufficiently large to fracture the solid. Given a constant injection rate, this leads to the formation of fractures preceded by an invasion front, as in the viscous fracturing transition regime, but with a more uniform saturation front (due to lower viscous stresses) and less solid compaction (hence narrower fractures). We note that the crossover between each of the four regimes is continuous, meaning that systems with $N_{vF}$ or $N_{cF}\sim1$ can share elements of neighboring regimes. 

\begin{figure}[t!]
\begin{center}
\includegraphics[width=0.983\textwidth]{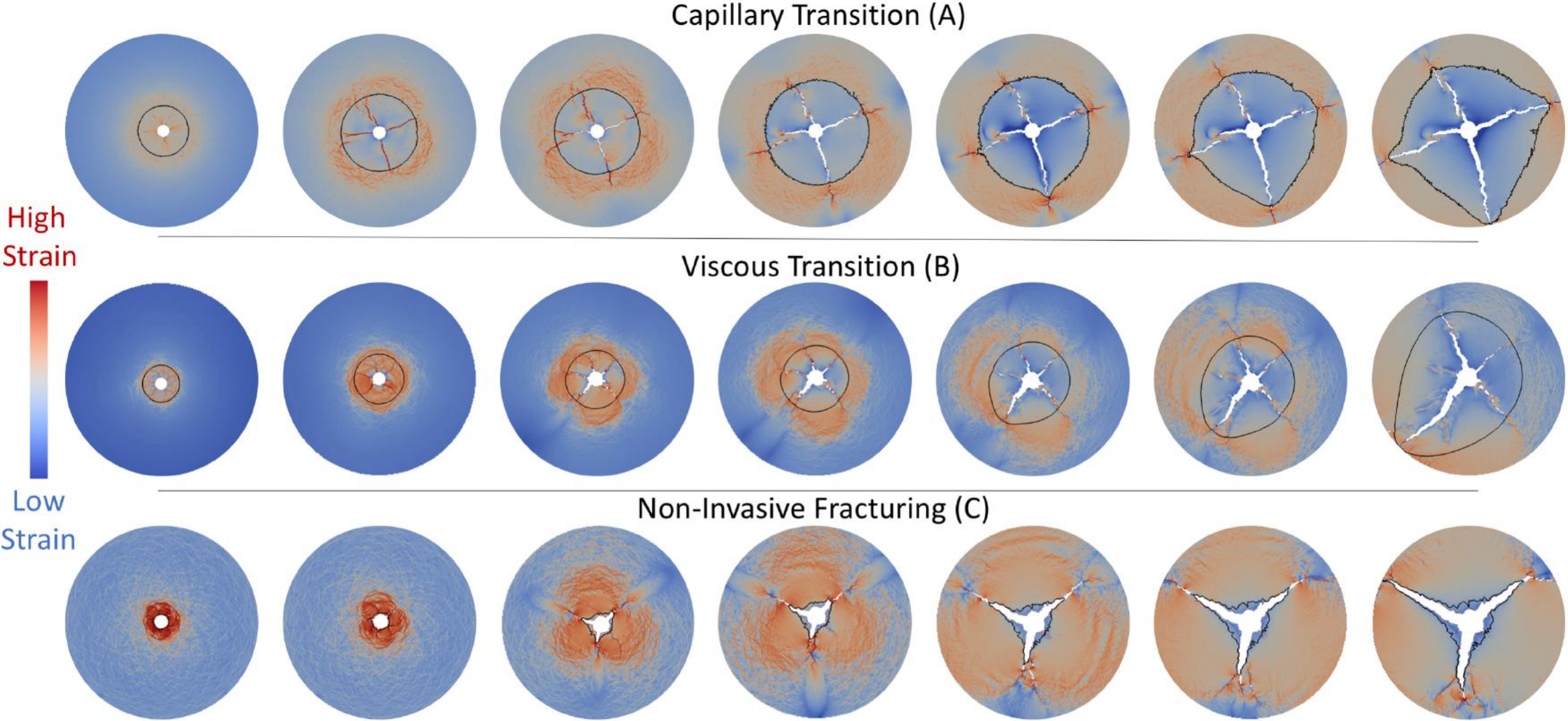}
\caption[Dynamic fracture formation mechanisms]{\label{fig:strain} Dynamic fracture formation mechanisms. Each row represents the time-dependant fracture formation process for each fracturing type, where time advances from left to right. Here, the red-blue color scheme represents the log-normalized strain-rate magnitude specific to each simulated case, fractures are shown in white and the advancing fluid-fluid interface is shown as a thin black line.}
\end{center}
\end{figure}

Although $N_{vF}$ and $N_{cF}$ are fairly intuitive numbers, their impacts on fracture propagation mechanisms are not. For this reason, we also studied the dynamics of fracture nucleation and growth and the evolution of the solid's strain for all three fracturing regimes. As seen in Fig. \ref{fig:strain}, fracturing in the two transition zones is characterized by the initial formation of non-flow-bearing failure zones (hereafter referred to as cracks), which function as nucleation sites for propagating flow-bearing fractures. These cracks are formed by the simultaneous movement of large contiguous sections of the porous medium in different directions, a process induced by uniform fluid invasion into the porous medium. However, the similarities between both transition zones end here. In the viscous fracturing transition regime, fractures quickly become the dominant deformation mechanism, localizing the majority of the stresses and solid compaction around the advancing fracture tip. Conversely, in the capillary fracturing transition regime, fluid-invasion continues to serve as the main flow mechanism and source of deformation, where fractures and cracks are slowly expanded due to the more evenly-distributed capillarity-induced stresses localized at the advancing invasion front. Finally, non-invasive fracturing follows a different process, where there is little-to-no crack formation and fracture propagation is the main source of deformation and flow. Here, the co-advancing fracture and saturation fronts uniformly compress the solid around and in front of them until this deformation reaches the outer boundary of the simulated system (see the ``jet" like-structures at fracture tips in Fig. \ref{fig:strain}C.)

\section{Influence of Localized and Uniform Deformation}

So far we have explored how independently changing $k$, $p_c$, and $\tau_{yield}$ (among others) can affect the fracturing of plastic materials. However, our results also have implications for situations in which these variables are all varied simultaneously, such as during the compaction of soils, sediments, or viscoplastic sedimentary rocks (i.e. mudstones or clay-shales). In such situations, with increasing compaction, $k^{-1}$, $p_c$, and $\tau_{yield}$ should all increase, although at different rates. As such, we now study the effects of local and uniform deformation on the outlined fracturing regimes. 

\subsection{Localized Deformation}

\begin{figure}[t!]
\begin{center}
\includegraphics[width=0.9\textwidth]{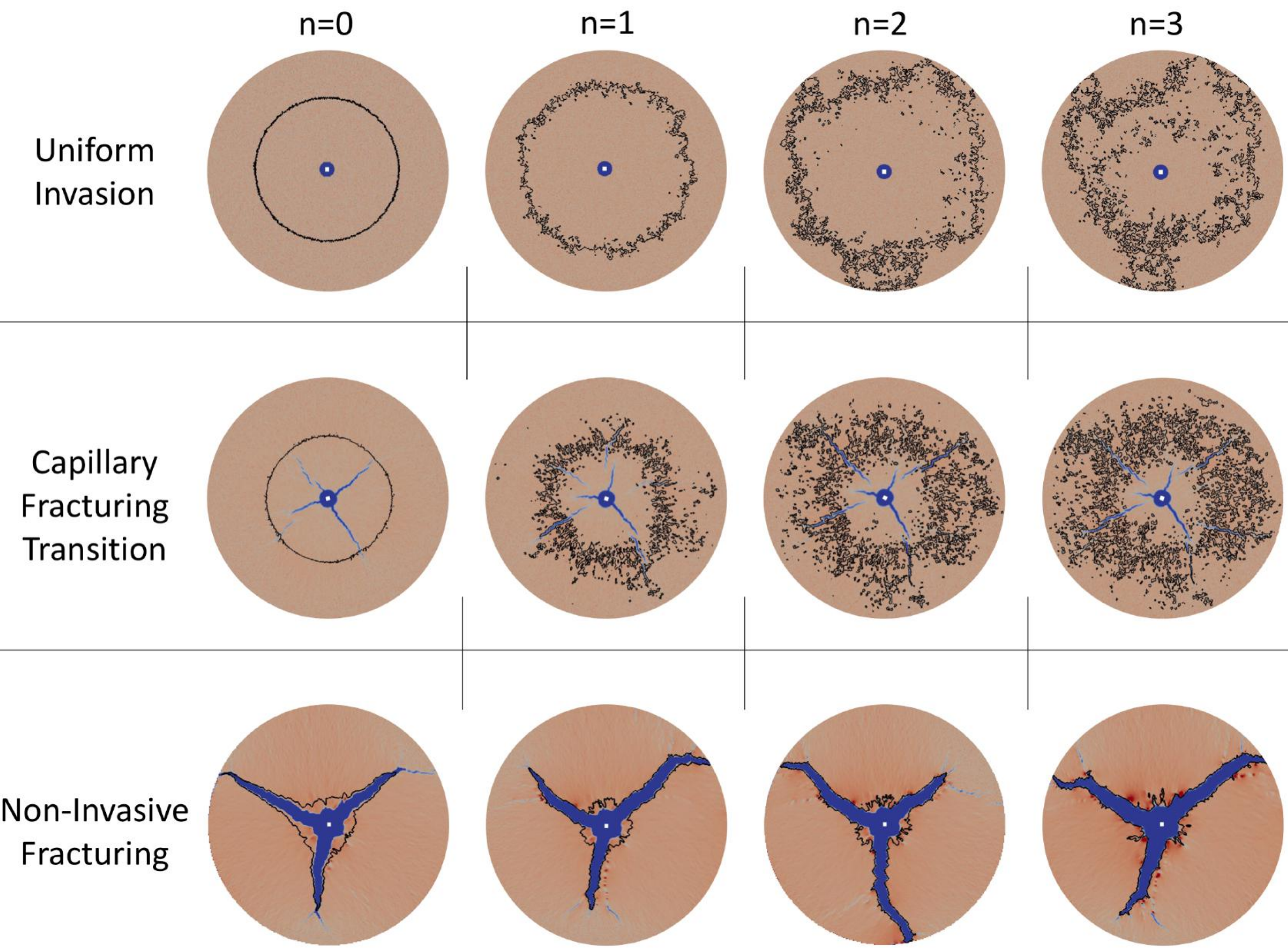}
\caption[Influence of the porosity-dependence of capillary pressure on fluid invasion and fracturing patterns]{\label{fig:variable_pc} Influence of the $\phi_f$-dependence of $p_c$ on fluid invasion and fracturing patterns for all three fracturing regimes. Here, $n$ represents the sensitivity parameter in the Leverett J-function analogue presented above. The color scheme is the same as in Fig. \ref{fig:Huang_Frac}.}
\end{center}
\end{figure}

The simulations presented above were carried out using the simplifying assumption that $p_c$ is invariant with $\phi_f$ (whereas $k$ and $\tau_{yield}$ are not). To evaluate the impact of this simplification on the results shown in Figs. \ref{fig:phase_diagrams}-\ref{fig:Final_phase}, we carried out additional simulations for all four regimes with a deformation-dependent capillary entry pressure based on a simplified form of the Leverett J-function where $p_{c,0} = p_{c,0}^*(\phi_s/\phi_s^{avg})^n$, $p_{c,0}^*$ is the capillary pressure at $\phi_s=\phi_s^{avg}$, and $n>0$ is a sensitivity parameter \citep{CLeverett,Li2015}. The results show that non-zero values of $n$ promote the creation of finger-like instabilities and the nucleation of cracks at the fluid invasion front, particularly in the capillary fracturing transition regime. Simulation predictions with different \textit{n} values are shown in Fig. \ref{fig:variable_pc}.

Despite the additional complexity of the resulting fluid invasion and fracturing patterns, results with $n>0$ conform to the overall phase diagram presented in Fig. \ref{fig:Final_phase}. The results at $n=0$ are therefore highlighted in the previous sections due to the greater simplicity of their fluid and solid distribution patterns.

\subsection{Uniform Deformation}

\begin{figure}[t!]
\begin{center}
\includegraphics[width=1\textwidth]{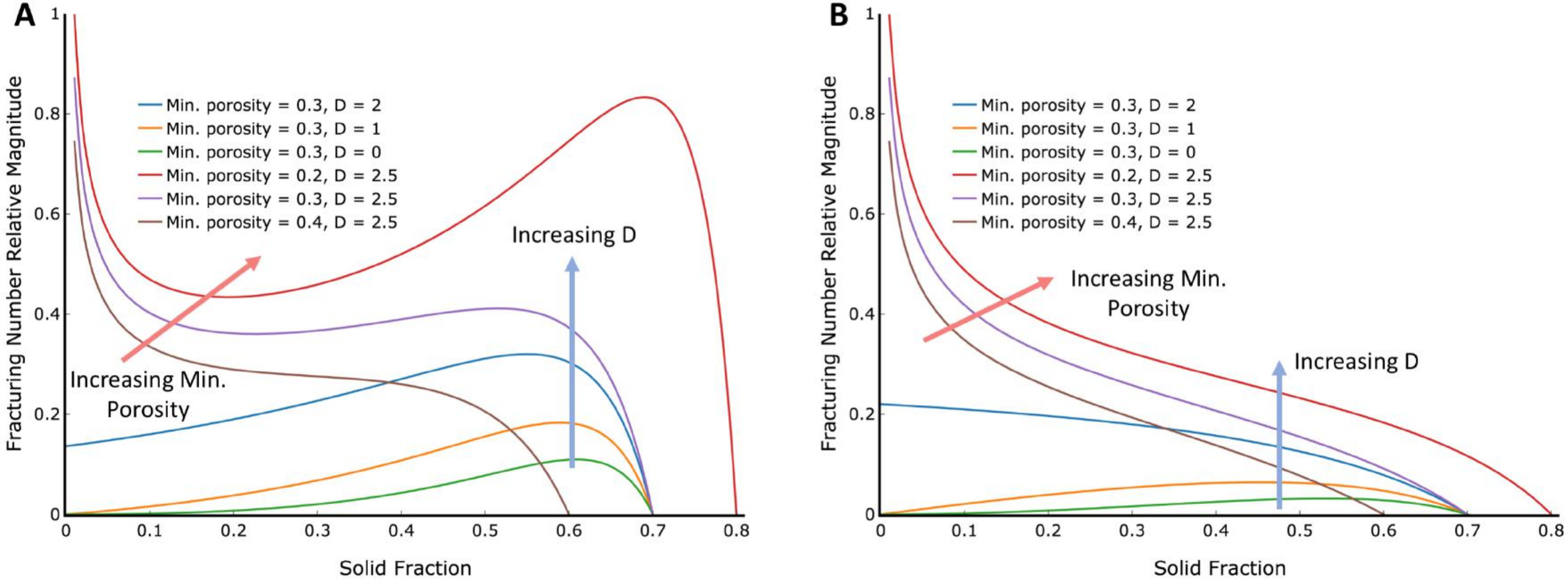}
\caption[Fracturing numbers' dependence on porosity]{\label{fig:NDep} Fracturing numbers' dependence on Porosity. A) For $N_{vF}$ B) For $N_{cF}$. Colored arrows represent the overall trends in the fracturing number behaviours when changing the fractal parameter $D$ and the minimum porosity parameter $\phi_{f,min}$ in Eqns. \ref{eq:N_vis_compaction}-\ref{Eq:N_cap_compaction}. Note the appearance of ``valleys" at high values of D in sub-figure A.}
\end{center}
\end{figure}

Having verified that the applicability of the fracturing numbers holds for systems were $k$, $\tau_{yield}$, and $p_c$ all vary with $\phi_f$, we now examine the effects of uniform compaction on said numbers. A direct analysis using the widely-used porosity-parameter relationships implemented above (the Kozeny-Carman relation for $k$, Leverett J-Function for $p_c$, and Quemada model for $\tau_{yield}$ \citep{CLeverett,Quemada1977,Spearman2017}) yields the following fracturing number - porosity dependence:

\begin{equation}\label{eq:N_vis_compaction}
    N_{vF} \propto \frac{(1-\phi_f)^{2-D}(1-\phi_{f,min}/\phi_f)}{\phi_f^2} 
\end{equation}

\begin{equation} \label{Eq:N_cap_compaction} 
N_{cF} \propto (1-\phi_f)^{2-D}(1-\phi_{f,min}/\phi_f)  
\end{equation} 

\noindent where $D$ is a rheological parameter based on the solid's fractal dimension (common values range for 1.7-2.9 for different clayey sediments \citep{Spearman2017}) and $\phi_{f,min}$ is the maximum possible degree of compaction. Through these relations, we can see that uniform compaction (or expansion) has a highly non-linear effect on fracturing. Equations \ref{eq:N_vis_compaction}-\ref{Eq:N_cap_compaction} indicate that whereas $N_{cF}$ tends to consistently decrease with increasing compaction, $N_{vF}$ is considerably more susceptible to changes in $\phi_f$ and exhibits multiple changes in the sign of its first derivative when $D > 2$, non-intuitively suggesting that fracturing can be either induced or suppressed through uniform compression. Plots of $N_{vF}$ and $N_{cF}$ as a function of solid fraction are shown in Fig. \ref{fig:NDep}.

\section{Conclusions}

In this chapter, we used the Multiphase DBB modeling framework to create a phase diagram that identifies two non-dimensional parameters that categorize the crossover between viscously-stable fluid drainage and fracturing as a function of wettability, solid deformability, and hydrodynamics. To the best of our knowledge, our results are the first to relate all three of these properties to characterize multiphase flow in viscoplastic porous media. As expected intuitively, we observe that fracturing occurs if the viscous and/or capillary stresses are sufficient to overcome the solid's structural forces. Thus, when it comes to systems with multiple fluids, it is necessary to consider the effects of surface tension, wettability, and pore size on the fluids' propensity to fracture or invade the permeable solid. Furthermore, we found that the two non-dimensional fracturing numbers described above delineate the existence of three fracturing regimes with distinct fracture propagation mechanisms. Lastly, we examined how uniform compression or expansion affect said non-dimensional numbers and a system's propensity to fracture . 
\begin{savequote}[75mm]
Science without religion is lame, religion without science is blind \qauthor{Albert Einstein}
\end{savequote}

\chapter{Simulation and Prediction of Stochastic Clogging Processes in Porous Media} \label{chapter:clogging}

\newthought{In this chapter,} we take a small detour from modeling multiphase flow in deformable porous media to focus on a conceptually simple, yet physically complex phenomenon: Clogging mechanics in porous media. This work was done as part of Princeton's Center for Statistics and Machine Learning (CSML) graduate certificate program. Although this chapter has not been published yet, we hope to do so in the near future. 

\section{Introduction}

The erosion, transport, and eventual deposition of fluid-suspended particles is ubiquitous within both natural and engineered porous systems. These processes control the evolution of sedimentary formations, the distribution of contaminants in the environment, and the clogging of porous media, pipes, and arteries \citep{Molnar2015,Phenrat2009,RobertDeSaintVincent2016,Zuriguel2014}. Although material transport in porous media is relatively well understood, the topics of particle deposition, accumulation, and clogging are still relatively new in the field.

The difficulty of predicting clogging in porous media arises from the complexity, heterogeneity, and stochastic nature of the relevant systems. Clogging in porous media is notoriously hard to probe and characterize, as it is necessary to account for each system's 3D geometry, grain size distribution, pore size distribution, rock/fluid chemistry, particle size, and even surface charge \citep{Ding2015,Sahimi1991a,Liu1995,Gerber2018,Mirabolghasemi2015,Pham2017}. Added to these spatial challenges is the fact that clogging is also a temporal process. Clogging mechanisms often don't reach a steady state: Clogs can form, redirect fluid fields, change pressure gradients, break, and form again downstream \citep{Bizmark2020}.  

Experimentalists have taken two separate routes in addressing these challenges. The first approach attempts to characterize particle deposition and clogging by reducing the complexity of the studied systems through the creation of simplified micro-models \citep{Wyss,Agbangla2012,Auset2006,Gerber2019}. This allows a measure of control of a handful of relevant variables (fluid flow rate, particle size, pore size, flow time) while significantly constraining others (flow path complexity and material heterogeneity). These studies have yielded noteworthy results. In particular, they have shown that clogging scales as a function of the pore to particle size ratio and that, under certain conditions, it is independent of particle injection rate and system porosity \citep{Wyss}. However, it is not clear if these conclusions still hold for more complex natural systems. 

The second approach relies on using advanced imaging techniques to probe real porous systems (or close approximations to them). These often rely on computed X-ray micro-tomography (XCT) and/or confocal microscopy. The former has taken a key role in characterizing particle aggregate formation in natural rock samples and identifying their dependence on fluid chemistry and rock geometry \citep{Liu1995,Li2006,Li2006a,Chen2009}. However, although highly informative, this technique suffers from not having large enough fields-of-view or high enough temporal resolutions. Confocal microscopy takes advantage of the fact that it is possible to match the refractive index of an artificial porous medium with the index of its permeating fluid phase in order to ``see" through the whole solid-fluid mixture. This technique has allowed scientists to characterize particle deposition and clogging as a function of time while having a high level of control of the relevant flow variables and material characteristics \citep{Bizmark2020,Mays2011}. 

However, all experimental studies suffer from the same practical problem: they cannot probe the full parametric phase-space required to properly generalize their results to naturally-occurring porous media due to the difficulty of creating and running an experiment with more than 3-5 independent variables. The necessity to investigate a larger range of systems is evidenced by the fact that experimental studies often reach conflicting conclusions. For example, some studies maintain that particle transport suppresses fluid flow by reducing the permeability of porous media \citep{Civan2010,Liu1995,Wiesner1996}, while others state that particles can actually help enhance flow \citep{Weber2009a,Kersting1999,Ryan1996,Schneider2021}. Even so, assuming that we could obtain a large-enough data set, there is no guarantee that we could derive general relationships between variables: some conclusions may only hold at low flow rates and/or low grain heterogeneity, while others may only hold at low fluid viscosities and/or high particle concentrations. One way to address this problem is through the application of Machine Learning (ML) algorithms to analyze such large data sets. These approaches have been shown to learn and predict complex relations in a wide variety of fields, from predicting fluid turbulence in jet engines \citep{Sirignano2020} to enabling image recognition in autonomous vehicles \citep{Kacic2019}. These models, however, require highly extensive data sets to become effective predictors.

Fortunately, due to the rise of large-scale parallelized cloud computing, it is now possible to run thousands of particle transport simulations in order to systematically study 10-20 dimensional parametric phase spaces. However, to the best of our knowledge, there have not been any studies that have used numerical models for this purpose. Current popular approaches rely on continuum-level (i.e. volume averaging) approximations such as filtration theory to study particle deposition \citep{Molnar2015,Messina2016,boccardo2018} and erosion \citep{hilpert2018,jager2017}. Contrastingly, coupled CFD-DEM approaches \citep{Zhao2013,Mirabolghasemi2015,Zuriguel2014,Natsui2012} attempt to model particle deposition directly through careful consideration of coupled particle-fluid mechanics and direct numerical simulations (DNS). However, due to their complexity and relatively-high computational cost, very few studies have attempted to model particle mechanics through these means.

In this chapter, we leverage Princeton's University's large computational resources in order to explore the underlying principles that control clogging mechanics. In particular, we explore the feasibility of using ML approaches to create a computational tool that can predict clogging \textit{a-priori} and that can be used to improve and streamline the design process of engineered porous systems. To do so, we ran 2000 CFD-DEM simulations in randomly-generated porous geometries while systematically varying 12 of the system’s design parameters (pore size, pore size heterogeneity, grain size heterogeneity, particle size, particle-particle attraction, particle-wall attraction, particle flux, porosity, system size, fluid velocity, and fluid viscosity). The resulting cases where labeled and analyzed in order to train and evaluate several different ML classifiers and regressors, which we then used to predict clogging in equivalent systems. Then, through model optimization, we identified which training features were most indicative of clogging in heterogeneous porous media. To the best of our knowledge, this is the first time ML approaches have been successfully used to reliably predict and control clogging processes. 


\section{Methods}

\subsection{CFD-DEM}

Our numerical simulations where performed on the CFDEM{\circledR} computational modeling framework, which couples Computational Fluid Dynamics (CFD) simulations performed in OpenFOAM{\circledR} with Discrete Element Models (DEM) performed in LIGGGHTS{\circledR}. Both C++ libraries are free-to-use, open-source, and parallelizable platforms. OpenFOAM{\circledR} uses the Finite Volume Method to discretize and solve partial differential equations to describe fluid motion in complex 3-D grids. In turn, LIGGGHTS{\circledR} uses Lagragian approaches and discrete particle models to describe the motion and interaction of a large number of individual particles within a grid-free environment. The object-oriented structure of both codes and multitude of supporting libraries allows the user to easily customize each simulation's setup with different numerical discretization schemes, time-stepping procedures, matrix-solution algorithms, and supporting physical models.

\begin{figure}[htb!] 
\begin{center}
\includegraphics[width=1\textwidth]{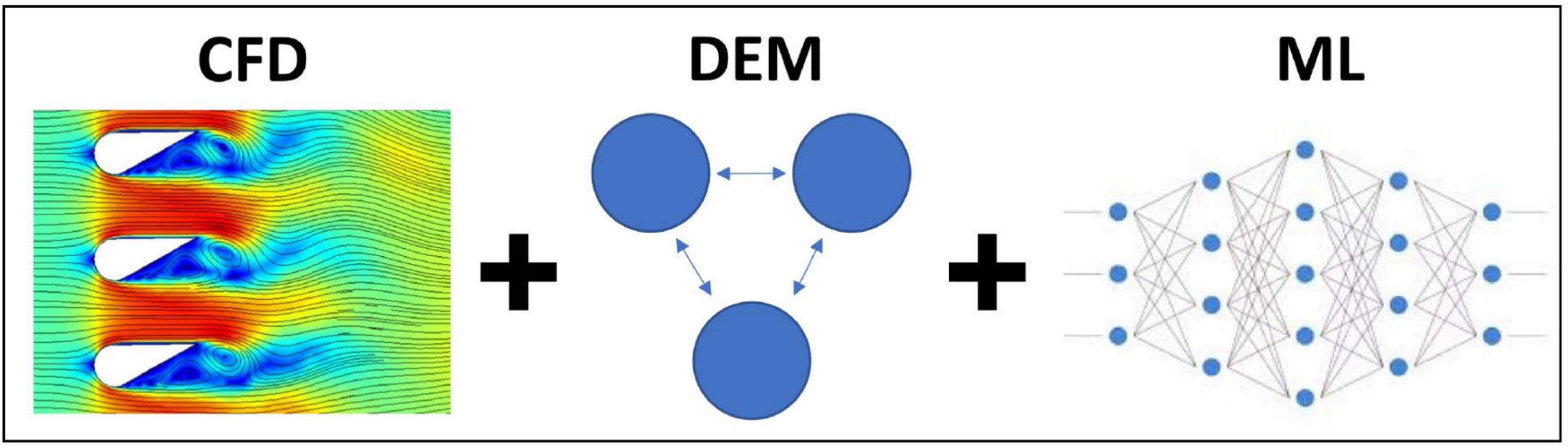}
\caption[Conceptual representation of the models used for clogging simulations and predictions]{\label{fig:clogging_models} Conceptual representation of the models we used for clogging simulation and prediction.}
\end{center}
\end{figure}

CFDEM{\circledR} successfully couples both models through an iterative algorithm that super-imposes their physical domains. Here, an additional particle-drag term is added to the fluid's momentum conservation equation to account for particle movement, and an additional fluid-drag term is added to each particle's equation of motion to account for fluid-induced motion (i.e. flow and buoyancy). Each model can probe its counterpart at any given moment in time, at which point the CFDEM{\circledR} algorithm iterates over both solver's solutions until the desired convergence metric is reached. Just as in its two base models, CFDEM{\circledR} counts with several customizable options regarding the coupling frequency, numerical schemes, and coupling-physical models. Our specific numerical implementation will be discussed in Section \ref{cfdem_setup}.

\subsection{Machine Learning}

The statistical and ML analysis performed in this study was carried out through Scikit-learn, a free-to-use, python-based library. Sckikit-learn was selected for its ability to allow users to easily train, optimize, and interpret a large number of supervised and unsupervised machine learning algorithms. More complex and customizable ML libraries such as TensorFlow and pyTorch would have been equally adequate. In our case, since we worked exclusively with labeled data, we only focused on training and testing a representative sample of ML classifiers and regressors (see Tables \ref{fig:table_classifiers} and \ref{fig:table_regressor}).

\section{Workflow}
   
Having established the numerical simulation and ML basis of our investigation, we now describe our project workflow: \textbf{First}, we 
created randomized porous systems. In this step, we generated the geometry and computational mesh on which we performed our numerical simulations. A combination of randomization and systematic-variation of the relevant geometric variables (e.g. pore size) was necessary for the eventual generalization of our results. 
\textbf{Second}, we ran CFDEM simulations. During this step, we ran thousands of clogging processes over a large parameter space by varying the specified fluid and particle variables (e.g. fluid flow rate and particle size). \textbf{Third}, we identified clogged systems by developing a standardized metric for labeling and characterizing our completed numerical simulations. \textbf{Fourth}, we identified and engineered features for training. In this step, we converted the varied system parameters into a set of standardized non-dimensionalized variables. This was done for two reasons: it reduces redundancy and creates more predictive features (e.g. the ratio between particle size and pore size is a better predictor for clogging than either variable by itself) while also allowing us to apply the resulting ML model to any potential system that can be similarly non-dimensionalized. \textbf{Fifth}, we trained the ML models. Here, we trained, optimized, and identified the best ML classifier and regressor for predicting our simulated clogging processes. \textbf{Finally}, we interpreted and analysed the models. In other words we identified the most predictive features for particle clogging in porous media. In addition, we also evaluated the efficacy of different models, their shortcomings, and their potential improvements. 

The rest of the chapter will focus on developing and discussing each of these individual steps. 

\section{Experimental Setup}

\subsection{Creating Randomized Porous Systems}

\begin{figure}[htb!] 
\begin{center}
\includegraphics[width=0.7\textwidth]{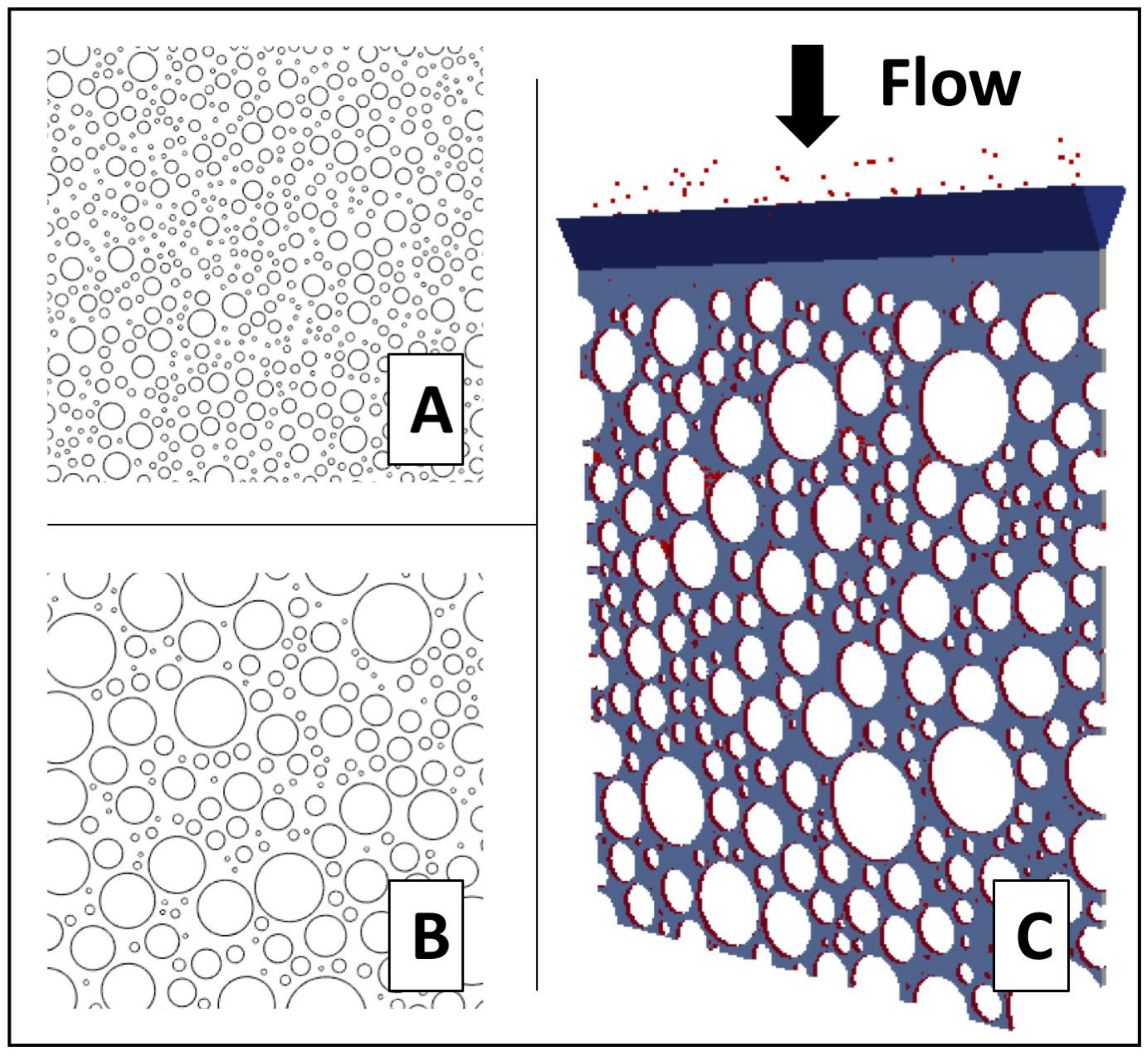}
\caption[Examples of geometries used for clogging simulations.]{\label{fig:clogging_geometry} A,B) Examples of randomly-generated porous media configurations with different grain sizes and porosities. C) Example of the numerical mesh and simulation setup. Here, the areas within the cylinders (i.e. grains) are removed during meshing to making them inaccessible to fluids and particles. Also note the direction of flow and the fact that the only inlets and outlets are at the upper and lower boundaries of the geometry, respectively.}
\end{center}
\end{figure}

The randomized porous geometries used in this study were generated through a custom Matlab{\circledR} script which allowed us to specify the following criteria: average grain size, grain size standard deviation, average pore size (i.e. shortest distance between two given grains), pore size standard deviation, and porosity (see Figure \ref{fig:clogging_geometry}). Note that, for simplicity, all grains where assumed to be cylindrical in shape and the domain's length and width where kept constant at 50 cm. Given a fixed domain space, the porosity and pore size distribution become coupled variables, which is why only one of these could be specified for a given geometry (the other one was measured at the end of geometry generation). Furthermore, to allow for a measurable particle flux into the geometry, a funnel-type feature was added to the top of the generated geometry during the numerical simulation setup (see Figure \ref{fig:clogging_geometry}C). The result is an algorithm that can generate well-characterized, yet randomized porous systems on which we can perform our numerical simulations. 

\subsection{CFDEM Simulations}\label{cfdem_setup}

We now present our numerical simulation setup. As discussed earlier, the main advantage that numerical simulations have over conventional experiments is their ability to probe large parameter spaces with ease. As such, we systematically varied the following 12 variables over the following ranges in 2000 CFDEM simulations:

\begin{enumerate}
    \item Porosity ($\phi$): $0.26$ to $0.98$
    \item Average pore size ($D$): ${10}^{-4} $ to ${10}^{-2}$ m
    \item Pore size standard deviation ($D_{std}$): $5\times{10}^{-4} $ to $ 5\times{10}^{-3}$ m 
    \item Average grain size ($G$): ${10}^{-3} $ to $ 5\times{10}^{-2}$ m
    \item Grain size standard deviation ($G_{std}$): ${10}^{-3} $ to ${10}^{-2}$ m
    \item Geometric thickness ($T$): ${10}^{-4} $ to ${10}^{-2}$ m
    \item Particle diameter ($d$): ${10}^{-4} $ to ${10}^{-2}$ m
    \item Particle flux ($F$): ${10}^{2} $ to $ 1.2\times{10}^{4}$ particles/s
    \item Particle-particle attraction ($PP$): $0 $ to $ 5\times{10}^{6}$ J/$\mathrm{m^3}$
    \item Particle-wall attraction ($PW$):  $0 $ to $ 5\times{10}^{6}$ J/$\mathrm{m^3}$
    \item Fluid velocity ($U$): $0 $ to $ 1.5$ m/s
    \item Fluid kinematic viscosity ($\nu$): ${10}^{-2} $ to ${10}^{-8}$ $\mathrm{m^2}$/s.

\end{enumerate}

All other parameters were kept constant. In particular, fluid density was 1000 $\mathrm{kg}/\mathrm{m^3}$, particle density was 1200 $\mathrm{kg}/\mathrm{m^3}$, and gravity ($g=9.8 m/s^2$) was set to be constant in the direction of flow. Particle-Particle/Wall attraction was modeled through the SJKR model, Particle-Particle/Wall collisions where captured through the Granular Hertz model with a Poisson Ratio of 0.22, and the fluid-particle drag coupling was calculated through the DiFelice Drag model every 0.001 seconds \citep{Barthel2008}. These models imply the following: A) the attraction force is normal to the inter-particle contact area and is only activated once particles are in direct contact with each other. B) Inter-particle collisions are modeled by defining both a normal force (spring + damping forces) and a tangential force (shear + damping forces). This allows particles to both bounce and roll around obstacles and other particles. C) Fluid-particle interactions are governed by a drag force that is proportional to the relative velocity between a given particle and the fluid, the solid volume fraction in the specified control volume, and the square of the Reynolds number \citep{ZhouCFDEMDrag}. 

\begin{figure}[htb!] 
\begin{center}
\includegraphics[width=0.7\textwidth]{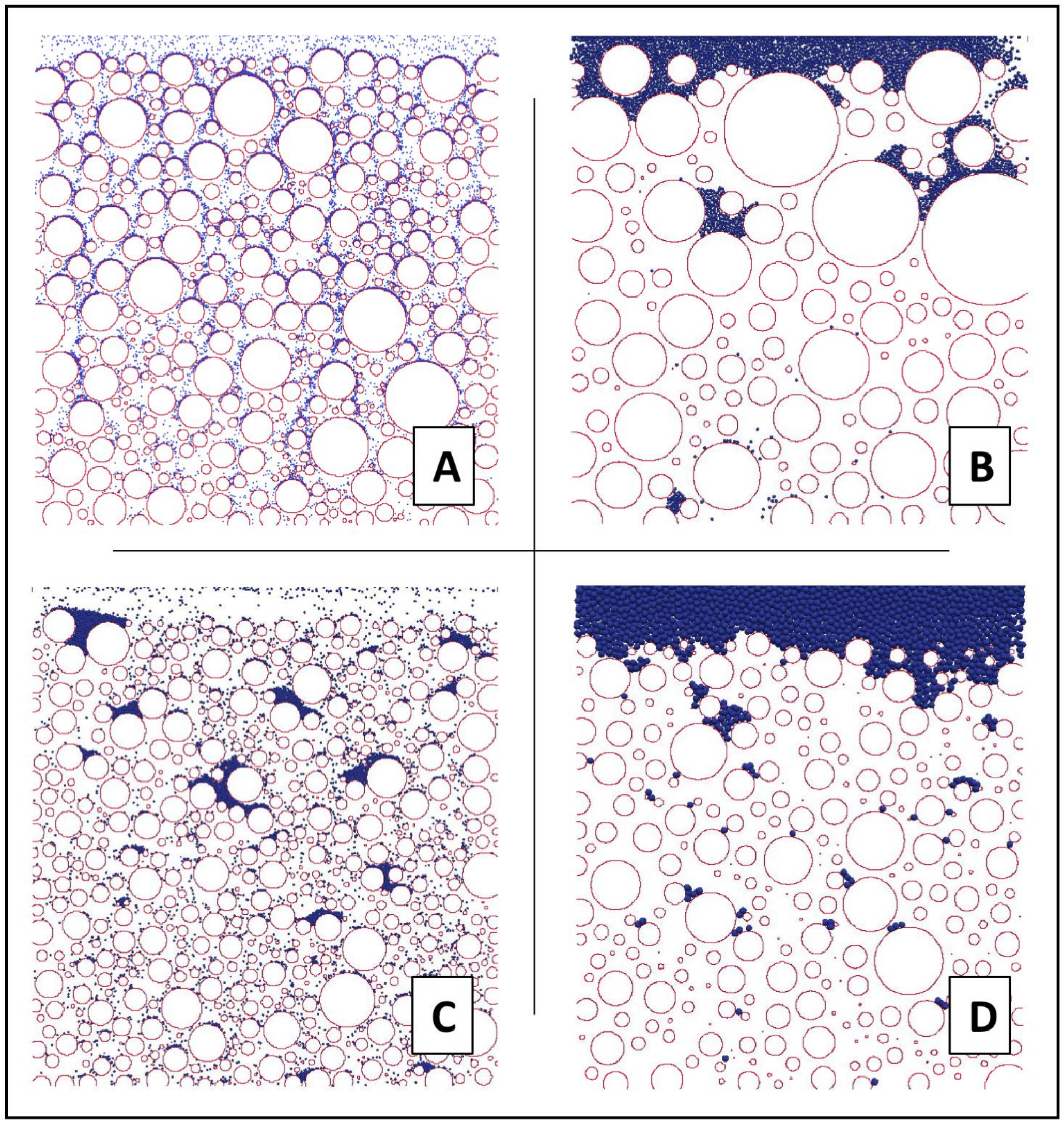}
\caption[Examples of different clogging simulations]{\label{fig:clogging_simulations} Examples of different clogging simulations. Note the wide range of particle sizes (A and D), porosities (B and C), pore size distributions (C and D), and grain size distributions (B and C).}
\end{center}
\end{figure}

In order to avoid modeling trivial clogging cases dictated by size exclusion, we exclusively considered systems where the particles' diameter was less than the pore size and geometric thickness (See Figure \ref{fig:clogging_simulations}). Each simulation was set to run for 5 hydraulic residence times (0.5 m / fluid velocity), which took on average about 5.5 hours of computing time on a 28-core Broadwell Xeon node.


\section{Data Analysis and Model Training}

\subsection{Identifying Clogged Systems}

\begin{figure}[htb!] 
\begin{center}
\includegraphics[width=1\textwidth]{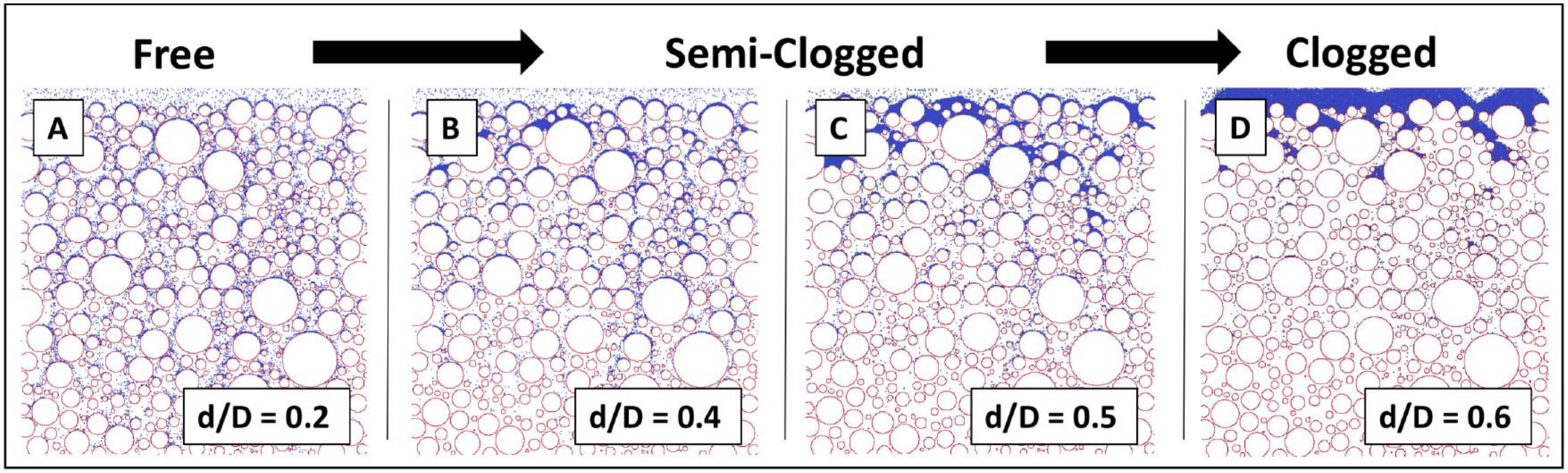}
\caption[Examples of different clogging levels]{\label{fig:clogging_clogging} Examples of different clogging levels. Each panel in this figure shows the final state of four separate clogging simulations with identical geometries and flow conditions. Their only difference is the size of the simulated particles (shown here as the particle size to pore size ratio, d/D, which increases from left to right). A) Free geometry. B, C) Semi-clogged geometries D) Fully-clogged geometry.}
\end{center}
\end{figure}

Accurate characterisation of our final configurations was crucial for the training and testing of our ML algorithms. This process was complicated by the fact that clogging in a heterogeneous porous medium is not necessarily a discrete state. As shown Figure \ref{fig:clogging_clogging}, clogging might only occur at certain pore throats, might not occur at all, or might occur throughout the porous medium \cite{Bizmark2020}. As such, in order to properly label our clogging geometries for ML classification (or obtain a continuous prediction variable for ML regression) we had to define an objective metric that could characterize the degree of clogging in each system. We called this variable the ``Clogging Number" (CN). 

The Clogging Number (Eqn. \ref{eq:CN}) is the product of two independent factors obtained at the end of each simulation : 1) The average relative distance between particles (DP). 2) The symmetry of the particle's final velocity distribution (SYMM). The significance of the first factor is fairly intuitive; systems with particles that end up closer together tend to have a higher degree of clogging. The second factor describes the relative behaviour of all the particles throughout the porous medium, where relatively high symmetry indicates homogeneous particle behaviour (i.e. the system is either fully clogged or fully unclogged) and low symmetry indicates heterogeneous behaviour (i.e. some particles are moving and some are stationary).

\begin{equation}\label{eq:CN}
    \mathrm{CN} = \mathrm{DP}*\mathrm{SYMM}
\end{equation}

\begin{figure}[htb!] 
\begin{center}
\includegraphics[width=0.8\textwidth]{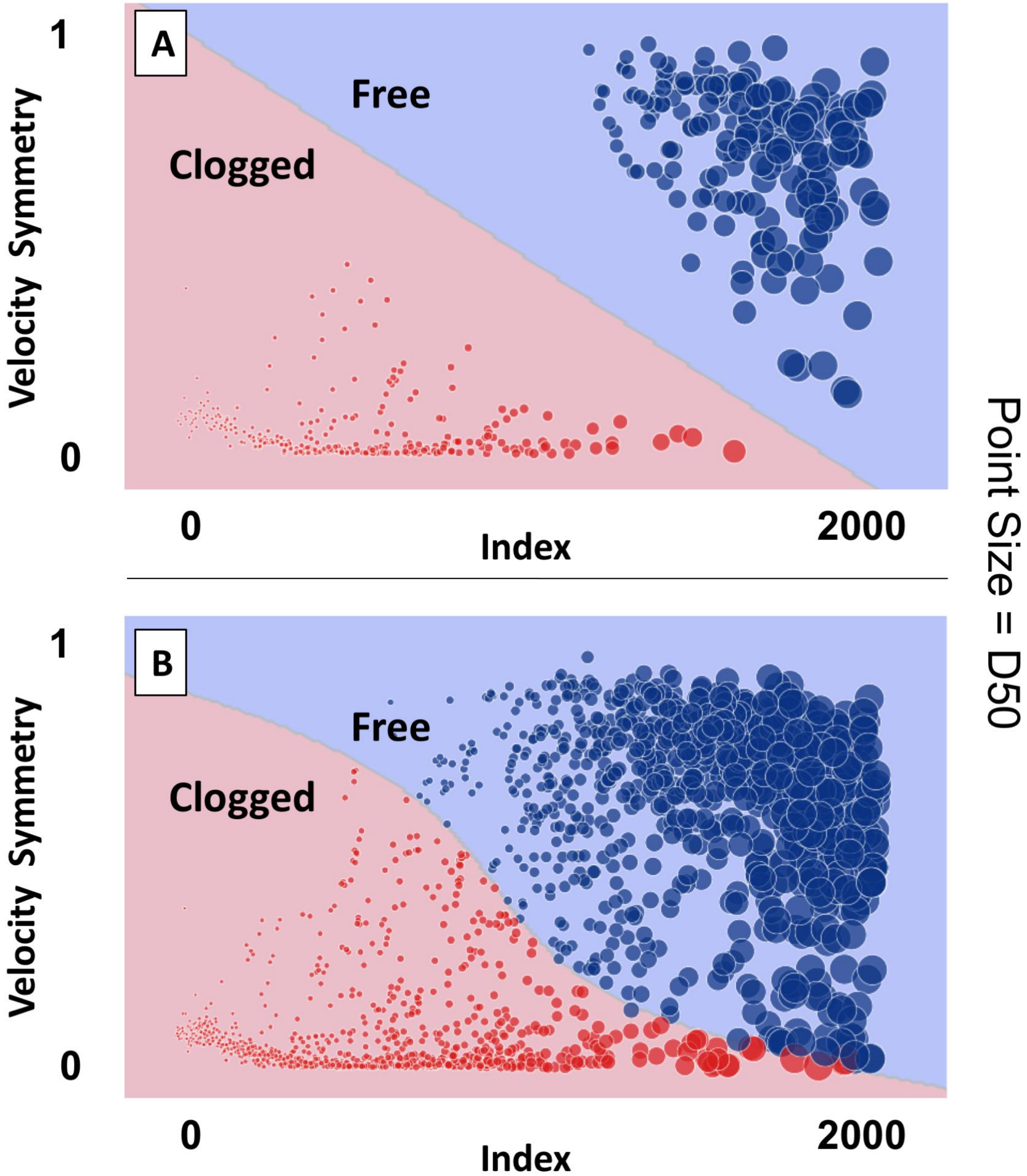}
\caption[Clogging as a function of the average final distance between particles and the symmetry of the particles' final velocity distribution.]{\label{fig:clogging_graphs} Clogging sample distribution as a function of the average final distance between particles (DP) and symmetry of the particles' final velocity distribution. Here, red and blue colors represent clogged and unclogged samples, respectively. Note that point size is directly proportional to DP and that all cases where plotted from lowest DP to highest DP values for visualization purposes. A) Result of using CN <1 to separate 300 manually-labelled fully clogged and unclogged samples. B) Result of using CN <1 to label all data samples. The color contours in both graphs represent the phase-space prediction of the optimized MLP classifier trained with their respective samples and discussed in Section \ref{Training_classifiers}.}
\end{center}
\end{figure}

Therefore, multiplying both factors together into the CN allows us to characterize each simulation's degree of clogging, where DP tells us if the system tends to be clogged or unclogged, and SYMM tells us the degree to which they are one or the other. After testing several different averaging procedures and symmetry measures, the following two metrics yielded the best performance when used to create a CN that could separate fully clogged from fully unclogged systems (see Figure \ref{fig:clogging_graphs}A). DP was calculated by averaging the average distance between each particle and its closest 50 neighbors, dividing by the particle diameter, and subtracting 1 from said value (to avoid double counting). In turn, SYMM was determined by calculating the ratio between the particles' median velocity and their mean velocity. This ratio describes the symmetry of the distribution by using the fact that the median of a sample is independent of changes in its extreme values, while the mean of a sample is. Therefore, whenever a group of particles form a blockage in an otherwise unclogged geometry, the median of the velocity will change at a much slower rate than the mean velocity and their ratio will decrease. Conversely, if we have a fully clogged or fully unclogged system, the velocity distribution will have a higher degree of symmetry, and this ratio will be close to one. 

Figure \ref{fig:clogging_graphs} shows that CN now provides us with a suitable way of characterizing the final state our clogging simulations. It also gives us a straightforward way to form discrete labels for classification out of continuous variables, where all samples with CN < $1$ will be labeled as ``Clogged" ($1$) and all others as ``Unclogged" ($0$). This threshold was chosen by identifying the best linear classifier that could separate a sample of 300 manually-labeled fully clogged and fully unclogged systems (without considering any semi-clogged samples). Applying this threshold to the complete data set yields 1012 samples classified as Clogged and 988 classified as Unclogged. 

\subsection{Feature Standardization and Non-dimensionalization}

In order to generalize our simulations and improve our models' prediction power, we now non-dimensionalize and/or re-scale our training features. The objective is to create a standardized set of features that can be readily quantified from any system involving the flow of solid bodies through a set of static obstacles (such as cars through a road, sheep through a gate, contaminants through soil, ext...). Nondimensionalization also allows us to add information to the model by explicitly dictating important relationships between variables. The following features are all system characteristics than can be measured \textit{a-priori}, meaning that, if trained correctly, a ML model will be able to use these features to predict clogging without actually having to run any simulations or experiments.

\begin{enumerate}
    \item Particle - pore size ratio ($d/D$): $d/D$
    \item Particle - geometric thickness ratio ($d/T$): $d/T$
    \item Particle - grain size ratio ($d/G$): $d/G$
    \item Geometric thickness - pore size ratio ($T/D$): $T/D$
    \item Grain size - pore size ratio ($G_{nd}$): $G/D$
    \item Non-dimensional grain size standard deviation ($G_{std,nd}$): $G_{std}/D$
    \item Non-dimensional pore size standard deviation ($D_{std,nd}$): $D_{std}/D$
    \item Non-dimensional particle-wall attraction ($PW_{nd}$): $\frac{PW \times \pi \times (D/2)^2}{ (\mathrm{Particle Mass} \times g)}$
    \item Non-dimensional particle-particle attraction ($PP_{nd}$): $\frac{PP \times \pi \times (D/2)^2}{ (\mathrm{Particle Mass} \times \textbf{g})}$ 
    \item Standardized particle flux ($F_{nd}$): $(F \times D )/ \phi $
    \item Standardized Fluid velocity ($U_{std}$): $U/D$
    \item Logarithm of kinematic viscosity ($log(\nu)$): $log_{10}(\nu)$.

\end{enumerate}

The resulting 11 variables where used to train the classifiers and the regressors showcased in the following sections. 

\subsection{Training and Identifying Best Classifiers} \label{Training_classifiers}

The ML classifiers shown in Table \ref{fig:table_classifiers} were trained and tested by using k-folds cross-validation (k=5) on all 2000 cases, resulting in training sample of 1600 and a testing sample of 400. These classifiers were chosen by virtue of their differing approaches, as we had little-to-no intuition of which one would work best for our purposes (no one has ever used ML to predict clogging mechanisms). Classifier performance was evaluated by calculating their testing accuracy, precision, the area under their Receiving Operating Characteristic (ROC) curve, and quantifying the total number of true positives (TP), false positives (FP), true negatives (TN), and false negatives (FP). In this case, a true positive pertains to correctly labeling a case as ``Clogged". These metrics where chosen due to their ability to quantify overall classifier performance (accuracy), while also testing for its ability to correctly predict positive clogging labels (precision + ROC). Furthermore, we implemented a grid-search algorithm in order to optimize our choice of model hyperparameters and thus improve the predictive power of each type of classifier. Table \ref{fig:table_classifiers} shows the performance metrics of each classifier as a result of average precision grid-search optimization. 

\begin{table}[htb!] 
\begin{center}
\includegraphics[width=1\textwidth]{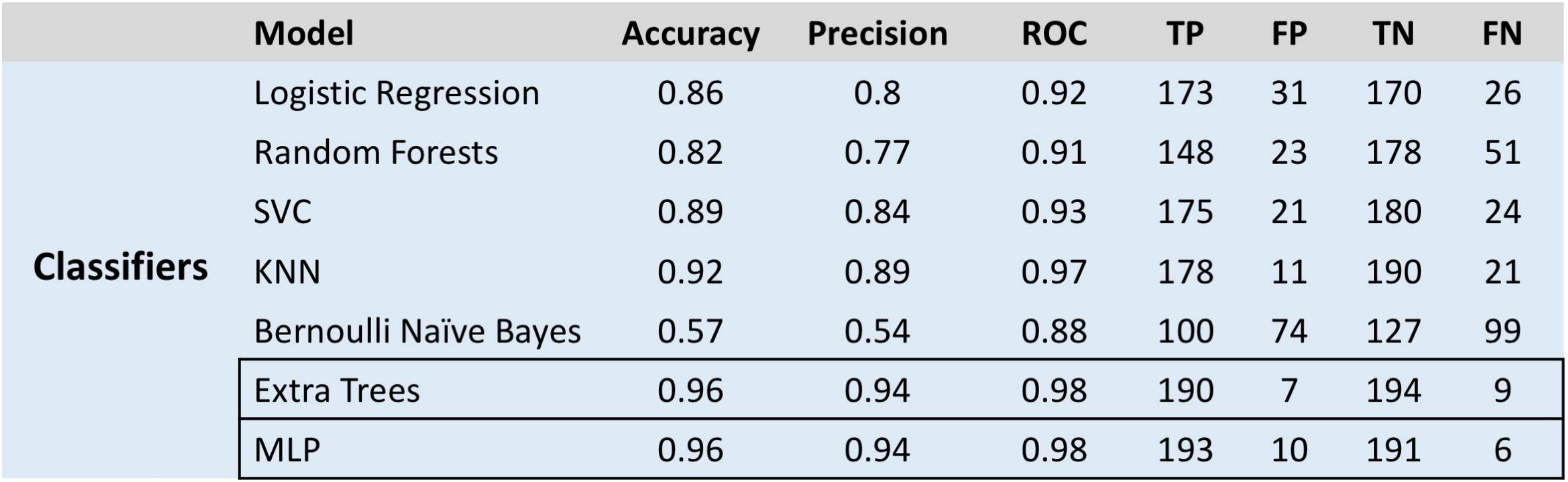}
\caption[Clogging prediction performance of the chosen ML classifiers.]{\label{fig:table_classifiers} Clogging prediction performance of the chosen ML classifiers. The highlighted sections represent the classifiers with the best performance.}
\end{center}
\end{table}

The results in Table \ref{fig:table_classifiers} are very encouraging, showing that the best two classifiers, Extra Trees and the Multi-Layer Perceptron (MLP), reliably achieve a classifying precision of 0.94 and an accuracy of 0.96, meaning their false positive/negative rates are practically indistinguishable. Further analysis into the misclassified samples show that all of them are partially-clogged samples that lie within the classification decision boundary shown in Figure \ref{fig:clogging_graphs}B. Please refer to Section \ref{clogging_conclusion} for a description of the optimized models. 

\subsection{Classification Feature Analysis} \label{class_feature_analysis}

Although the testing metrics of our models are relatively high, it might not be practical for a potential user to have to obtain all 11 training variables for every system of interest that he/she wishes to test. It might be more efficient to reduce a model's accuracy in favor of reducing its complexity and number of features. As such, we now aim to identify the most predictive features of the two most succesful models identified above. 

To do so, we implemented a permutation feature importance algorithm, where we quantified the decrease in the models' precision as a result of shuffling the values of a single feature. This effectively disassociates the feature values from each sample's labels. The relative magnitude of the drop in the model's precision is proportional to the relative importance of the shuffled feature. This drop is then normalized with respect to all other shuffled features, and a feature importance ranking is created. In our case, we chose to shuffle and measure the score effect of each feature 50 times before continuing on to the next one. 

\begin{table}[htb!] 
\begin{center}
\includegraphics[width=1\textwidth]{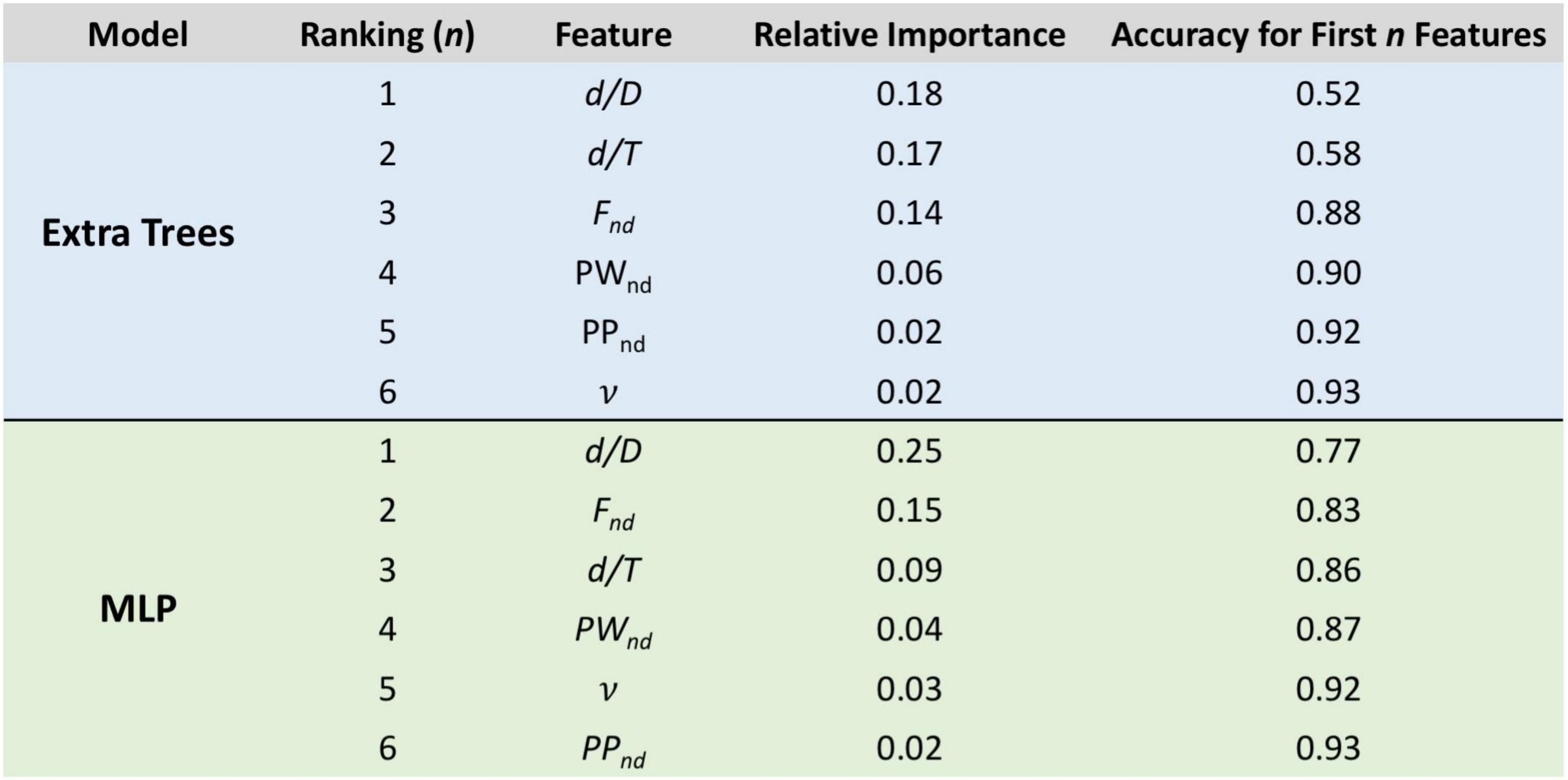}
\caption[Classifier Feature Relative Importance]{\label{fig:classifier_features} Tabulated relative feature importance for the top 6 classifier features. The rightmost column identifies the accuracy of a model trained and tested exclusively with the top ``n" features. }
\end{center}
\end{table}

Table \ref{fig:classifier_features} shows the top 6 features for our optimized Extra Trees and MLP classifiers. Not surprisingly, both cases have the ratio between particle diameter and pore size ($d/D$) as their top feature, followed by the standardized particle flux ($F_{nd}$) and the ratio between the particle diameter and the geometric thickness ($d/T$). These are then followed by the non-dimensionalized attractive forces and the fluid viscosity. What is surprising however, is the relatively small predictive power that $d/D$ has, specially in the Extra Trees classifier, where a model trained and tested with only said feature cannot predict clogging any better than a coin flip. The addition of the next two features increases prediction accuracy to about 0.88 for both classifiers, meaning that the bulk of the clogging cases can be explained/predicted by just 2 complementary physical processes: particle size exclusion and particle flux. The remaining cases rely on more complex physics, where it is important to consider particle-particle-wall attractions and viscous flow-particle couplings. 

Porosity, pore size and grain heterogeneity, and the normalized fluid velocity do not seem to have as much as an influence on clogging. This might be explained by two different factors: 1) We did not probe the sections of the 11-dimensional parameter space for which these features become significant. 2) These features really do not affect clogging in a significant way. Since fluid viscosity (i.e. drag) did appear to have a quantifiable influence on clogging mechanics, we also believe that fluid velocity should play a part as well. For this reason we are inclined to believe the first explanation is more likely, which is why this will be the first point to be addressed in our future work. 

\subsection{Training and Identifying Best Regressors} \label{Training_regressors}

Training and testing ML regressors followed the same procedure outlined in Section \ref{Training_classifiers}. The only difference was the fact that we are now aiming to predict the CN associated with each sample, as opposed to its discrete label. Given that the degree of clogging is a non-discrete latent variable, it can be argued that using regressors to predict a CN might actually be more useful than binary classification. However, it is important to consider that the CN is an abstract metric; it is not obvious how this number's magnitude reflects actual physical processes. For this reason we make no value-judgement as to which approach might be more useful and instead present the results of both types of predictors.

\begin{figure}[htb!] 
\begin{center}
\includegraphics[width=1\textwidth]{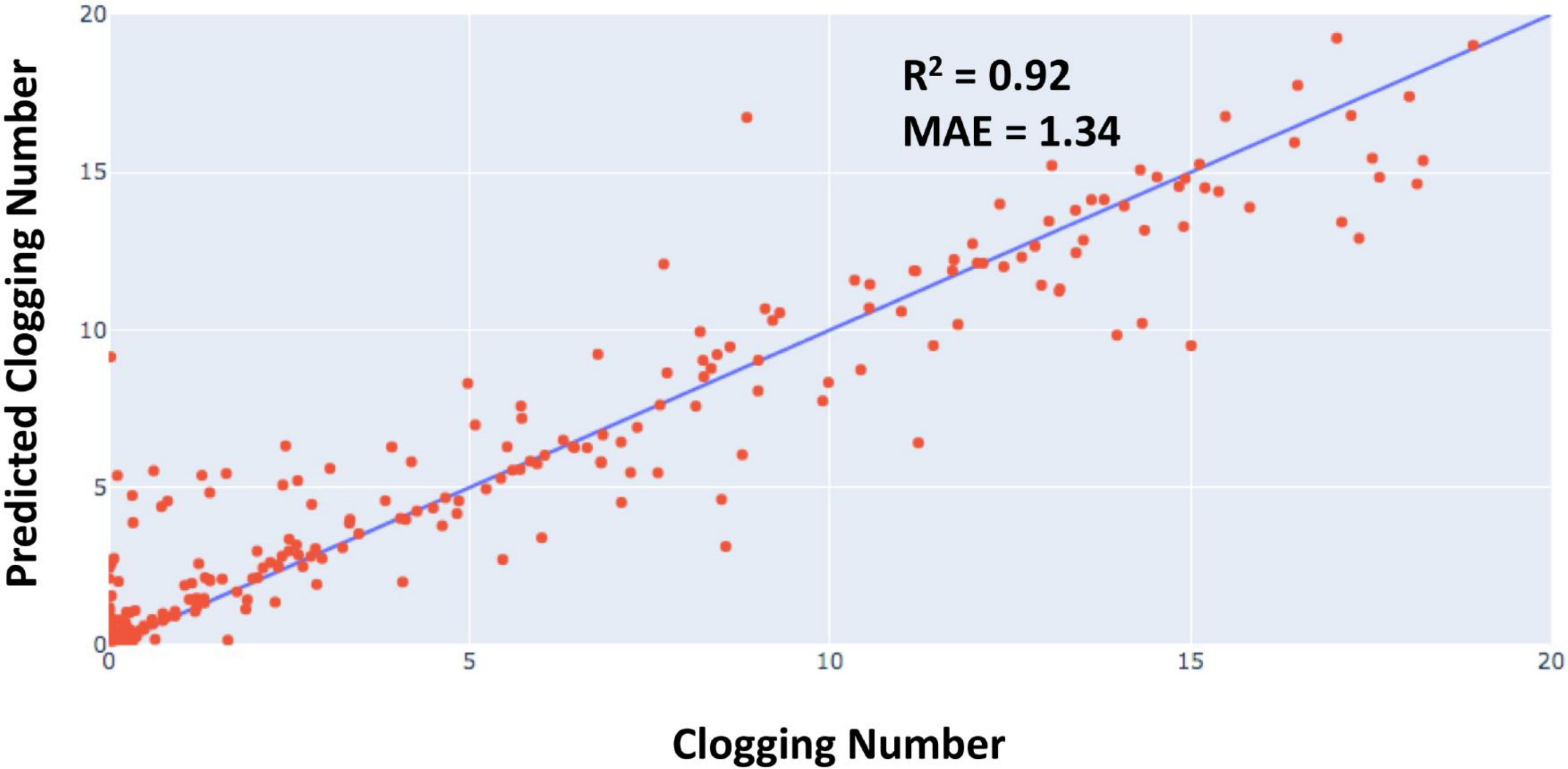}
\caption[Optimized Random Forest regression results]{\label{fig:regression_curve} Optimized Random Forest regression results.}
\end{center}
\end{figure}

The results of our training, testing, and optimization procedure are shown in Table \ref{fig:table_regressor}, where we evaluate regressor performance by quantifying the $\mathrm{R^2}$ and Mean Absolute Error (MAE) obtained from plotting a ``predicted CN" vs ``actual CN" curve (Figure \ref{fig:regression_curve}). 

\begin{table}[htb!] 
\begin{center}
\includegraphics[width=0.7\textwidth]{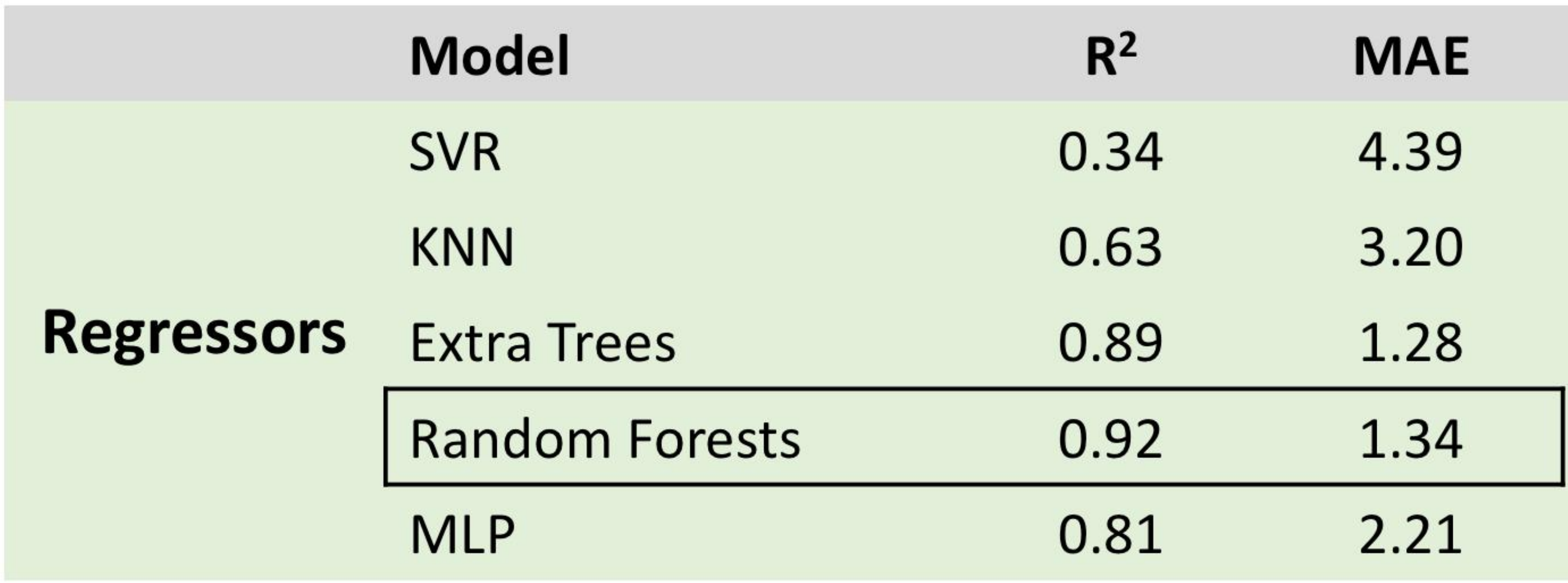}
\caption[Clogging prediction performance of the chosen ML regressors]{\label{fig:table_regressor} Clogging prediction performance of the chosen ML regressors. The highlighted section represents the regressor with the best performance.}
\end{center}
\end{table}

Table \ref{fig:table_regressor} clearly shows that decision tree-based ML models are, once again, the best predictors for clogging mechanisms in porous media, closely followed by neural networks (MLPs). A description of the best optimized model (Random Forest) can be found in Section \ref{clogging_conclusion} .

\subsection{Regression Feature Analysis} \label{Reg_feature_analysis}

Just as we did for classifiers, we now turn to evaluate feature importance for our best regression algorithm (Random Forest) through a permutation feature importance algorithm. The results shown in Table \ref{fig:regressor_features} are surprising, especially when compared to the ones presented in Section \ref{class_feature_analysis}. The top two features follow the same pattern we saw before, where $d/D$ and $F_{nd}$ are once again the most significant features. However, this time, the ratio between particle size and grain size ($d/G$) takes the third spot, followed by the particle-wall attraction ($PW_{nd}$), the particle-geometric thickness ratio ($d/T$), and the porosity ($\phi$). The previously-unseen correlation between $d/G$ might be explained by the fact that we used the symmetry of the particle velocity distribution to construct the CN. Larger grains/obstacles produce larger deviations in the particles' velocity than their smaller counterparts, effectively changing the CN without actually signifying the presence of clogging in the porous medium. A similar thing can be said about the effect of porosity on the CN, where less porous systems may exhibit more velocity fluctuations than more porous ones. 

\begin{table}[htb!] 
\begin{center}
\includegraphics[width=1\textwidth]{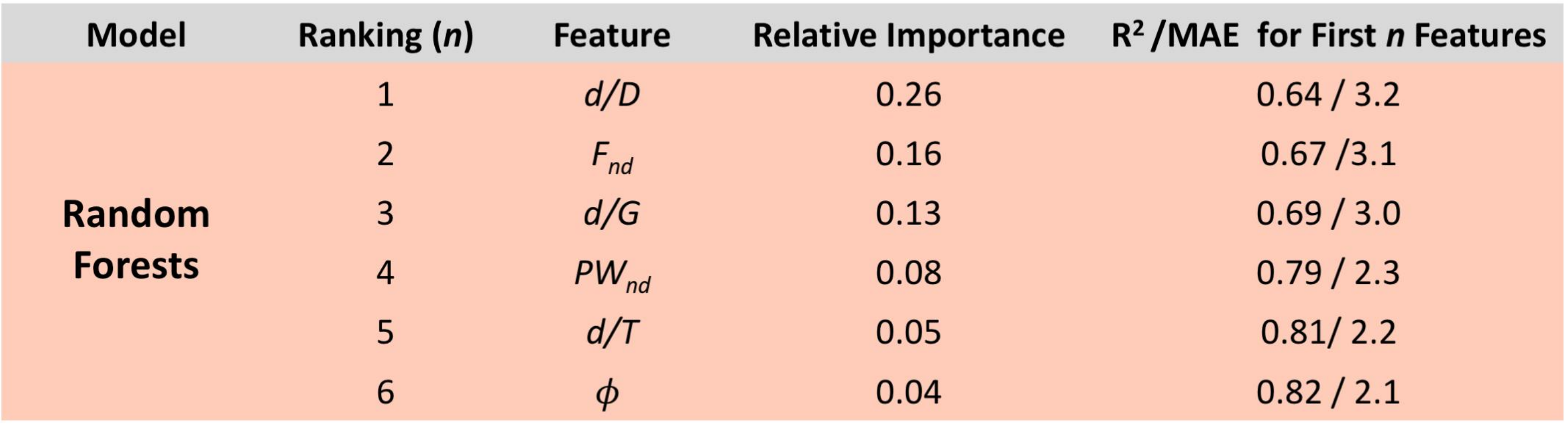}
\caption[Classifier Feature Relative Importance]{\label{fig:regressor_features} Tabulated relative feature importance for the top 6 regression features. The rightmost column identifies the score metrics of the model if it is trained and tested exclusively with the top ``n" features.}
\end{center}
\end{table}

This analysis then begs the question of whether the importance of $d/G$ and $\phi$ is actually predictive of clogging mechanisms or just an artifact of our simulation characterization procedure. We believe that the answer is the later, as the marginal increase in the prediction accuracy of these two feature seen in Table \ref{class_feature_analysis} is very small compared to their counterparts (about 0.01 each). Furthermore, if we train and test our model without $d/G$ and $\phi$ we only reduce the overall $\mathrm{R^2}$ score by 0.02 ($\mathrm{R^2} = 0.90$), while also replicating the original feature rankings obtained by the classifiers. This implies that even though $d/G$ and $\phi$ can be seen as significant by our permutation feature ranking algorithm, most of their effects on the CN can be captured by other features. 

\section{Conclusions} \label{clogging_conclusion}

In this chapter, we combined direct numerical simulations with machine learning approaches to predict clogging in heterogeneous porous media. In order to obtain the necessary data for training said algorithms we developed a computational workflow designed to simulate particle flow through randomly-generated porous media over thousands of different experimental and parametric conditions. One particular challenge that arose was the fact that clogging is a continuous process, not a discrete one. For this reason, we also developed the concept of a ``Clogging Number" in order to properly characterize the level of clogging in our computational simulations. Furthermore, in order to standardize our results we developed 11 non-dimensional training features that can be measured \textit{a-priori} from any system involving the flow of solid particles through a static obstacle field. 

After training, testing, and optimizing several classifiers and regressors we concluded that the best performing classifier was a Multi-Layer Perceptron with a logistic activation function, 175 hidden layers and an L2 regularization term of 0.05. This neural network was able to achieve 0.96 and 0.94 labeling accuracy and precision, respectively. In turn, the best regressor was a Random Forest regressor with 100 trees, a Mean Squared Error split criteria, minimum sample split of 2, and no maximum depth. This regressor achieved an $\mathrm{R^2}$ value of 0.92 when used to predict the samples' CN.

Feature importance analysis of both Machine Learning approaches showed that the most predictive features for clogging in porous media are, in order of importance: the ratio between particle size and pore size, the flux of particles through the porous medium, particle-particle/wall attraction forces, and fluid viscosity. They also showed that fluid velocity and grain heterogeneity did not play a significant role in determining whether a system clogs or not, a point that we attributed to the fact that we did not probe the 11-dimensional parameter space where these features are relevant. 

The result of this investigation is a generalized clogging prediction algorithm that can accurately predict clogging in heterogeneous porous media, a tool which we hope can help improve the design process of engineered porous systems. This is the first time that anyone has probed such a large parametric phase space and/or applied Machine Learning to try and describe this difficult problem, to the best of our knowledge. However, additional work would be required the evaluate the ability of our models to predict clogging in most physical systems. To do so we would need to fine-tune our clogging characterization algorithm and expand our parameter space into different length scales, different particle shapes, different grain geometries, and a larger range of fluid velocities and simulation times. Lastly, and arguably the most important step, we would need to use actual experimental data to test (or maybe even train) these models. We hope that this investigation spurns further work into integrating Machine Learning approaches with numerical simulations in order to probe and characterize stochastic physical phenomena. 
\begin{savequote}[75mm]
In science one tries to tell people, in such a way as to be understood by
everyone, something that no one ever knew before. But in poetry, it's the
exact opposite. \qauthor{Paul Dirac}
\end{savequote}

\chapter{Conclusions}
\label{conclusion}

\newthought{In this dissertation}, we derived, implemented, benchmarked, and showcased a novel CFD approach for simulation of multiscale multiphase flow within and around deformable porous media. This micro-continuum modeling framework is based on elementary physics and was rigorously derived in Chapter \ref{FullDerivation} through the method of volume averaging and asymptotic matching. The result is a set of partial differential equations that approximate the multiphase Volume of Fluid equations in solid-free regions and multiphase Biot Theory in porous regions. These equations are valid in every simulated grid cell within a multiphase porous system, regardless of content, which obviates the need to define different meshes, domains, or complex boundary conditions within the simulation. The solver's numeric and algorithmic development were presented in Chapter \ref{numerical_impl}. The equations were implemented into \textit{hybridPorousInterFoam} and \textit{hybridBiotInterFoam}, two open-source packages accessible \href{https://github.com/Franjcf}{here} for free to any interested party. 

Throughout this thesis we show that the Multiphase DBB model can be readily used to model a large variety of systems, from multiphase flow in static porous media, to elastic systems under compression, to viscosity- or capillarity-dominated fracturing systems, all the way up to multiscale wave propagation in poroelastic coastal barriers (Chapters \ref{chp:singlePhaseDBB} to \ref{chapter:PRL}). In particular, we used this model to investigate and obtain parametric relationships for: A) the permeability-clay content relationship in sedimentary rocks (Chapter \ref{chp:singlePhaseDBB}), and B) the crossover between fluid drainage and fracturing as a function of wettability, solid deformability, and hydrodynamics (Chapter \ref{chapter:PRL}).

We note, however, that the solver presented here cannot be liberally applied to any porous system, as it comes with the following inherent limitations. First, closure of the model's system of equations requires appropriate constitutive and parametric relations that describe fluid pressure, permeability, capillarity, and rheology within volume averaged porous regions. Therefore, the assumptions present in each of these models should be carefully considered. Second, volume averaging imposes important length scale restrictions in order to conform to the scale separation hypothesis: the pore sizes within the averaging volume must be substantially smaller than the chosen REV, and the REV must be substantially smaller than the macroscopic length scale. Third, as implemented here, the multiphase DBB framework only represents continuum-level elastic or plastic solid mechanics that can be described from an Eulerian frame of reference. As such, it cannot be used to model large elastic deformations or phenomena originating from sub-REV heterogeneities such as fluidization or granular mechanics \citep{Meng2020}, except insofar as they are captured in an averaged manner at the REV scale. Fourth, the use of the CSF as a representation of capillary forces within solid-free regions enforces mass conservation, but it creates a diffuse fluid-fluid interface that may generate spurious and parasitic currents.

Finally, although the modeling framework developed here opens up significant new possibilities in the simulation of coupled fluid-solid mechanics, it also creates a need for the development of constitutive relations describing the coupling between multiphase flow and poromechanics. Of particular importance is the formulation of saturation and deformation-dependent solid rheological models (both plastic and elastic), as well as the rigorous derivation of the interfacial condition between solid-free and deformable porous regions. In this study we proposed a suitable approximation for said boundary condition based on our single-field formulation, the implementation of a wettability boundary condition, and the previous work done by \citet{Neale1974} and \citet{Zampogna2019}. However, the accuracy and validity of such an approximation is still an open question, one that is at the frontier of our modeling and characterization capabilities \citep{Qin2020}. The derivation and implementation of said boundary condition, along with the addition of erosion and chemical reactions into this modeling framework, will be the focus of subsequent investigations.  

We hope that the results developed of this thesis will inspire more work in this direction, as suitable and accurate CFD models will be key pieces in our path to meet the world's steadily increasing water and energy demand.  

\begin{appendices}
    \chapter{Relative Permeability and Capillary Pressure Models}
\label{multiphase_models}

\section{Relative Permeability Models} \label{relative_perm}

The two relative permeability models used in this paper and implemented in the accompanying code depend on the definition of an effective saturation in order to account for the presence of irreducible saturations within a porous medium,

\[{\alpha }_{w,eff}\ \ =\ \frac{{\alpha }_w-{\alpha }_{w,irr}}{1-{\alpha }_{w,irr}-{\alpha }_{w,irr}\ }\] 

\noindent Here, ${\alpha }_{w,eff}$ is the wetting fluid's effective saturation, which is the wetting fluid's saturation normalized by each fluid's irreducible saturation ${\alpha }_{i,irr}.\ $The \citet{RHBrooks1964} model relates each phase's relative permeability to saturation through the following expressions:

\[k_{r,n}=\ {\left(1-{\alpha }_{w,eff}\right)}^m\]

\[k_{r,w}\ ={\left({\alpha }_{w,eff}\right)}^m\]

\noindent where $m$ is a non-dimensional coefficient that controls how sensitive the relative permeability is with respect to saturation. The \citet{VanGenutchen1980} model calculates relative permeabilities in the following way:

\[k_{r,n}\ =\ {\left(1-{\alpha }_{w,eff}\ \right)}^{\frac{1}{2}}{\left({\left(1-{\alpha }_{w,eff}\right)}^{\frac{1}{m}}\right)}^{2m}\] 

\[k_{r,w}={\left({\alpha }_{w,eff}\right)}^{\frac{1}{2}}\ {\left(1-{\left(1-{\left({\alpha }_{w,eff}\right)}^{\frac{1}{m}}\right)}^m\right)}^2\]

In this case, $m$ controls how wetting (or non-wetting) the porous medium is to a given wetting (or non-wetting) fluid. High values of $m$ indicate high relative permeabilities for the non-wetting fluid, while low values of $m$ indicate very low relative permeabilities for the same fluid. 

\section{Capillary Pressure Models} \label{cap_pressure}

The implemented capillary pressure models also depend on the definition of an effective wetting-fluid saturation ${\alpha }_{w,pc}$,

\[{\alpha }_{w,pc}\ =\ \frac{{\alpha }_w-{\alpha }_{pc,irr}}{{\alpha }_{pc,max}-{\alpha }_{pc,irr}}\ \] 

\noindent Here, ${\alpha }_{pc,max}$ is the maximum saturation of the wetting fluid and ${\alpha }_{pc,irr}$ is its irreducible saturation. The \citet{RHBrooks1964} model uses the following expression to calculate the capillary pressures within a porous medium:

\[p_c\ =\ p_{c,0}{\left({\alpha }_{w,pc}\right)}^{-\beta }\] 

\noindent where $p_{c,0}$ is the entry capillary pressure, and $\beta $ is a parameter depending on the pore size distribution. Conversely, the \citet{VanGenutchen1980} model calculates the capillary pressure with the following relation:

\[p_c\ ={p_{c,0}\left({\left({\alpha }_{w,pc}\right)}^{-\frac{1}{m}}-1\right)}^{1-m}\ \] 
    \chapter{Solid Rheology Models}
\label{rheology_mod}

\section{Hershel-Bulkley Plasticity} \label{rheology_Herschel}

A Bingham plastic is a material that deforms only once it is under a sufficiently high stress. After this yield stress is reached, it will deform viscously and irreversibly. The Herschel-Bulkley rheological model combines the properties of a Bingham plastic with a power-law viscosity model, such that said plastic can be shear thinning or shear thickening during deformation. In OpenFOAM{\circledR} this model is implemented as follows: 

\[\boldsymbol{\sigma }={\mu }^{eff}_s\left(\ \nabla {\boldsymbol{U}}_s+{ \ \left(\nabla {\boldsymbol{U}}_s\right)}^T-\frac{2}{3}\nabla \cdot\left({\boldsymbol{U}}_s\mathrm{\ }\boldsymbol{I}\right)\right)\mathrm{\ }\] 

\noindent where ${\mu }^{eff}_s$ is the effective solid plastic viscosity, which is then modeled through a power law expression: 

\[{\mu }^{eff}_s={\mathrm{min} \left({\mu }^0_s\mathrm{\ ,\ \ \ }\frac{\tau }{\eta }+{\mu }_s{\eta}^{n-1}\right)\ }\]      

\noindent where$\ {\mu }^0_s$ is the limiting viscosity (set to a very large value), $\tau \ $is the yield stress, ${\mu }_s$ is the viscosity of the solid once the yield stress is overcome, $n$ is the flow index ($n=1$ for constant viscosity), and $\eta $ is the shear rate.       

\section{Quemada Rheology Model} \label{rheology_Quemada}

The Quemada rheology model \citep{Quemada1977,Spearman2017} is a simple model that accounts for the fact that the average yield stress and effective viscosity of a plastic are functions of its solid fraction. These two quantities are large at high solid fractions and small at low solid fractions, as described by the following equations:

\[\tau ={\tau }_0{\left(\frac{\left({{\phi }_s}/{{\phi }^{max}_s}\right)}{\left(1-{{\phi }_s}/{{\phi }^{max}_s}\boldsymbol{\ }\right)}\right)}^D\] 

\[{\mu }_s=\frac{{\mu }_0}{{\left(1-\frac{{\phi }_s}{{\phi }^{max}_s}\right)}^2}\] 

\noindent here, ${\phi }^{max}_s$ is the maximum solid fraction possible (perfect incompressible packing),$\ {\tau }_0$ is the yield stress at ${\phi }_s={\phi }^{max}_s/2$ ,$\ {\mu }_0$ is the viscosity of the fluid where the solid would be suspended at low solid fractions (high fluid fractions), and $D$ is a scaling parameter based on the solid's fractal dimension. 

\section{Linear Elasticity} \label{rheology_Elasticity}

A linear elastic solid assumes that a solid exhibits very small reversible deformations under stress. Linear elasticity is described by the following relation:

\[\boldsymbol{\sigma }={\mu }_s\nabla {\boldsymbol{u}}_{\boldsymbol{s}}+{{\mu }_s\left(\nabla {\boldsymbol{u}}_{\boldsymbol{s}}\right)}^T+{\lambda }_str\left(\nabla {\boldsymbol{u}}_{\boldsymbol{s}}\right)I\]

\noindent where $u_s\ $is the solid displacement vector (not to be confused with solid velocity $U_s = \frac{\partial u_s}{\partial t}$), and ${\mu }_s$ and ${\lambda }_s$ are the Lam\'{e} coefficients. The implementation of linear elasticity in OpenFOAM{\circledR} follows the procedure outlined in \citet{Jasak2000}.
    \chapter{Fracturing Instabilities}\label{fracturing_instabilities}

The following figures demonstrate how different fracturing patterns can result from different solid fraction initializations. Here we set up two sets of four identical experiments. In the first set, the only difference between cases is the value of the standard deviation of their respective normally-distributed solid fraction field (all centered at ${\phi }_s=0.64$). These experiments follow the same simulation setup used for the fracturing case shown in Figure \ref{fig:hele-shaw}K. 

\begin{figure}[htb]
\begin{center}
\includegraphics[width=0.75\textwidth]{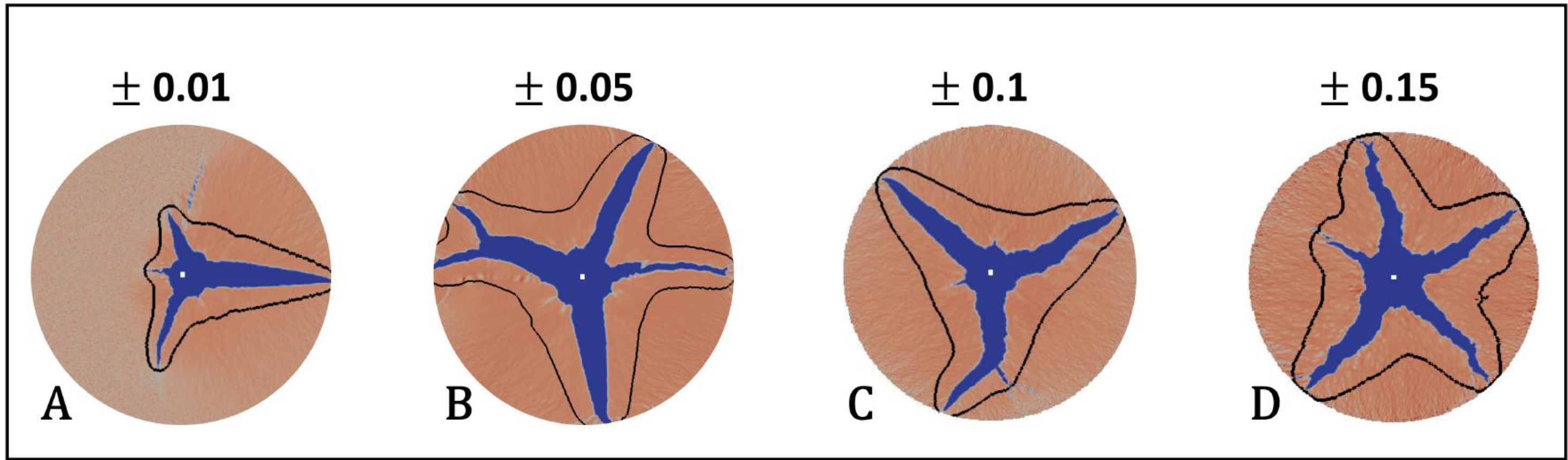}
\caption[Effects of the solid fraction field's standard deviation on fracturing]{Effects of the solid fraction field's standard deviation on fracturing. \label{frac_1}}
\end{center} 
\end{figure}

In the second set of experiments we simulated the base case presented in Figure \ref{fig:surface} with different solid fraction profiles picked from the same normal distribution ${\phi }_s=0.6\ \pm 0.05$. 

\begin{figure}[htb]
\begin{center}
\includegraphics[width=0.75\textwidth]{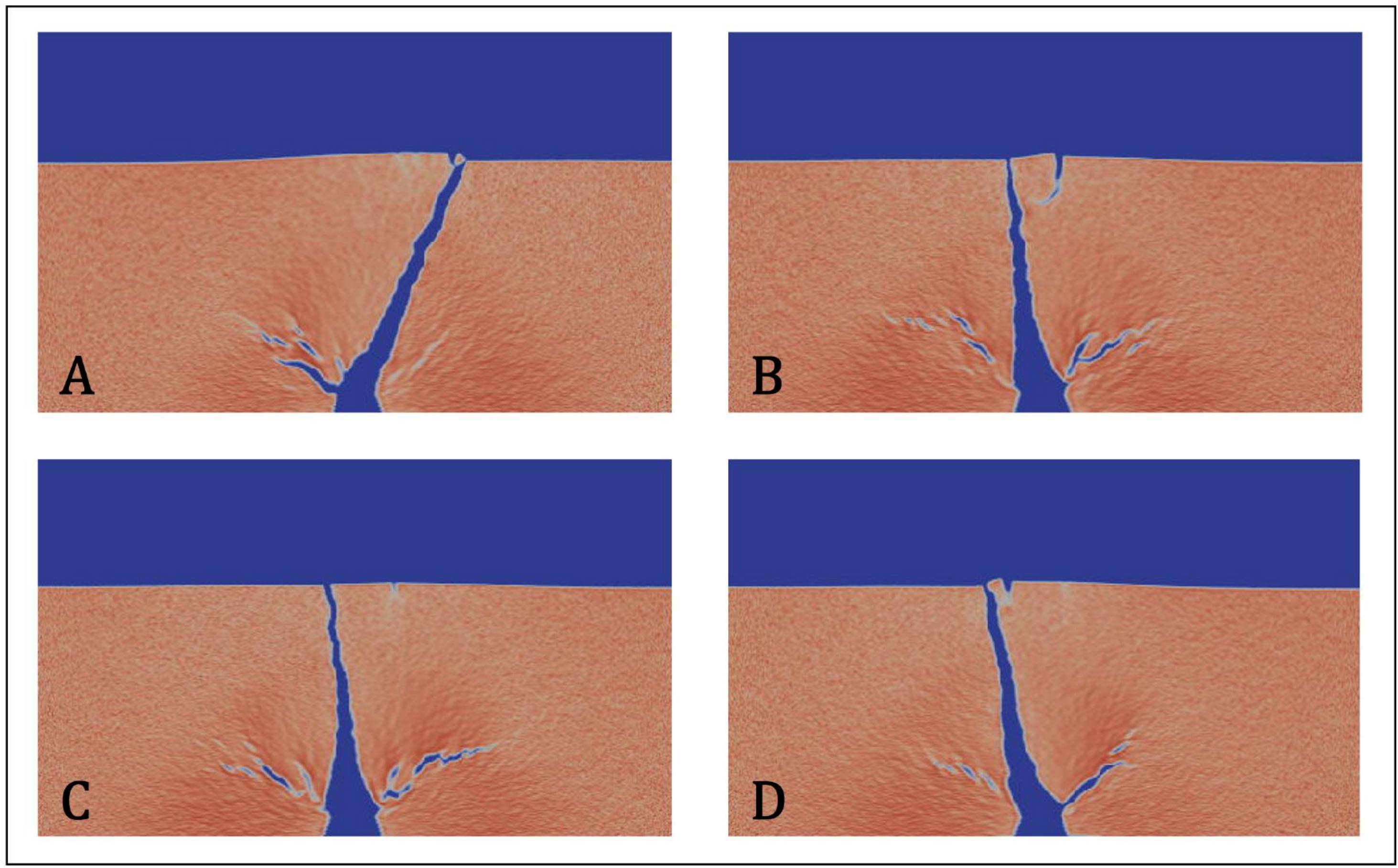}
\caption[Effects of different solid fraction field initializations on fracturing]{Effects of different solid fraction field initializations on fracturing. \label{frac_2}}
\end{center} 
\end{figure}

Figures \ref{frac_1} and \ref{frac_2} clearly show that the shape of the created fractures is dependent on the initial solid fraction distribution. 
    \chapter{Additional Derivation Steps}\label{APP:derivation_steps}

\section{Skipped Steps in the Derivation of the Fluid Continuity Equation} \label{App:Skipped_Fluid_Continuity}

We start with a basic continuity equation for fluid phase $i$,

\begin{equation}\frac{\partial {\rho}_i}{\partial t}+\nabla\cdot\left({\rho}_i\boldsymbol{U}_i\right)=0\end{equation} 

\noindent where ${\rho}_i$ and $\boldsymbol{U}_i$ are the fluid's density and velocity, respectively. Applying the averaging operators for a volume containing a solid phase ($s$) and an additional fluid phase ($j$) we get

\begin{equation}\overline{\frac{\partial {\rho}_i}{\partial t}}+\overline{\nabla\cdot\left({\rho}_i\boldsymbol{U}_i\right)}=0\end{equation} 

Expanding the averaged functions

\begin{multline}
\frac{\partial \overline{{\rho}_i}}{\partial t}-\frac{1}{V}\int_{A_{i,s}}{{\rho}_i{\boldsymbol{U}}_s \cdot {\boldsymbol{n}}_{i,s}dA}-\frac{1}{V}\int_{A_{i,j}}{{\rho}_i{\boldsymbol{U}}_i \cdot {\boldsymbol{n}}_{i,j}dA} \\
+\nabla\cdot\left(\overline{{\rho}_i{\boldsymbol{U}}_i}\right)+\frac{1}{V}\int_{A_{i,s}}{{\rho}_i{\boldsymbol{U}}_i\cdot {\boldsymbol{n}}_{i,s}dA}+\frac{1}{V}\int_{A_{i,j}}{{\rho}_i{\boldsymbol{U}}_i\cdot {\boldsymbol{n}}_{i,j}dA}=0
\end{multline} 

Canceling like terms we then obtain

\begin{equation}\frac{\partial \overline{{\rho}_i}}{\partial t}+\nabla\cdot\left(\overline{{\rho}_i{\boldsymbol{U}}_i}\right)+\frac{1}{V}\int_{A_{i,s}}{{\rho}_i{(\boldsymbol{U}}_i-{\boldsymbol{U}}_s)\cdot {\boldsymbol{n}}_{i,s}dA}=0\end{equation} 

Now, given that the fluid and solid velocities are equal at the fluid-solid interface (${\boldsymbol{U}}_i={\boldsymbol{U}}_s$ at $A_{i,s}$) we can cancel the terms within the integral. This results in Equation \ref{eq:fluid_mass} presented in the main text.

\begin{equation}\frac{\partial \overline{{\rho}_i}}{\partial t}+\nabla\cdot\left(\overline{{\rho}_i{\boldsymbol{U}}_i}\right)=0\end{equation} 

\section{Derivation of the Single-Field Viscous Stress Tensor} \label{App:Viscous_Stress}

Following the main derivation in Section \ref{Sect:FluidMomeqn}, the single-field viscous stress tensor should have the following form:
\begin{equation}\overline{\boldsymbol{S}}={\alpha}_w{\mu}_w\left(\nabla{\overline{\boldsymbol{U}}}^w_w+{\left(\nabla{\overline{\boldsymbol{U}}}^w_w\right)\ }^T\right)+{\alpha}_n{\mu}_n\left(\nabla{\overline{\boldsymbol{U}}}^n_n+{\left(\nabla{\overline{\boldsymbol{U}}}^n_n\right)}^T\right)\end{equation} 

Additionally, the definition of the single field and relative velocities (see Sections \ref{Sect:FluidMass} and \ref{Sect:saturationEqn}) state that
\begin{equation}{\overline{\boldsymbol{U}}}^w_w={\phi}^{-1}_f{\boldsymbol{U}}_f+{\alpha}_n{\boldsymbol{U}}_r\end{equation} 
\begin{equation}{\overline{\boldsymbol{U}}}^n_n={\phi}^{-1}_f{\boldsymbol{U}}_f-{\alpha}_w{\boldsymbol{U}}_r\end{equation}

Taking the gradient within the free fluid (where the viscous stress tensor is dominant), we get,

\begin{equation}{\nabla\overline{\boldsymbol{U}}}^w_w=\nabla{\boldsymbol{U}}_f+{\alpha}_n\nabla{\boldsymbol{U}}_r-{\boldsymbol{U}}_r\cdot \nabla{\alpha}_w\end{equation} 
\begin{equation}\nabla{\overline{\boldsymbol{U}}}^n_n=\nabla{\boldsymbol{U}}_f-{\alpha}_w\nabla{\boldsymbol{U}}_r-{\boldsymbol{U}}_r\cdot \nabla{\alpha}_w\end{equation} 

Inserting these definitions into the first equation and expanding we then obtain the following equation.  

\begin{multline}
\overline{\boldsymbol{S}}=({\alpha}_w{\mu}_w+{\alpha}_n{\mu}_n)\left(\nabla{\boldsymbol{U}}_f+ {\left(\nabla{\boldsymbol{U}}_f\right)}^T\right) 
+  {\alpha}_w{\alpha}_n\left({\mu}_w-{\mu}_n\right)(\nabla{\boldsymbol{U}}_r+ \\{\left(\nabla{\boldsymbol{U}}_r\right)}^T)  -\left({\alpha}_w{\mu}_w+{\alpha}_n{\mu}_n\right){\boldsymbol{U}}_r \cdot \left(\nabla{\alpha}_w+{\left(\nabla{\alpha}_w\right)}^T\right)
\end{multline} 

We can further simplify this expression by defining the single-field fluid viscosity ${\mu}_f={\alpha}_w{\mu}_w+{\alpha}_n{\mu}_n$. The result is a complete expression for the single-field viscous stress tensor.
\begin{multline}
\overline{\boldsymbol{S}}={\mu}_f\left(\nabla{\boldsymbol{U}}_f+{\left(\nabla{\boldsymbol{U}}_f\ \right)}^T\right) \\ +{\alpha}_w{\alpha}_n\left({\mu}_w-{\mu}_n\right)\left(\nabla{\boldsymbol{U}}_r+{\left(\nabla{\boldsymbol{U}}_r\right)}^T\right)-{\mu}_f{\boldsymbol{U}}_r \cdot \left(\nabla{\alpha}_w+{\left(\nabla{\alpha}_w\right)}^T\right)
\end{multline} 

If ${\boldsymbol{U}}_r\ll {\boldsymbol{U}}_f$, as argued in \citet{Fleckenstein2015}, this expression reduces to the equation presented in Section \ref{Sect:FluidMomeqn}.

\begin{equation}\overline{\boldsymbol{S}}\approx {\mu}_f\left(\nabla{\boldsymbol{U}}_f+{\left(\nabla{\boldsymbol{U}}_f\ \right)}^T\right)\end{equation} 

\section{Derivation of the Single-Field Material Derivative} \label{App:material_derivative}

We start with the standard material derivative for a given fluid phase $i$.
\begin{equation}\frac{\partial {\rho}_i{\boldsymbol{U}}_i}{\partial t}+\nabla\cdot \left({\rho}_i{\boldsymbol{U}}_i{\boldsymbol{U}}_i\right)\end{equation} 

We then apply the averaging operators for a volume that contains an additional fluid phase and a solid phase.
\begin{equation}\overline{\frac{\partial {\rho}_i{\boldsymbol{U}}_i}{\partial t}}+\overline{\nabla\cdot \left({\rho}_i{\boldsymbol{U}}_i{\boldsymbol{U}}_i\right)}\end{equation} 

Expanding the terms according to the averaging theorems, we obtain:

\begin{multline}
\frac{\partial \overline{{\rho}_i{\boldsymbol{U}}_i}}{\partial t}-\frac{1}{V}\int_{A_{i,s}}{{\rho}_i{\boldsymbol{U}}_i{\boldsymbol{U}}_{i,s} \cdot {\boldsymbol{n}}_{i,s}dA}-\frac{1}{V}\int_{A_{i,j}}{{\rho}_i{\boldsymbol{U}}_i{\boldsymbol{U}}_{i,j} \cdot {\boldsymbol{n}}_{i,j}dA} \\
+\nabla\cdot \left(\overline{{\rho}_i{\boldsymbol{U}}_i{\boldsymbol{U}}_i}\right)+\frac{1}{V}\int_{A_{i,s}}{{\rho}_i{\boldsymbol{U}}_i{\boldsymbol{U}}_i\cdot {\boldsymbol{n}}_{i,s}dA}\ \ \ +\frac{1}{V}\int_{A_{i,j}}{{\rho}_i{\boldsymbol{U}}_i{\boldsymbol{U}}_i\cdot {\boldsymbol{n}}_{i,j}dA}
\end{multline} 

Given that the velocities of two immiscible fluids are equal to each other at the fluid-fluid interface (i.e. ${\boldsymbol{U}}_i={\boldsymbol{U}}_j={\boldsymbol{U}}_{i,j}$ at $A_{i,j})$ and that the fluid and solid velocities are equal at the fluid-solid interface (i.e. ${\boldsymbol{U}}_i={\boldsymbol{U}}_s={\boldsymbol{U}}_{i,s}$ at $A_{i,s})$, we can obtain the following equation \citep{Higuera2015}:

\begin{equation}\frac{\partial \overline{{\rho}_i{\boldsymbol{U}}_i}}{\partial t}+\nabla\cdot \left(\overline{{\rho}_i{\boldsymbol{U}}_i{\boldsymbol{U}}_i}\right)=0\end{equation} 

We can further simplify the convective term by separating the fluid's velocity into its intrinsic average and deviation terms ${\boldsymbol{U}}_i={\overline{\boldsymbol{U}}}^i_i+{\widetilde{\boldsymbol{U}}}_i$ \citep{Whitaker2013}, which results in

\begin{equation}\nabla\cdot \left(\overline{{\rho}_i{\boldsymbol{U}}_i{\boldsymbol{U}}_i}\right)=\nabla\cdot \left({{\phi}_f{\alpha}_i\rho}_i\overline{{\boldsymbol{U}}_i{\boldsymbol{U}}_i}\right)=\nabla\cdot \left({\phi}_f{\alpha}_i{\rho}_i{\overline{\boldsymbol{U}}}^i_i{\overline{\boldsymbol{U}}}^i_i\right)+\nabla\cdot \left({\rho}_i\overline{{\widetilde{\boldsymbol{U}}}_i{\widetilde{\boldsymbol{U}}}_i}\right)\end{equation} 

For low \textit{Re} number simulations, we can neglect the deviation term to obtain the averaged material derivative for a single fluid phase.  

\begin{equation}\frac{\partial {\rho}_i{\phi}_f{\alpha}_i{\overline{\boldsymbol{U}}}^i_i}{\partial t}+\nabla\cdot \left({\phi}_f{\alpha}_i{\rho}_i{\overline{\boldsymbol{U}}}^i_i{\overline{\boldsymbol{U}}}^i_i\right)\end{equation} 

To obtain the equivalent single-field expression we add the material derivatives of a wetting ($w$) and non-wetting ($n$) fluid, 

\begin{equation}\label{D20}\frac{\partial {{\phi}_f(\rho}_w{\alpha}_w{\overline{\boldsymbol{U}}}^w_w+\ {\rho}_n{\alpha}_n{\overline{\boldsymbol{U}}}^n_n)}{\partial t}+\nabla\cdot \left({\phi}_f\left({\alpha}_w{\rho}_w{\overline{\boldsymbol{U}}}^w_w{\overline{\boldsymbol{U}}}^w_w+{\alpha}_n{\rho}_n{\overline{\boldsymbol{U}}}^n_n{\overline{\boldsymbol{U}}}^n_n\right)\right)\end{equation} 

From the definitions of the single-field and relative velocities we can write the following relations:

\begin{equation}{\overline{\boldsymbol{U}}}^w_w={\phi}^{-1}_f{\boldsymbol{U}}_f+{\alpha}_n{\boldsymbol{U}}_r\end{equation} 
\begin{equation}{\overline{\boldsymbol{U}}}^n_n={\phi}^{-1}_f{\boldsymbol{U}}_f-{\alpha}_w{\boldsymbol{U}}_r\end{equation} 

Based on these definitions, the terms within the time derivative and within the divergence operator in Eqn. \ref{D20} can be expressed as

\begin{equation}{{\phi}_f(\rho}_w{\alpha}_w{\overline{\boldsymbol{U}}}^w_w+\ {\rho}_n{\alpha}_n{\overline{\boldsymbol{U}}}^n_n)=\left({{\alpha}_w\rho}_w+{{\alpha}_n\rho}_n\right){\boldsymbol{U}}_f+{\phi}_f{\alpha}_w{\alpha}_n{\boldsymbol{U}}_r({\rho}_w-{\rho}_n)\end{equation} 

and, 
\begin{multline}
{\phi}_f\left({\alpha}_w{\rho}_w{\overline{\boldsymbol{U}}}^w_w{\overline{\boldsymbol{U}}}^w_w+{\alpha}_n{\rho}_n{\overline{\boldsymbol{U}}}^n_n{\overline{\boldsymbol{U}}}^n_n\right)= \\
{\phi}_f{({\alpha}_w\rho}_w+{\alpha}_n{\rho}_n)\left({\phi}^{-2}_f{\boldsymbol{U}}_f{\boldsymbol{U}}_f+{\boldsymbol{U}}_r{\boldsymbol{U}}_r\right)+2{\phi}^{-1}_f{\alpha}_w{\alpha}_n{\boldsymbol{U}}_f{\boldsymbol{U}}_r({\rho}_w-{\rho}_n)
\end{multline} 

To simplify things, we now define the single-field density ${\rho}_f={\alpha}_w{\rho}_w+{\alpha}_n{\rho}_n$, and the viscosity difference $\Delta\rho={\rho}_w-{\rho}_n$. Combining the previous equations we obtain the complete averaged single-field expression for the material derivative

\begin{multline}\frac{\partial {\rho}_f{\boldsymbol{U}}_f}{\partial t}+\frac{\partial {\phi}_f{\alpha}_w{\alpha}_n\Delta\rho{\boldsymbol{U}}_r}{\partial t}+\nabla\cdot \left(\frac{{\rho}_f}{{\phi}_f}{\boldsymbol{U}}_f{\boldsymbol{U}}_f\right)+ \\ \nabla\cdot \left({\phi}_f{\rho}_f{\boldsymbol{U}}_r{\boldsymbol{U}}_r\right)+\nabla\cdot \left(2{\alpha}_w{\alpha}_n\Delta\rho{\boldsymbol{U}}_f{\boldsymbol{U}}_r\right)\end{multline} 

Finally, if ${\boldsymbol{U}}_r\ll {\boldsymbol{U}}_f$, as argued in \citet{Fleckenstein2015}, this expression reduces to the expression used in the derivation of Equation \ref{eq:complete_fluid}. 

\begin{equation}\frac{\partial {\rho}_f{\boldsymbol{U}}_f}{\partial t}+\nabla\cdot \left(\frac{{\rho}_f}{{\phi}_f}{\boldsymbol{U}}_f{\boldsymbol{U}}_f\right)\end{equation}

\section{Recovery of Biot Theory from the Fluid and Solid Momentum Equations}\label{App:Biot_Theory}

We start by adding together the final fluid and solid momentum equations (Eqns. \ref{Eq:Final_fluid} and \ref{Eq:Final_Solid}) under the assumption of low Reynolds numbers and low permeability. This is analogous to the steps used to obtain Eqn. \ref{added} in Section \ref{App:Biot_Terms}.

\begin{equation}-\nabla\cdot\overline{\boldsymbol{\sigma}}={\phi}_s\nabla\cdot{\overline{\boldsymbol{\tau}}}^s-{\phi}_f\nabla p+\left({{\phi}_s\rho}_s+{\phi}_f{\rho}_f\right)\boldsymbol{g}+{\boldsymbol{F}}_{c,2}\end{equation} 

Now, assuming uniform confining pressure and no swelling pressure (i.e $\nabla\cdot{\overline{\boldsymbol{\tau}}}^s\boldsymbol{=-}\nabla p)$ we obtain:

\begin{equation}-\nabla\cdot\overline{\boldsymbol{\sigma}}=-{\phi}_s\nabla p-{\phi}_f\nabla p+\left({{\phi}_s\rho}_s+{\phi}_f{\rho}_f\right)\boldsymbol{g}+{\boldsymbol{F}}_{c,2}\end{equation} 

We can then set ${\rho}^*=({{\phi}_s\rho}_s+{\phi}_f{\rho}_f\boldsymbol{)}$ and input the definition of the capillary force term (Eqn. \ref{Fc2_def}), such that

\begin{equation}-\nabla\cdot\overline{\boldsymbol{\sigma}}=-\nabla p+{\rho}^*\boldsymbol{g}-p_c\nabla{\alpha}_w\end{equation} 

The resulting expression is the momentum conservation equation used in multiphase Biot Theory \citep{Jha02014,Kim2013}. In said papers, the corresponding fluid mass conservation equation is

\begin{equation}\frac{\partial m}{\partial t}+\nabla\cdot {\boldsymbol{U}}_f=0\end{equation} 

\noindent where $m$ is the mass of fluid per control volume. The time derivative is often expressed in terms of the fluid pressure and volumetric strain ($\epsilon$) by applying the following a pressure-strain relation \citep{Coussy2010,Kim2011}:

\begin{equation}\frac{1}{{\rho}_f}\left(m-m_0\right)=b\boldsymbol{\epsilon}+\frac{1}{M}(p-p_0)\end{equation} 

\noindent Therefore, the continuity equation becomes,
\begin{equation}{\rho}_f\left(\frac{1}{M}\frac{\partial p}{\partial t}+b\frac{\partial \epsilon }{\partial t}\right)+\nabla\cdot {\boldsymbol{U}}_f=0\end{equation} 

\noindent where $M$ and $b$ are the Biot modulus and coefficient, respectively. In our modelling framework, however, we note that the initial continuity equation is equal to the averaged fluid continuity equation presented in the main text by considering that $m={\phi}_f{\rho}_f$.
\begin{equation}\frac{\partial {\phi}_f{\rho}_f}{\partial t}+\nabla\cdot {\boldsymbol{U}}_f=0\end{equation} 

\section{Integral Geometric Relation for Vectors}

The following geometric relation holds for vector values within the averaging integrals; note its similarity to the geometric relation for scalars shown in the main text \citep{Whitaker1986} 
\begin{equation}\frac{1}{V}\int_{A_{i,s}}{\boldsymbol{I}\cdot {\boldsymbol{n}}_{f,s}dA}\ =\ -\boldsymbol{I} \cdot \nabla{\phi}_f\end{equation}  
    \chapter{Semi-Analytic Solution for the Seismic Stimulation of a Poroelastic Core}\label{analytical_occilation}

Here we present the analytical solution used to describe the oscillating elastic system in Section \ref{occilation}. Given a Biot coefficient of unity and incompressible fluids, the fractional change in an oscillating poroelastic core's fluid content $\Omega\ $as a function of time $t$ is given by \citep{Lo2012}:

\begin{equation}\label{infinite_sum}
    \Omega\left(t\right)=-a_1\upsilon +a_2{\alpha }_wp_a+0.5\left(a_2p_0 sin\mathrm{}(\omega t\right)-a_2{\alpha }_wp_a)+\ \sum^{\infty }_{n=1}{A} 
\end{equation}

\begin{equation}
    A={\left(n\pi \right)}^{-2}2{cos \left(n\pi \right)\ }a_2p_0\left({sin \left(\omega t\right)+\frac{{\omega }^2_n{sin \left(\omega t+{\delta }_n\right)\ }}{{{{((\omega }^2-{\omega }^2_n)}^2+D^2{\omega }^2)}^{0.5}}\ }\right)\left(1-{cos \left(n\pi \right)\ }\right)
\end{equation}

\noindent where $\upsilon$ is the uniaxial confining pressure, $p_a$ is the fixed pressure at the left boundary, $p_0$ is the amplitude of the oscillating pressure at the right boundary, and $\omega =2\pi f$ is the angular frequency of the pressure variation. The summation terms ${\omega }_n$ and ${sin \left({\delta }_n\right)\ }$ are defined as

\begin{equation}
    {\omega }_n={\left(\frac{Cn\pi }{Length}\right)}^2
\end{equation} 

\begin{equation}
    {sin \left({\delta }_n\right)\ }=\frac{D\omega }{{{{((\omega }^2-{\omega }^2_n)}^2+D^2{\omega }^2)}^{0.5}}
\end{equation}

\begin{equation}
    {cos \left({\delta }_n\right)\ }=\frac{{\omega }^2-{\omega }^2_n}{{{{((\omega }^2-{\omega }^2_n)}^2+D^2{\omega }^2)}^{0.5}}
\end{equation}

Furthermore, the dissipation constant $D$, the wave speed $C$, and the compressibility constants $a_1$ and $a_2$ are defined as follows

\begin{equation}
    D=\frac{1}{k_0}\ \frac{1}{\frac{T}{{\phi }_f}\left(\frac{{{\rho }_wM}_w}{{\alpha }_w}+\frac{{{\rho }_nM}_n}{{\alpha }_n}\right)-\left({{\rho }_wM}_w+{{\rho }_nM}_n\right)}
\end{equation}
 
\begin{equation}
    C^2=\left(K_b+\frac{4}{3}G\right)\frac{M}{\frac{T}{{\phi }_f}\left(\frac{{{\rho }_wM}_w}{{\alpha }_w}+\frac{{{\rho }_nM}_n}{{\alpha }_n}\right)-\left({{\rho }_wM}_w+{{\rho }_nM}_n\right)}
\end{equation} 

\begin{equation}
    a_1={\left(3K_b\right)}^{-1} 
\end{equation}

\begin{equation}
    a_2=K^{-1}_b 
\end{equation}

\noindent where $T=0.5\left(1+{\phi }^{-1}_f\right)$ is the tortuosity, $K_b$ is the bulk modulus of the solid matrix, $G$ is the shear modulus of the solid matrix, and the rest of the variables are defined as in the main manuscript. The infinite sum in Eqn. \ref{infinite_sum} was calculated through a python script, where it was truncated at the point where the last sum term represented 0.01\% of the previous term. 

\end{appendices}

\singlespacing

\clearpage
\bibliography{ref,refCyp,cloggingRefs}
\addcontentsline{toc}{chapter}{References}
\bibliographystyle{apalike2}

\newpage


\vspace*{200pt}

\begin{center}
\parbox{200pt}{\lettrine[lines=3,slope=-2pt,nindent=-4pt]{\textcolor{SchoolColor}{T}}{his thesis was typeset} using \LaTeX, originally developed by Leslie Lamport and based on Donald Knuth's \TeX. The body text is set in 11 point Egenolff-Berner Garamond, a revival of Claude Garamont's humanist typeface. A template that can be used to format a PhD thesis with this look and feel has been released under the permissive \textsc{mit} (\textsc{x}11) license, and can be found online at \href{https://github.com/suchow/Dissertate}{github.com/suchow/Dissertate} or from its author, Jordan Suchow, at \href{mailto:suchow@post.harvard.edu}{suchow@post.harvard.edu}.}
\end{center}

\end{document}